\documentclass[a4paper,11pt]{article}
\pdfoutput=1
\usepackage[utf8]{inputenc}
\usepackage{jheppub}
\usepackage{graphicx}
\usepackage{caption}
\usepackage{subcaption}
\usepackage{hyperref}
\usepackage[T1]{fontenc}

\newcommand{\be}{\begin{equation}}
\newcommand{\ee}{\end{equation}}
\newcommand{\bea}{\begin{eqnarray}}
\newcommand{\eea}{\end{eqnarray}}

\newcommand{\nn}{\nonumber\\}

\def\tp{\tau_{\Pi}}

\def\CA{\mathcal{A}}
\def\CB{\mathcal{B}}

\def\CG{\mathcal{G}}

\def\CN{\mathcal{N}}

\def\CW{\mathcal{W}}
\def\CZ{\mathcal{Z}}

\def\qfr{\mathfrak{q}}
\def\wfr{\mathfrak{w}}

\def\lgb{\lambda_{\scriptscriptstyle GB}}

\def\taur{\tau_{\scriptscriptstyle R}}
\def\taumft{\tau_{\scriptscriptstyle mft}}
\def\gammagb{\gamma_{\scriptscriptstyle GB}}
\def\viscb{\hbar/4\pi k_B}
\def\ggb{\gamma_{\scriptscriptstyle GB}}
\def\re{\mbox{Re}}
\def\im{\mbox{Im}}
\title{From strong to weak coupling in holographic models of thermalization}

\author[a]{Sa\v{s}o Grozdanov,}
\author[a]{Nikolaos Kaplis}
\author[b]{and Andrei O. Starinets}

\affiliation[a]{Instituut-Lorentz for Theoretical Physics, Leiden University, \\ Niels Bohrweg 2,  Leiden 2333 CA, The Netherlands }
\affiliation[b]{Rudolf Peierls Centre for Theoretical Physics, University of Oxford, \\ 1 Keble Road,  Oxford OX1 3NP, United Kingdom }

\emailAdd{grozdanov@lorentz.leidenuniv.nl}
\emailAdd{kaplis@lorentz.leidenuniv.nl}
\emailAdd{andrei.starinets@physics.ox.ac.uk}

\abstract{
We investigate the analytic structure of thermal energy-momentum tensor correlators at large but finite coupling in quantum field theories with gravity duals. We compute corrections to the quasinormal spectra of black branes due to the presence of higher derivative $R^2$ and $R^4$ terms in the action, focusing on the dual to $\CN=4$ SYM theory and Gauss-Bonnet gravity. We observe the appearance of new poles in the complex frequency plane at finite coupling. The new poles interfere with hydrodynamic poles of the correlators leading to the breakdown of hydrodynamic description at a coupling-dependent critical value of the wave-vector. The dependence of the critical wave vector on the coupling implies that the range of validity of the hydrodynamic description increases monotonically with the coupling. The behavior of the quasinormal spectrum at large but finite coupling may be contrasted with the known properties of the hierarchy of relaxation times determined by the spectrum of a linearized kinetic operator at weak coupling. We find that the ratio of a transport coefficient such as viscosity to the relaxation time determined by the fundamental non-hydrodynamic quasinormal frequency changes rapidly in the vicinity of infinite coupling but flattens out for weaker coupling, suggesting an extrapolation from strong coupling to the kinetic theory result. We note that the behavior of the quasinormal spectrum is qualitatively different depending on whether the ratio of shear viscosity to entropy density is greater or less than the universal, infinite coupling value of $\hbar/4\pi k_B$. In the former case, the density of poles increases, indicating a formation of branch cuts in the weak coupling limit, and the spectral function shows the appearance of narrow peaks. We also discuss the relation of the viscosity-entropy ratio to conjectured bounds on relaxation time in quantum systems.
}

\preprint{OUTP-16-11P, INT-PUB-15-076}

\begin{document}
\maketitle
\flushbottom

\section{Introduction}
\label{sec:intro}
Nuclear matter produced in heavy ion collisions at RHIC and LHC appears to be well described by relativistic fluid dynamics at the time shortly after the collision, i.e. for $t>\tau_H$, where the ``hydrodynamization'' time $\tau_H$ is  of the order of $1 - 2$ fm/c \cite{Teaney:2000cw,Heinz:2013wva,Luzum:2008cw,Luzum:2009sb,Schenke:2010rr,Song:2010mg}. The hydrodynamic description fits the available experimental data well provided the shear viscosity - entropy density ratio of the resulting nuclear fluid is low, $\eta/s \sim \hbar/4 \pi k_B$. An interesting and not fully understood question is how the matter reaches the hydrodynamic stage of its evolution so quickly and which physical mechanisms are responsible for such a rapid thermalization at intermediate values of QCD coupling. The regime of intermediate coupling can in principle be approached from either the weak or the strong coupling side and accordingly, issues related to thermalization have been studied in kinetic theory at weak coupling and in gauge-string duality (holography) at strong coupling. While the kinetic theory approach and the holographic methods are very different, it is clear that in one and the same theory (e.g. in ${\cal N}=4$ supersymmetric $SU(N_c)$ Yang-Mills (SYM) theory at infinite $N_c$) one should expect an interpolation between strong and weak coupling results for observables describing thermalization, similar to the coupling constant dependence of the shear viscosity - entropy density ratio \cite{Kovtun:2004de,Buchel:2004di} or pressure \cite{Gubser:1998nz,Blaizot:2006tk}. The goal of this paper is to investigate such a dependence for a number of models where corrections to known holographic results at infinitely strong coupling can be computed by using higher derivative terms in the dual gravity action.

Among relevant observables, we focus on the hierarchy of times characterizing the approach to thermal equilibrium. In simple models of kinetic theory, the appropriate time scales emerge as eigenvalues of the linearized collision operator, with the largest eigenvalue, $\taur$, essentially (within a specified approximation scheme) setting the time scale for transport phenomena \cite{ford-book,gross-1959,grad-1963,liboff-book} (see Section \ref{sec:relaxation} for details). In particular, for the shear viscosity in the non-relativistic kinetic theory one typically obtains \cite{chapman-book}
\begin{equation}
\eta = \taur\, n\, k_B\, T\,,
\label{eq:rel-visc-nr}
\end{equation}
where $n$ is the particle density. The relativistic analogue of Eq.~(\ref{eq:rel-visc-nr}) is
\begin{equation}
\eta = \taur\, s\, T\,,
\label{eq:rel-visc-rel}
\end{equation}
where $s$ is the volume entropy density\footnote{To get the factors of $k_B$ right, one may consult the equation carved on Boltzmann's tombstone.}. In kinetic theory, the relaxation time $\taur$ is simply proportional to the (equilibrium) mean free time for corresponding particles or quasiparticles and thus the internal time scale associated with the kinetic operator acquires a transparent physical meaning. In the regime of validity of Eq.~(\ref{eq:rel-visc-rel}), the dependence of $\eta/s$ on e.g. the coupling is the same as the dependence on the coupling of $\taur T$ and thus we expect the ratio $\eta/s \taur T$ to be (approximately) constant in that regime. Another interesting feature of kinetic theory models is the breakdown of the hydrodynamic description for sufficiently large values of the wave vector $q>q_c$ and the appearance of the strongly damped Knudsen modes \cite{boltzmann-book}. We shall see that these phenomena have their counterparts in the regime of strong coupling despite the fact that kinetic theory is not applicable in that regime.

It is believed that the quark-gluon plasma created in heavy ion collisions at energies available at RHIC or LHC is a strongly interacting system, for which a direct or effective (via a suitable quasiparticle picture) application of kinetic theory is difficult to justify. Instead, insights into the time-dependent processes at strong coupling are obtained by studying qualitatively similar strongly coupled theories having a dual holographic description in terms of higher-dimensional semiclassical gravity. Holography \cite{jorge-book, Ammon:2015wua, nastase-book, natsuume-book, zaanen-book} provides a convenient framework for studying non-equilibrium phenomena in strongly interacting systems. The dynamics and evolution of non-equilibrium states in a strongly interacting quantum many-body system is mapped (in the appropriate limit) into the dynamics and evolution of gravitational and other fields of a dual theory. Holography should in principle be capable of encoding all types of non-equilibrium behavior. In particular, evolution of the system towards thermal equilibrium is expected to be described by the dynamics of gravitational collapse. Numerical and analytical studies of processes involving strong gravitational fields including black holes and neutron stars mergers resulting in black hole formation and particles falling into black holes show a characteristic scenario in which a primary signal (strongly dependent on the initial conditions) is followed by the quasinormal ringdown (dependent on the final state parameters only) and then a late-time tail (see e.g. \cite{Frolov:1998wf}, \cite{Berti:2009kk}). A holographic description of fully non-equilibrium quantum field theory states via dual gravity has been developed over the last several years and the results suggest that the quasinormal spectrum (i.e. the eigenvalues of the linearized Einstein's equations of the dual black brane background) and in particular the fundamental (the least damped non-hydrodynamic) quasinormal frequency play a significant role in the description of relaxation phenomena. Recent studies (including sophisticated numerical general relativity approaches) of equilibration processes in the dual gravity models \cite{Chesler:2008hg, Chesler:2009cy,Chesler:2015wra,Chesler:2015fpa,Casalderrey-Solana:2013aba,Casalderrey-Solana:2013aba,Heller:2011ju,Bantilan:2014sra,Buchel:2015saa,Jankowski:2014lna,Keranen:2015mqc} reveal that the hydrodynamic stage of  evolution is reached by a strongly coupled system long before the pressure gradients become small and that the relevant time scales are essentially determined by the lowest quasinormal frequency, even for non-conformal backgrounds \cite{Buchel:2015saa,Janik:2015waa,Janik:2015iry,Attems:2016ugt,Janik:2016btb,Gursoy:2016ggq}. The characteristic time scale here is set by the inverse Hawking temperature of the dual equilibrium black hole.

A seemingly natural question to ask is whether the relation between transport phenomena and the relaxation time(s) familiar from kinetic theory exists also at strong coupling and if yes, how it changes as a function of coupling. Is there a limiting value of the wave vector beyond which hydrodynamic description breaks down at large but finite coupling? Extrapolating kinetic theory results to the regime of intermediate coupling was the subject of recent investigation by Romatschke \cite{Romatschke:2015gic}. In holography, these questions can be studied by computing coupling constant corrections to the full quasinormal spectra using the appropriate higher derivative terms in dual gravity. Recently, such corrections have been studied in Refs.~\cite{Stricker:2013lma}, \cite{Waeber:2015oka}.

In this paper, we compute the quasinormal spectra of metric perturbations of the gravitational background with $R^4$ higher derivative term (dual to $\CN=4$ SYM at finite temperature and large but finite  't Hooft coupling), and for the background with $R^2$ terms including Gauss-Bonnet gravity in $d=5$ dimensions. Normally, higher derivative terms are treated as infinitesimally small corrections to the second order equations of motion of Einstein gravity, otherwise one is doomed to encounter the Ostrogradsky instability and related problems. Accordingly, extrapolating results from infinitesimal to finite values of the corresponding parameters requires caution. Gauss-Bonnet and more generally Lovelock gravity are good laboratories since their equations of motion are of second order and thus can handle finite values of the parameters multiplying higher derivative terms. However, such theories appear to suffer from internal inconsistencies for any finite value of the parameters \cite{Camanho:2014apa} (for an apparently dissenting view, see \cite{Reall:2014pwa}). The passage between Scylla and Charybdis of those two difficulties may be hard to find, if it exists at all. We find some solace in the fact that our results show a qualitatively similar picture regardless of the exact form of higher derivative terms used.

The paper is organized as follows: Our main results are summarized in Section \ref{sec:relaxation}, where we also review some facts about the relaxation times in quantum critical, kinetic and gravitational systems, adding a number of new observations along the way. In Section \ref{sec:SYM}, we compute the (inverse) 't Hooft coupling corrections to the quasinormal spectrum of gravitational fluctuations in AdS-Schwarzschild black brane background modified by the higher derivative terms and discuss the relaxation time behavior, the density of poles and the inflow of extra poles from infinity. In Sections \ref{sec:GB} and \ref{sec:r2}, correspondingly, a similar procedure is applied to Gauss-Bonnet gravity and to the background with generic curvature squared terms. We briefly discuss the results in the concluding Section \ref{sec:discussion}. Some technical issues and comments about our numerical procedures appear in the Appendices.
\section{Relaxation times at weak and strong coupling}
\label{sec:relaxation}
In this Section, we briefly review the appearance of the hierarchy of relaxation times in kinetic theory, holography and some models of condensed matter physics, emphasizing their similarities and adding some new observations. In this context, at the end of the Section, we list the main results of the present paper.

\begin{figure}[htbp]
\centering
\includegraphics[width=0.65\textwidth]{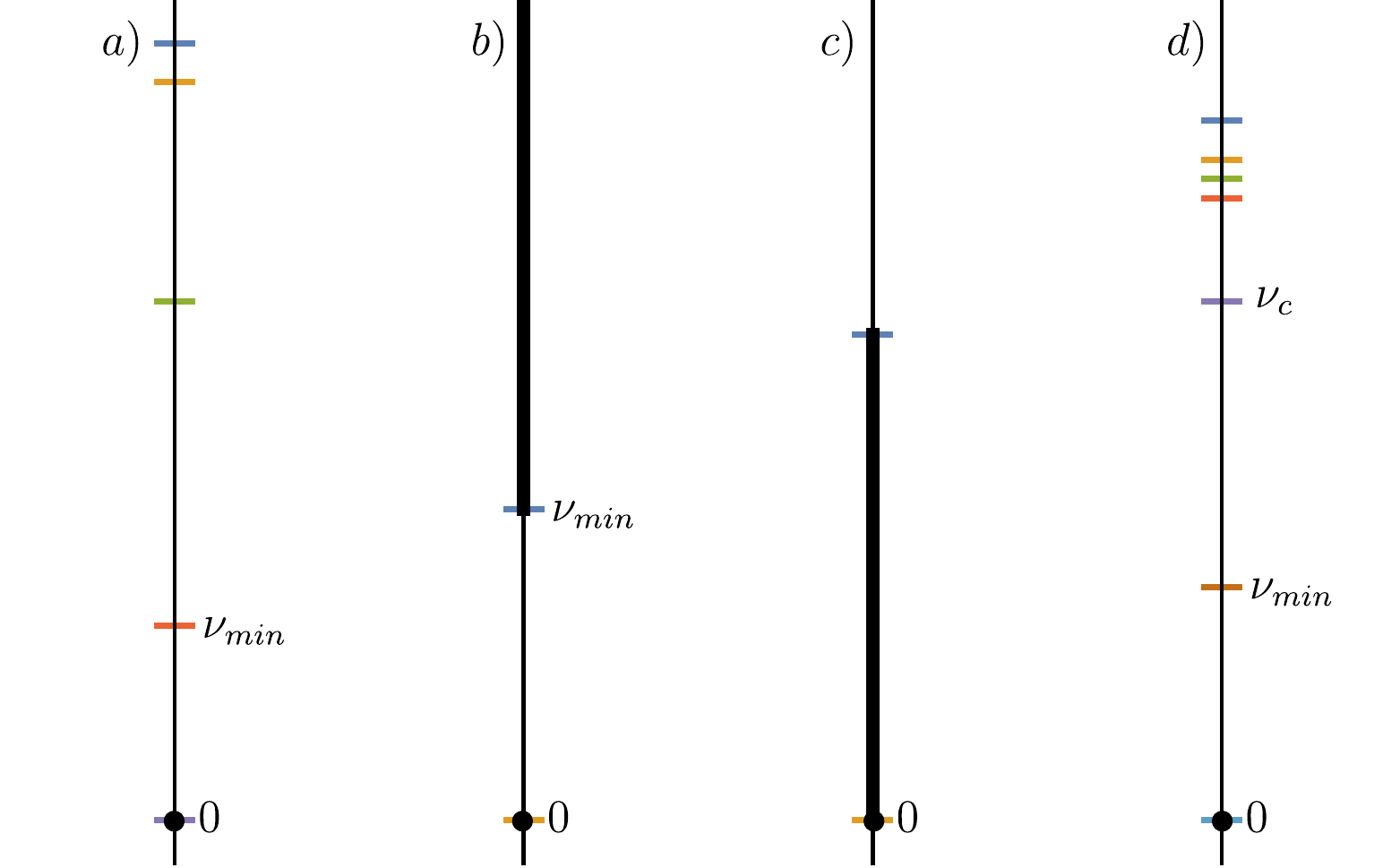}
\caption{The spectrum of a linear collision operator: a) discrete spectrum, b) continuous spectrum with a gap, realized for the interaction potential $U=\alpha/r^n$, $n>4$, c) gapless continuous spectrum, realized for the interaction potential $U=\alpha/r^n$, $n<4$, d) Hod spectrum (see text): $0 \leq \nu_{min} \leq \nu_c$. In all cases, $\nu =0$ is a degenerate eigenvalue corresponding to hydrodynamic modes (at zero spatial momentum).}
\label{fig:spectrum_kinetic}
\end{figure}

In kinetic theory, transport coefficients and relaxation time(s) are intimately related. To be clear, by the relaxation time we mean the characteristic time interval during which a local thermal equilibrium (e.g. a local Maxwell-Boltzmann equilibrium) is formed everywhere in the system. We are not interested in the momentum-dependent equilibration time-scales of the densities of conserved charges (these densities always relax hydrodynamically) which are, strictly speaking, infinite in the limit of vanishing spatial momentum. Consider, for illustration, non-relativistic Boltzmann equation obeyed by the one-particle distribution function $F(t,{\bf r},{\bf p})$
\begin{equation}
\frac{\partial F}{\partial t} + \frac{p_i}{m}\, \frac{\partial F}{\partial r^i}  -\frac{\partial U (r)}{\partial r^i}\, \frac{\partial F}{\partial p_i} = C[F]\,,
\end{equation}
where $U(r)$ is the external potential and $C[F]$ is the Boltzmann collision operator containing details of the interactions. For small deviations from the local thermal equilibrium described by the distribution function $F_0({\bf r},{\bf p})$, the kinetic equation can be linearized by the ansatz
\begin{equation}
F(t,{\bf r},{\bf p}) = F_0({\bf r},{\bf p})\left[1 + \varphi (t, {\bf r},{\bf p})\right]\,,
\label{eq:anz}
\end{equation}
where $\varphi \ll 1$. The ansatz (\ref{eq:anz}) leads to the evolution equation
\begin{equation}
\frac{\partial \varphi}{\partial t} =  - \frac{p_i}{m}\, \frac{\partial \varphi}{\partial r^i} + \frac{\partial U (r)}{\partial r^i}\, \frac{\partial \varphi}{\partial p_i} + L_0 [\varphi ]\,,
\label{eq:linear_collision_op}
\end{equation}
where $L_0$ is a linear integral operator resulting from linearization of $C[F]$. Formal solution to Eq.~(\ref{eq:linear_collision_op}) with the initial condition $\varphi (0, {\bf r},{\bf p}) = \varphi_0 ({\bf r},{\bf p})$ can be written in the form \cite{ferziger-kaper-book}
\begin{equation}
\varphi (t, {\bf r},{\bf p}) = e^{t L} \, \varphi_0 ({\bf r},{\bf p}) = \frac{1}{2\pi i} \int\limits_{\gamma - i \infty}^{\gamma+i \infty} \, e^{s t} \, R_s d s\, \varphi_0 ({\bf r},{\bf p})\,,
\end{equation}
where $R_s = \left( s I - L\right)^{-1}$ is the resolvent whose analytical structure in the complex $s$-plane determines the relaxation properties. In some simple cases, such as e.g. the relaxation of a low-density gas of light particles in a gas of heavy particles, the resolvent can be constructed explicitly and the time dependence fully analyzed \cite{silin-book}. Generically, however, the time evolution is not known explicitly. For spatially homogeneous equilibrium distributions and perturbations, a simple ansatz $\varphi (t, {\bf p}) = e^{-\nu t} h ({\bf p})$ reduces the linearized kinetic equation to the eigenvalue problem for the linear collision operator:
\begin{equation}
- \nu h  = L_0 [h]\,.
\end{equation}
The eigenvalues of $L_0$ determine the spectrum of (inverse) relaxation times in the system. One can then write a general solution of the linearized kinetic equation in the form
\begin{equation}
\varphi (t, {\bf p}) = \sum\limits_n C_n e^{-\nu_n t} h_n ({\bf p})\,,
\label{eq:sum-eigen}
\end{equation}
where the coefficients $C_n$ are determined by the initial conditions and the sum should be replaced by an integral if the spectrum turns out to be continuous. The hierarchy $\{ \nu_n \}$ in Eq.~(\ref{eq:sum-eigen}) is clearly reminiscent of the hierarchy of imaginary times of the quasinormal modes in the dual gravity treatment of near-equilibrium processes at strong coupling. The spectrum of the operator $L_0$ for (classical) particles interacting via the potential $U(r) = \alpha/r^n$ has been investigated by Wang Chang and Uhlenbeck \cite{ford-book} and by Grad \cite{grad-1963}. The spectrum consists of a five-fold degenerate null eigenvalue, corresponding to conserved quantities and the rest of the spectrum which can be discrete (for $n=4$) or continuous \cite{ford-book, grad-1963,liboff-book}, with or without a gap, depending on $n$ (see Fig.~\ref{fig:spectrum_kinetic}). The time dependence is obviously sensitive to the type of the spectrum: discrete spectrum leads to a clear exponential relaxation, whereas continuous spectrum implies a more complicated pattern including a pure power-law fall-off in the gapless case.
Assuming the spectrum is discrete and denoting $\taur = 1/\nu_{min}$, in the relaxation time approximation, when the sum in (\ref{eq:sum-eigen}) is dominated by a single term with $\nu_n = \nu_{min}$, we find
\begin{equation}
\frac{\partial F}{\partial t} = - \frac{F - F_0}{\taur}\,.
\end{equation}
Generalization to weakly inhomogeneous systems gives \cite{kvasnikov-book,liboff-book}
\begin{equation}
\frac{\partial F}{\partial t} + \frac{p_i}{m}\, \frac{\partial F}{\partial r^i}  -\frac{\partial U (r)}{\partial r^i}\, \frac{\partial F}{\partial p_i} =  - \frac{F - F_0}{\taur}\,.
\label{eq:relax-term-eq}
\end{equation}
The equation (\ref{eq:relax-term-eq}) has been remarkably successful in describing transport phenomena in systems with a kinetic regime\footnote{The equation (\ref{eq:relax-term-eq}) with a semi-phenomenological $\tau_{\scriptscriptstyle R} = \taur (v)$ is sometimes called the Krook-Gross-Bhatnagar (KGB) equation \cite{KGB}.} \cite{gross-1959}, \cite{ferziger-kaper-book}. In particular, assuming $\taur = const$, for the shear viscosity one obtains the result (\ref{eq:rel-visc-nr}). Estimates of $\taur$ based on Ritz variational method relate the relaxation time to the mean free time: $\tau_{\scriptscriptstyle R} = 15/8\, \taumft \sim \sqrt{m}/\sqrt{T} n \sigma$, where $\sigma$ is the interaction cross-section. The account above may look too schematic but a more detailed treatment is available in the standard kinetic theory \cite{ferziger-kaper-book} (including relativistic and quantum cases \cite{liboff-book}, \cite{groot-book}), in the mathematical theory of Boltzmann equation \cite{saint-raymond-book} and in thermal gauge theory \cite{Arnold:2002zm,Arnold:2003zc}.

Do the relations between transport coefficients and relaxation time(s) similar or identical to the ones in Eqs.~(\ref{eq:rel-visc-nr}) and (\ref{eq:rel-visc-rel}) hold beyond the regime of applicability of kinetic theory and in the absence of quasiparticles? One may appeal to dimensional analysis and the uncertainty principle \cite{Kovtun:2003wp} or "general wisdoms'' \cite{zaanen-book} when arguing for an affirmative answer\footnote{Indeed, the characteristic time scale in the kinetic regime is the mean free path $\tau \sim t_{mfp}$ and in the regime of strong coupling it is the inverse temperature of a dual black hole, $\tau \sim \hbar/k_B T$. Assuming $\eta/s \sim \tau k_B T$, we have $\eta/s \sim \sqrt{m T}/n \sigma$ in the first case and $\eta/s \sim \hbar /k_B$ in the second.} but in all cases the concept of weakly interacting quasiparticles seems to be lurking behind such reasoning. At the same time, the concepts of relaxation time and transport are meaningful irrespective of whether or not the kinetic theory arguments are applicable.
\begin{figure}[tbp]
\begin{center}
\includegraphics[width=.7\textwidth]{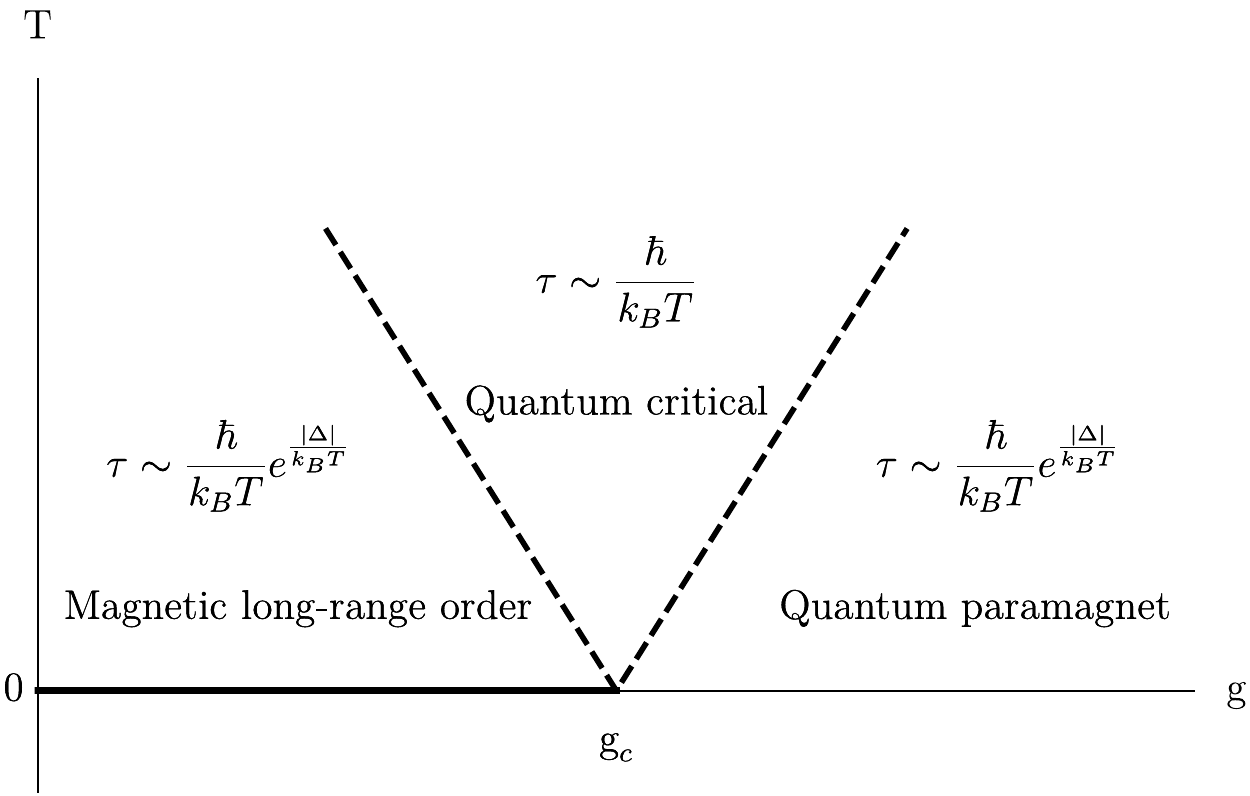}
\caption{\label{fig:quantum_critical_diag} Phase diagram of the $d=1+1$ quantum Ising model \cite{sachdev-book-2}. The relaxation time in the quantum critical region is determined by the lowest quasinormal frequency of the BTZ black hole.}
\end{center}
\end{figure}
In particular, in condensed matter physics, considerable attention has been drawn to the studies of quantum critical regions \cite{sachdev-book-2}, where the characteristic time scales of strongly interacting theories at finite temperature are of the order of $\tau \sim \hbar/k_B T$ (see Fig.~\ref{fig:quantum_critical_diag}). Moreover, estimates of thermal equilibration time $\taur$ in relevant models suggest that \cite{sachdev-book-2}
\begin{equation}
\taur \geq {\cal C} \, \frac{\hbar}{k_B T}\,,
\label{eq:sachdev-const}
\end{equation}
where ${\cal C}$ is a constant of order one, with the inequality saturated in the quantum critical region. In some models, the constant ${\cal C}$ can be computed analytically. For the quantum Ising model in $d=1+1$ dimension serving as one of the main examples illustrating quantum critical behavior in \cite{sachdev-book-2}, the relaxation time of the order parameter $\hat{\sigma}_z$ having the anomalous dimension $\Delta = 1/8$ in the quantum critical region is determined by the correlation function of a $1+1$-dimensional CFT at finite temperature. The (equilibrium) retarded two-point correlation function of an operator of (non-integer) conformal dimension $\Delta$ in momentum space is given by \cite{Son:2002sd}
\begin{align}
 G^{R}(\omega , q) & = {{\cal C}_\Delta \over \pi\, \Gamma^2 ( \Delta  -1)\sin{\pi\Delta}}\left| \Gamma \left( \frac{\Delta}{2}+  {i (\omega - q)\over 4 \pi T }\right)\Gamma \left(\frac{\Delta}{2}+  {i (\omega + q)\over 4 \pi T }\right)\right|^2 \nonumber \\
&\quad \times \Biggl[ \cosh{{q\over 2 T}} -\cos{\pi\Delta}\cosh{ {\omega\over 2 T}  }  + i \sin{\pi\Delta}\sinh{{\omega\over 2 T} }\Biggr]\,,
\label{eq:full_green_ni}
\end{align}
where ${\cal C}_\Delta$ is the normalization constant and we put $T_L=T_R=T$. The correlator has a sequence of poles at
\begin{equation}
\omega = \pm q - i 4 \pi T \left( n +\frac{\Delta}{2}\right)\,,
\end{equation}
where $n=0,1,2,...$. Note that these are precisely the quasinormal frequencies of the dual BTZ black hole \cite{Birmingham:2001pj}, \cite{Son:2002sd}. At zero spatial momentum, the lowest quasinormal frequency determines the relaxation time
\begin{equation}
\taur = \frac{1}{2\pi \Delta}\, \frac{\hbar}{k_B T}\,,
\label{eq:rel-time-scaling-dim}
\end{equation}
and thus the constant ${\cal C}$ in Eq.~(\ref{eq:sachdev-const}) is ${\cal C} = 4/\pi \approx 1.273$ for the Ising model considered\footnote{In \cite{sachdev-book-2}, the relaxation time was determined by expanding the denominator of the correlation function in Taylor series around $\omega =0$. This approximates the singularity of the correlator rather crudely giving $\taur = \frac{\hbar}{2 k_B T} \cot{[\frac{\pi}{16}]}$ and ${\cal C} = \frac{1}{2} \cot{[\frac{\pi}{16}]} \approx 2.514$.} in \cite{sachdev-book-2}. Curiously, inserting $\Delta=2$ (the scaling dimension of the energy-momentum tensor) into Eq.~(\ref{eq:rel-time-scaling-dim}) and using Eq.~(\ref{eq:rel-visc-rel}), one formally\footnote{The shear viscosity is not defined in $d=1+1$.} finds $\eta/s = 1/4\pi$.

\begin{figure}[htbp]
\centering
\includegraphics[width=0.45\textwidth]{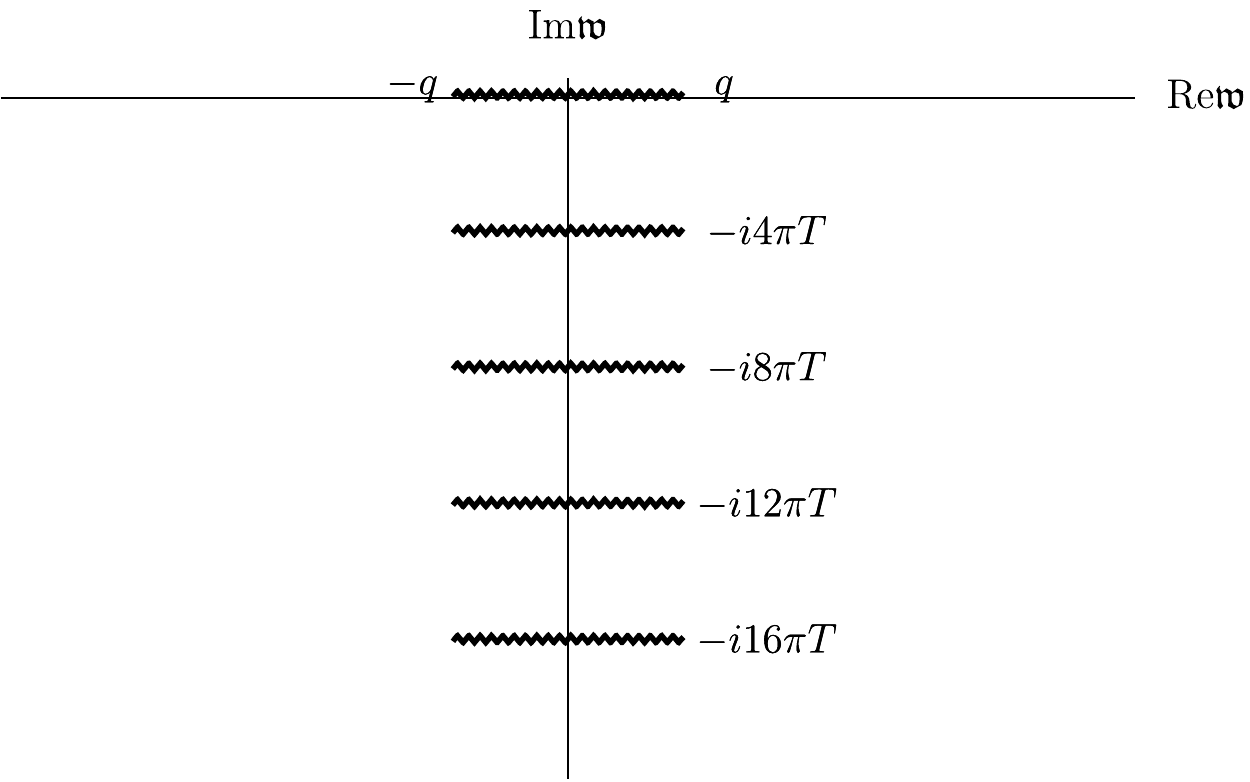}
\includegraphics[width=0.45\textwidth]{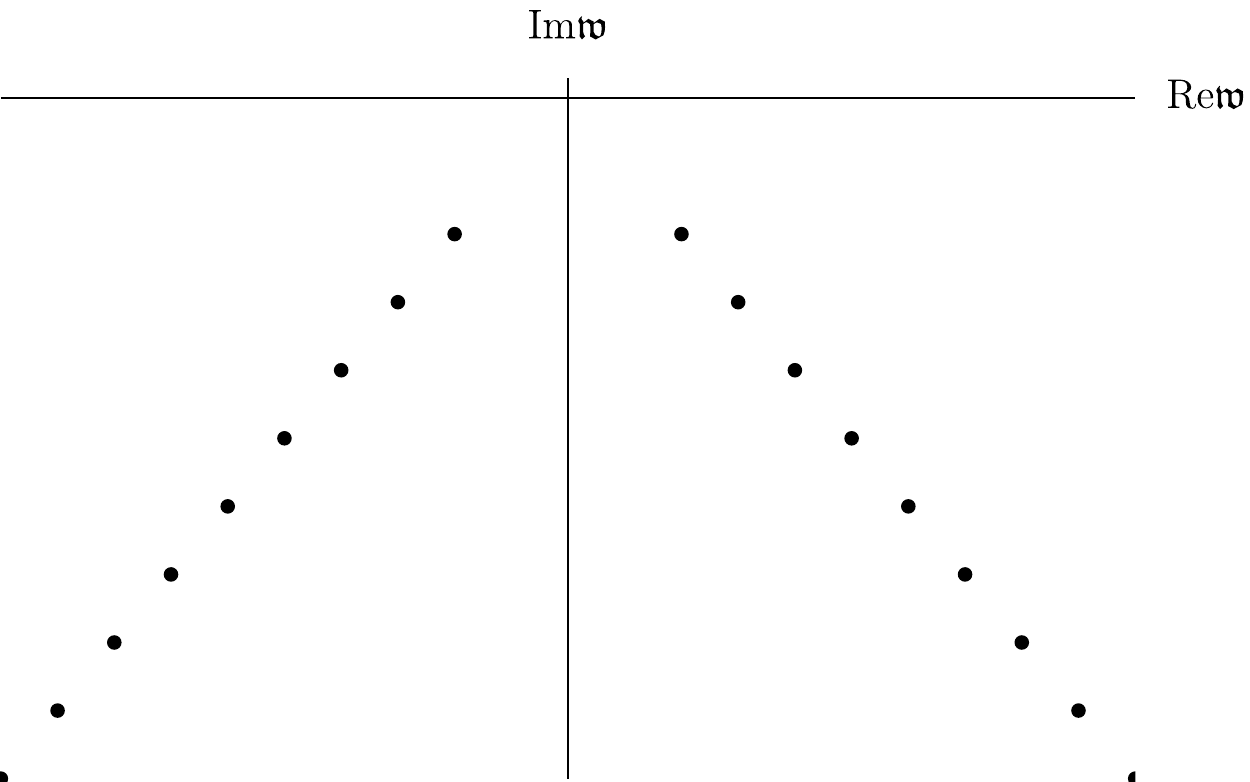}
\caption{Singularities of a thermal two-point correlation function in the complex frequency plane at (vanishingly) small \cite{Hartnoll:2005ju} (left panel) and infinitely large \cite{Starinets:2002br} (right panel) values of the coupling.}
\label{fig:cuts_poles}
\end{figure}

In holography, the importance  of the quasinormal spectrum  as the fundamental characteristic feature of near-equilibrium phenomena in a dual field theory has been recognized early on \cite{KalyanaRama:1999zj}, \cite{Horowitz:1999jd}, \cite{Danielsson:1999fa} and later it was observed \cite{Birmingham:2001pj} and shown \cite{Son:2002sd}, \cite{Kovtun:2005ev} that the quasinormal frequencies correspond to poles of the dual retarded correlators. A typical distribution of poles in the complex frequency $\omega$ plane at fixed spatial momentum $q$ of an equilibrium retarded correlator computed via holography in the supergravity approximation (e.g. at infinite 't Hooft coupling and infinite $N_c$ in $\CN=4$ SYM) is shown in Fig.~\ref{fig:cuts_poles} (right panel), where the spectrum of a scalar fluctuation is shown \cite{Starinets:2002br}.
\begin{figure}[htbp]
\centering
\includegraphics[width=0.45\textwidth]{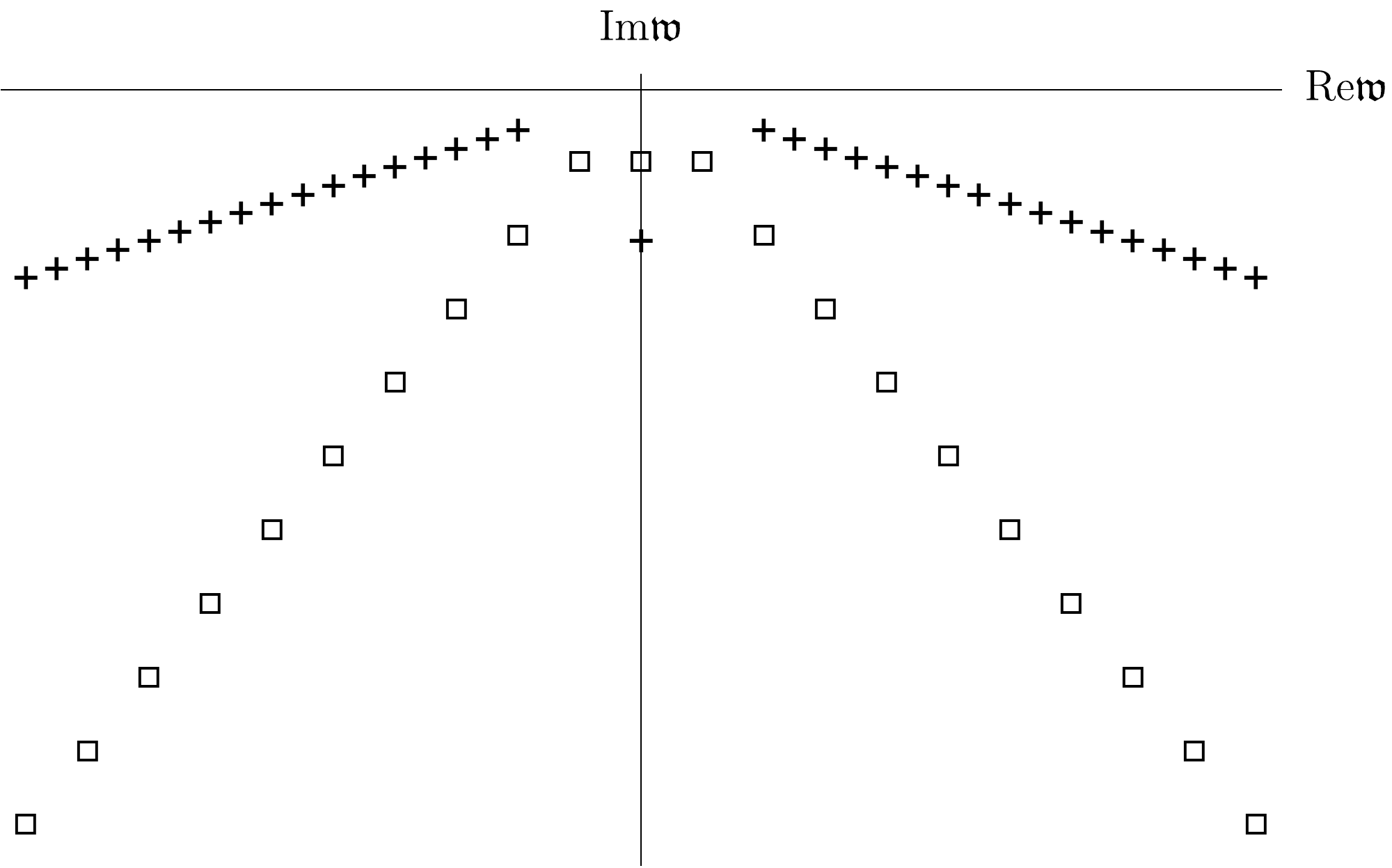}
\includegraphics[width=0.45\textwidth]{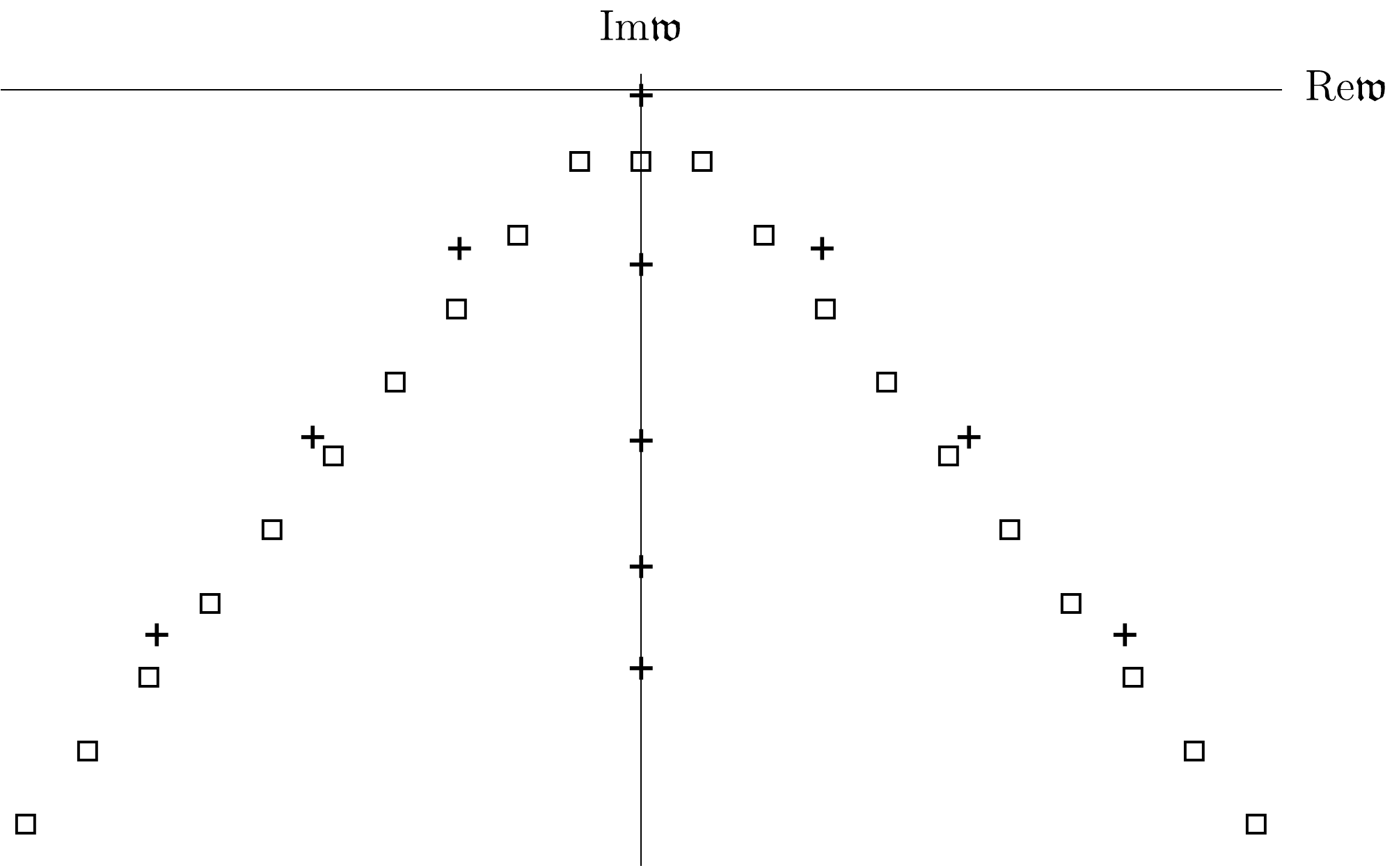}
\caption{Singularities of thermal two-point correlation function of the energy-momentum tensor in the shear channel in the complex frequency plane at large coupling at $\eta/s>\viscb$ (left panel) and at $\eta/s<\viscb$ (right panel). Poles at infinitely large coupling are indicated by squares. At large but finite coupling, their new positions are shown by crosses.}
\label{fig:poles_finite_coupling}
\end{figure}
For correlators of conserved quantities such as the energy-momentum tensor, the spectrum, in addition to an infinite tower of gapped strongly damped modes $\omega_n = \omega_n (q)$, contains also a sector of gapless  hydrodynamic modes $\omega = \omega (q)$ with the property $\omega (q) \rightarrow 0$ for $q\rightarrow 0$ \cite{Starinets:2002br,Nunez:2003eq,Kovtun:2005ev}. Asymptotics of these spectra were computed in Refs.~\cite{Cardoso:2004up,Natario:2004jd} (for large $n$) and in Ref.~\cite{Festuccia:2008zx} (for large $q$). Curiously, at weak coupling the correlators at finite spatial momentum $q$ seem to have branch cuts stretching from $-q$ to $q$ rather than poles \cite{Hartnoll:2005ju} (see left panel of Fig.~\ref{fig:cuts_poles}). At zero spatial momentum, the branch cuts reduce to a sequence of poles on the imaginary axis \cite{Hartnoll:2005ju}. These issues are further discussed in \cite{Romatschke:2015gic} and in the present paper.

Finite coupling corrections to the quasinormal spectrum can be computed by using higher derivative terms in the appropriate supergravity action. Such corrections for gravitational backgrounds involving $R^4$ higher derivative term were recently computed in Refs.~\cite{Stricker:2013lma,Waeber:2015oka}. In this paper, we consider $R^4$ and $R^2$ terms, including Gauss-Bonnet gravity. We find a number of novel features in addition to those reported in
Refs.~\cite{Stricker:2013lma,Waeber:2015oka}. Our observations can be summarized as follows (see Sections \ref{sec:SYM}, \ref{sec:GB}, \ref{sec:r2} for full details):
\begin{itemize}
\item The positions of all poles change with the coupling. In the shear channel in particular, two qualitatively different trends are seen depending on whether $\eta/s >\viscb$ or $\eta/s <\viscb$ (see Fig.~\ref{fig:poles_finite_coupling}). In the first case (realized, for example, in $\CN=4$ SYM), the symmetric branches of non-hydrodynamic poles lift up towards the real axis\footnote{In their motion toward the real axis, the branches remain essentially straight, in agreement with earlier observations in Ref.~\cite{Waeber:2015oka}. We do not observe the phenomenon of poles with large imaginary parts bending toward the real axis reported in Ref.~\cite{Stricker:2013lma}.} and the diffusion pole moves deeper down the imaginary axis. In the second case (corresponding to known examples of the dual gravity actions with curvature squared corrections, in particular, Gauss-Bonnet gravity with positive coupling), the branches move up only very slightly and the diffusion pole comes closer to the origin.

\item For $\eta/s >\viscb$, the density of poles in the symmetric branches increases monotonically with the coupling changing from strong to weak values as shown schematically in Fig.~\ref{fig:poles_finite_coupling}. Qualitatively, this seems to be compatible with the poles merging and eventually forming branch cuts $(-\infty,q]$ and $[q,\infty)$, where $q$ is the spatial momentum, in the complex frequency plane at vanishing coupling. For $\eta/s <\viscb$, however, the density of poles decreases and they seem to disappear from the finite complex plane completely in the limit of vanishing viscosity.

\item In the holographic models we considered, the function $\eta / s\, \taur T$ is a slowly varying function of the coupling, with an appreciable change in the vicinity of infinite coupling only, suggesting that approximations of the type $\eta /s \sim const\, \taur T$ are not unreasonable in the strongly coupled regime even though they cannot possibly follow from kinetic theory arguments.

\item In view of the relation between $\eta/s$ and relaxation time, a bound on quasinormal frequencies of black branes similar to the one proposed by Hod for black holes \cite{Hod:2006jw} may imply a bound on $\eta/s$. This is further discussed in Section \ref{sec:discussion}.

\item As $\eta/s$ increases well beyond $\viscb$ and the poles approach the real axis, we expect them to be visible as clear quasiparticle-like excitations (i.e. well-defined, high in amplitude and very narrow peaks) in the appropriate spectral function of the dual field theory, well known from weakly coupled theories. This is indeed the case (see Section \ref{sec:SpectFunGBShear} for a calculation of the shear channel spectral function in the Gauss-Bonnet theory where these feature can be seen explicitly).

\item An inflow of new poles from complex infinity is observed at finite coupling. The new poles ascend from the negative infinity towards the origin along the imaginary axis as the coupling changes. The behavior of these new poles as a function of coupling also depends on whether $\eta/s > \viscb$ or $\eta /s < \viscb$. In the Gauss-Bonnet model with $\eta /s < \viscb$ (i.e. with positive values of the Gauss-Bonnet coupling), the poles reach the
asymptotic values known analytically \cite{GBNesojen}, without interfering with the hydrodynamic pole. However, in models with $\eta/s > \viscb$ ($\CN=4$ SYM or Gauss-Bonnet holographic liquid with negative coupling), a qualitatively different picture is observed. In this case, in the shear channel, the least damped new pole reaches the hydrodynamic pole at a certain value of the coupling (for each fixed $q$), the two poles merge and then move off the imaginary axis. Furthermore, as the coupling constant varies at fixed $q$, the poles previously describing the hydrodynamic excitations (diffusion and sound) become the leading (i.e. having the smallest $\im |\omega|$) poles of the two symmetric branches. We interpret these phenomena as the breakdown of the hydrodynamic gradient expansion at some value of the coupling (for each $q$). Phrased differently, at each value of the coupling $\lambda$, there exists a critical value of the wave vector $q_c (\lambda)$ such that for $q > q_c (\lambda)$ the hydrodynamic description becomes inadequate. In the holographic models we considered, the function $q_c (\lambda)$ is a monotonically increasing function of the coupling suggesting that the range of validity of the hydrodynamic description is larger at strong coupling. Details are reported in Sections \ref{sec:SYM} and \ref{sec:GB}. This is reminiscent of the weak coupling kinetic theory behavior mentioned earlier \cite{boltzmann-book} and also the one described in \cite{Romatschke:2015gic}, although our interpretation is somewhat different from the one in Ref.~\cite{Romatschke:2015gic}.
\end{itemize}

The reported observations (admittedly, made only for a few holographic models and suffering from various limitations mentioned above) seem to suggest the following picture: First, the relations such as (\ref{eq:rel-visc-rel}) may still hold in the regime of the coupling where the kinetic theory approach used to derive them can no longer be justified. This may explain why using the kinetic theory formally outside its regime of applicability can still give results compatible with experimental data. Second, it seems that for a fixed value of the coupling, there exist critical length- and time-scales beyond which the hydrodynamic approximation fails. The dependence of these critical scales on coupling extracted from the holographic models suggests that hydrodynamics has a wider range of applicability at strong coupling in comparison to weaker coupling. This appears to be compatible with the widely reported ``unreasonable effectiveness of hydrodynamics'' in models of strongly coupled plasma.

\section{Coupling constant corrections to equilibrium energy-momentum tensor correlators in strongly interacting $\CN=4$ SYM theory}
\label{sec:SYM}
For $\CN=4$ supersymmetric $SU(N_c)$ Yang-Mills (SYM) theory in $d=3+1$ (flat) dimensions, corrections in inverse powers of the 't Hooft coupling $\lambda=g^2_{YM} N_c$ at infinite $N_c$ to thermodynamics \cite{Gubser:1998nz,Pawelczyk:1998pb} and transport \cite{Buchel:2004di,Buchel:2008sh,Benincasa:2005qc,Buchel:2008ac,Buchel:2008bz,Buchel:2008kd,Saremi:2011nh,Grozdanov:2014kva} have been computed using the higher derivative $R^4$ term \cite{Grisaru:1986px,Gross:1986iv} in the effective low-energy type IIB string theory action\footnote{The full set of $\alpha'^3$ terms in the ten-dimensional effective action is currently unknown. Corrections involving the self-dual Ramond-Ramond five-form were considered in Refs.~\cite{Green:2003an,deHaro:2003zd,Green:2005qr,Paulos:2008tn}. Following the arguments in \cite{Myers:2008yi}, in this paper we assume that the (unknown) corrections to fields whose background values vanish to leading order in $\alpha'^3$ for a given supergravity solution will not modify the quasinormal spectrum at order $\alpha'^3$ and thus can be neglected. We thank A.~Buchel and K.~Skenderis for discussing these issues with us.}. In particular, for the shear viscosity to entropy density ratio the coupling constant correction to the universal infinite coupling result is positive \cite{Buchel:2004di,Buchel:2008sh}:
\begin{equation}
\frac{\eta}{s} = \frac{1}{4\pi} \left( 1 + 15\zeta (3) \lambda^{-3/2}+ \ldots \right)\,.
\label{eq:eta-s-correction}
\end{equation}
The result (\ref{eq:eta-s-correction}) can be found, in particular, by computing the $\lambda^{-3/2}$ correction to the hydrodynamic (gapless) quasinormal frequency in the shear channel of gravitational perturbations of the appropriate background. Coupling constant corrections to the full quasinormal spectrum of gravitational perturbations of the AdS-Schwarzschild black brane background, dual to finite-temperature $\CN=4$ SYM were previously computed by Stricker \cite{Stricker:2013lma} (see also \cite{Waeber:2015oka}). In this Section, we reproduce those results and find some new features focusing on the relaxation time and the behavior of the old and new poles.
\subsection{Equations of motion}
The source of finite 't Hooft coupling corrections is the ten-dimensional low-energy effective action of type IIB string theory
\begin{align}
S_{IIB} = \frac{1}{2\kappa_{10}^2} \int d^{10} x \sqrt{-g} \left( R - \frac{1}{2} \left(\partial \phi\right)^2 - \frac{1}{4\cdot 5!} F_5^2 + \gamma e^{-\frac{3}{2} \phi} \CW + \ldots \right)\,,
\label{eq:10DAct}
\end{align}
where $\gamma = \alpha'^3 \zeta(3) / 8$ and the term $\CW$ is proportional to the contractions of the four copies of the Weyl tensor
\begin{align}
\label{eq:Wterm}
\CW = C^{\alpha\beta\gamma\delta}C_{\mu\beta\gamma\nu} C_{\alpha}^{~\rho\sigma\mu} C^{\nu}_{~\rho\sigma\delta} + \frac{1}{2} C^{\alpha\delta\beta\gamma} C_{\mu\nu\beta\gamma} C_{\alpha}^{~\rho\sigma\mu} C^\nu_{~\rho\sigma\delta}\,.
\end{align}
Considering corrections to the AdS-Schwarzschild black brane background and its fluctuations, potential  $\alpha'$ corrections to supergravity fields other than the metric and the five-form field have been argued to be irrelevant \cite{Myers:2008yi}. Moreover, as discussed in \cite{Buchel:2008ae}, for the purposes of computing the corrected quasinormal spectrum one can use the Kaluza-Klein reduced five-dimensional action
\begin{align}
S = \frac{1}{2\kappa_5^2} \int d^5 x \sqrt{-g} \left(R  + \frac{12}{L^2} + \gamma \CW \right)\,,
\label{eq:hd-action}
\end{align}
where $\CW$ is given by Eq.~(\ref{eq:Wterm}) in $5d$. The parameter $\gamma$ is related to the value of the 't Hooft coupling $\lambda$ in $\CN=4$ SYM via $\gamma  = \lambda^{-3/2}\zeta (3) L^6/8$ (we set $L=1$ in the rest of this Section). Higher derivative terms in the equations of motion are treated as perturbations and thus any reliable results are restricted to small values of the parameter $\gamma$. The effective five-dimensional gravitational constant is connected to the rank of the gauge group by the expression $\kappa_5 = 2\pi /N_c$.

To leading order in $\gamma$, the black brane solution to the equations of motion following from (\ref{eq:hd-action}) is given by  \cite{Gubser:1998nz,Pawelczyk:1998pb}
\begin{align}
ds^2 = \frac{r_0^2}{u} \left( - f(u) Z_t dt^2 + dx^2 +dy^2 +dz^2 \right) + Z_u \frac{du^2}{4u^2 f}\,,
\label{eq:corrected_metric}
\end{align}
where $f(u) = 1 - u^2$, $r_0$ is the parameter of non-extremality of the black brane geometry and the functions $Z_t$ and $Z_u$ are given by
\begin{align}
Z_t = 1 - 15\gamma\left(5u^2+5u^4-3 u^6 \right) , && Z_u = 1 + 15\gamma \left(5u^2 + 5 u^4 - 19 u^6 \right) .
\end{align}
The $\gamma$-corrected Hawking temperature corresponding to the solution (\ref{eq:corrected_metric}) is $T = r_0 (1+15\gamma)/\pi$. For the isotropic $\CN=4$ SYM medium, we now consider fluctuations of the metric of the form $g_{\mu\nu} = g_{\mu\nu}^{(0)} +  h_{\mu\nu}(u,t,z)$, where $g_{\mu\nu}^{(0)}$ is the background (\ref{eq:corrected_metric}). We Fourier transform the fluctuations with respect to $t$ and $z$ to introduce $h_{\mu\nu}(u,\omega,q)$, choose the radial gauge with $h_{ u \nu} = 0$ and follow the recipes in \cite{Kovtun:2005ev} to write down the equations of motion for the three gauge-invariant modes $Z_i = Z^{(0)}_i + \gamma Z^{(1)}_i$, $i=1,2,3$, in the scalar, shear and sound channels, respectively. Explicitly, the three modes and the corresponding equations of motion are given by the following expressions\footnote{We note that there seems to be a typo in Eq.~(23) of Ref.~\cite{Stricker:2013lma} describing metric fluctuations in the shear mode.}:
\paragraph{\bf Scalar channel}
\begin{align}
&Z_1 = \frac{ u}{\pi^2 T_0^2} h_{xy},
\label{eq:Ginv4g1} \\
&\partial^2_u Z_1 - \frac{1+u^2}{u\left(1-u^2\right)} \partial_u Z_1  + \frac{\wfr^2 - \qfr^2 \left(1-u^2 \right)}{u \left(1-u^2\right)^2} Z_1  = \gamma \, \CG_1 \left[ Z_1   \right] \,. \label{ScalarEqN4}
\end{align}
\paragraph{\bf Shear channel}
\begin{align}
&Z_2 =\frac{ u }{\pi^2 T_0^2} \left( q h_{tx} + \omega h_{xz} \right), \label{eq:Ginv4g2} \\
&\partial^2_u Z_2 - \frac{\left(1+u^2\right) \wfr ^2-\qfr^2 \left(1-u^2\right)^2}{u \left(1-u^2\right) \left(\wfr^2-\qfr^2 \left(1-u^2\right)\right)} \partial_u Z_2  +\frac{\wfr ^2 - \qfr^2 \left(1-u^2\right)}{u \left(1-u^2\right)^2} Z_2  = \gamma \, \CG_2\left[Z_2\right]\,. \label{ShearEqN4}
\end{align}
\paragraph{\bf Sound channel}
\begin{align}
&Z_3 =  - \frac{u}{2 \pi^2 T_0^2} \left[1 - \frac{q^2}{\omega^2} \left(1+u^2 + 15 \gamma u^2 \left(21 u^6-40 u^4+5\right)\right)  \right] \left(h_{xx} + h_{yy} \right) \nonumber \\
& \qquad \;\; +\frac{u}{\pi^2 T_0^2} \,\left[   \frac{q^2}{\omega^2} h_{tt}+ h_{zz} + \frac{2q}{\omega} h_{tz} \right]\,,   \label{Ginv4g3} \\
&\partial^2_u Z_3 - \frac{3 \left(1+u^2\right) \wfr ^2 -  \qfr^2 \left(3-2 u^2 +3 u^4\right)}{u \left(1-u^2\right) \left(3 \wfr^2 - \qfr^2 \left(3-u^2\right)\right)}  \partial_u Z_3  \nonumber\\
&+\frac{3 \wfr ^4 - 2  \left(3-2 u^2\right) \wfr ^2 \qfr^2  -  \qfr^2 \left(1-u^2\right) \left(4 u^3+\qfr^2 \left(u^2-3\right)\right)}{u \left(1-u^2\right)^2 \left(3 \wfr ^2 -  \qfr^2 \left(3-u^2\right)\right)}    Z_3  = \gamma \, \CG_3\left[Z_3\right]\,.\label{eq:SoundEqN4}
\end{align}
The functions $\CG_1$,  $\CG_2$ and $\CG_3$ appearing on the right hand side of the equations can be found in Appendix \ref{sec:appendix-N4}. Here and in the rest of the paper we use the dimensionless variables
\begin{align}
\wfr = \frac{\omega}{2\pi T},  \qquad  \qfr = \frac{q}{2\pi T}\,.
\label{eq:gothic}
\end{align}
The equations of motion are solved numerically and the quasinormal spectrum is extracted using the standard recipes \cite{Starinets:2002br,Nunez:2003eq,Kovtun:2005ev,Buchel:2004di,Buchel:2008sh,Benincasa:2005qc}. Our numerical approach is described in Appendix \ref{sec:Numerics}.
\subsection{The spectrum of the metric fluctuations}
\label{sec:N4Results}
Given the smooth dependence of the equations of motion on the parameter $\gamma$, we may expect the eigenvalues to shift somewhat in the complex frequency plane with respect to their $\gamma=0$ positions. This is indeed the case, as noted previously in Refs.~\cite{Stricker:2013lma,Waeber:2015oka} and the details of this shift are interesting. In addition to this, we observe an inflow of new poles from complex infinity along the imaginary axis.
\begin{figure}[ht]
\centering
\begin{subfigure}[t]{0.45\linewidth}
\includegraphics[width=1\linewidth]{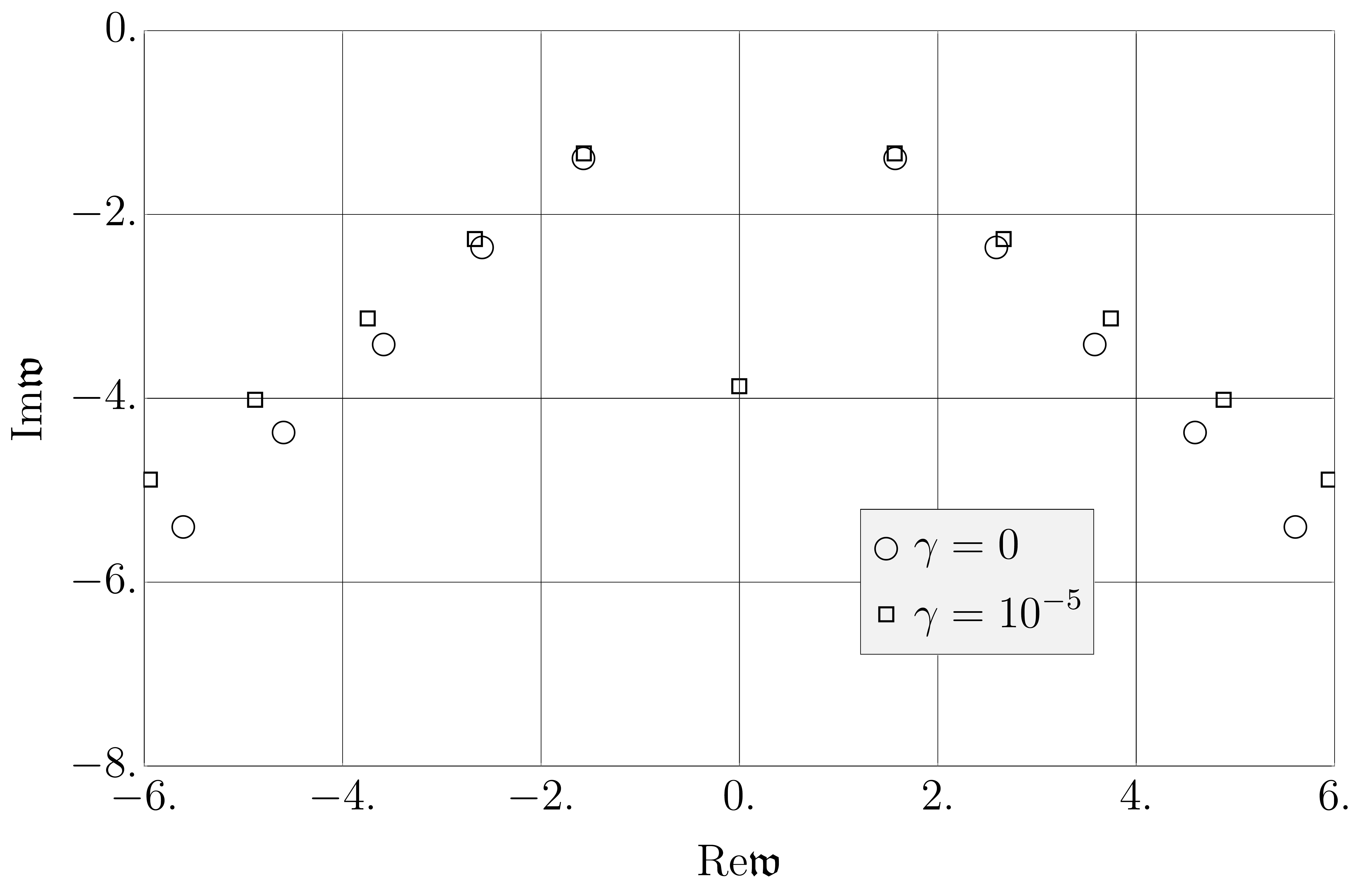}
\end{subfigure}
\qquad
\begin{subfigure}[t]{0.45\linewidth}
\includegraphics[width=1\linewidth]{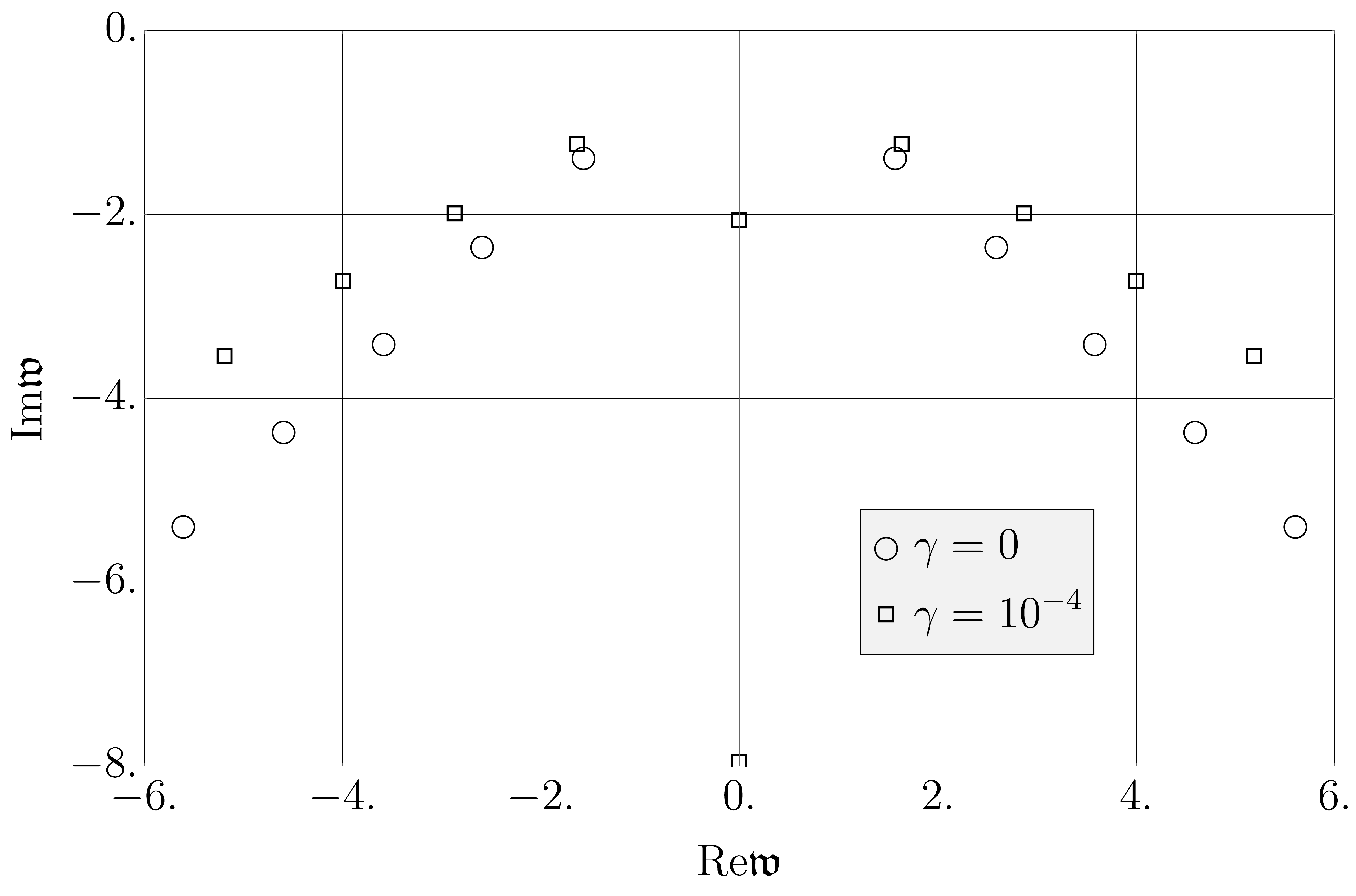}
\end{subfigure}
\\
\begin{subfigure}[b]{0.45\linewidth}
\includegraphics[width=1\linewidth]{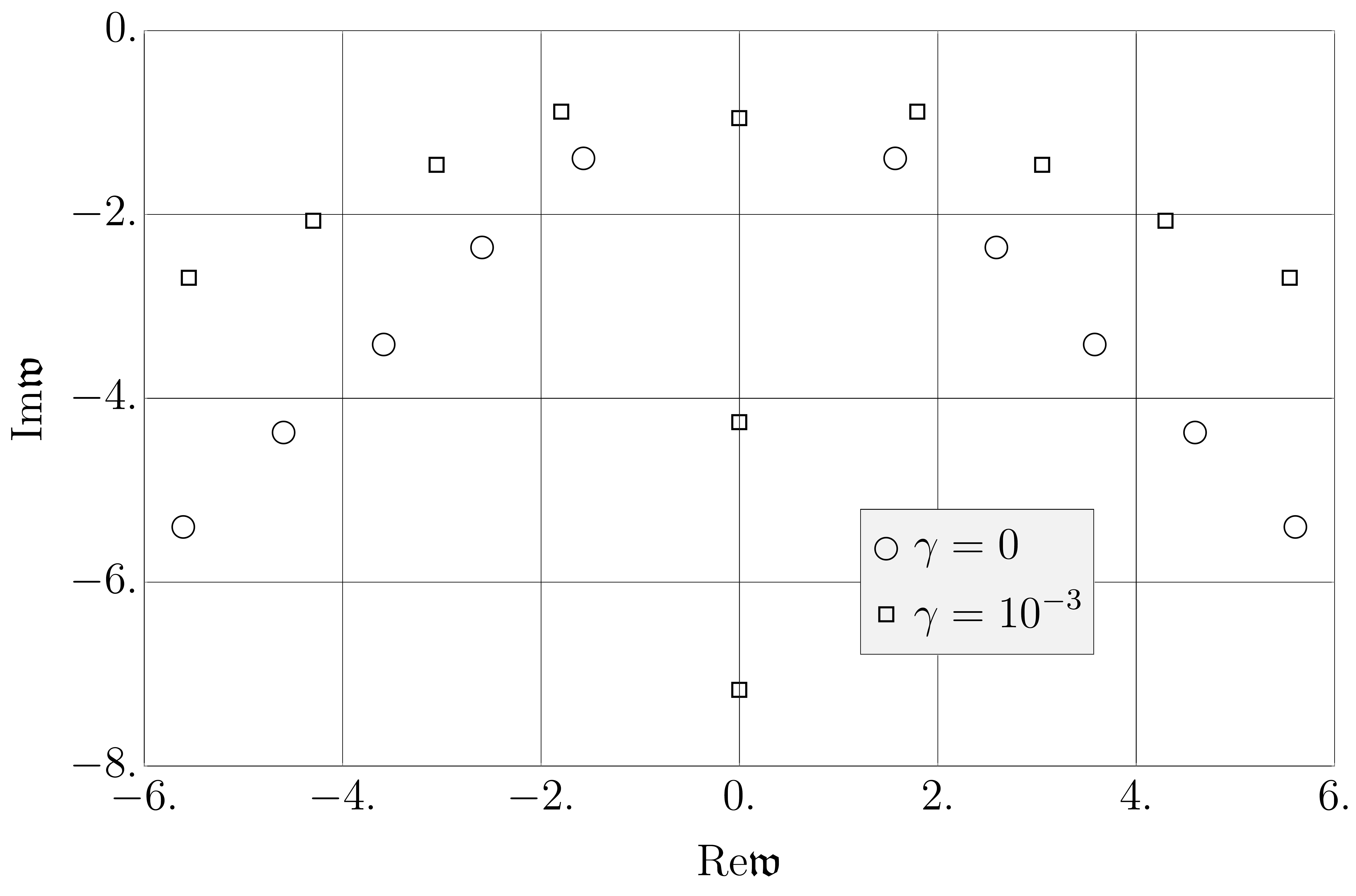}
\end{subfigure}
\qquad
\begin{subfigure}[b]{0.45\linewidth}
\includegraphics[width=1\linewidth]{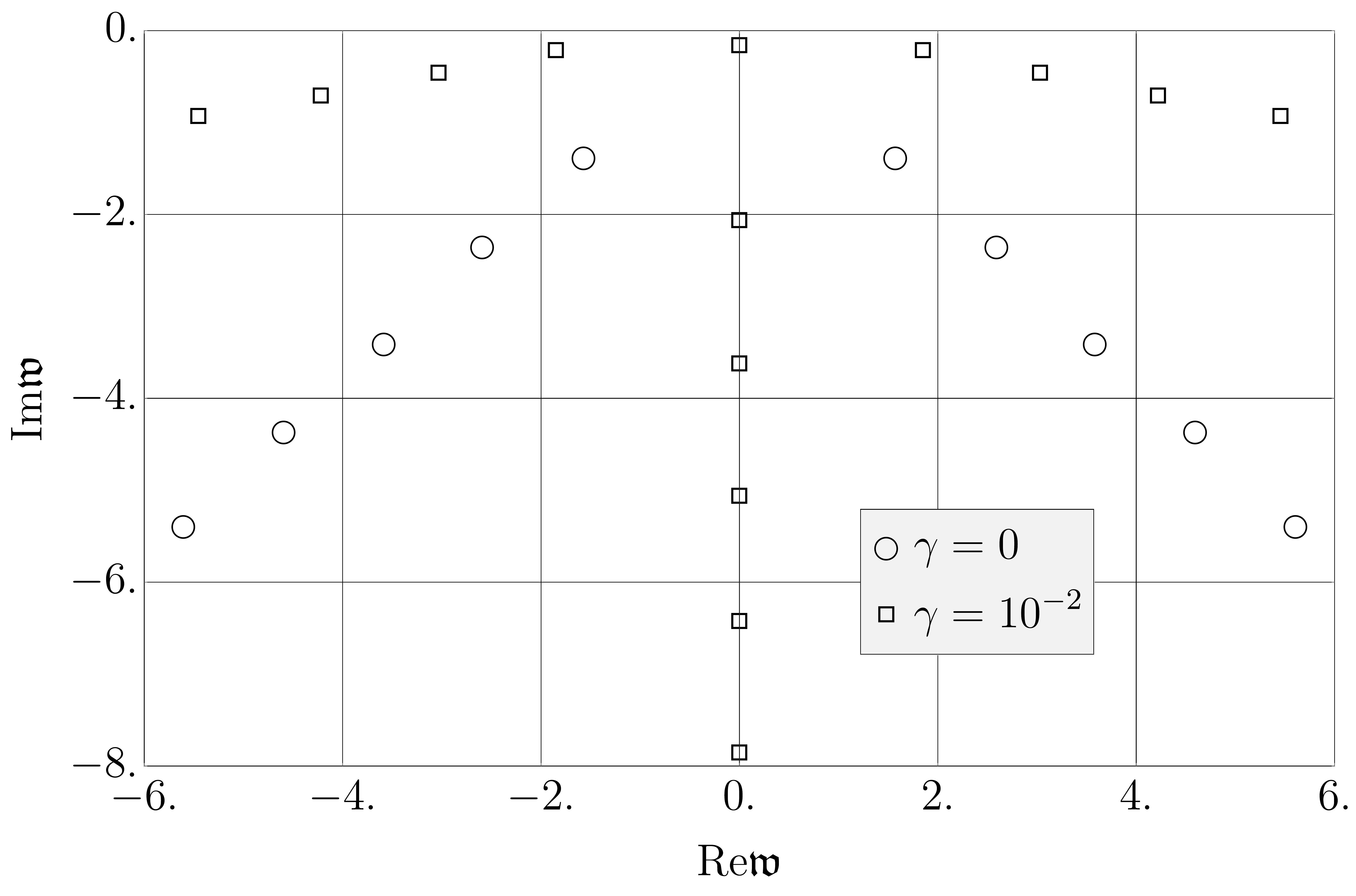}
\end{subfigure}
\caption{Poles (shown by squares) of the energy-momentum retarded two-point function of $\mathcal{N}=4$ SYM in the scalar channel, for various values of the coupling constant and $\qfr=0.1$. From top left: $\gamma = \{10^{-5},\, 10^{-4}, \,10^{-3},\, 10^{-2}\} $ corresponding to values of the 't Hooft coupling $\lambda \approx \{609,\, 131, \, 28,\, 6\} $. Poles at $\gamma = 0$ ($\lambda\rightarrow \infty$) are shown by circles.}
\label{fig:N=4+gamma-Scalar-channel}
\end{figure}
These poles are non-perturbative in $\gamma$ (the relevant quasinormal frequencies scale as $1/\gamma$) but under certain conditions they are visible in the finite complex frequency plane and can even be approximated by analytic expressions. The new poles appear in all three channels of perturbations. In the shear and sound channels, they interfere with the hydrodynamic poles and effectively destroy them at still sufficiently small, $q$-dependent values of  $\gamma$. A qualitatively similar phenomenon is observed in Gauss-Bonnet gravity where the equations of motion are second order and fully non-perturbative (see Section \ref{sec:GB}).
\subsubsection{Scalar channel}
The scalar equation of motion  \eqref{ScalarEqN4} is solved numerically with the incoming wave boundary condition at $u=1$ and Dirichlet condition at $u=0$ for fixed small values of $\gamma > 0$. A typical distribution of the quasinormal frequencies (poles of the scalar components of the energy-momentum retarded two-point function of $\mathcal{N}=4$ SYM) in the complex frequency plane is shown in Fig.~\ref{fig:N=4+gamma-Scalar-channel}.

The two symmetric branches of the modes move up towards the real axis relative to their $\gamma = 0$ position. Here and in all subsequent calculations, we do not observe the bending of the quasinormal modes with large real and imaginary parts towards the real axis reported earlier in  Ref.~\cite{Stricker:2013lma}. Rather, our findings agree with the results of Ref.~\cite{Waeber:2015oka}, where the two branches lift up without bending. The two branches become more and more horizontal with the 't Hooft coupling decreasing and move closer to the real axis.
\begin{figure}[ht]
\centering
\begin{subfigure}[t]{0.45\linewidth}
\includegraphics[width=1\linewidth]{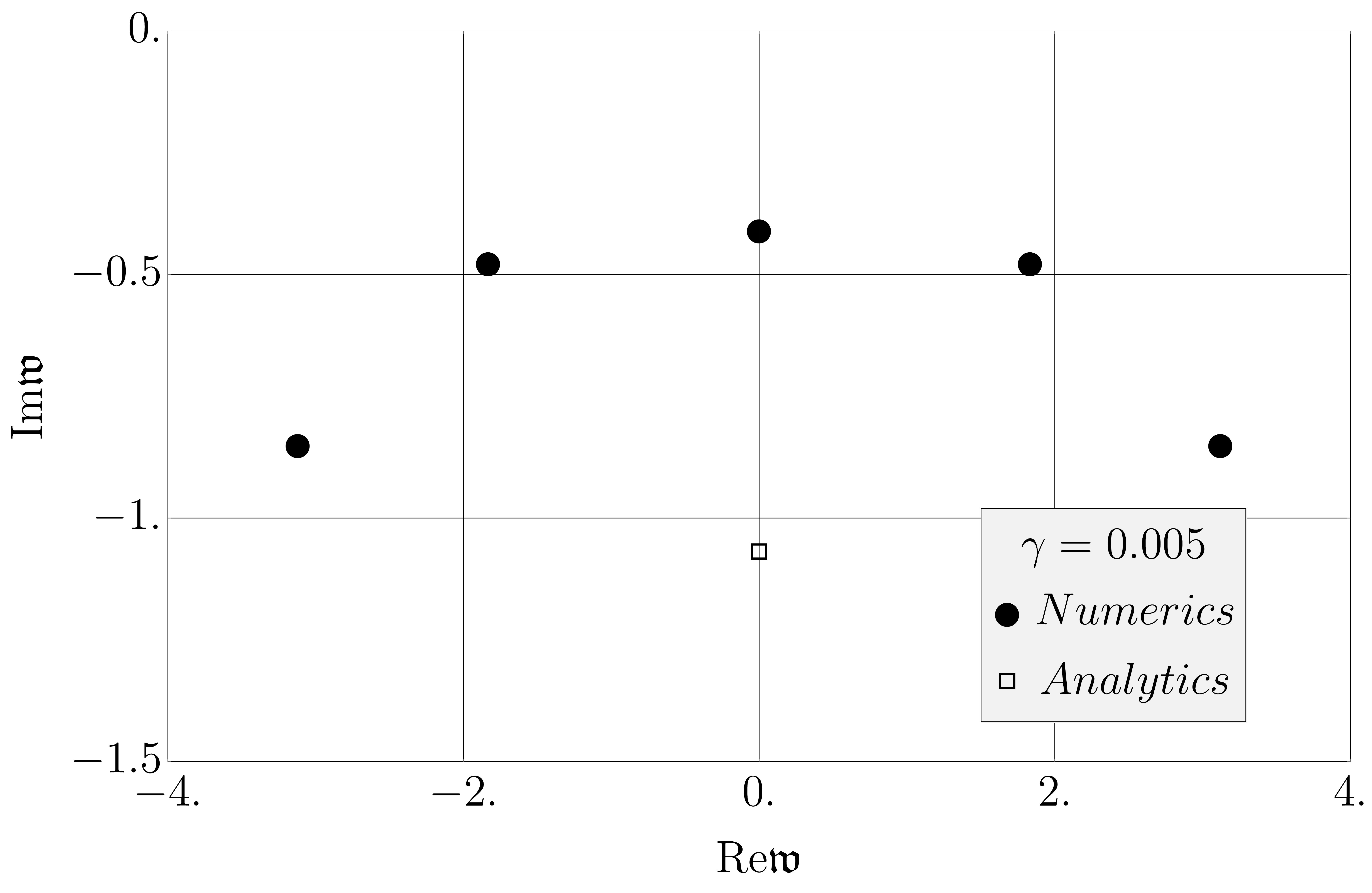}
\end{subfigure}
\qquad
\begin{subfigure}[t]{0.45\linewidth}
\includegraphics[width=1\linewidth]{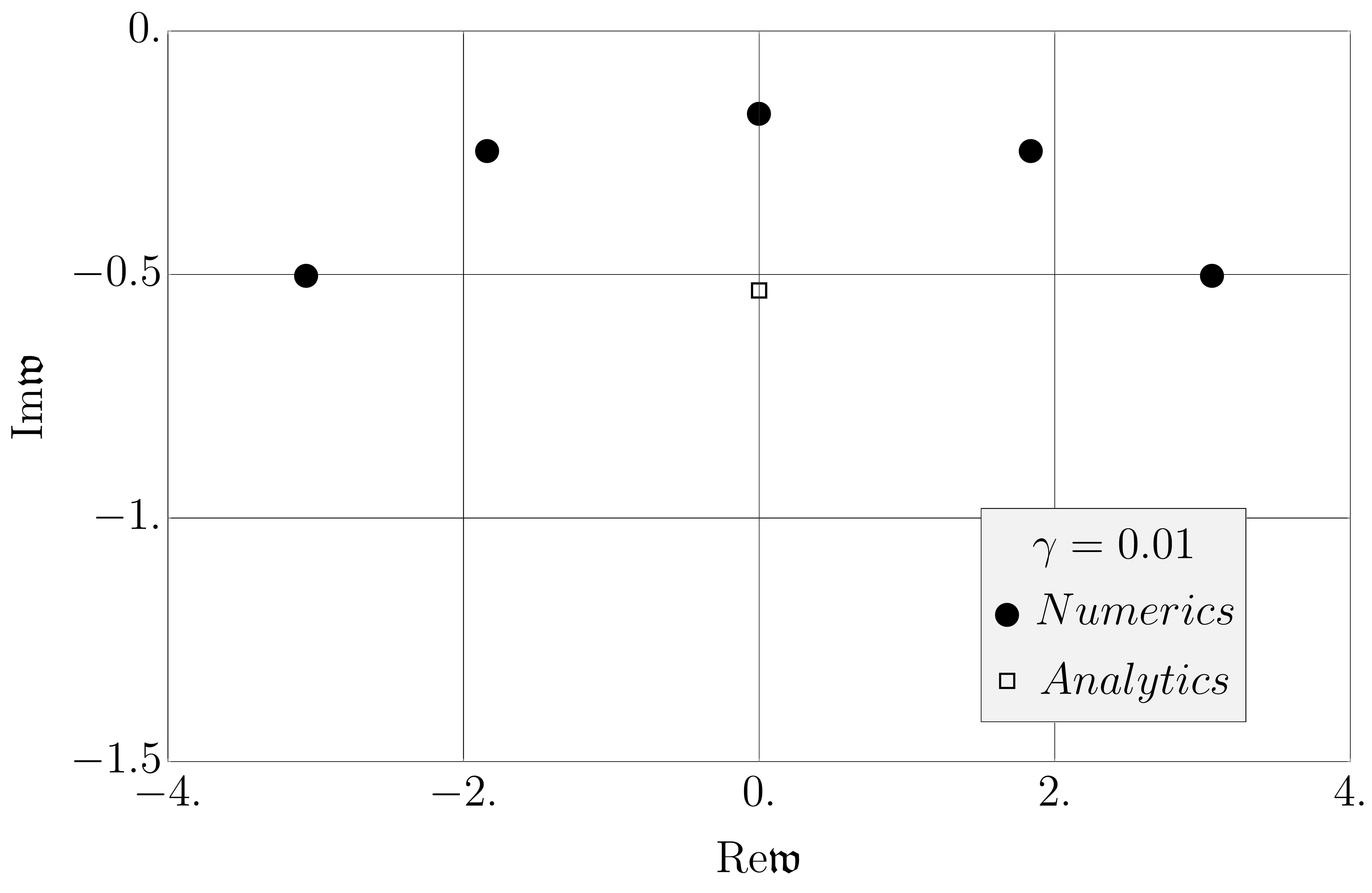}
\end{subfigure}
\\
\begin{subfigure}[b]{0.45\linewidth}
\includegraphics[width=1\linewidth]{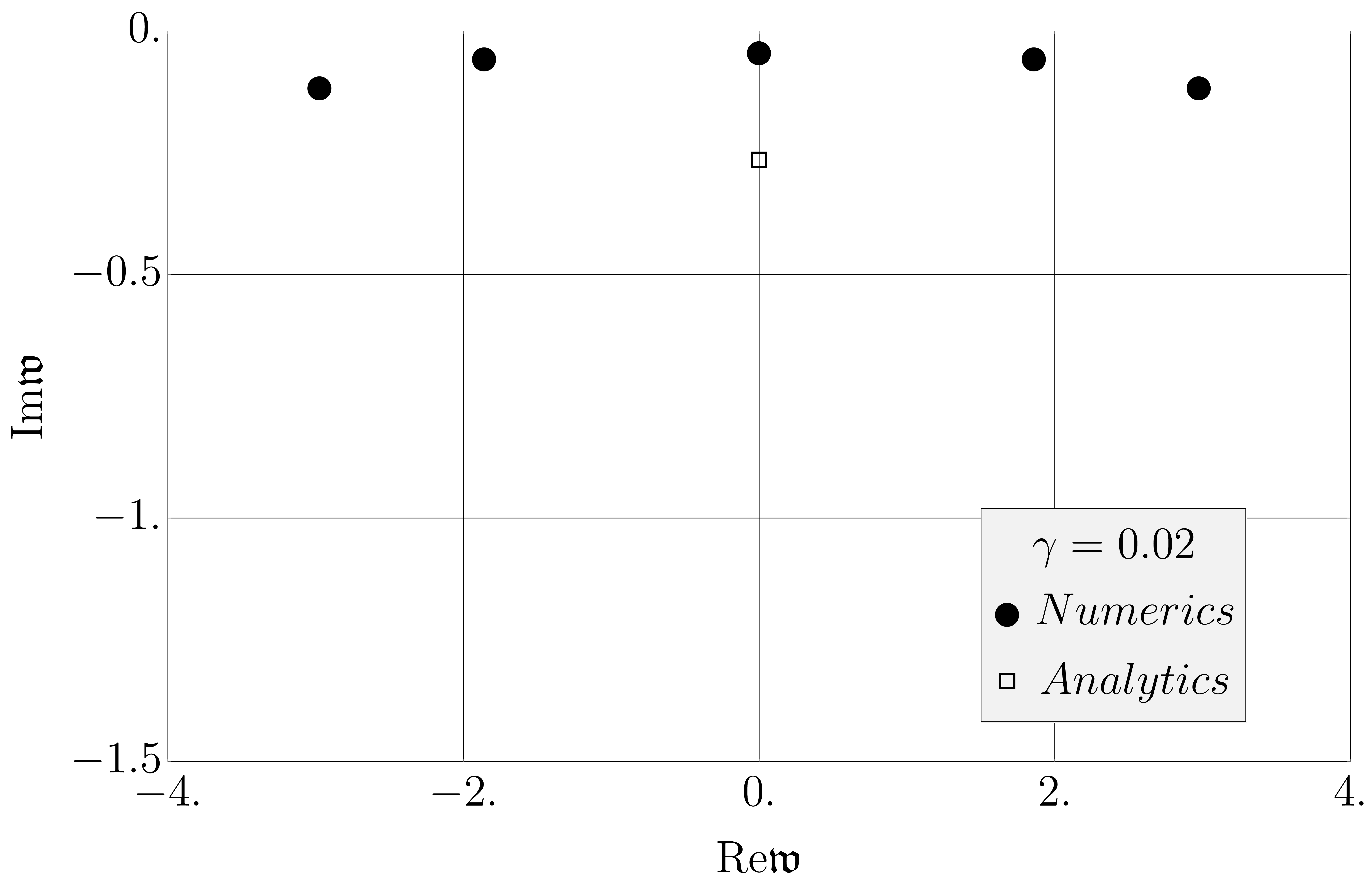}
\end{subfigure}
\qquad
\begin{subfigure}[b]{0.45\linewidth}
\includegraphics[width=1\linewidth]{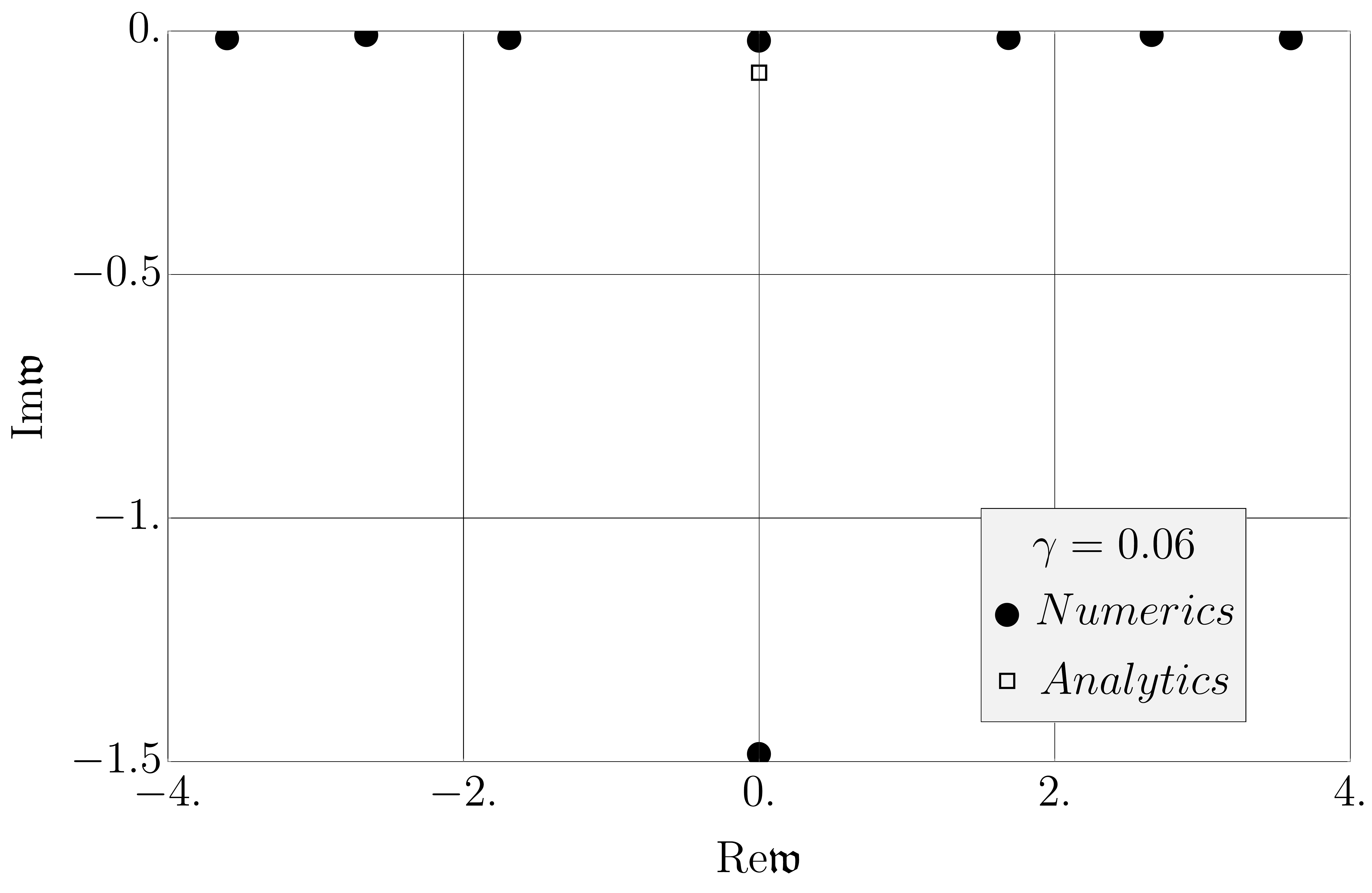}
\end{subfigure}
\caption{The closest to the origin poles (shown by black dots) of the energy-momentum retarded two-point function of $\mathcal{N}=4$ SYM in the scalar channel, for various values of the coupling constant and $\qfr=0.1$. From top left: $\gamma = \{0.005,\, 0.010, \,0.020,\, 0.060\} $ corresponding to values of the 't Hooft coupling $\lambda \approx \{10,\, 6, \, 4,\, 2\}$. The crude analytical approximation \eqref{eq:ScalarN4newpole} to the new pole on the imaginary axis becomes more accurate for larger $\gamma$.}
\label{fig:N=4+gamma-Scalar-zoom}
\end{figure}

At the same time, the distance between the poles in the branches decreases: in a sense, there is an inflow of new poles from complex infinity along the branches. This last effect is too small to be noticeable e.g. in  Fig.~\ref{fig:N=4+gamma-Scalar-channel} because in $\mathcal{N}=4$ SYM we are restricted to the $\gamma\ll 1$ regime. Extrapolating to larger values of $\gamma$ (smaller values of 't Hooft coupling) would not be legitimate with the $R^4$ corrections treated perturbatively but it is conceivable that in the limit of vanishing 't Hooft coupling the poles in the two branches merge forming two symmetric branch cuts $(-\infty,-q]$ and $[q,\infty)$. We shall see more evidence for this behavior in Gauss-Bonnet gravity, where the equations of motion are second-order and the coupling dependence is fully non-perturbative (see Section \ref{sec:GB}). The closeness of the two branches of poles to the real axis at intermediate and small values of the 't Hooft coupling raises the question of the behavior of the spectral function and the appearance of quasiparticles. Again, this is investigated in detail in Gauss-Bonnet gravity in Section \ref{sec:GB}, where we are not constrained by the smallness of the perturbation theory parameter.

We also observe a novel phenomenon: a sequence of new poles ascends along the imaginary axis towards the origin as $\gamma$ increases from zero to small finite values. The first of these poles reaches the vicinity of the origin at $\gamma \sim 0.01$. One can find a crude analytic approximation for this top pole by solving the equation in the regime $|\wfr| \ll 1$ (for simplicity, we also take $|\qfr| \ll 1$). We assume the scaling $\wfr \to \epsilon \wfr$ and $\qfr \to \epsilon \qfr$ for $\epsilon \ll 1$, so that to first order in $\epsilon$, the function  $Z_1 (u) = (1 - u)^{-i \wfr / 2 } \left( z_1^{(0)} + \epsilon z_1^{(1)} \right)$. The functions $z_1^{(0,1)}$ are found perturbatively to first order in $\gamma$. To find the quasinormal frequency, we solve the polynomial equation $Z_1(u=0,\wfr,\qfr) = 0$, looking for a solution of the form $\wfr(\qfr)$. To leading order in $\qfr$, we find a gapped pole on the imaginary axis with the dispersion relation
\begin{align}
\wfr = \wfr_{\mathfrak{g}} = -\frac{2 i}{373 \gamma -\ln 2} \approx -\frac{2 i}{373 \gamma}.
\label{eq:ScalarN4newpole}
\end{align}
As shown in Fig.~\ref{fig:N=4+gamma-Scalar-zoom}, the analytic approximation \eqref{eq:ScalarN4newpole} works better for larger values of $\gamma$. For $\gamma \rightarrow 0$, the pole recedes deep into the complex plane along the negative imaginary axis (the approximate formula \eqref{eq:ScalarN4newpole} is compatible with this observation but breaks down when $|\wfr|$ becomes large).
\subsubsection{Shear channel}
\begin{figure}[ht]
\centering
\begin{subfigure}[t]{0.45\linewidth}
\includegraphics[width=1\linewidth]{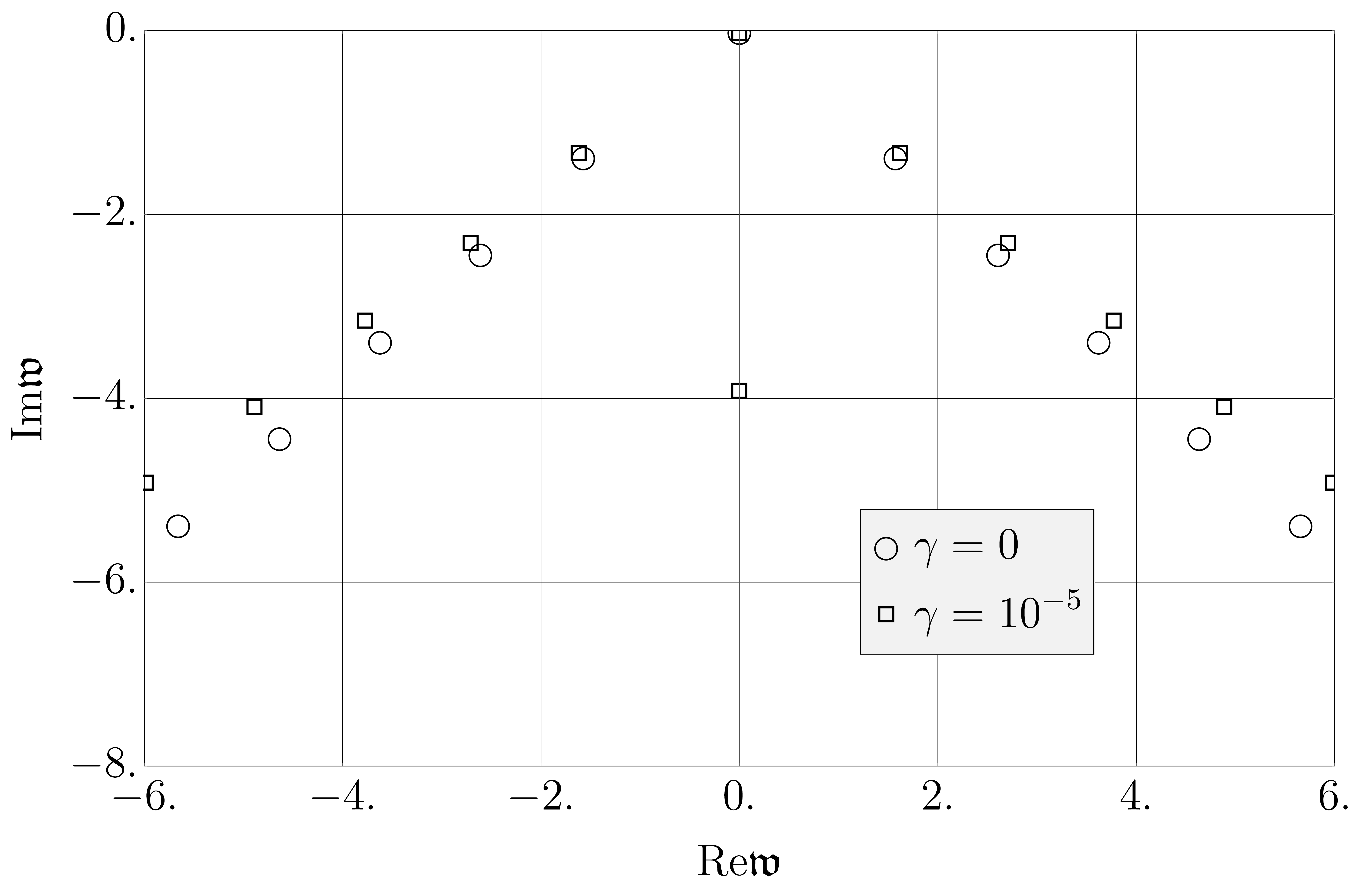}
\end{subfigure}
\qquad
\begin{subfigure}[t]{0.45\linewidth}
\includegraphics[width=1\linewidth]{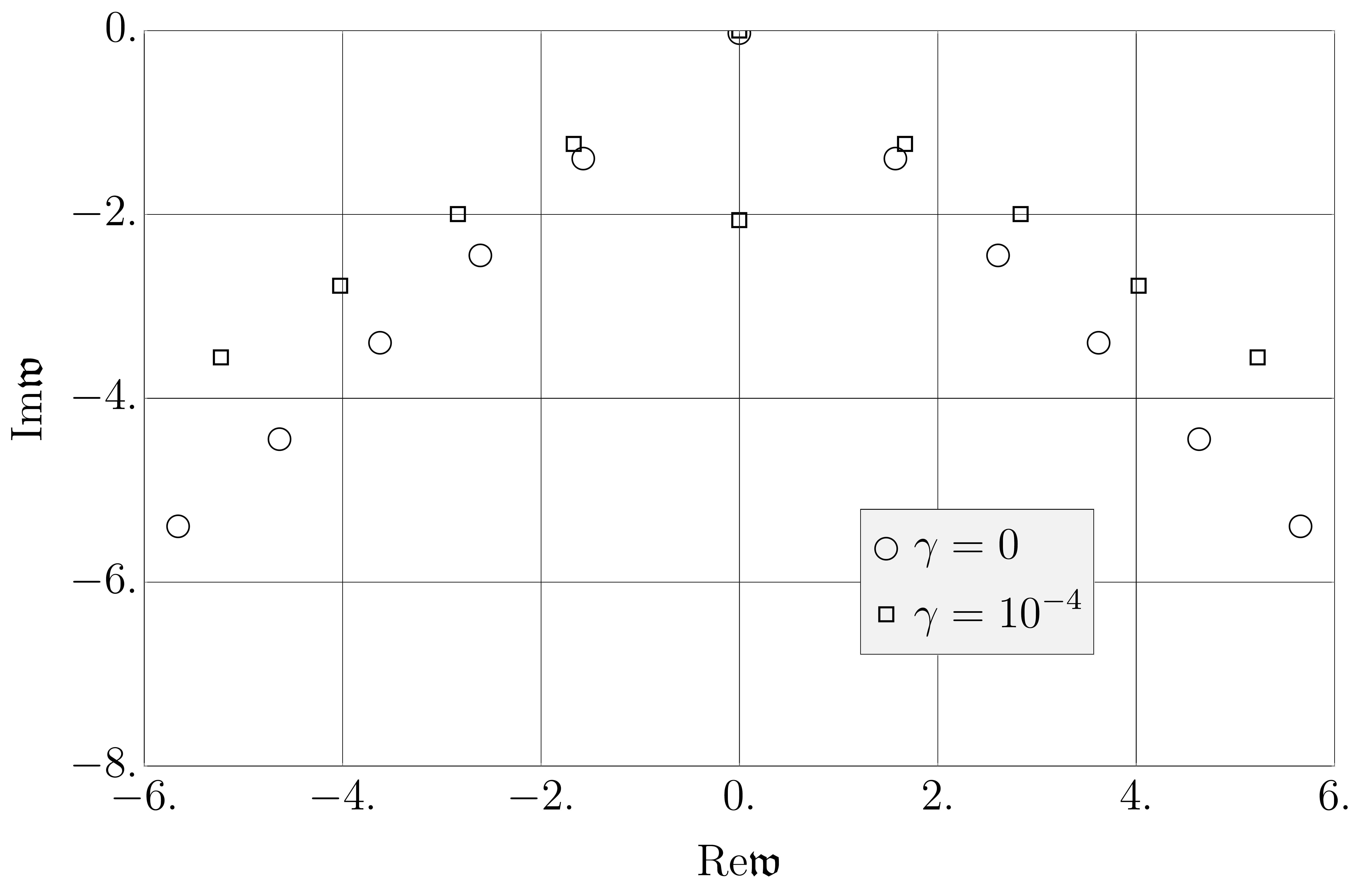}
\end{subfigure}
\\
\begin{subfigure}[b]{0.45\linewidth}
\includegraphics[width=1\linewidth]{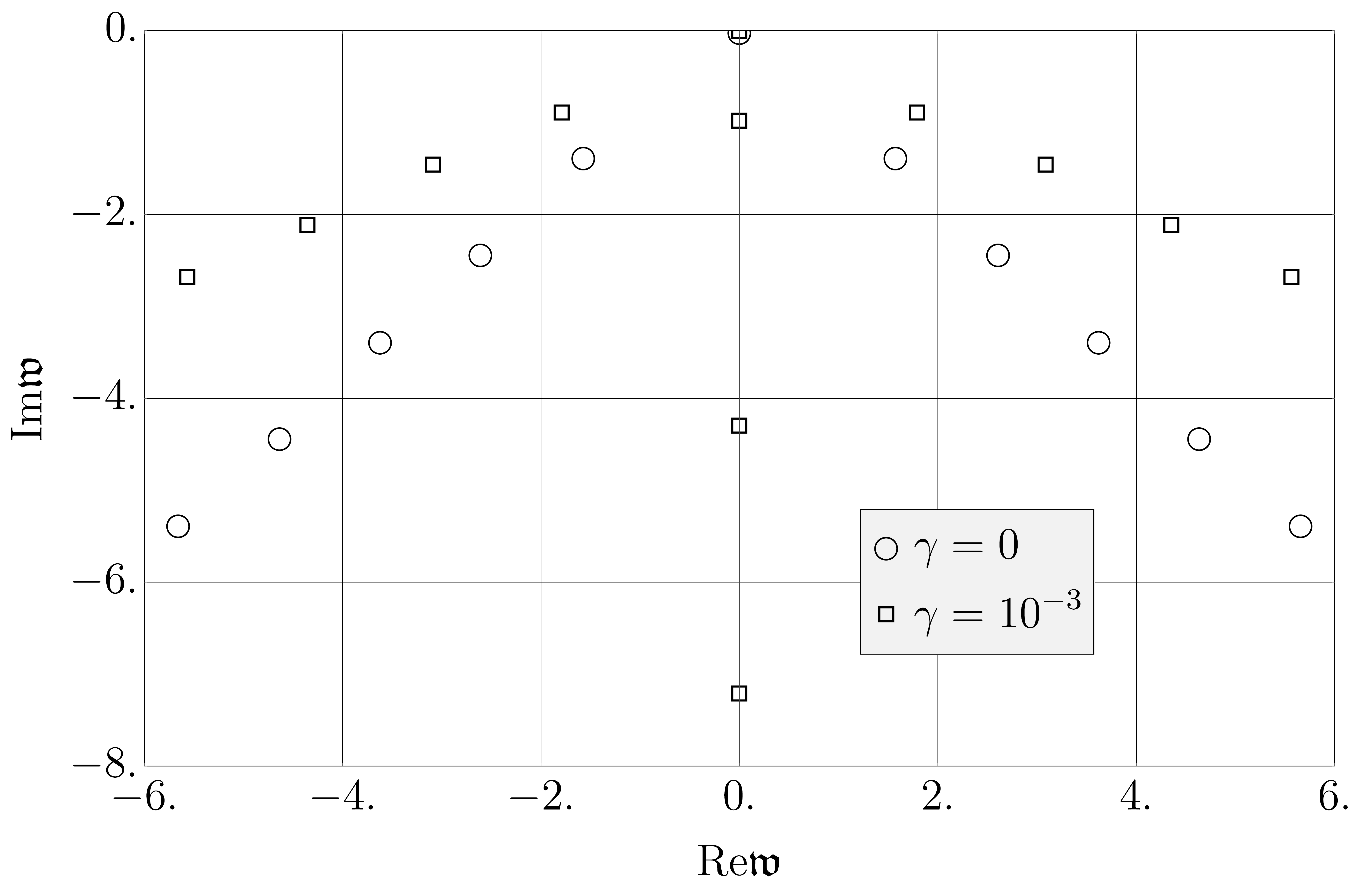}
\end{subfigure}
\qquad
\begin{subfigure}[b]{0.45\linewidth}
\includegraphics[width=1\linewidth]{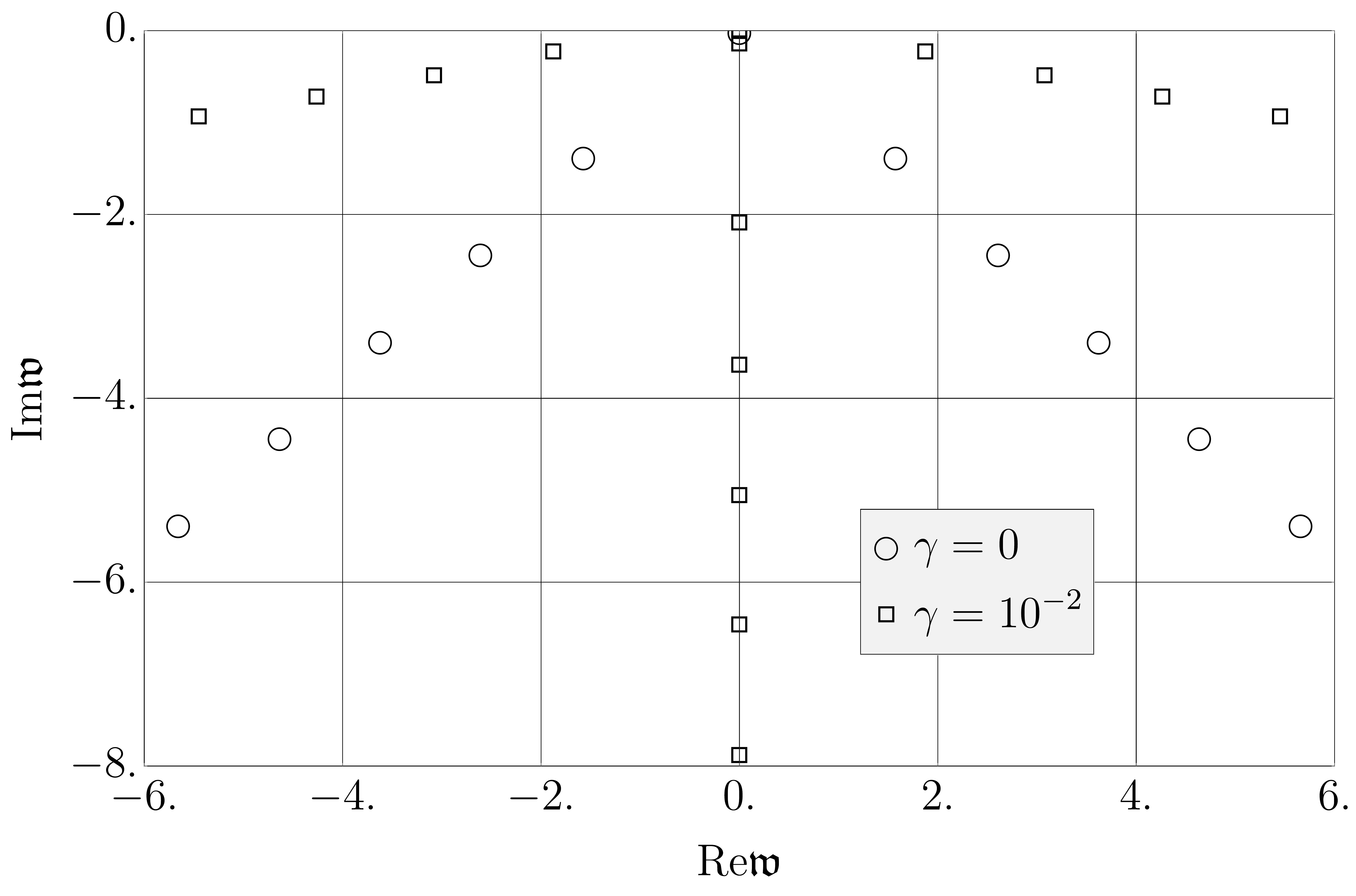}
\end{subfigure}
\caption{Poles (shown by squares) of the energy-momentum retarded two-point function of $\mathcal{N}=4$ SYM in the shear channel, for various values of the coupling constant and $\qfr=0.1$. From top left: $\gamma = \{10^{-5},\, 10^{-4}, \,10^{-3},\, 10^{-2}\} $ corresponding to values of the 't Hooft coupling $\lambda \approx \{609,\, 131, \, 28,\, 6\} $. Poles at $\gamma = 0$ ($\lambda\rightarrow \infty$) are shown by circles.}
\label{fig:N=4+gamma-Shear-channel}
\end{figure}

In the shear channel, the distribution of poles at finite coupling is similar to the one in the scalar channel. The exception is the gapless hydrodynamic pole on the imaginary axis responsible for the momentum diffusion. The poles are shown in Fig.~\ref{fig:N=4+gamma-Shear-channel} for several values of $\gamma$. General properties of non-hydrodynamic poles described in detail for the scalar channel are observed here as well. The new feature is the interaction between the diffusion pole and the first of the new poles rising up from complex infinity along the imaginary axis with increasing $\gamma$.

The dispersion relation for the diffusion pole is given by the formula \cite{Policastro:2002se,Baier:2007ix,Grozdanov:2015kqa}
\begin{align}
\omega = - i \frac{\eta}{\varepsilon + P}\, q^2 - i  \left[ \frac{\eta^2\tau_\Pi}{(\varepsilon + P)^2}-\frac{\theta_1}{2(\varepsilon + P)}\right] q^4 + \cdots \,,
\label{eq:shear_disp}
\end{align}
where in the absence of the chemical potential $\varepsilon + P = s T$. In $\CN=4$ SYM theory, one has \cite{Policastro:2001yc,Buchel:2004di,Buchel:2008sh,Baier:2007ix,Benincasa:2005qc,Bhattacharyya:2008jc,Buchel:2008bz,Buchel:2008kd,Buchel:2008bz,Grozdanov:2015kqa}
\begin{align}
\frac{\eta}{s} &= \frac{1}{4\pi} \left( 1 + 120 \gamma + \cdots \right)\,, \\
\tau_\Pi &= \frac{2 - \ln {2}}{2 \pi T} + \frac{375 \gamma}{4\pi T} + \cdots\,, \\
\theta_1 &= \frac{N_c^2 T}{32 \pi} + O(\gamma)\,.
\label{eq:coeffi-n=4}
\end{align}
The coupling constant correction to the coefficient $\theta_1$ of the third-order hydrodynamics is currently unknown. However, for $\qfr \ll 1$, the $\qfr^2$ term in Eq.~(\ref{eq:shear_disp}) dominates and the pole moves down the imaginary axis with $\gamma$ increasing, in agreement with our numerical findings.

\begin{figure}[ht]
\centering
\begin{subfigure}[t]{0.45\linewidth}
\includegraphics[width=1\linewidth]{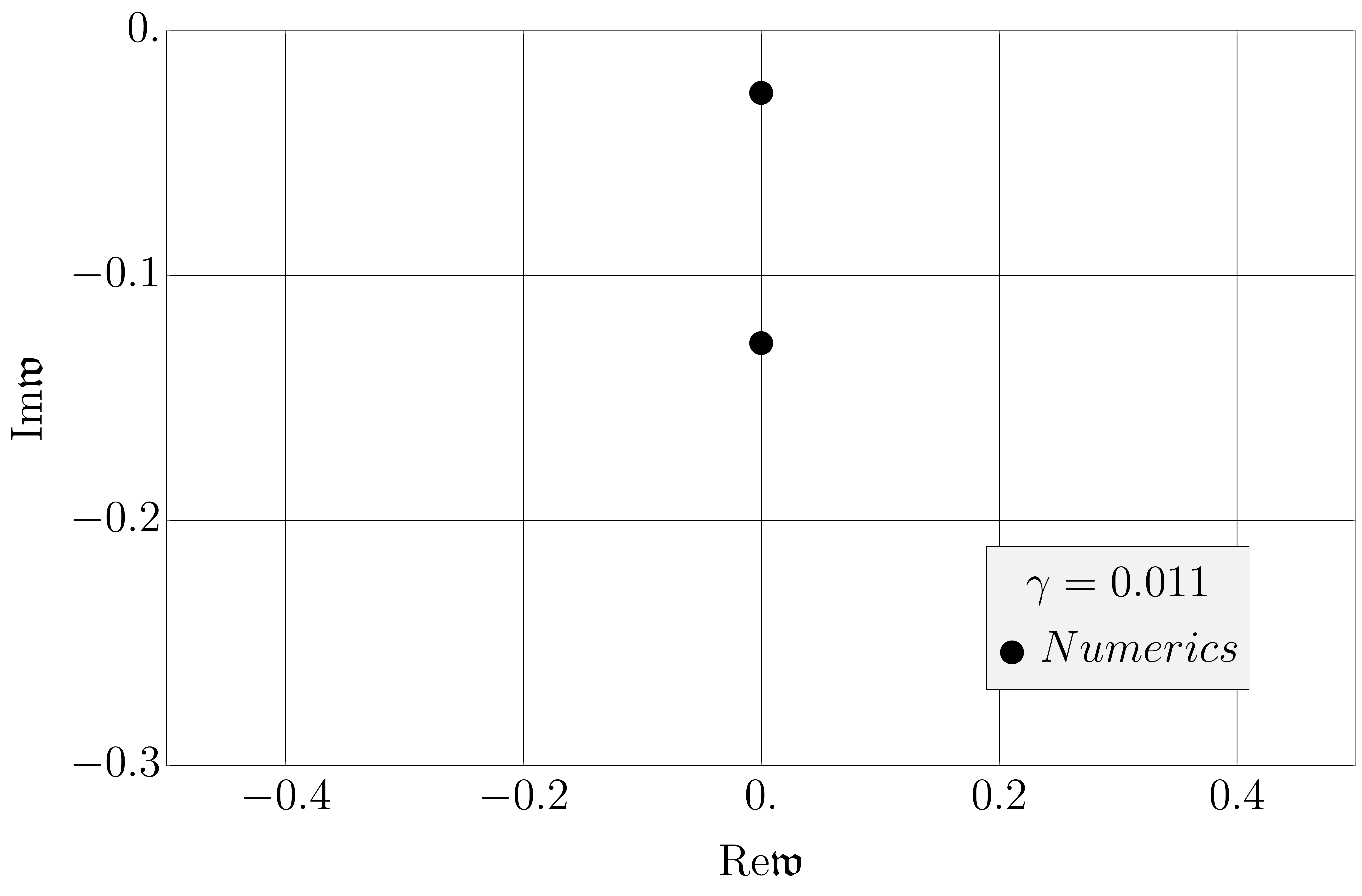}
\end{subfigure}
\qquad
\begin{subfigure}[t]{0.45\linewidth}
\includegraphics[width=1\linewidth]{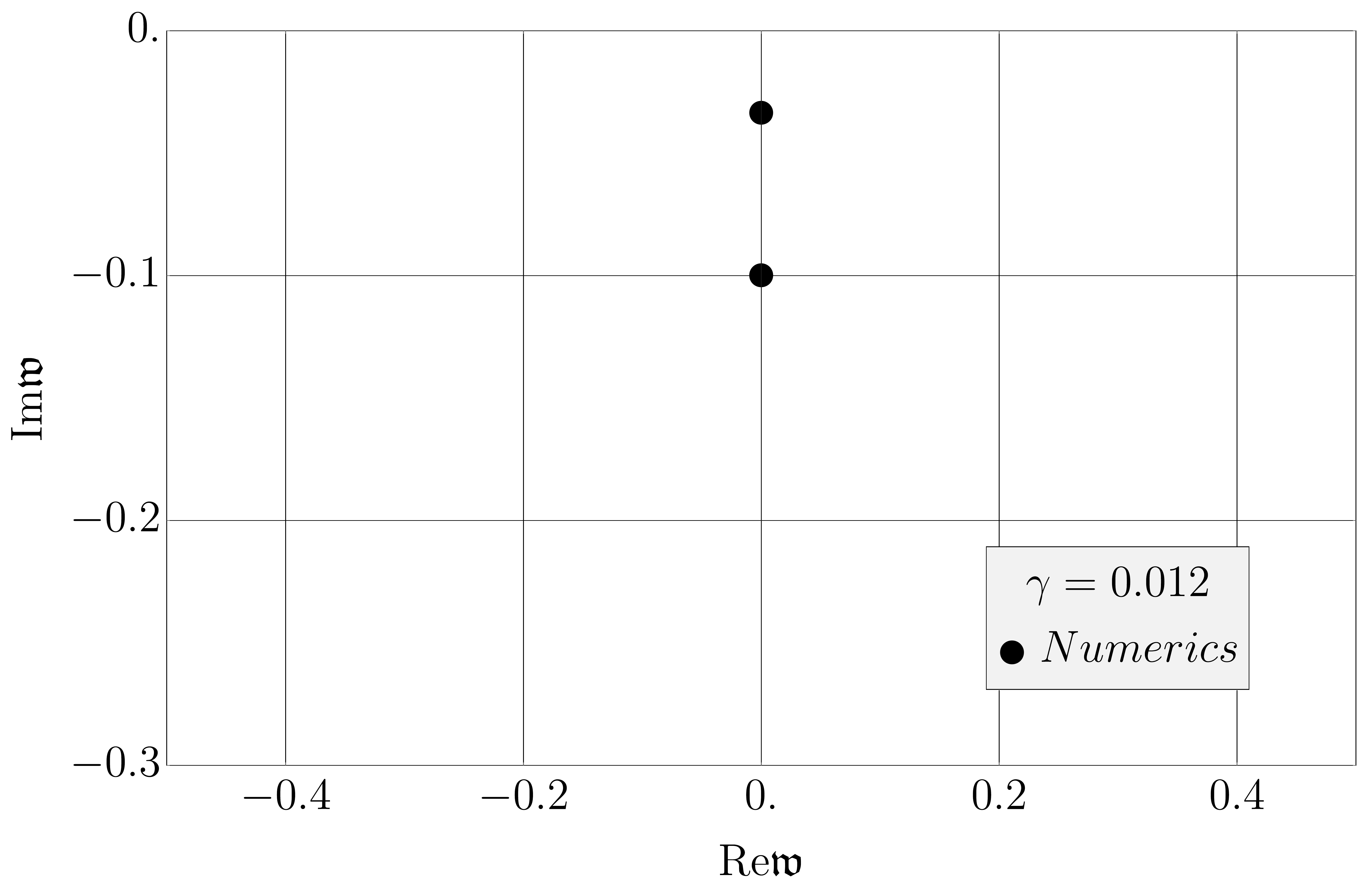}
\end{subfigure}
\\
\begin{subfigure}[b]{0.45\linewidth}
\includegraphics[width=1\linewidth]{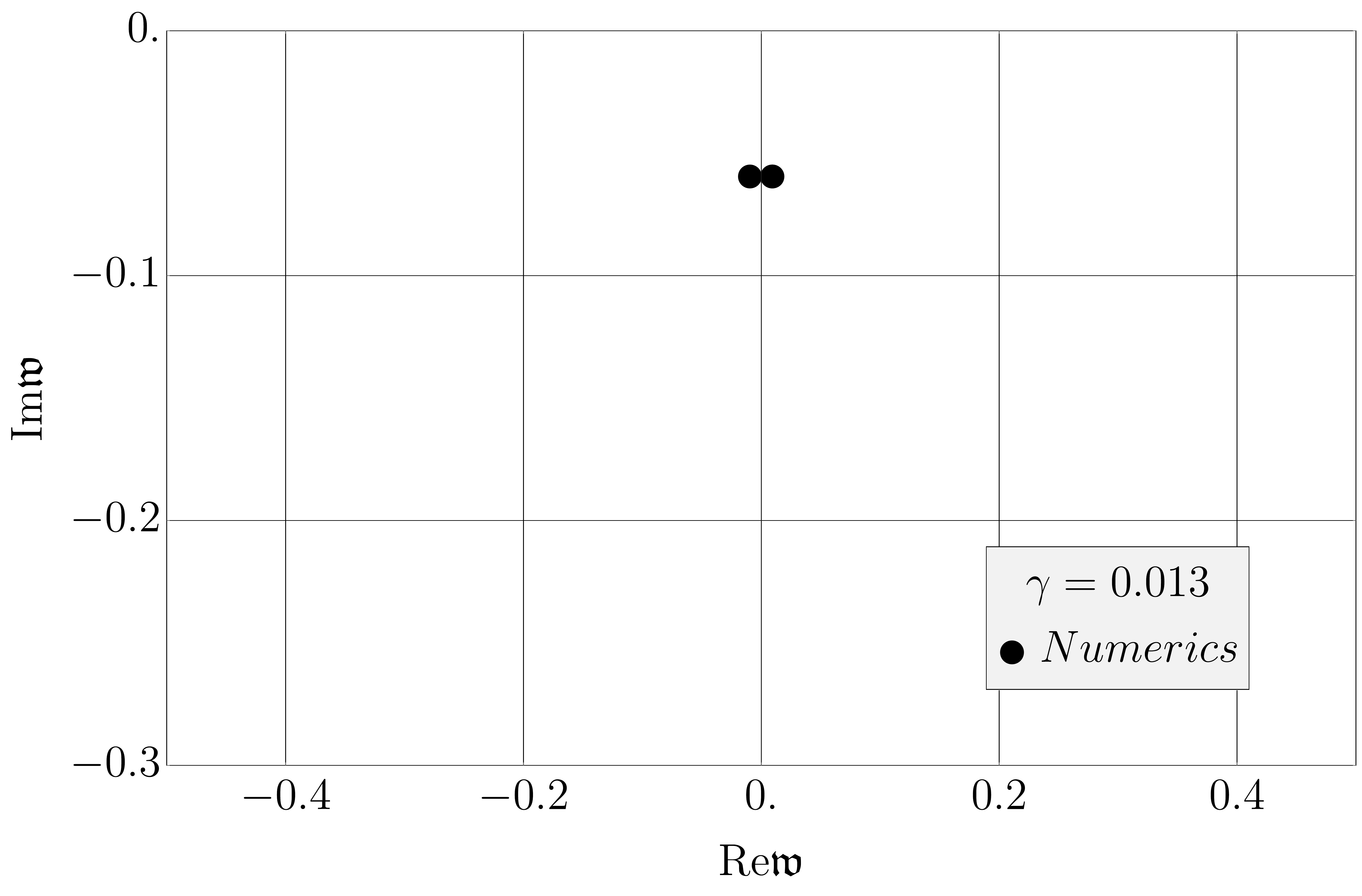}
\end{subfigure}
\qquad
\begin{subfigure}[b]{0.45\linewidth}
\includegraphics[width=1\linewidth]{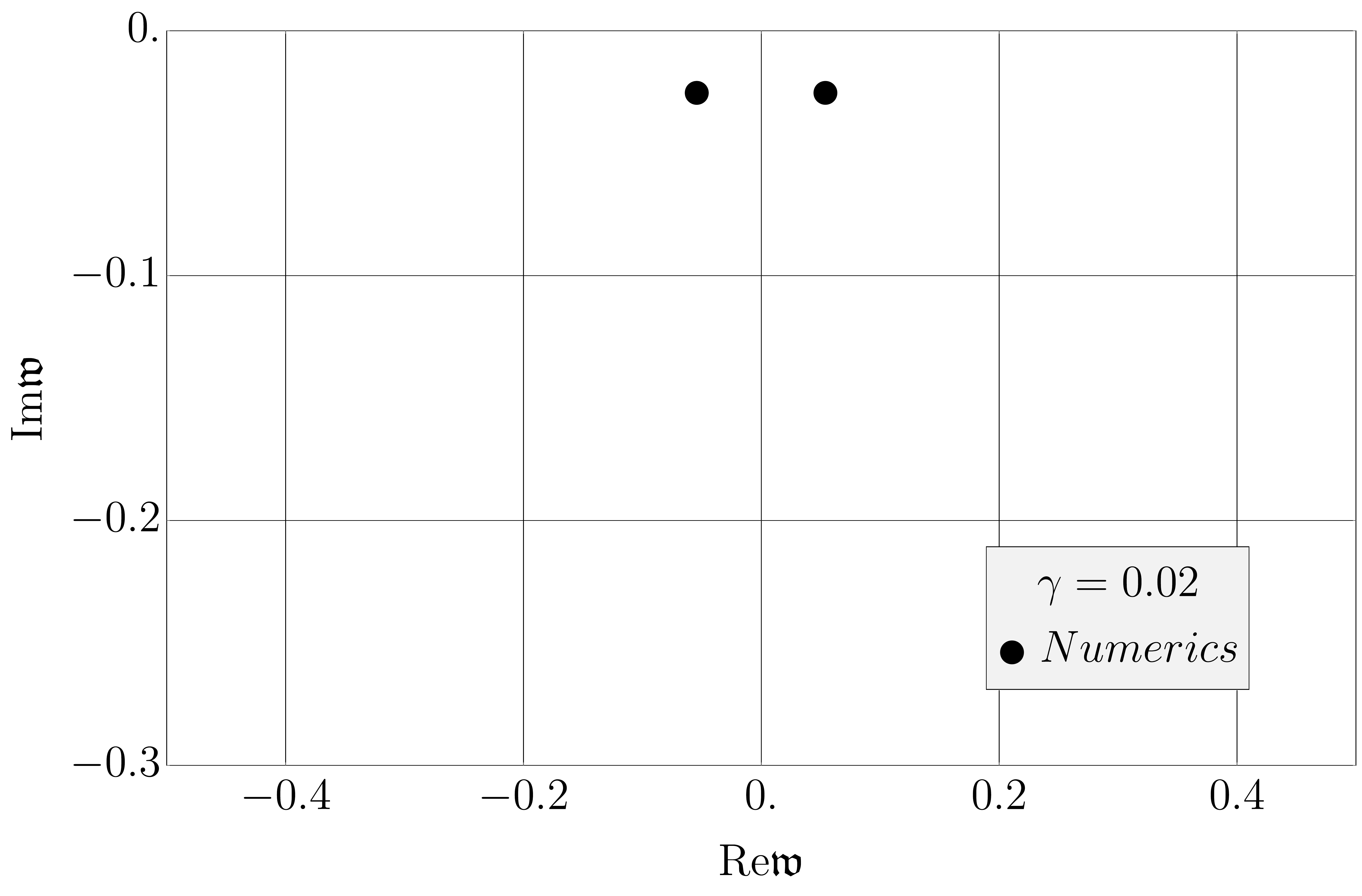}
\end{subfigure}
\caption{The closest to the origin poles (shown by black dots) of the energy-momentum retarded two-point function of $\mathcal{N}=4$ SYM in the shear channel, for various values of the coupling constant and $\qfr=0.1$. From top left:  $\gamma = \{0.011,\, 0.012, \,0.013,\, 0.020\} $ corresponding to values of the 't Hooft coupling $\lambda \approx \{5.7,\, 5.4, \, 5.1,\, 3.8\}$. The hydrodynamic pole moving down the imaginary axis and the new gapped pole moving up the axis merge and move off the imaginary axis. All other poles are outside the range of this plot.}
\label{fig:N=4+gamma-Shear-zoom}
\end{figure}

\begin{figure}[ht]
\centering
\includegraphics[width=0.7\linewidth]{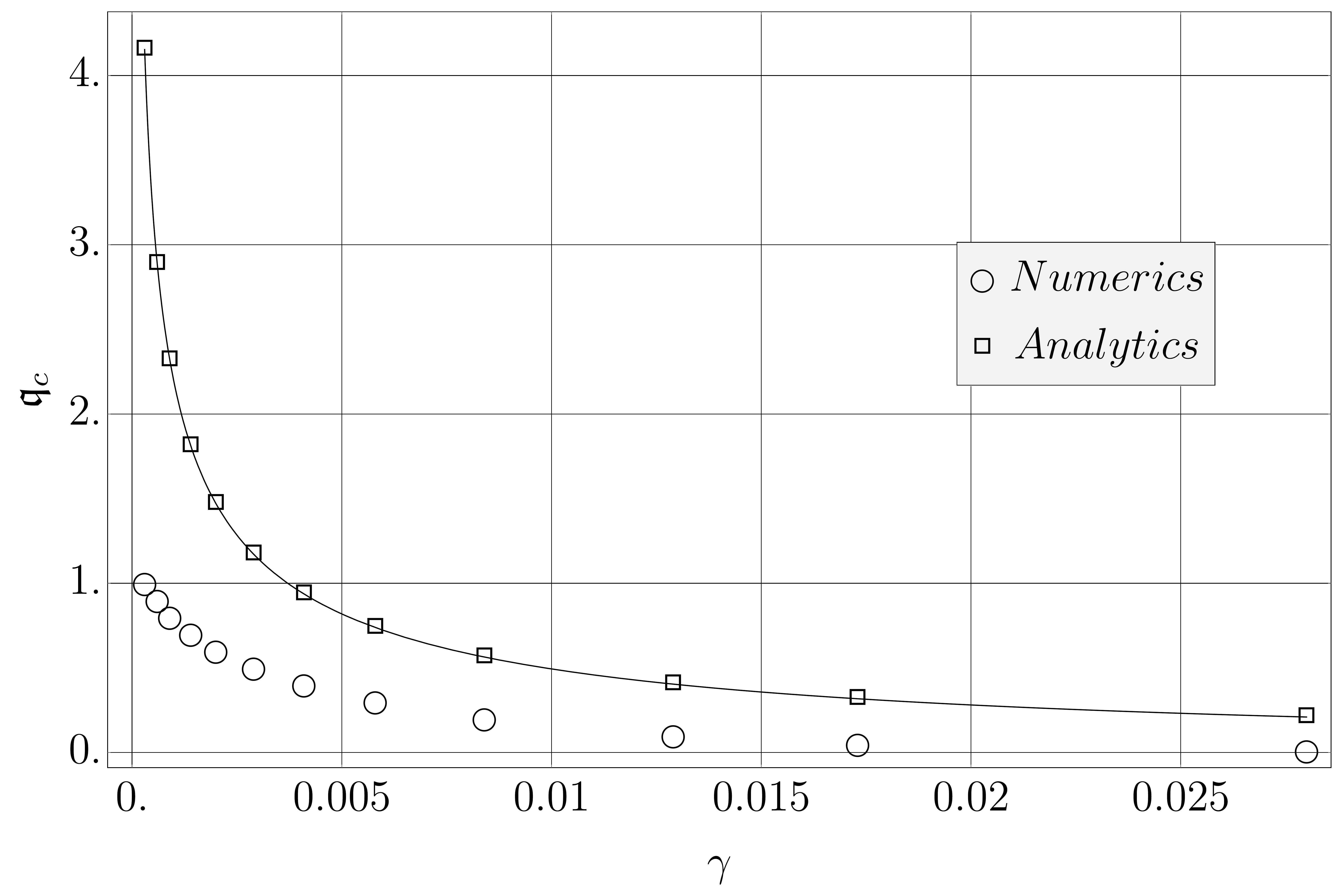}
\caption{Critical value of the spatial momentum $\qfr_c$, limiting the hydrodynamic regime, as a function of higher derivative coupling $\gamma$ in the shear channel of $\mathcal{N}=4$ SYM. Hydrodynamics has a wider range of applicability in $\qfr$ at smaller $\gamma$ (larger 't Hooft coupling).}
\label{fig:N=4+gamma-Shear-critical}
\end{figure}

For certain values of $\gamma \ll 1$, the leading new pole ascending the imaginary axis approaches the hydrodynamic pole. The two poles collide on the imaginary axis at some critical value of $\gamma$ at fixed $\qfr$ (or equivalently, at some $\qfr = \qfr_c (\gamma)$ at fixed $\gamma$) and then for larger $\gamma$ they symmetrically move off the imaginary axis, both having acquired non-zero real parts (see Fig.~\ref{fig:N=4+gamma-Shear-zoom}). At this point, the hydrodynamic pole (\ref{eq:shear_disp}) ceases to exists and for $\qfr > \qfr_c$ the hydrodynamic description appears to be invalid. We interpret this as the breakdown of hydrodynamics at sufficiently large, coupling-dependent value of the wave-vector. The function $\qfr_c(\gamma)$ is shown in Fig.~\ref{fig:N=4+gamma-Shear-critical}. It is monotonically decreasing with $\gamma$ suggesting that hydrodynamics has a wider applicability range at larger 't Hooft coupling as far as the spatial momentum dependence is concerned.

The phenomenon just described can be approximated analytically in the region of small $\wfr$ and $\qfr$ (although this approximation is not very precise quantitatively, it captures the behavior of the poles correctly). Indeed, solving the equation \eqref{ShearEqN4} perturbatively in $\wfr \ll 1$ and $\qfr \ll 1$ (still with $\gamma \ll 1$) and imposing the Dirichlet condition $Z_2(u=0,\wfr,\qfr)=0$, we find a quadratic equation
\begin{align}
2 \wfr +  i \wfr^2 \log{2}  + i \qfr^2 + i 120  \gamma \qfr^2  - i 373  \gamma  \wfr^2  = 0\,.
\label{eq:shear-quad-eq}
\end{align}
This equation has two roots parametrized by $\gamma$ and $\qfr$,
\begin{align}
&\wfr_1 = \frac{- i +i \sqrt{-44760 \gamma ^2 \qfr^2-373 \gamma  \qfr^2+120 \gamma \qfr^2 \ln 2+\qfr^2 \ln 2+1}}{373 \gamma -\ln 2}, \label{eq:ShearNeww1}\\
&\wfr_2 = \frac{-i -i \sqrt{-44760 \gamma ^2 \qfr^2-373 \gamma  \qfr^2+120 \gamma \qfr^2 \ln 2+\qfr^2 \ln 2+1}}{373 \gamma -\ln 2}.\label{eq:ShearNeww2}
\end{align}
At fixed $\qfr$ and sufficiently small $\gamma$, the roots are purely imaginary, moving closer to each other with increasing $\gamma$. Finally, the two roots merge and then acquire non-zero real parts for larger $\gamma$.
The physical meaning of the solutions \eqref{eq:ShearNeww1} and \eqref{eq:ShearNeww2} becomes transparent from their small $\qfr$ expansions:
\begin{align}
&\wfr_1 = - \frac{1}{2} i \left( 1 + 120\gamma\right)\qfr^2 + \ldots \, , \label{eq:ShearN4dispQ2}\\
&\wfr_2 = \wfr_{\mathfrak{g}} + \frac{1}{2} i \left( 1 + 120\gamma\right)\qfr^2 + \ldots  \, , \label{eq:ShearN4disp2Q2}
\end{align}
where $\wfr_{\mathfrak{g}}$ is given by Eq.~\eqref{eq:ScalarN4newpole}. Here, the mode  \eqref{eq:ShearN4dispQ2} is the standard hydrodynamic momentum diffusion pole predicted by Eq.~(\ref{eq:shear_disp}), whereas the mode \eqref{eq:ShearN4disp2Q2} approximates the new gapped pole moving up the imaginary axis. Note that the gap $\wfr_{\mathfrak{g}}$ in the mode \eqref{eq:ShearN4disp2Q2} is the same as in the scalar channel. Using Eqs.~(\ref{eq:ShearN4dispQ2}) and (\ref{eq:ShearN4disp2Q2}), we can find an approximate analytic expression for the function $\qfr_c(\gamma)$ plotted in Fig.~\ref{fig:N=4+gamma-Shear-critical}:
\begin{align}
\qfr_c = \sqrt{\frac{2}{373 \gamma \left(1+120 \gamma \right)} } \sim 0.04 \, \lambda^{3/2}\,.
\label{eq:q-crit}
\end{align}
As is evident from Fig.~\ref{fig:N=4+gamma-Shear-critical}, the analytic approximation becomes more precise with larger $\gamma$.
\subsubsection{Sound channel}
\begin{figure}[ht]
\centering
\begin{subfigure}[t]{0.45\linewidth}
\includegraphics[width=1\linewidth]{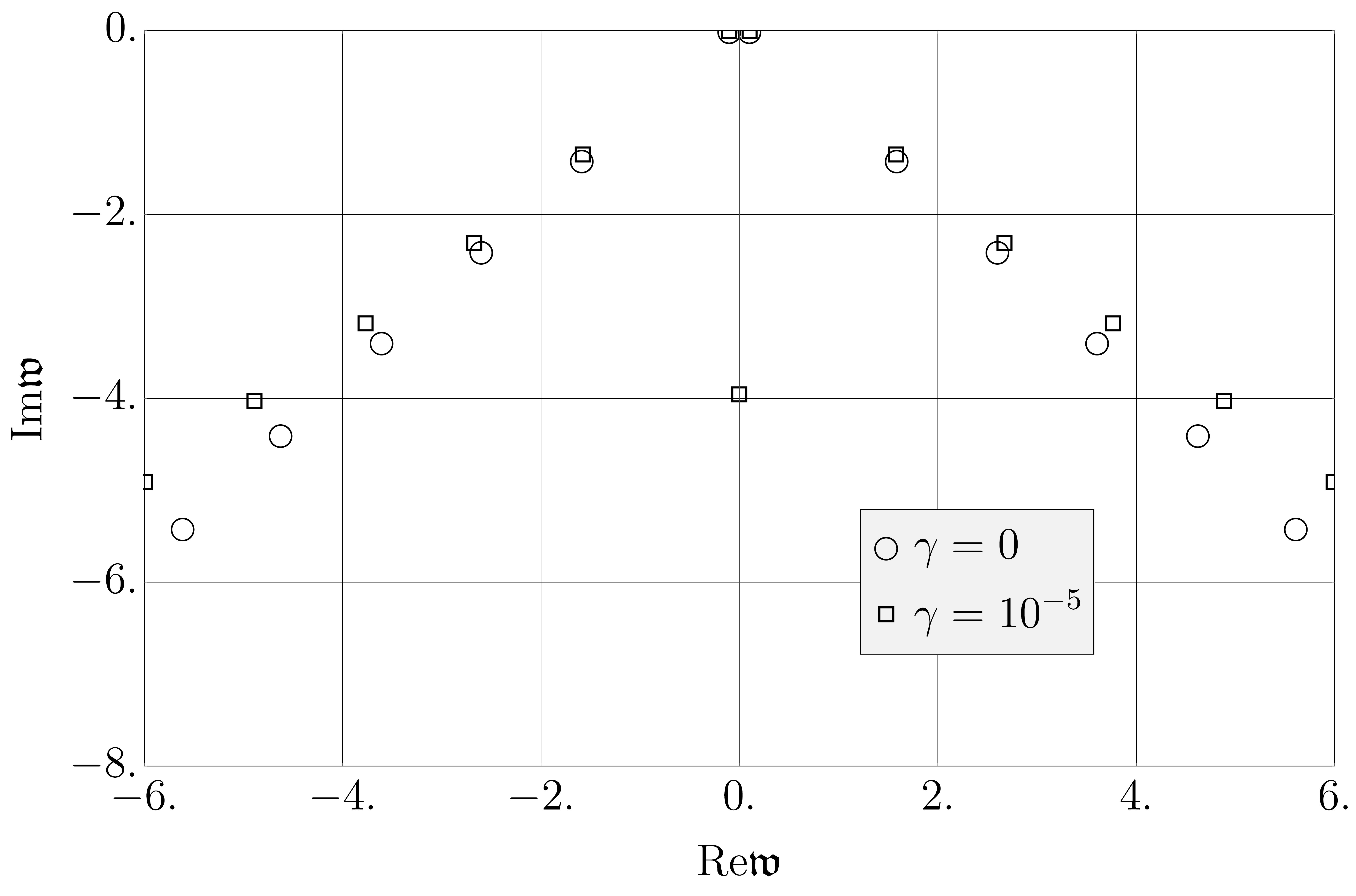}
\end{subfigure}
\qquad
\begin{subfigure}[t]{0.45\linewidth}
\includegraphics[width=1\linewidth]{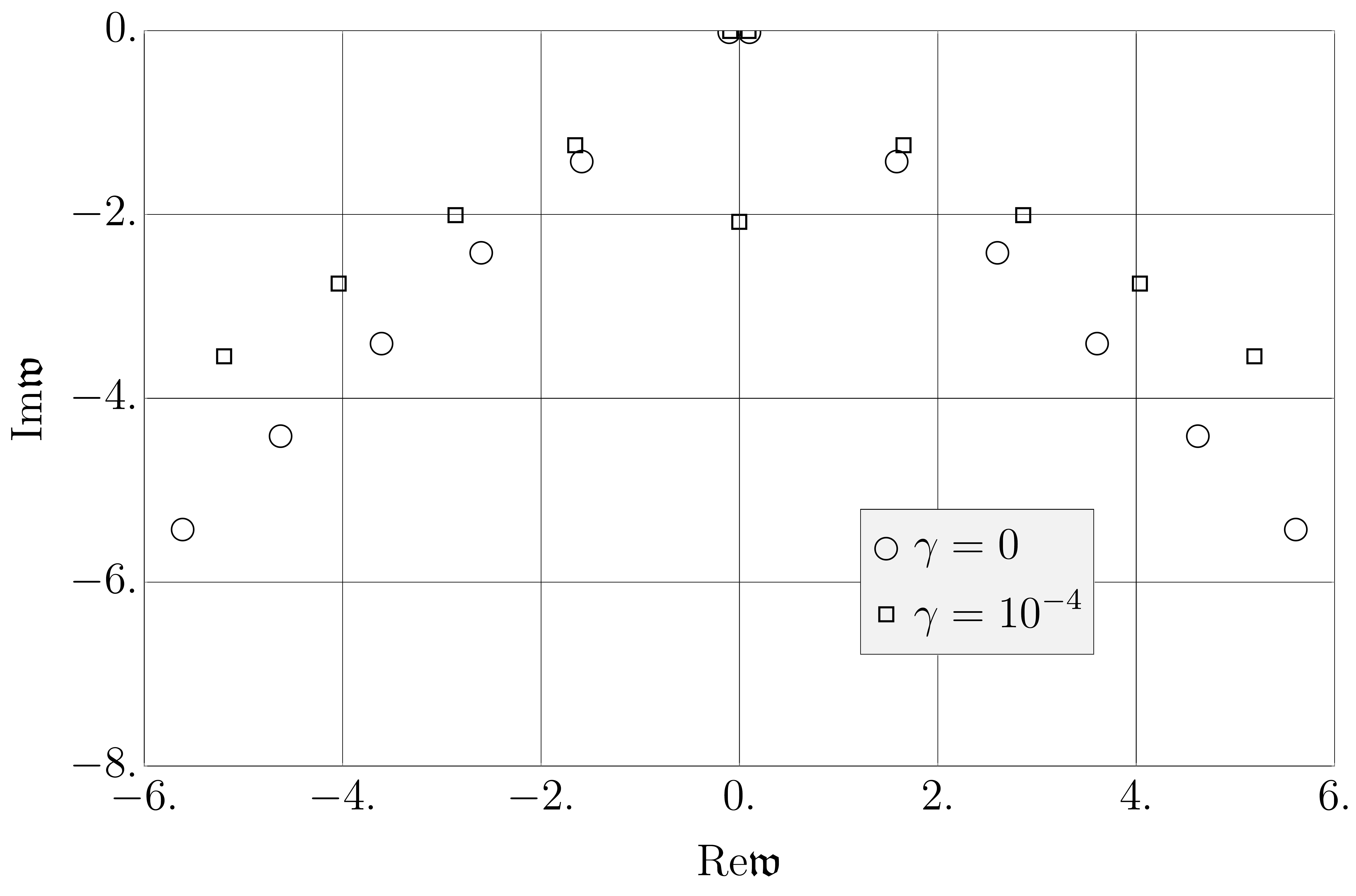}
\end{subfigure}
\\
\begin{subfigure}[b]{0.45\linewidth}
\includegraphics[width=1\linewidth]{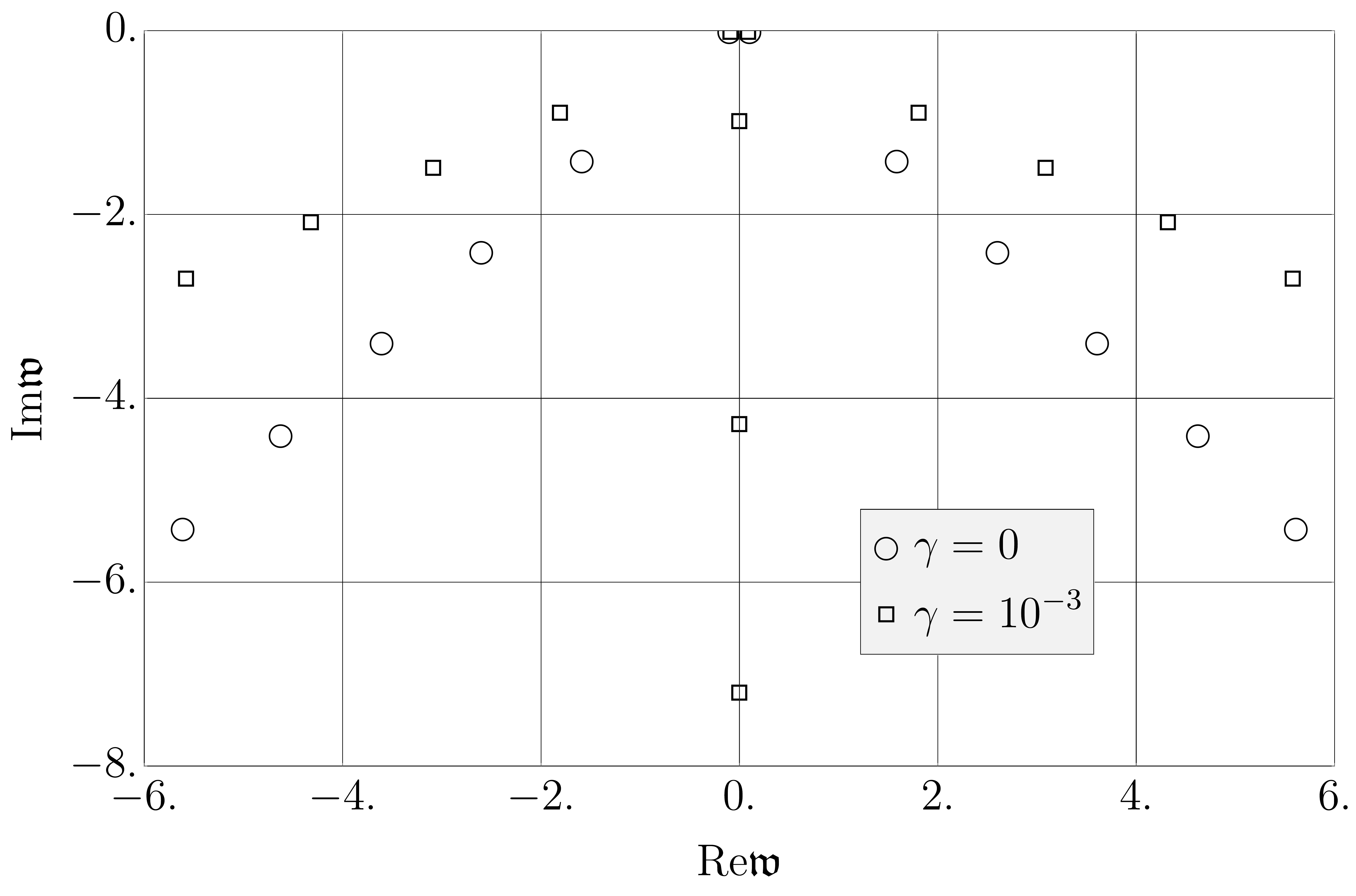}
\end{subfigure}
\qquad
\begin{subfigure}[b]{0.45\linewidth}
\includegraphics[width=1\linewidth]{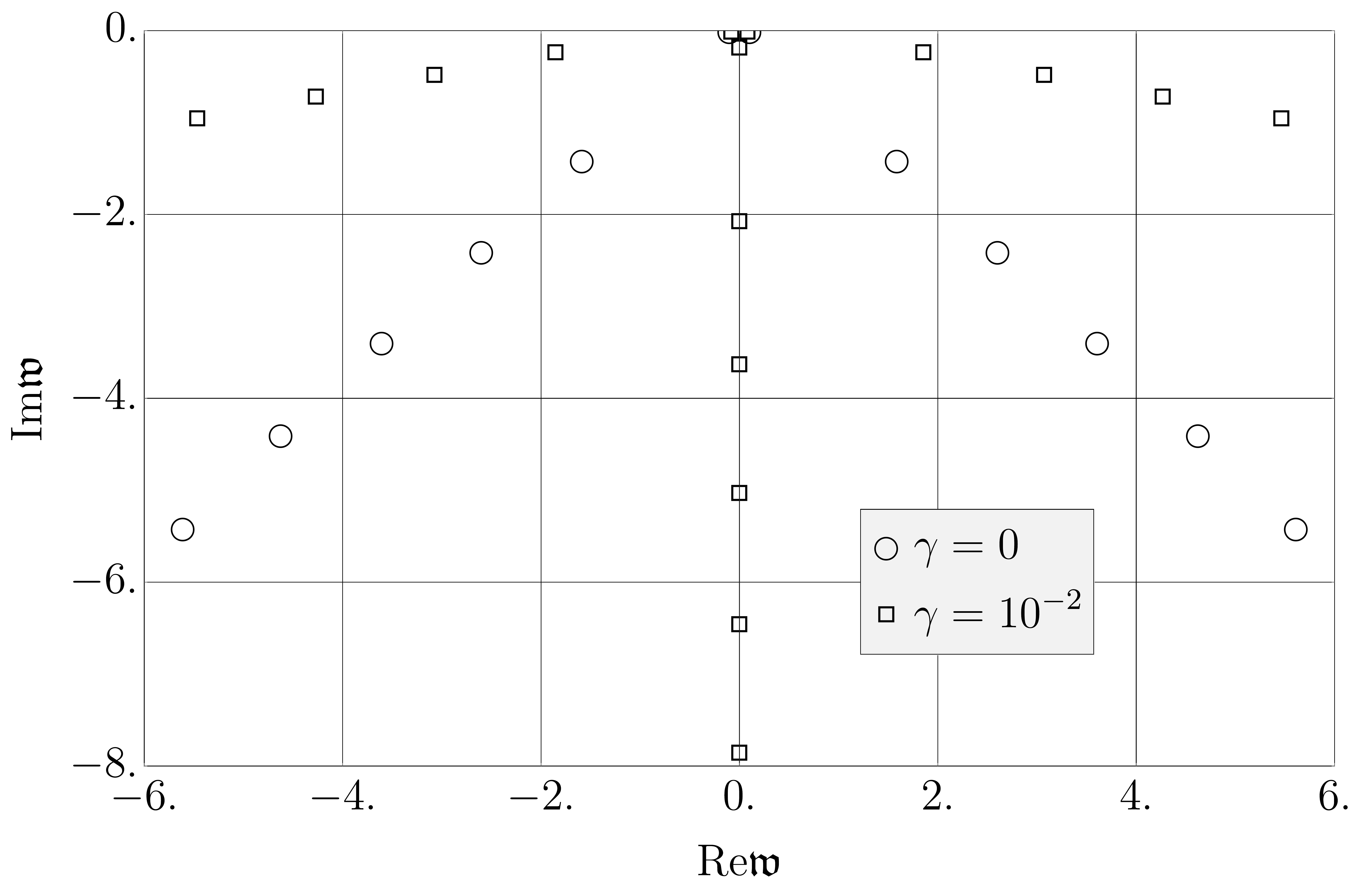}
\end{subfigure}
\caption{Poles (shown by squares) of the energy-momentum retarded two-point function of $\mathcal{N}=4$ SYM in the sound channel, for various values of the coupling constant and $\qfr=0.1$. From top left: $\gamma = \{10^{-5},\, 10^{-4}, \,10^{-3},\, 10^{-2}\} $ corresponding to values of the 't Hooft coupling $\lambda \approx \{609,\, 131, \, 28,\, 6\} $. Poles at $\gamma = 0$ ($\lambda\rightarrow \infty$) are shown by circles.}
\label{fig:N=4+gamma-Sound-channel}
\end{figure}

The quasinormal spectrum in the sound channel is found by solving Eq.~(\ref{eq:SoundEqN4}) and imposing the Dirichlet condition $Z_3 (u=0,\wfr,\qfr) = 0$.
The distribution of poles in the complex frequency plane at various values of the coupling is shown in Fig.~\ref{fig:N=4+gamma-Sound-channel}. The movement of the poles with varying coupling is qualitatively similar to the one observed in the scalar and shear channels. The two gapless sound poles symmetric with respect to the imaginary axis have the dispersion relation predicted by hydrodynamics \cite{Policastro:2002se,Baier:2007ix,Grozdanov:2015kqa}
\begin{align}
\omega =  \pm c_s \, q - i \Gamma\, q^2    \mp \frac{\Gamma}{2 c_s} \left( \Gamma - 2 c_s^2 \tau_\Pi \right)\, q^3 - i \left[ \frac{8}{9}\frac{\eta^2 \tau_\Pi}{(\varepsilon+P)^2} - \frac{1}{3} \frac{\theta_1+\theta_2}{\varepsilon +P}\right]\, q^4  + \cdots \,,
\label{eq:sound_disp}
\end{align}
where $c_s = 1/\sqrt{3}$ for conformal fluids in $d=3+1$ dimensions, $\Gamma = 2 \eta/3(\varepsilon +P)$ and $\varepsilon + P = s T$ at zero chemical potential. For $\CN=4$ SYM theory, the coefficients $\eta/s$, $\tau_\Pi$ and $\theta_1$ are given in Eq.~(\ref{eq:coeffi-n=4}) and 
\begin{align}
\theta_2 = \frac{N_c^2 T}{384 \pi} \left( 22 - \frac{\pi^2}{12} - 18 \ln{2} +\ln^2 2 \right)+ O(\gamma)\,.
\end{align}
The full $\gamma$-dependence of the quartic term in Eq.~(\ref{eq:sound_disp}) is currently unknown.

\begin{figure}[ht]
\centering
\begin{subfigure}[t]{0.45\linewidth}
\includegraphics[width=1\linewidth]{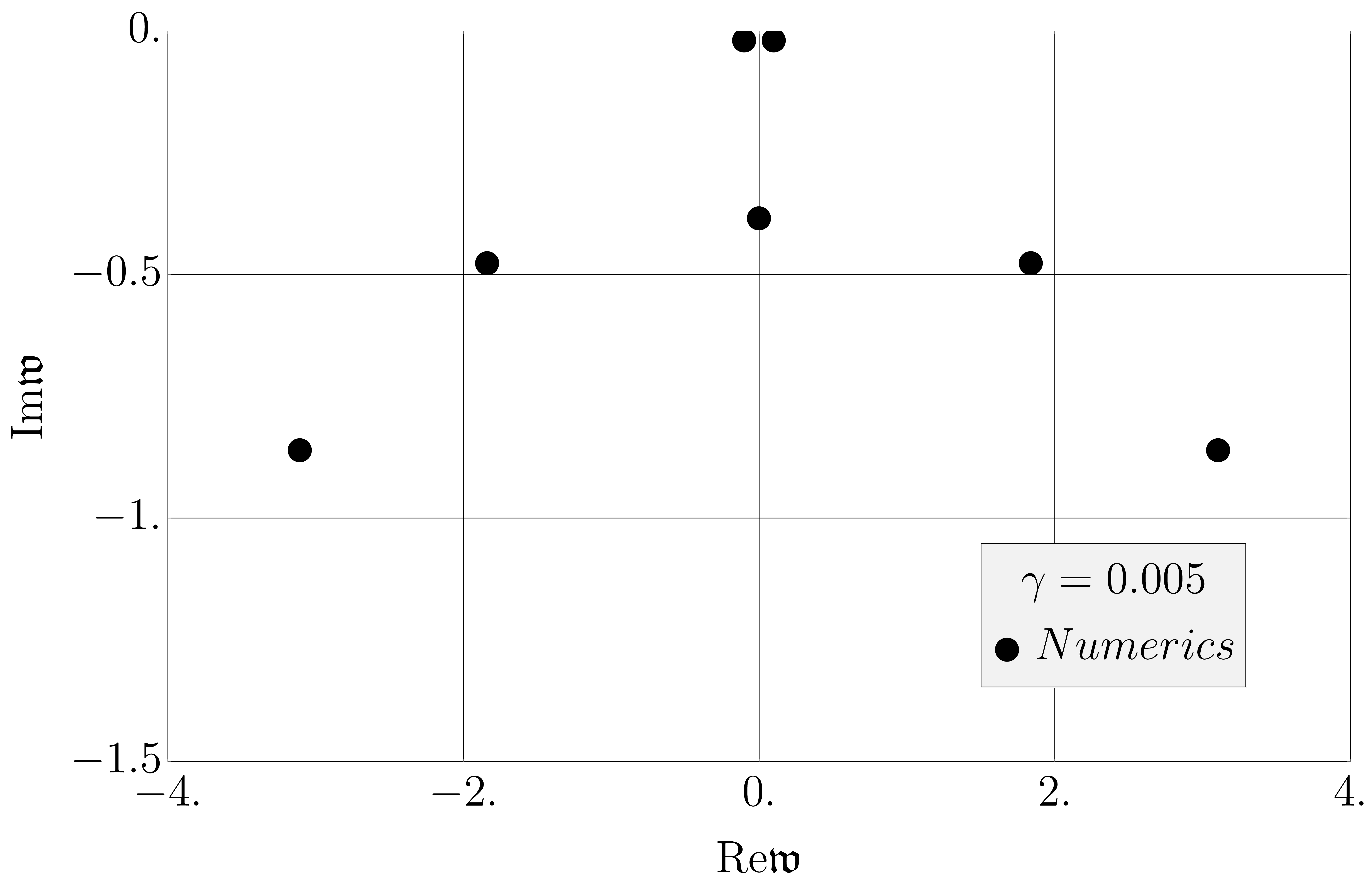}
\end{subfigure}
\qquad
\begin{subfigure}[t]{0.45\linewidth}
\includegraphics[width=1\linewidth]{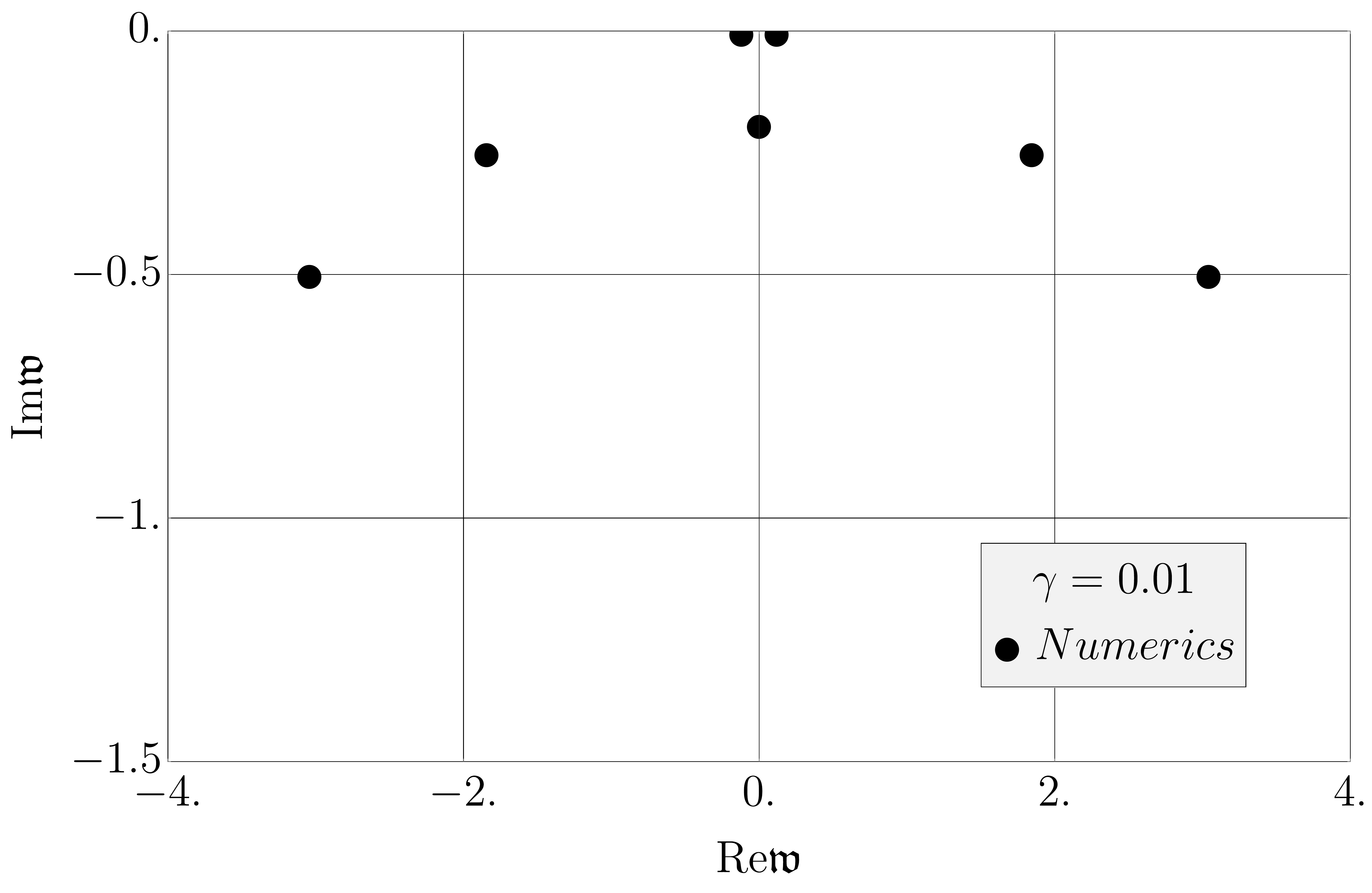}
\end{subfigure}
\\
\begin{subfigure}[b]{0.45\linewidth}
\includegraphics[width=1\linewidth]{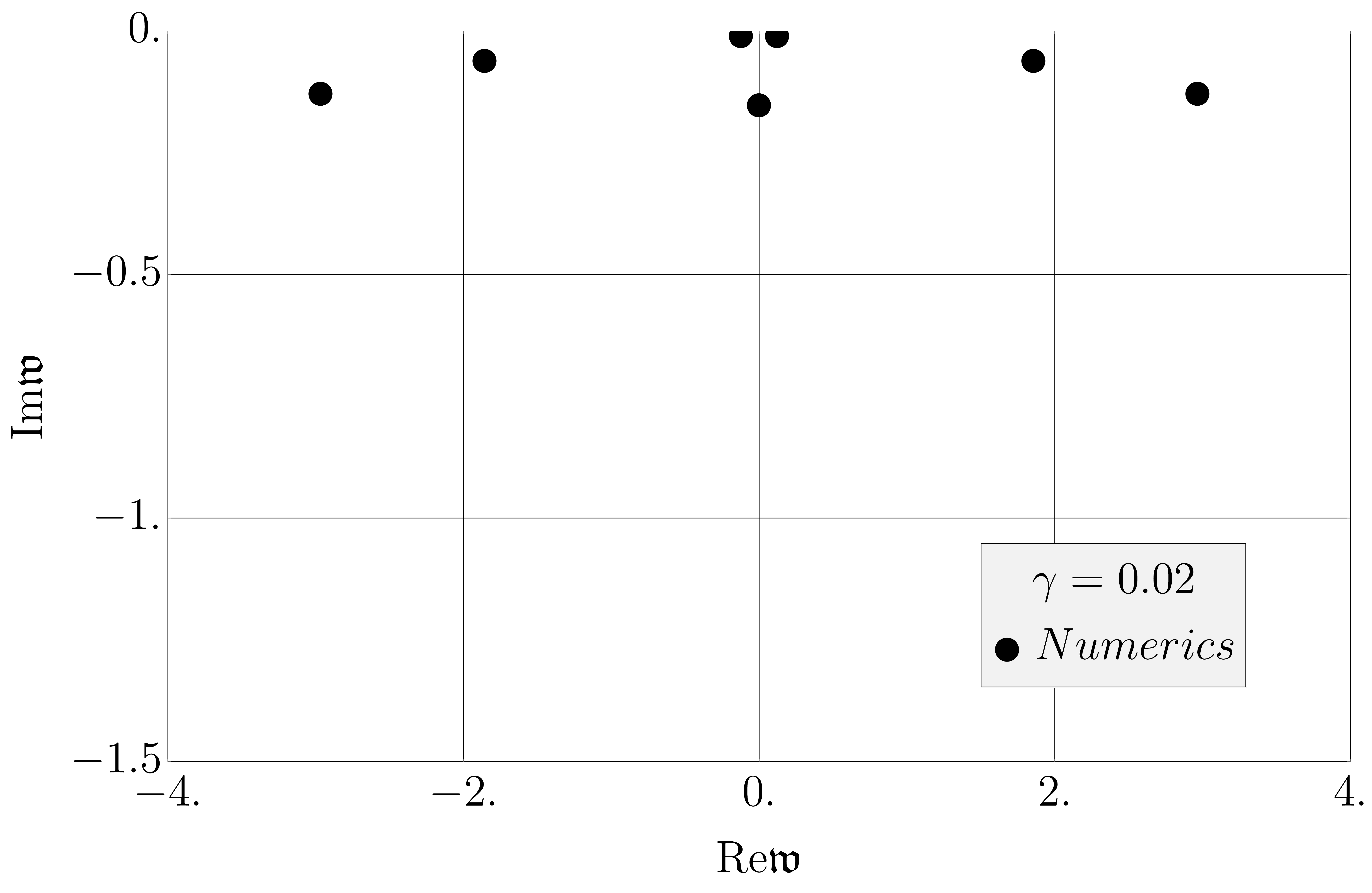}
\end{subfigure}
\qquad
\begin{subfigure}[b]{0.45\linewidth}
\includegraphics[width=1\linewidth]{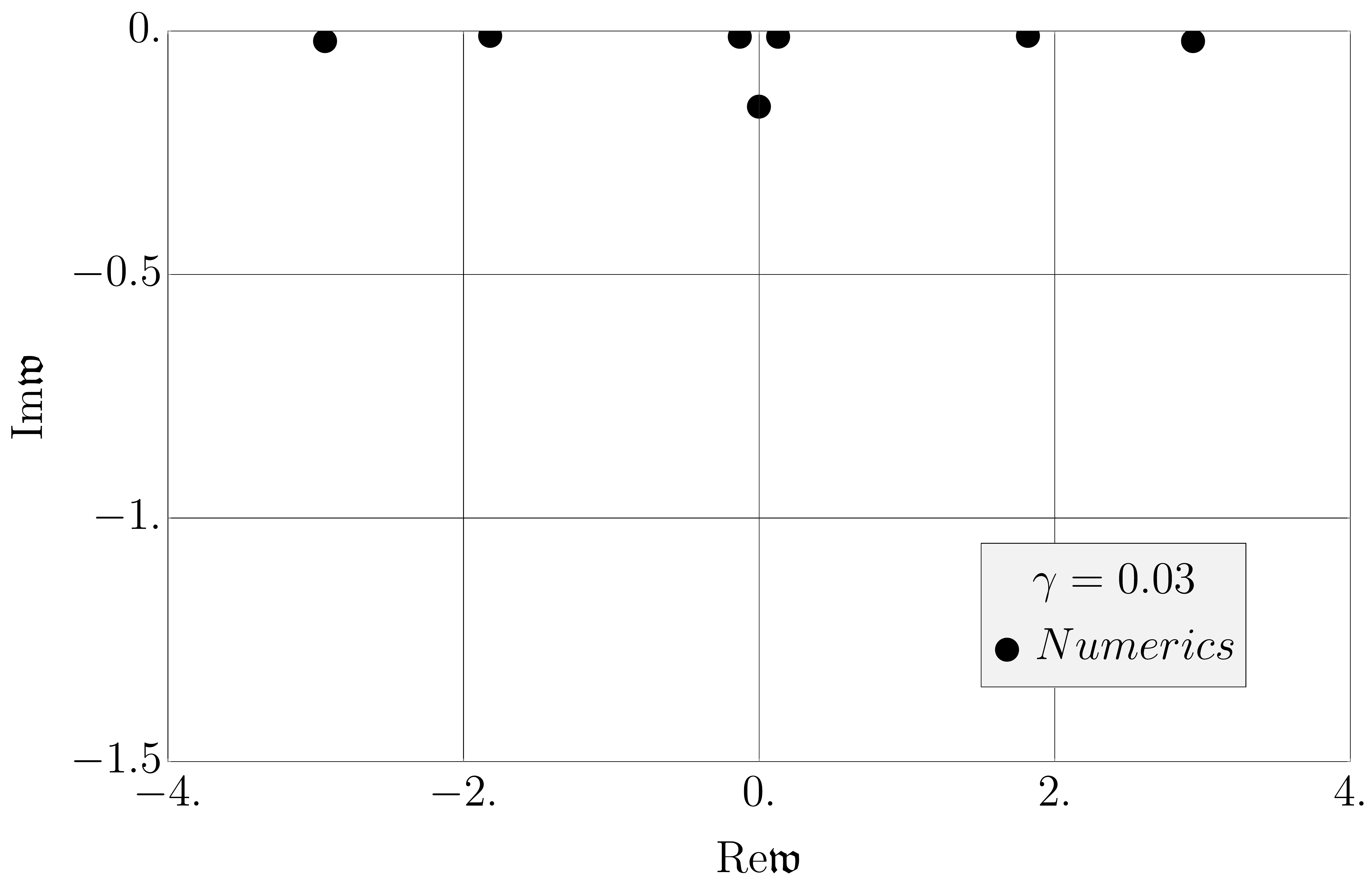}
\end{subfigure}
\caption{Closest to the origin poles (shown by black dots) of the energy-momentum retarded two-point function of $\mathcal{N}=4$ SYM in the sound channel, for various values of the coupling constant and $\qfr=0.1$. From top left:  $\gamma = \{0.005,\, 0.01, \,0.02,\, 0.03\} $ corresponding to values of the 't Hooft coupling $\lambda \approx \{10,\, 6, \, 4,\, 3\}$. All other poles are outside the range of this plot.}
\label{fig:N=4+gamma-Sound-zoom}
\end{figure}

With $\gamma$  increasing, the leading new gapless pole rising along the imaginary axis approaches the region of the sound poles (see Fig.~\ref{fig:N=4+gamma-Sound-zoom}). For $\wfr \ll 1$ and $\qfr \ll 1$, the equation (\ref{eq:SoundEqN4}) can be solved perturbatively and from the Dirichlet condition  $Z_3 (u=0,\wfr,\qfr) = 0$ one finds a quintic equation
\begin{align}
&420 \gamma  \qfr^4-2546 i \gamma  \qfr^4 \wfr-8 i \qfr^4 \wfr+4 \qfr^4+4797 i \gamma  \qfr^2 \wfr^3+12 i \qfr^2 \wfr^3 \nn
&-1260 \gamma  \qfr^2 \wfr^2-18 \qfr^2 \wfr^2-3357 i \gamma  \wfr^5+18 \wfr^4 =0.
\end{align}
Expanding further in $\gamma \ll 1$ and $\qfr \ll 1$, we obtain the following analytic expressions for the three closely located modes of interest:
\begin{align}
&\wfr_{1,2} = \pm \frac{1}{\sqrt{3}} \qfr - \frac{1}{3} i  (1 + 120\gamma) \qfr^2 + \ldots  \,, \label{eq:sound-gam}\\
&\wfr_3 = \wfr_{\mathfrak{g}} + \frac{2}{3} i (1 + 120\gamma) \qfr^2 + \ldots  \label{eq:sound-gam-gap}\,.
\end{align}
Here, the Eq.~(\ref{eq:sound-gam}) is the standard dispersion relation for the two sound modes as in Eq.~(\ref{eq:sound_disp}), and Eq.~\eqref{eq:sound-gam-gap} is the new gapped pole with $\wfr_{\mathfrak{g}}$ given by Eq. \eqref{eq:ScalarN4newpole}. Assuming, perhaps somewhat arbitrarily, that the hydrodynamic description fails when the imaginary part of the new gapped pole becomes equal to the one of the sound mode, from Eqs.~(\ref{eq:sound-gam}) and (\ref{eq:sound-gam-gap}) we find the critical value of the spatial momentum $\qfr_c$ which turns out to be exactly the same as in Eq. \eqref{eq:q-crit}.
\subsubsection{Coupling constant dependence of the shear viscosity - relaxation time ratio}
\label{sec:RatioN4}
\begin{figure}[ht]
\centering
\includegraphics[width=0.47\linewidth]{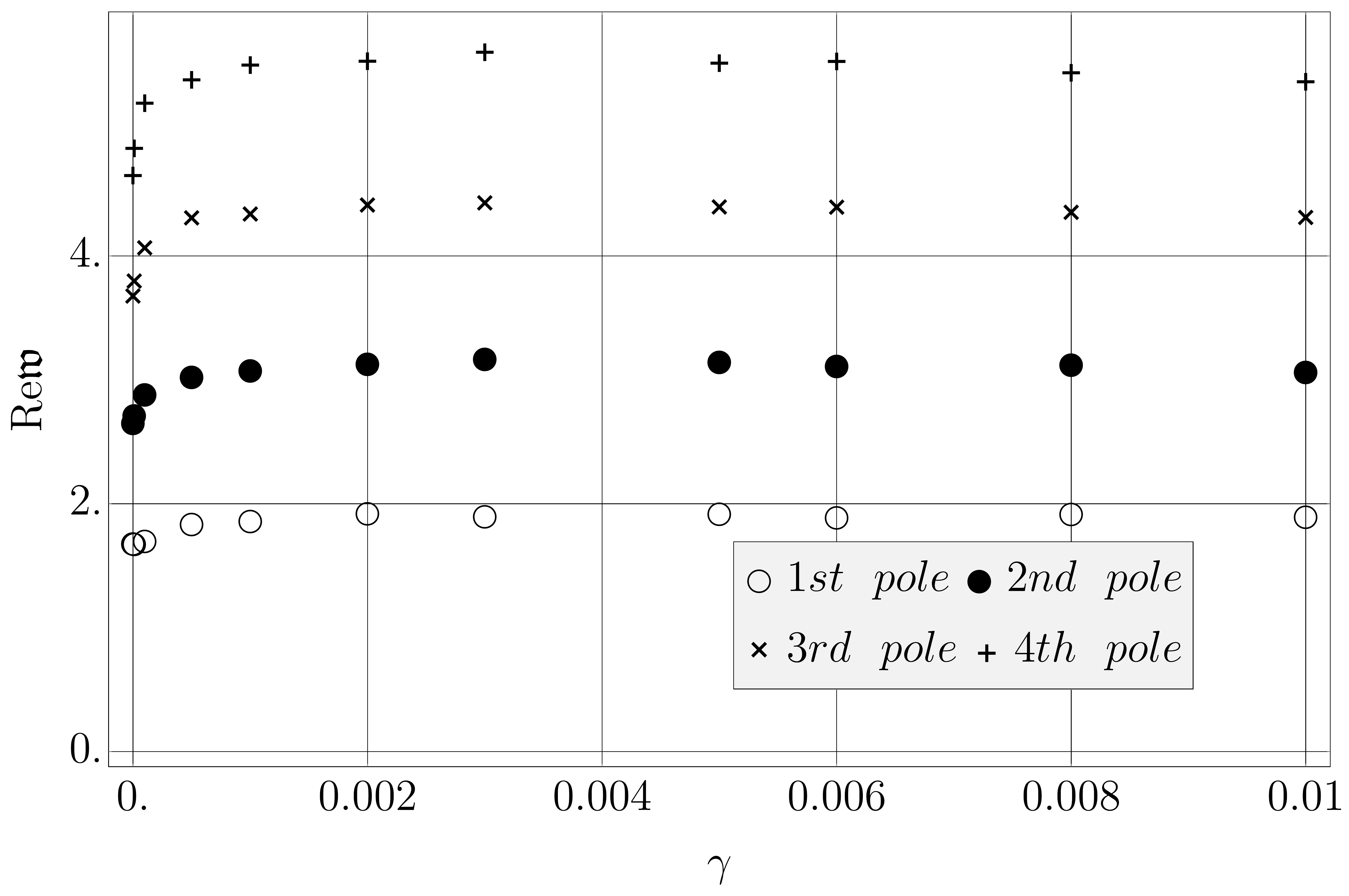}
\includegraphics[width=0.47\linewidth]{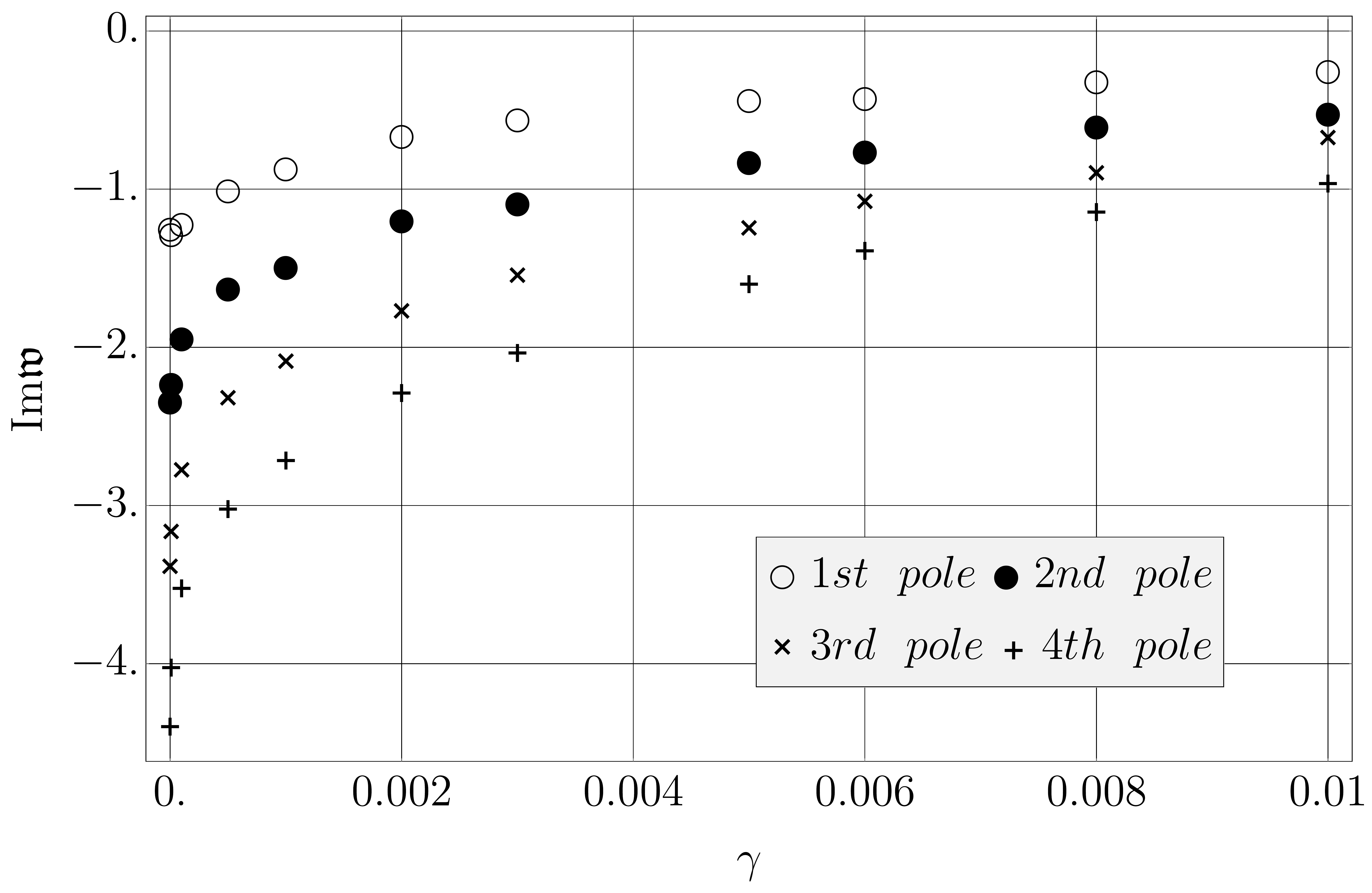}
\caption{$\CN=4$ SYM: Real (left panel) and imaginary (right panel) parts of the lowest four quasinormal frequencies in the scalar channel at $\qfr = 0.1$.}
\label{fig:N=4+gamma-Scalar-Poles-vs-gamma}
\end{figure}

\begin{figure}[ht]
\centering
\includegraphics[width=0.47\linewidth]{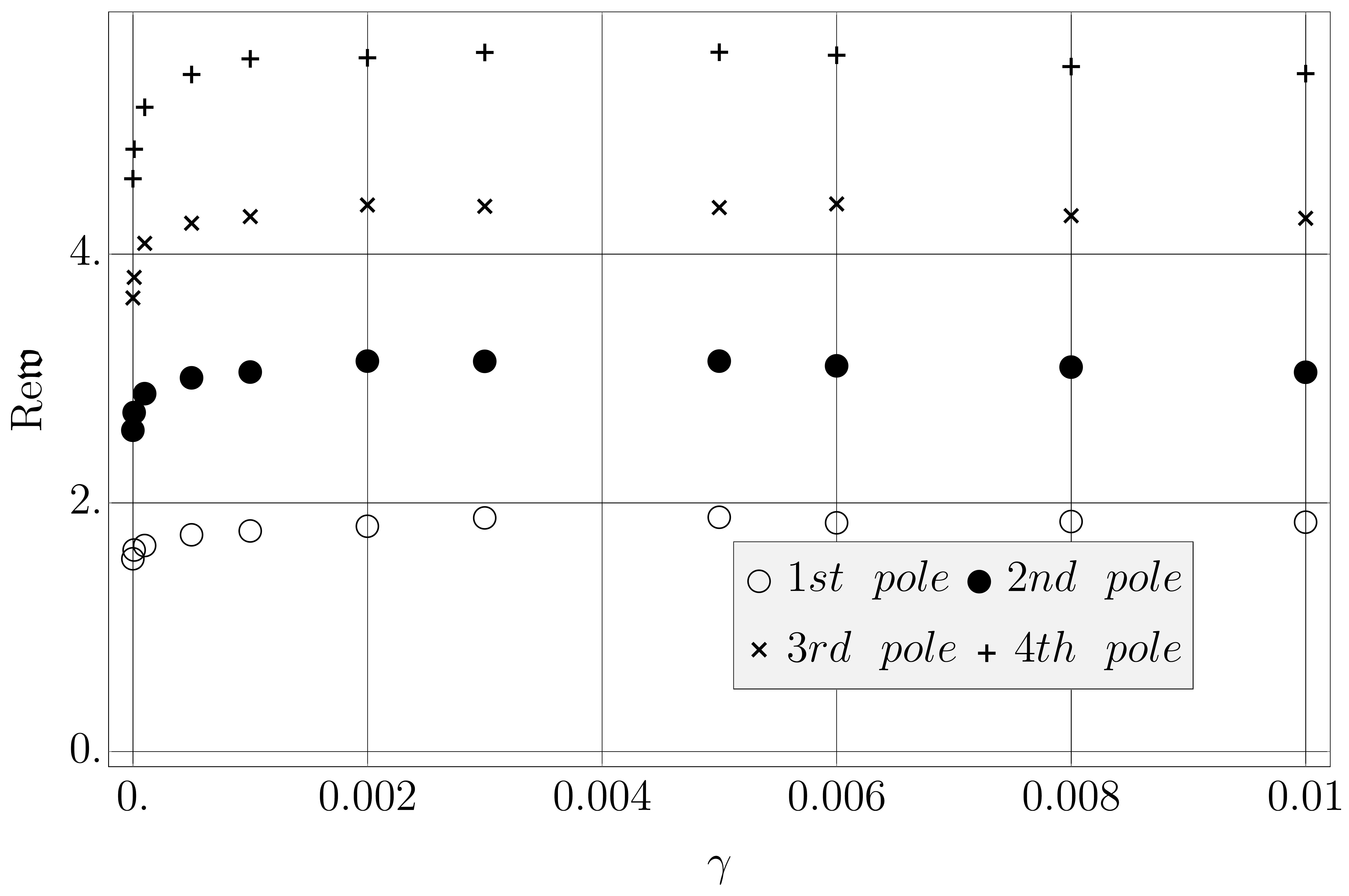}
\includegraphics[width=0.47\linewidth]{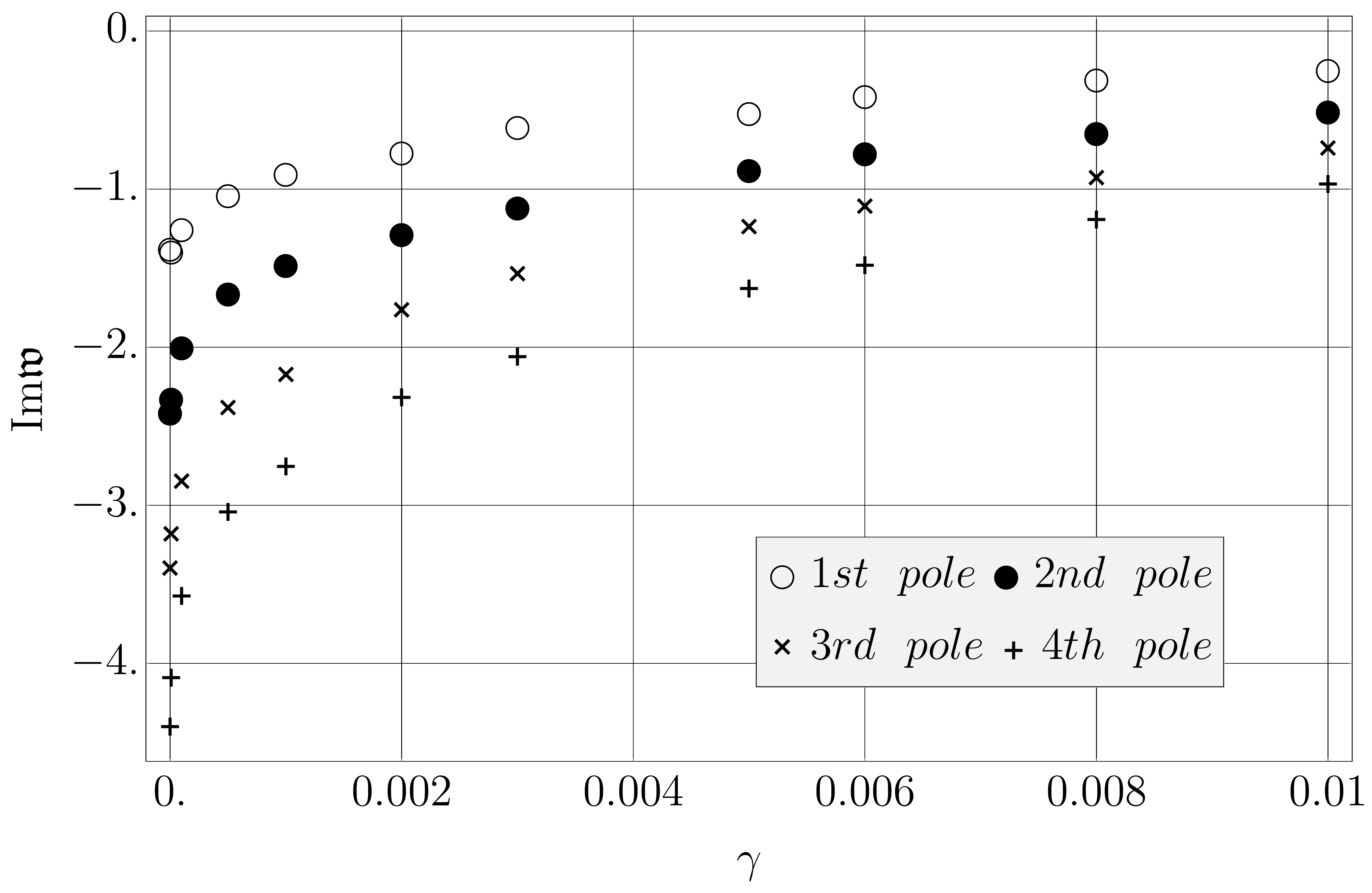}
\caption{$\CN=4$ SYM: Real (left panel) and imaginary (right panel) parts of the lowest four quasinormal frequencies in the shear channel at $\qfr = 0.1$.}
\label{fig:N=4+gamma-Shear-Poles-vs-gamma}
\end{figure}

\begin{figure}[ht]
\centering
\includegraphics[width=0.47\linewidth]{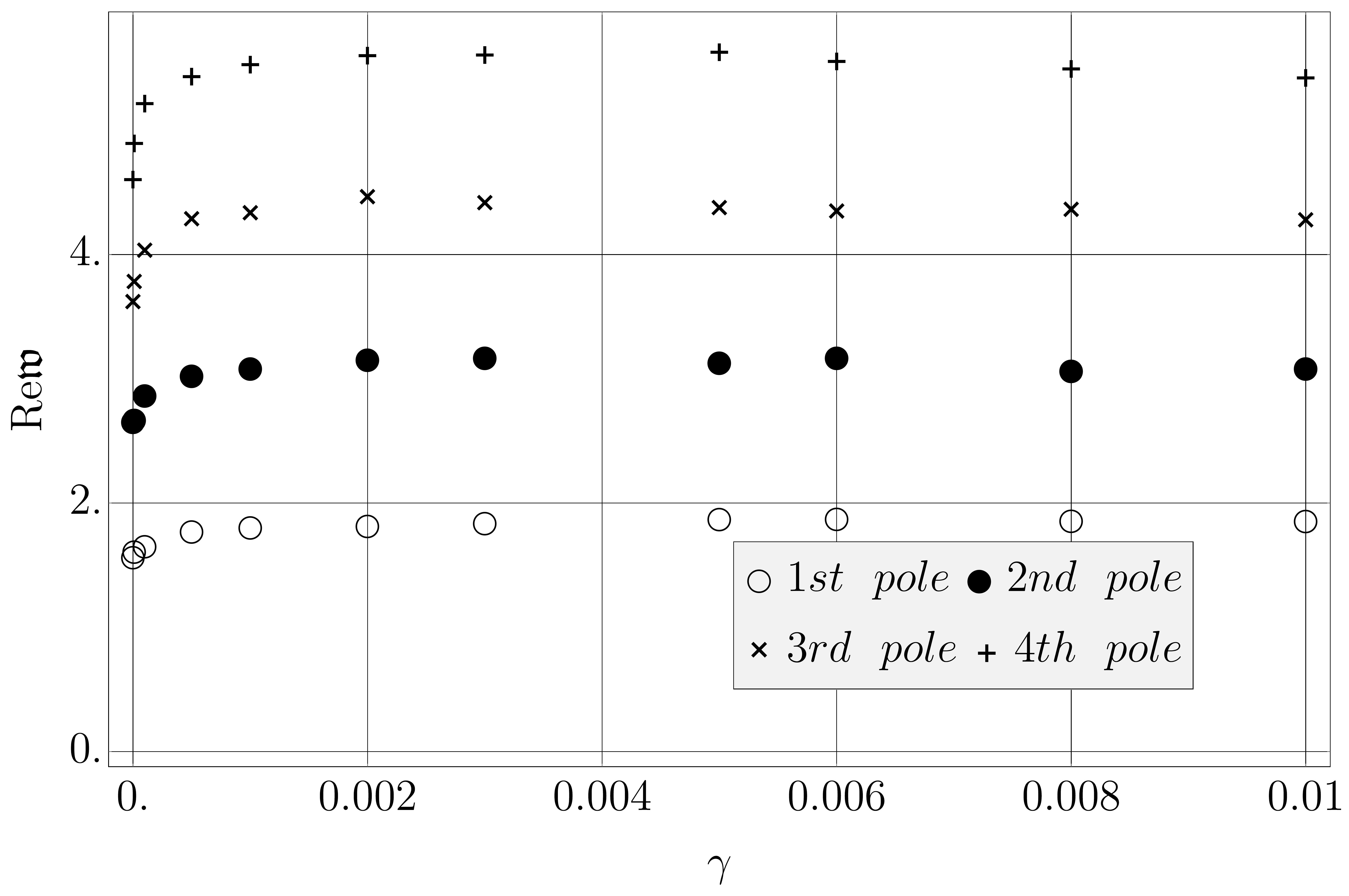}
\includegraphics[width=0.47\linewidth]{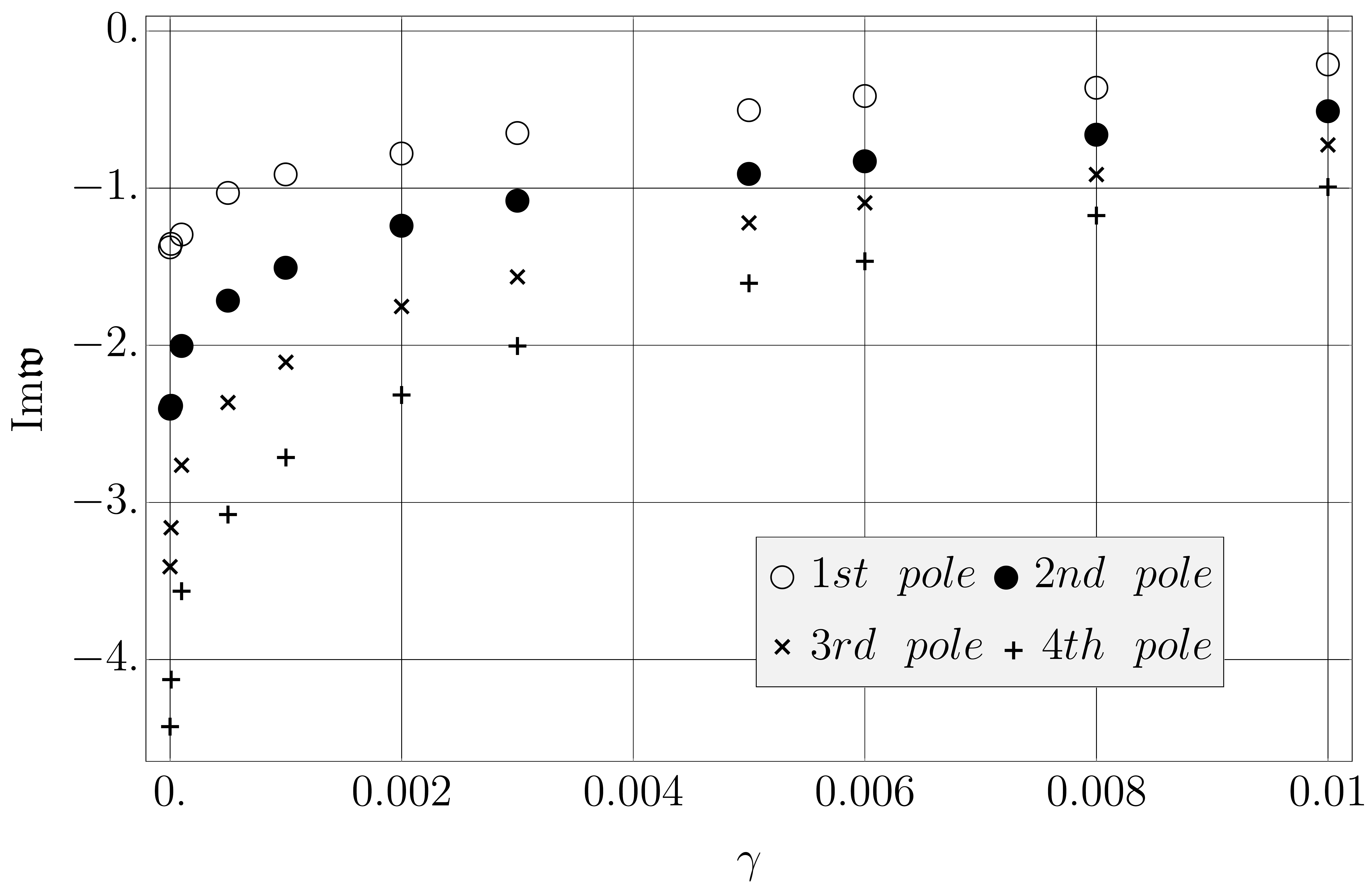}
\caption{$\CN=4$ SYM: Real (left panel) and imaginary (right panel) parts of the lowest four quasinormal frequencies in the sound channel at $\qfr = 0.1$.}
\label{fig:N=4+gamma-Sound-Poles-vs-gamma}
\end{figure}

\begin{figure}[htbp]
\centering
\includegraphics[width=0.7\linewidth]{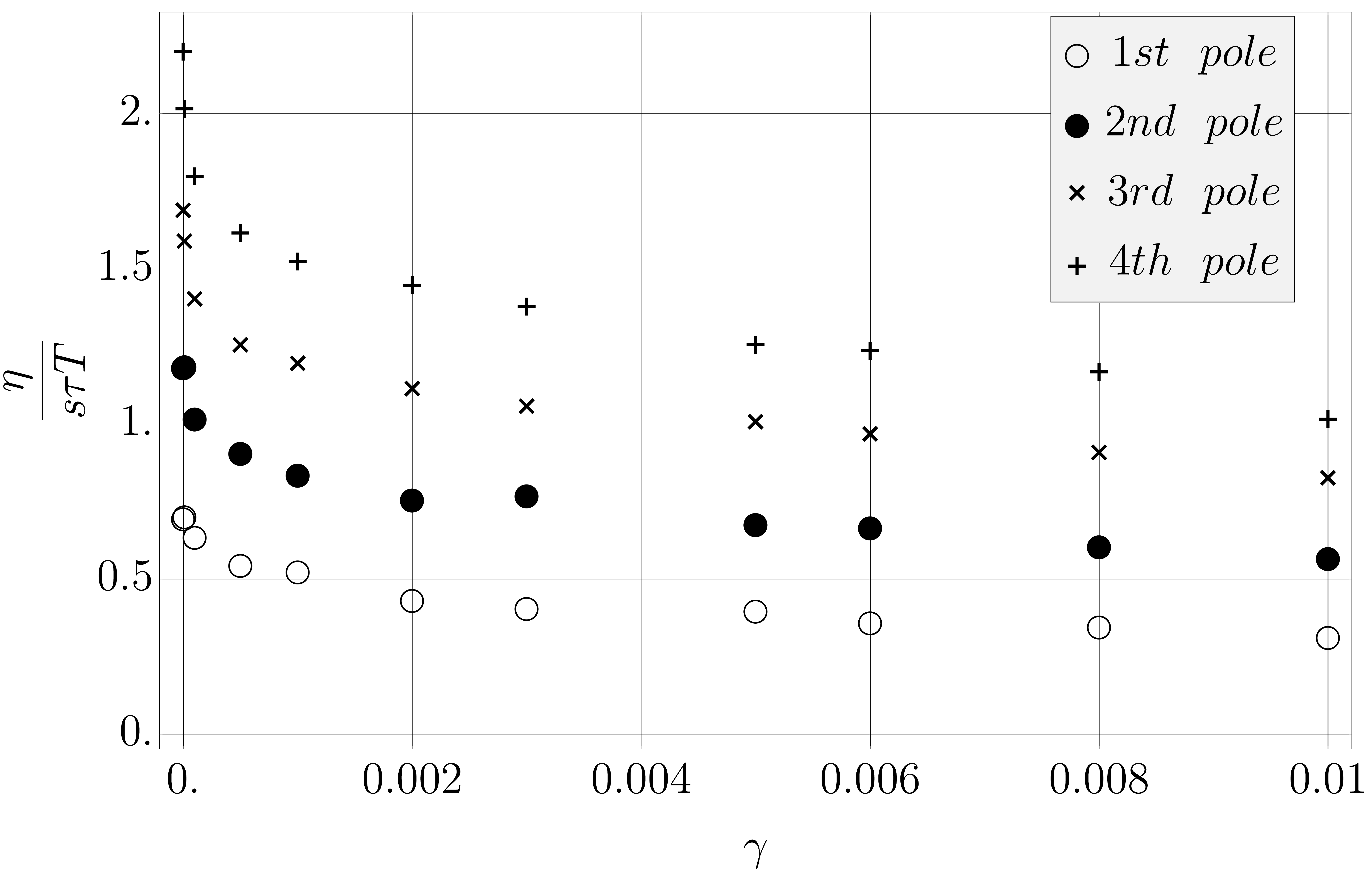}
\caption{The ratios $\eta/s \tau_k T $, $k=\{1,\,2,\,3,\,4\}$, as functions of $\gamma$ in $\CN=4$ SYM.}
\label{fig:N=4+gamma-Shear-etaOverStauT-vs-gamma}
\end{figure}

The dependence of real and imaginary parts of the smallest in magnitude quasinormal frequencies in the symmetric branches on $\gamma$ (at fixed $\qfr$) in the scalar, shear and sound channels, respectively, is shown in Figs.~\ref{fig:N=4+gamma-Scalar-Poles-vs-gamma}, \ref{fig:N=4+gamma-Shear-Poles-vs-gamma} and \ref{fig:N=4+gamma-Sound-Poles-vs-gamma}. In all three channels, a relatively strong dependence of the spectrum on $\gamma$ in the vicinity of $\gamma =0$ changes to a nearly flat behavior at larger values of $\gamma$. As discussed in the Introduction, these data can be used to test whether the relations between transport coefficients and the relaxation time typical for a kinetic regime of the theory may still hold at strong coupling. In kinetic theory, the hierarchy of relaxation times arises as the non-hydrodynamic part of the spectrum of a linearized Boltzmann operator (see Section \ref{sec:relaxation}). At strong coupling, it seems natural to associate this hierarchy with the (inverse) imaginary parts of the quasinormal spectrum frequencies. In particular, the relaxation time $\taur$ can be defined as
\begin{equation}
\taur (q, \lambda ) = \frac{2 \pi T}{\mbox{Im}\, \omega_F (q,\lambda)} = \frac{1}{\mbox{Im}\, \wfr_F (q,\lambda)}\,,
\end{equation}
where $\omega_F$ is the fundamental (lowest in magnitude) quasinormal frequency. The prediction of kinetic theory is that Eq.~(\ref{eq:rel-visc-rel}) holds at least at weak coupling, i.e. that the ratio $\eta / s\, \taur T$ is approximately constant. In Fig.~\ref{fig:N=4+gamma-Shear-etaOverStauT-vs-gamma}, we plot the ratios $\eta / s\, \tau_k T$, $k=1,2,3,4$, as functions of $\gamma$ using the data for $\tau_k=1/\mbox{Im} \, \wfr_k$ of the leading four non-hydrodynamic quasinormal frequencies (including the fundamental one) in the shear channel at $\qfr=0$. Curiously, although rapid decrease of all four functions is seen in the vicinity of $\gamma =0$, the dependence changes to a nearly flat one very quickly, already at $\gamma \approx 2 \times 10^{-3}$ (corresponding to the 't Hooft coupling $\lambda \sim 18$), which is well within the regime of small $\gamma$. Note that the 't Hooft coupling correction to $\eta / s$ for $\gamma \approx 2 \times 10^{-3}$ is approximately $25\%$. Thus, the naive use of kinetic theory expressions such as Eq.~(\ref{eq:rel-visc-rel}) may not be so disastrous at moderate or even strong coupling. We shall see in the next Sections that the features discussed here for the specific gravity dual with the higher derivative term of the type $R^4$ are also observed for gravity backgrounds with $R^2$ terms, in particular Gauss-Bonnet gravity.

\section{Relaxation time and poles of energy-momentum tensor correlators in a theory dual to Gauss-Bonnet gravity}
\label{sec:GB}
The action of Einstein-Gauss-Bonnet gravity in five space-time dimensions is given by
\begin{align}
\label{eq:GBaction}
S_{GB} = \frac{1}{2\kappa_5^2} \int d^5 x \sqrt{-g} \left[ R  + \frac{12}{L^2} + \frac{l^2_{GB}}{2} \left( R^2 - 4 R_{\mu\nu} R^{\mu\nu} + R_{\mu\nu\rho\sigma} R^{\mu\nu\rho\sigma} \right) \right],
\end{align}
where the scale $l^2_{GB}$ of the higher derivative term can be chosen to be set by a cosmological constant, $l^2_{GB} = \lgb L^2$, where $\lgb$ is the dimensionless parameter. The coefficients of the curvature-squared terms ensure that the equations of motion following from the action \eqref{eq:GBaction} are second order in derivatives. Thus, in the absence of Ostrogradsky instability and other difficulties usually induced by the dynamics with higher derivatives, Gauss-Bonnet and more generally Lovelock theories, are popular theoretical laboratories for studying non-perturbative effects of higher-derivative couplings. For example, the shear viscosity-entropy ratio in a (hypothetical) conformal fluid dual to five-dimensional Gauss-Bonnet gravity turns out to be \cite{Brigante:2007nu}
\begin{equation}
\frac{\eta}{s} = \frac{1 - 4 \lgb}{4\pi}\,,
\label{eq:gbviscosity}
\end{equation}
and this result is obtained without the assumption $|\lgb|\ll1$, i.e. non-perturbatively in the coupling. However, as pointed out and investigated in detail in Refs.~\cite{Brigante:2007nu,Brigante:2008gz,Buchel:2009tt,deBoer:2009pn,Camanho:2009vw,Buchel:2009sk}, for $\lgb$ outside of a certain interval, the dual theory suffers from pathologies associated with superluminal propagation of high momentum modes. More recently, Camanho {\it et al.} \cite{Camanho:2014apa} argued that Gauss-Bonnet theory suffers from causality problems in the bulk that can only be cured by adding higher spin fields. This would effectively imply that 
 Gauss-Bonnet and, most likely, general Lovelock theories\footnote{See Refs.~\cite{deBoer:2009gx,Camanho:2009hu,Camanho:2010ru} for relevant work in Lovelock theories.}  should loose their privileged non-perturbative status and be treated as any other theory with higher derivative terms, i.e. the coupling $\lgb$ in, for example, Eq.~(\ref{eq:gbviscosity}) should be seen as an infinitesimally small parameter (see, however, Refs.~\cite{Reall:2014pwa, Papallo:2015rna}). We note those difficulties but will not constrain $\lgb$ beyond its natural (here, limited by the existence of the black brane solution) domain $\lgb \in ( -\infty,1/4]$ in the following. 

Our goal is to compute the quasinormal spectrum of gravitational fluctuations of the Gauss-Bonnet black brane metric\footnote{Exact solutions and thermodynamics of black branes and black holes in Gauss-Bonnet gravity were considered in \cite{Cai:2001dz} (see also \cite{Nojiri:2001aj,Cho:2002hq,Neupane:2002bf,Neupane:2003vz}).}
\begin{align}
ds^2 = - f(r) N^2_{GB} dt^2 + \frac{1}{f(r)} dr^2 + \frac{r^2}{L^2} \left(dx^2 + dy^2 +dz^2 \right),
\label{eq:BB}
\end{align}
dual to a thermal state of a boundary CFT. Here
\begin{align}
f(r) = \frac{r^2}{L^2} \frac{1}{2\lgb} \left[1 - \sqrt{1-4\lgb \left(1 - \frac{r^4_0}{r^4} \right) } \right]
\label{eq:BBf}
\end{align}
and the constant $N_{GB}$ can be chosen to normalize the speed of light at the boundary to $c=1$:
\begin{align}
N_{GB}^2 = \frac{1}{2} \left(1+\sqrt{1-4\lgb} \right).
\label{eq:NGBDef}
\end{align}
The Hawking temperature corresponding to the solution \eqref{eq:BB} is given by
\begin{align}
T =  \frac{N_{GB} r_0}{\pi L^2} = \frac{r_0\sqrt{ 1+\gammagb}}{\sqrt{2} \pi L^2 }\,,
\label{eq:GBTemperature}
\end{align}
where we introduced the notation $\gammagb \equiv \sqrt{1-4\lgb}$. We shall use $\lgb$ and $\gammagb$ interchangeably in the following. The range $\lgb <0 $ corresponds to $\gammagb \in [1,\infty)$ and the interval $\lgb \in [0,1/4]$ maps into $\gammagb \in [0,1]$, with $\lgb =0$  corresponding to  $\gammagb=1$.
\subsection{Equations of motion}
Fluctuations $h_{\mu\nu}(r,t,z)$ of the Gauss-Bonnet black brane metric \eqref{eq:BB} can be  decomposed into the scalar, shear and sound channels in the standard way \cite{Son:2002sd,Kovtun:2005ev}. The corresponding gauge-invariant combinations $Z_1$, $Z_2$, $Z_3$ of the metric fluctuations $h_{\mu\nu}(r,\omega,q)$ (Fourier transformed in the variables along the brane directions) in the three channels are given by
\begin{align}
&\text{\bf Scalar:}&  &Z_1 = h^x_{~y}\,, \label{eq:GinvZ1}   \\
&\text{\bf Shear:}&  &Z_2 = \frac{q }{r^2} h_{tx} + \frac{\omega}{ r^2}  h_{xz}\,, \label{eq:Ginv4g2} \\
&\text{\bf Sound:}& &Z_3 = \frac{2 q^2}{r^2 \omega^2} h_{tt} +\frac{4 q}{r^2 \omega} h_{tz}  - \left(  1 - \frac{q^2 N_{GB}^2 \left(4 r^3 - 2 r f(r)\right)}{2 r \omega^2 \left(r^2 - 2 \lgb f(r)\right)}   \right) \left( \frac{h_{xx}}{r^2} + \frac{h_{yy}}{r^2} \right) + \frac{2}{r^2} h_{zz}\,.  \label{eq:GinvZ3}
\end{align}
Introducing the new variable $u = r_0^2/r^2$, the equation of motion in each of the three channels can be written in the form of a linear second-order differential equation
\begin{align}
\partial_u^2 Z_i + A_i \partial_u Z_i + B_i Z_i = 0\,,
\label{eq:eom_GB_ginv}
\end{align}
where $i=1,2,3$ and the coefficients $A_i$ and $B_i$ are given in Appendix \ref{sec:appendix-GB}. To find the quasinormal spectrum in the three channels, we impose the "incoming wave" boundary conditions at the horizon at $u=1$ \cite{Son:2002sd},
\begin{align}
Z_i(u) = (1-u)^{-i\wfr/2} \CZ_i(u,\wfr,\qfr)\,,
\end{align}
where the functions $\CZ_i$ are regular at $u=1$. The quasinormal spectra $\wfr = \wfr (\qfr)$ are then solutions to the equations $Z_i (u=0,\wfr,\qfr) = 0$. They can be found numerically. In addition, in the regime  $\wfr \ll 1$ and $\qfr\ll 1$, some frequencies are determined analytically. In all three channels, it will be convenient to use a new variable
\begin{align}
v = 1 - \sqrt{ 1 - \left(1-u^2\right)\left(1 - \ggb^2 \right) },
\end{align}
so that the horizon is at $v = 0$ and the boundary at $ v = 1 - \ggb$. The new coordinate is singular at zero Gauss-Bonnet coupling, $\lgb = 0$ ($\ggb = 1$) and the results for that point, which are identical to those of  $\CN = 4$ SYM theory at infinite 't Hooft coupling, have to be obtained independently.
\begin{figure}[ht]
\centering
\begin{subfigure}[t]{0.45\linewidth}
\includegraphics[width=1\linewidth]{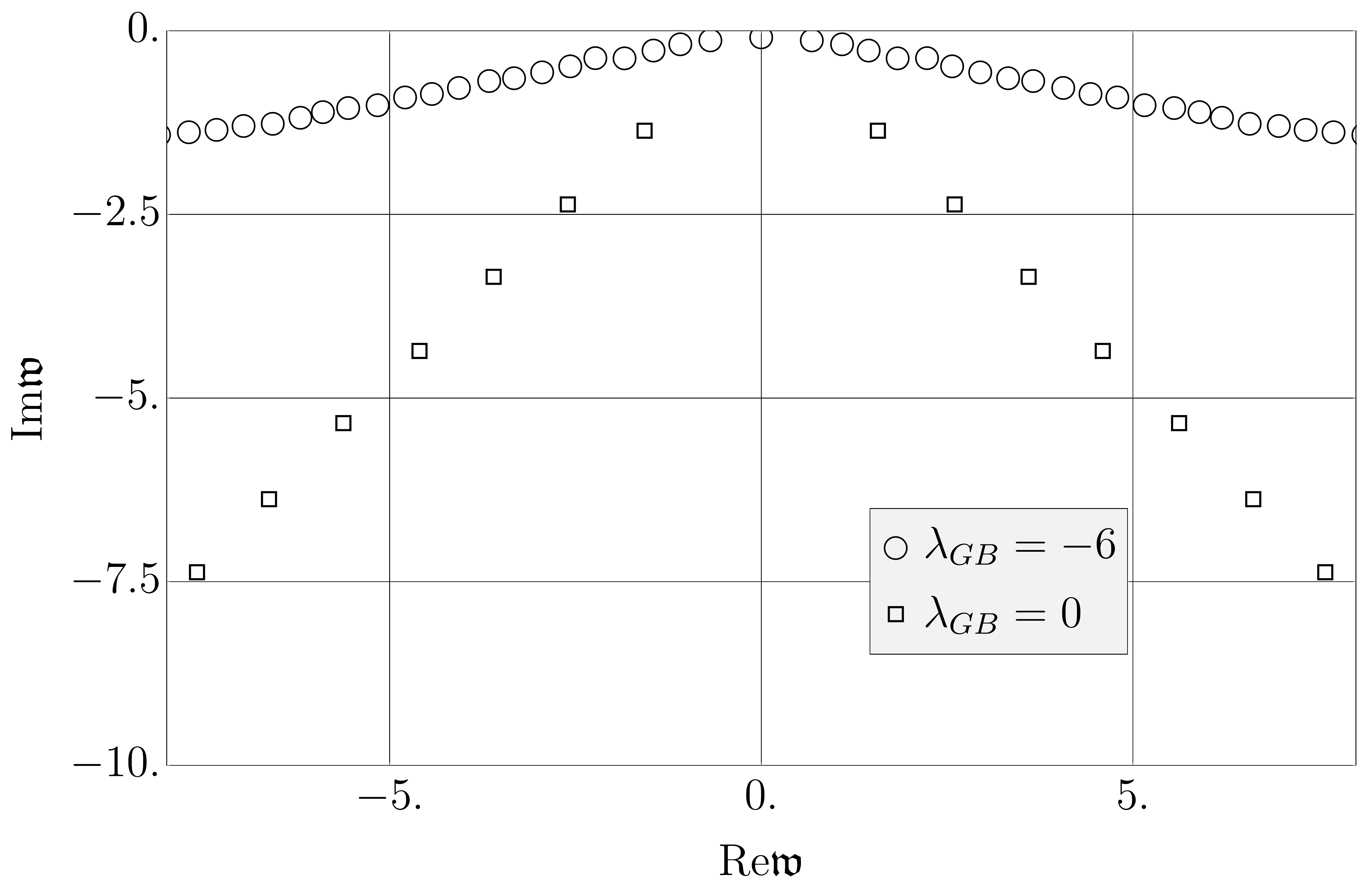}
\end{subfigure}
\qquad
\begin{subfigure}[t]{0.45\linewidth}
\includegraphics[width=1\linewidth]{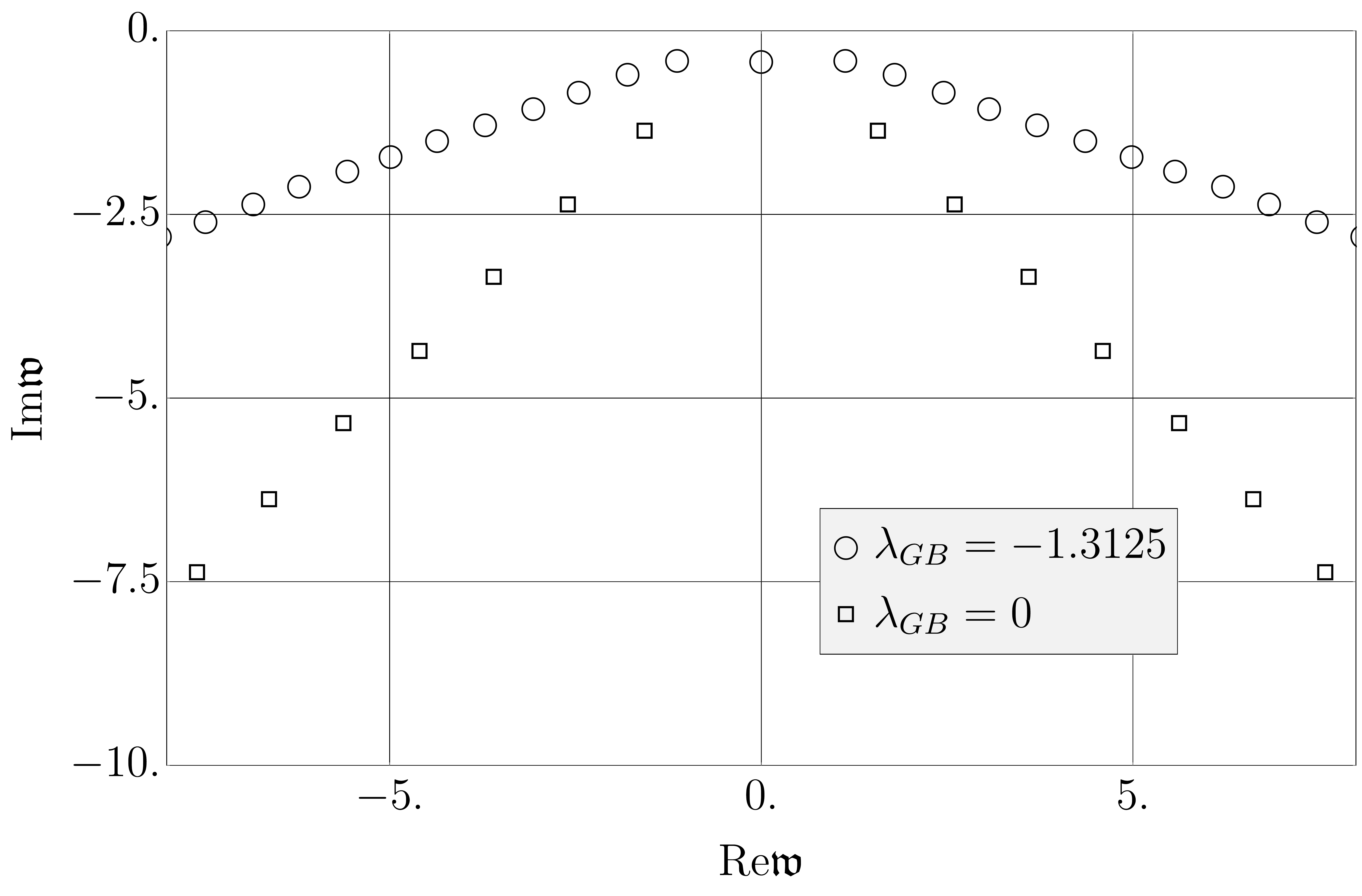}
\end{subfigure}
\\
\begin{subfigure}[b]{0.45\linewidth}
\includegraphics[width=1\linewidth]{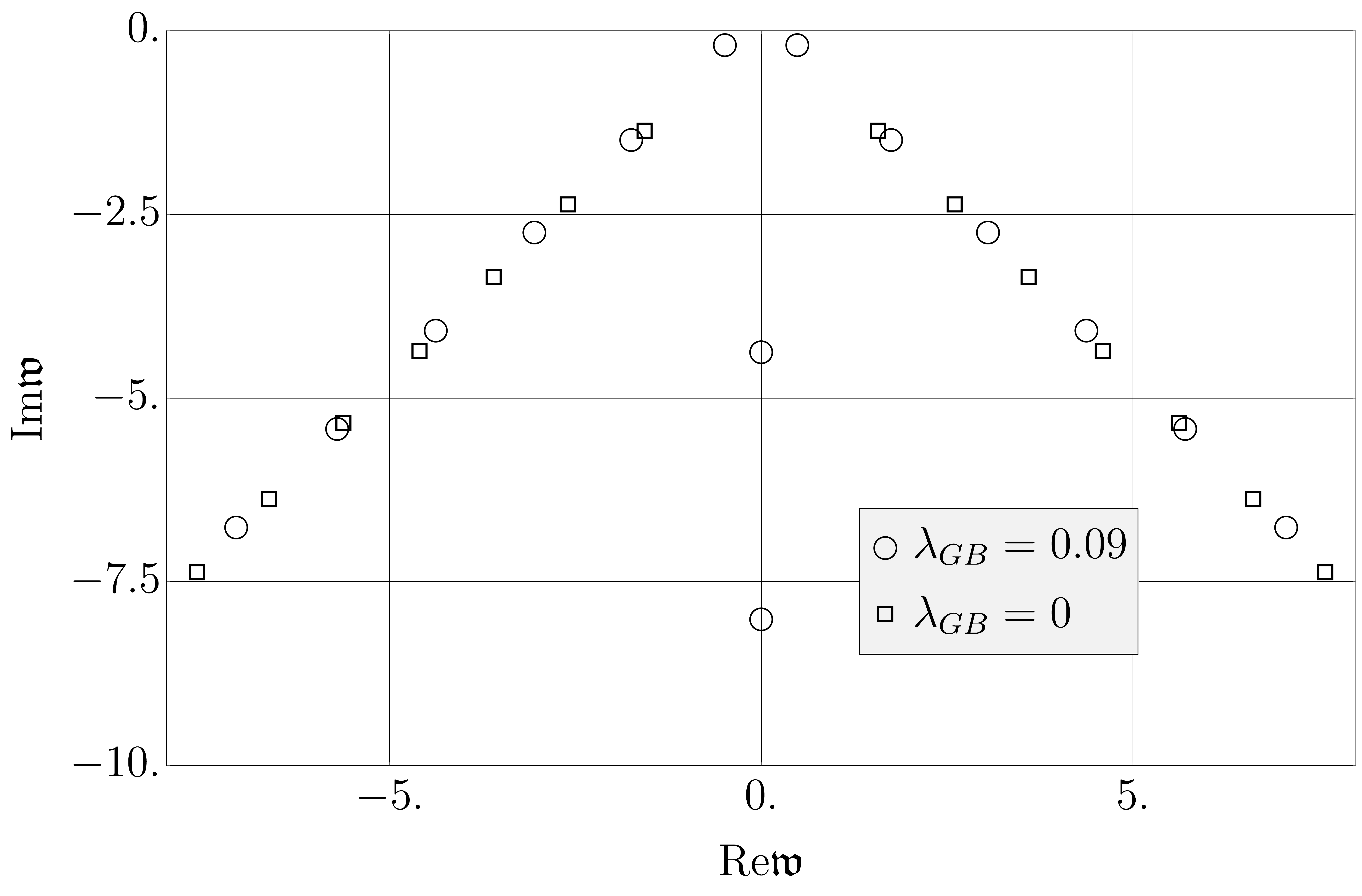}
\end{subfigure}
\qquad
\begin{subfigure}[b]{0.45\linewidth}
\includegraphics[width=1\linewidth]{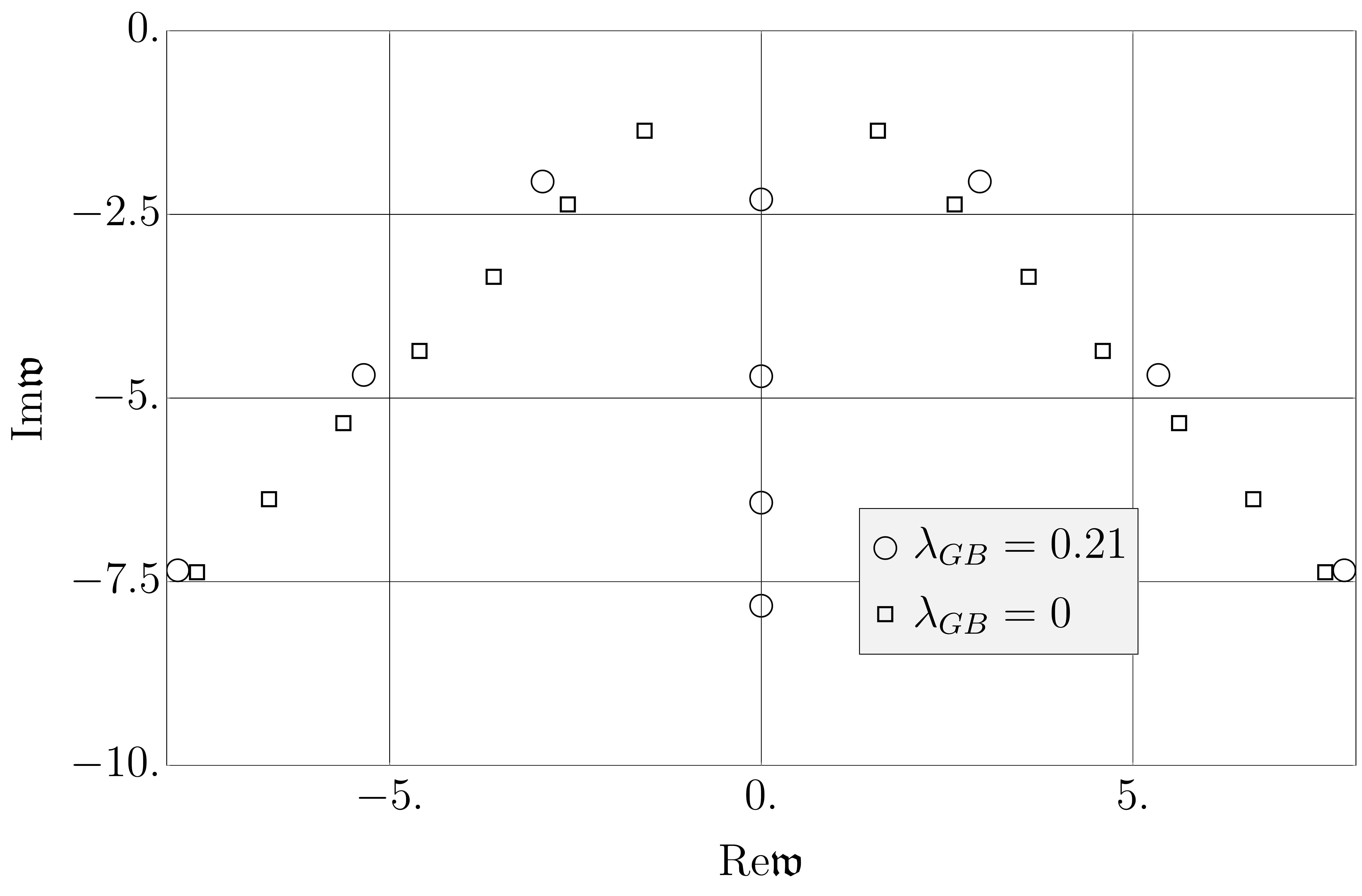}
\end{subfigure}
\caption{Quasinormal spectrum (shown by circles) of the scalar channel metric perturbations in Gauss-Bonnet gravity for various values of the coupling $\lgb$ and $\qfr=0.1$. From top left: $\lgb = \{-6,\, -1.3125, \, 0.09,\, 0.21\} $. For comparison, the spectrum at $\lgb = 0$ is shown by squares.}
\label{fig:GB-Scalar-channel}
\end{figure}
\subsection{The spectrum of the metric fluctuations}
\label{sec:GBResults}
The quasinormal spectra in Einstein-Gauss-Bonnet theory obtained non-perturbatively in $\lgb$ show the properties qualitatively similar to the ones discussed in Section  \ref{sec:N4Results} for the AdS-Schwarzschild background corrected by the $R^4$ term. In this Section we show the numerical results and analytic approximations for the spectra in the three channels, including the details of the breakdown of the hydrodynamic regime. There are some novelties in the Gauss-Bonnet case. First, not being restricted by the perturbative nature of the higher-derivative coupling, we are able to explore the coupling dependence to a fuller extent than in $\mathcal{N}=4$ SYM. In particular, we are able to say more about the spectral function and the density of poles in the complex plane than we could in $\mathcal{N}=4$ SYM owing to the restriction $\gamma \ll 1$. Second, Gauss-Bonnet gravity (and gravity with generic $R^2$ terms) provides an example of a holographic model, where the shear viscosity - entropy density ratio can be greater or less than $1/4\pi$, depending on the sign of $\lgb$. We find qualitatively different patterns in the behavior of relaxation time and other quantities in those two regimes.
\subsubsection{Scalar channel}
The spectrum of gravitational perturbations in the scalar channel is shown in Fig.~\ref{fig:GB-Scalar-channel}. Two different regimes are observed depending on the value of $\eta/s$.
\begin{figure}[ht]
\centering
\includegraphics[width=0.45\linewidth]{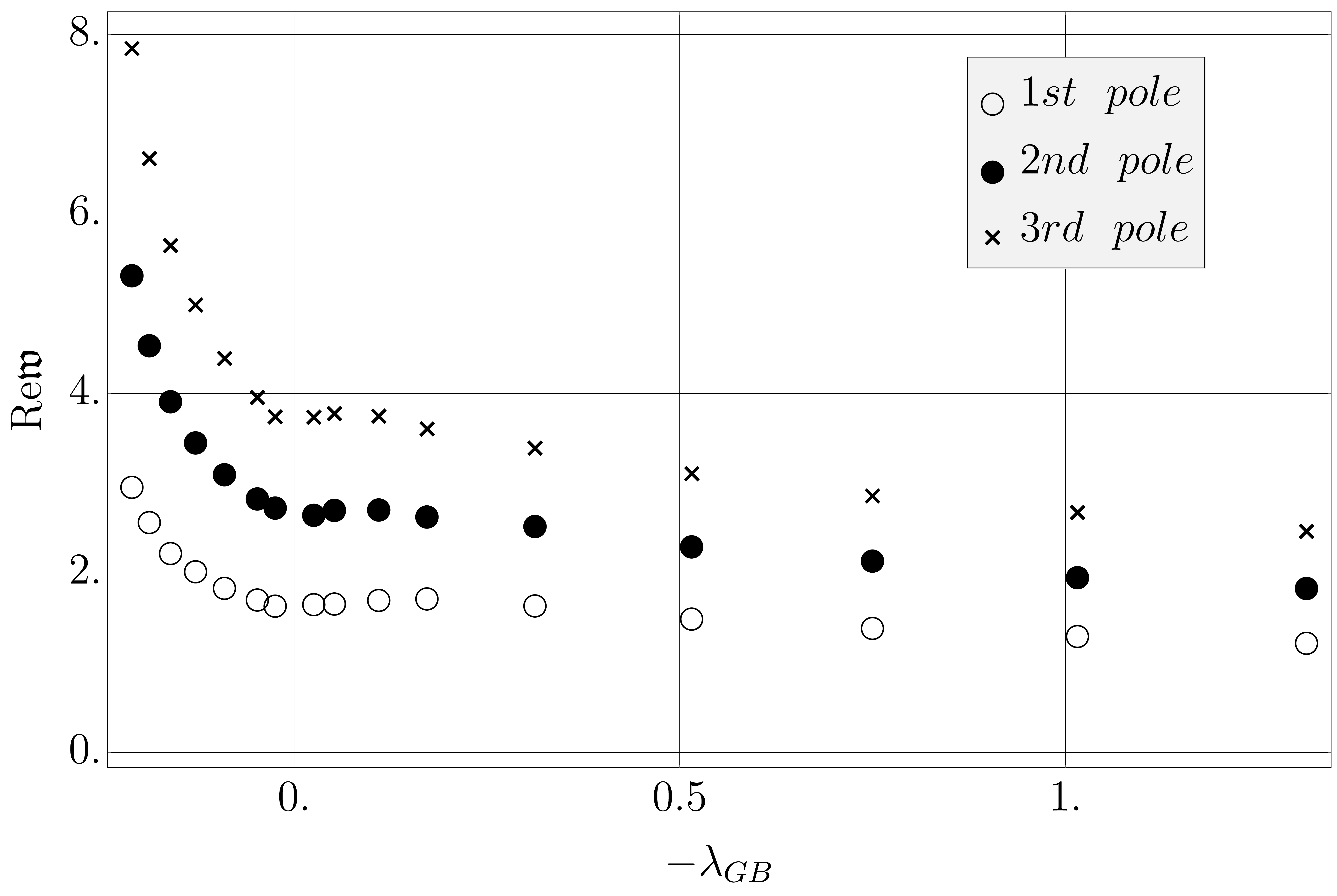}
\includegraphics[width=0.45\linewidth]{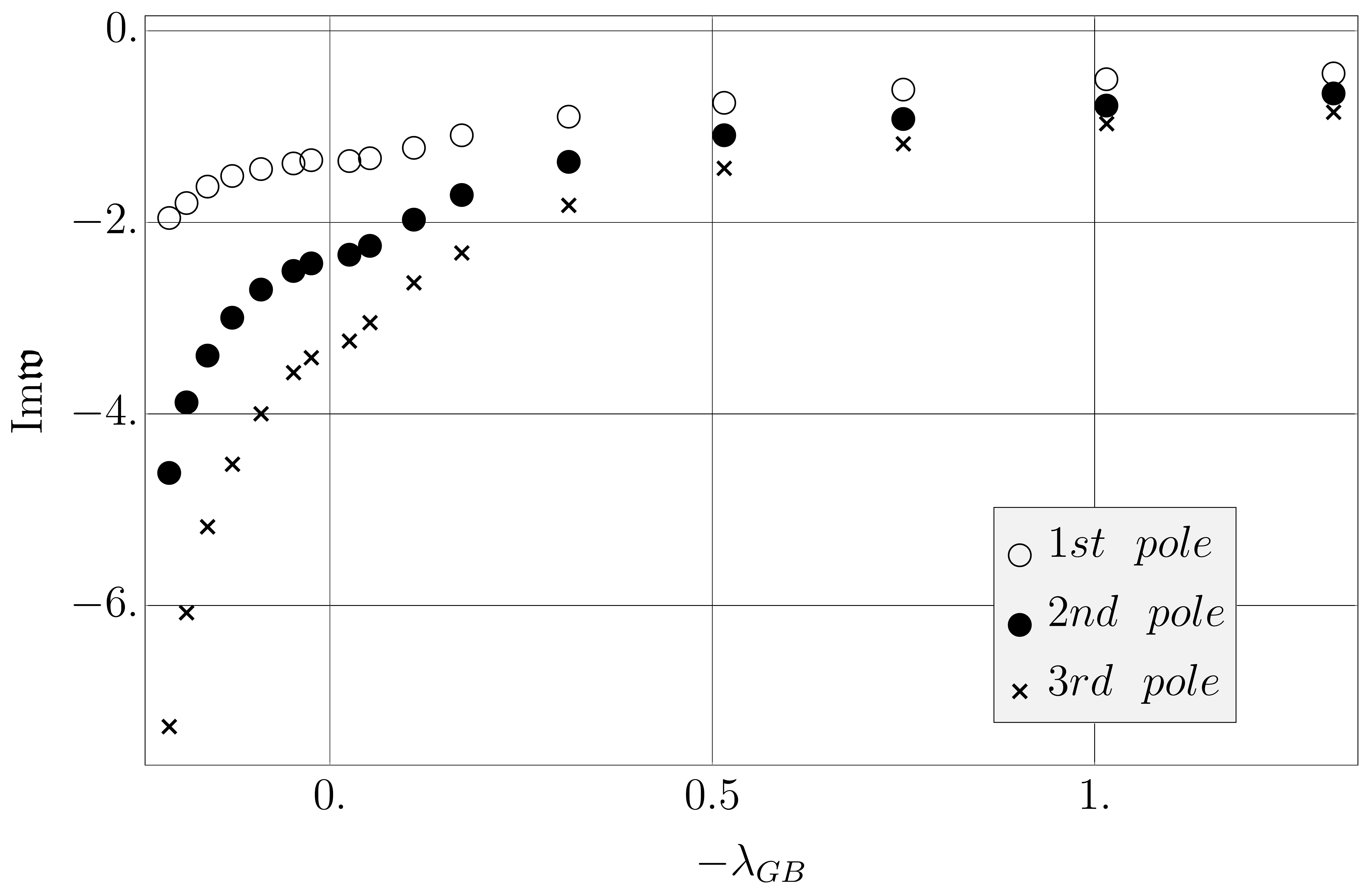}
\caption{Real (left panel) and imaginary (right panel) parts of the top three quasinormal frequencies of the symmetric branches in the scalar channel of Gauss-Bonnet gravity at $\qfr = 0.5$.}
\label{fig:GB-Scalar-Poles-vs-gamma}
\end{figure}
\begin{figure}[ht]
\centering
\includegraphics[width=0.8\linewidth]{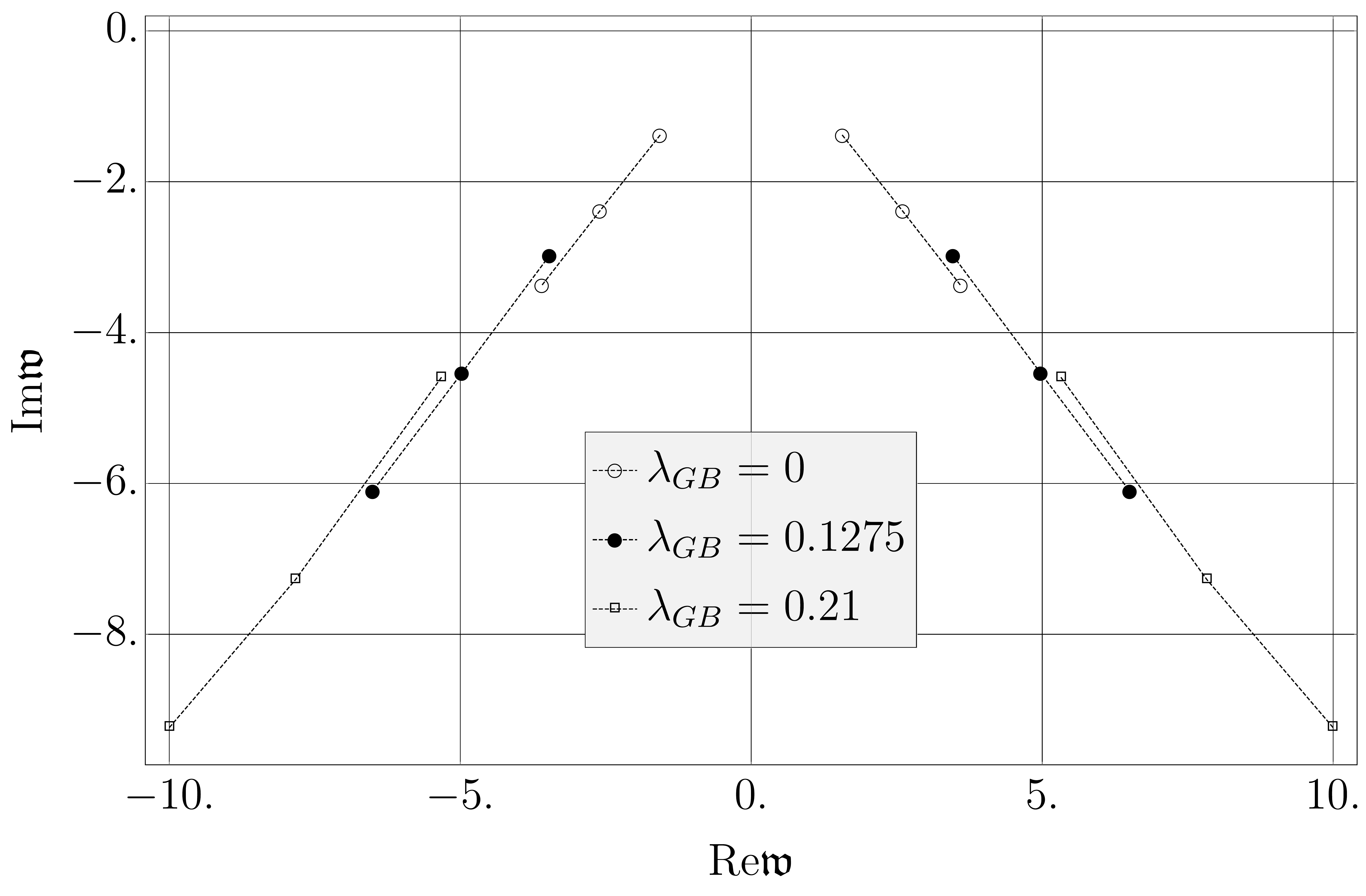}
\caption{Top three quasinormal frequencies (connected by lines for better visibility) of the symmetric branches in the scalar channel of gravitational perturbations in Gauss-Bonnet theory as functions of the coupling $\lgb > 0$ (i.e. in the regime $\eta/s < 1/4\pi$). The rest of the quasinormal spectrum is not shown in this figure.}
\label{fig:N=GB-Scalar-3-poles}
\end{figure}

For $\eta/s > 1/4\pi$ (corresponding to $\lgb < 0$), the behavior of the poles is qualitatively the same as in  $\mathcal{N}=4$ SYM: the two symmetric branches of gapped poles lift up towards the real axis monotonically with $|\lgb|$ increasing, the distance between the poles decreases suggesting a formation of branch cuts $(-\infty,-q]$ and $[q,\infty)$ in the limit $|\lgb|\rightarrow \infty$. Observing the motion of individual poles in the symmetric branches, one can say that there is an inflow of new poles from complex infinity along the branches with $|\lgb|$ increasing. The dependence of real and imaginary parts of the top three poles in the symmetric branches on $\lgb$ at $\qfr = 0.5$ is shown in Fig.~\ref{fig:GB-Scalar-Poles-vs-gamma}. Within the limits of numerical accuracy, this dependence is monotonic. One may notice that the functions become flat for large negative $\lgb$. When the poles in the two branches are sufficiently close to the real axis, we expect the spectral function to show the distinct quasiparticle peaks. We shall discuss this in detail for the shear channel, see subsection \ref{sec:SpectFunGBShear}.

There is a new pole rising up the imaginary axis from complex infinity\footnote{In contrast to the corresponding $\mathcal{N}=4$ SYM case, we observe only one new pole for $\lgb <0$, although it is difficult to make this conclusion with certainty using a numerical approach.}. The position of the new pole in the regime  $\wfr \ll 1$, $\qfr\ll 1$, $\ggb \gg 1$ can be determined analytically by solving the equation for  $\CZ_1$ perturbatively and imposing the condition $Z_1(u=0,\wfr,\qfr) = 0$:
\begin{align}
\wfr_1 = \wfr^{GB}_{\mathfrak{g}}  + \ldots = - \frac{4i}{\ggb  (\ggb +2) - 3 + 2 \ln \left( \frac{ 2}{\ggb + 1 }\right)} + \ldots  \approx -
\frac{i}{|\lgb|} \, .
\label{eq:GB-scalar-gap}
\end{align}
The mode remains purely on the negative imaginary axis and approaches the origin as $\ggb \to \infty$ ($\lgb \rightarrow -\infty$). This result is confirmed numerically. The residue vanishes in the limit $\ggb \to \infty$, and so the pole disappears in that limit \cite{GBNesojen}.
\begin{figure}[ht]
\centering
\begin{subfigure}[t]{0.45\linewidth}
\includegraphics[width=1\linewidth]{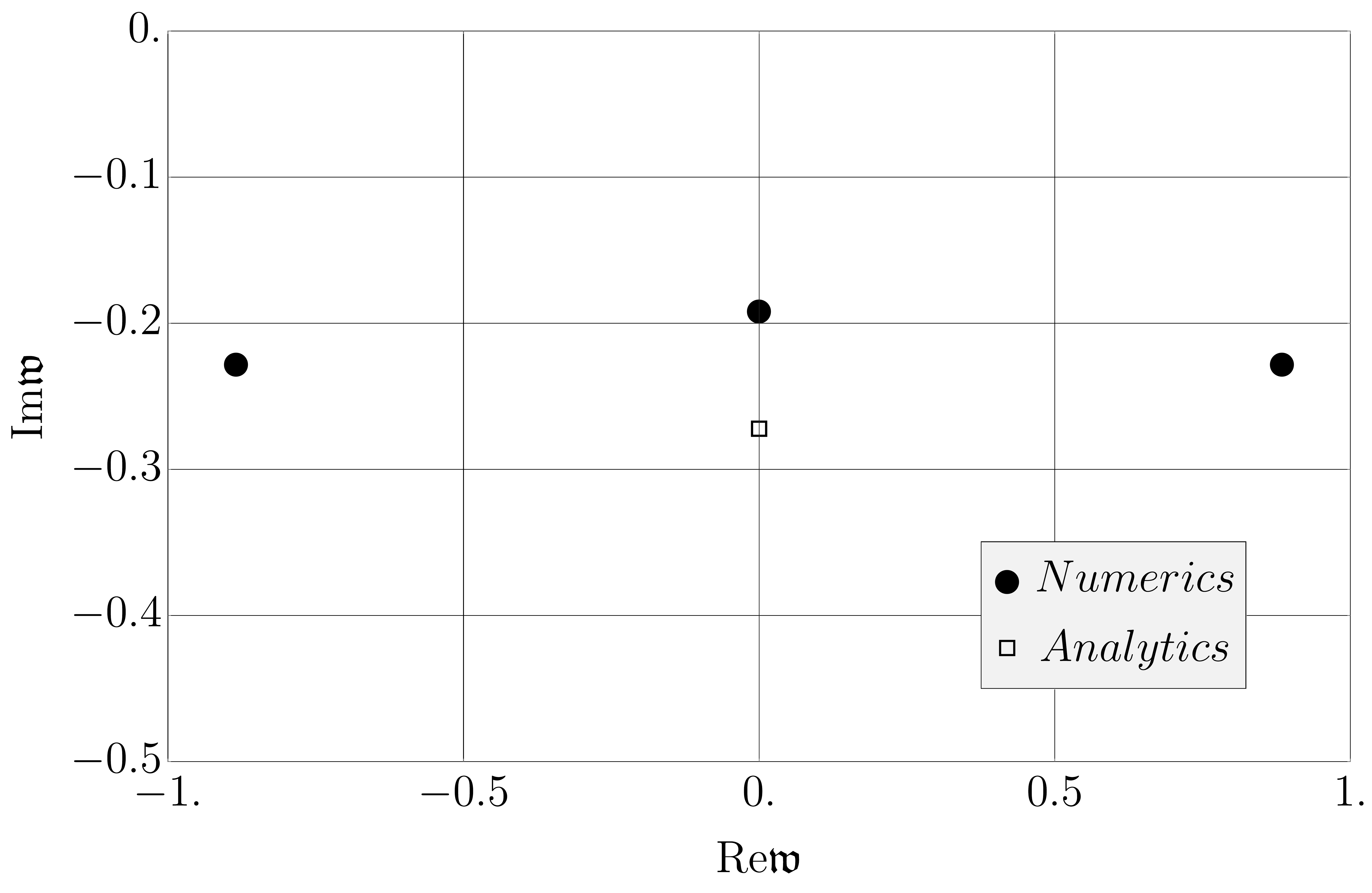}
\end{subfigure}
\qquad
\begin{subfigure}[t]{0.45\linewidth}
\includegraphics[width=1\linewidth]{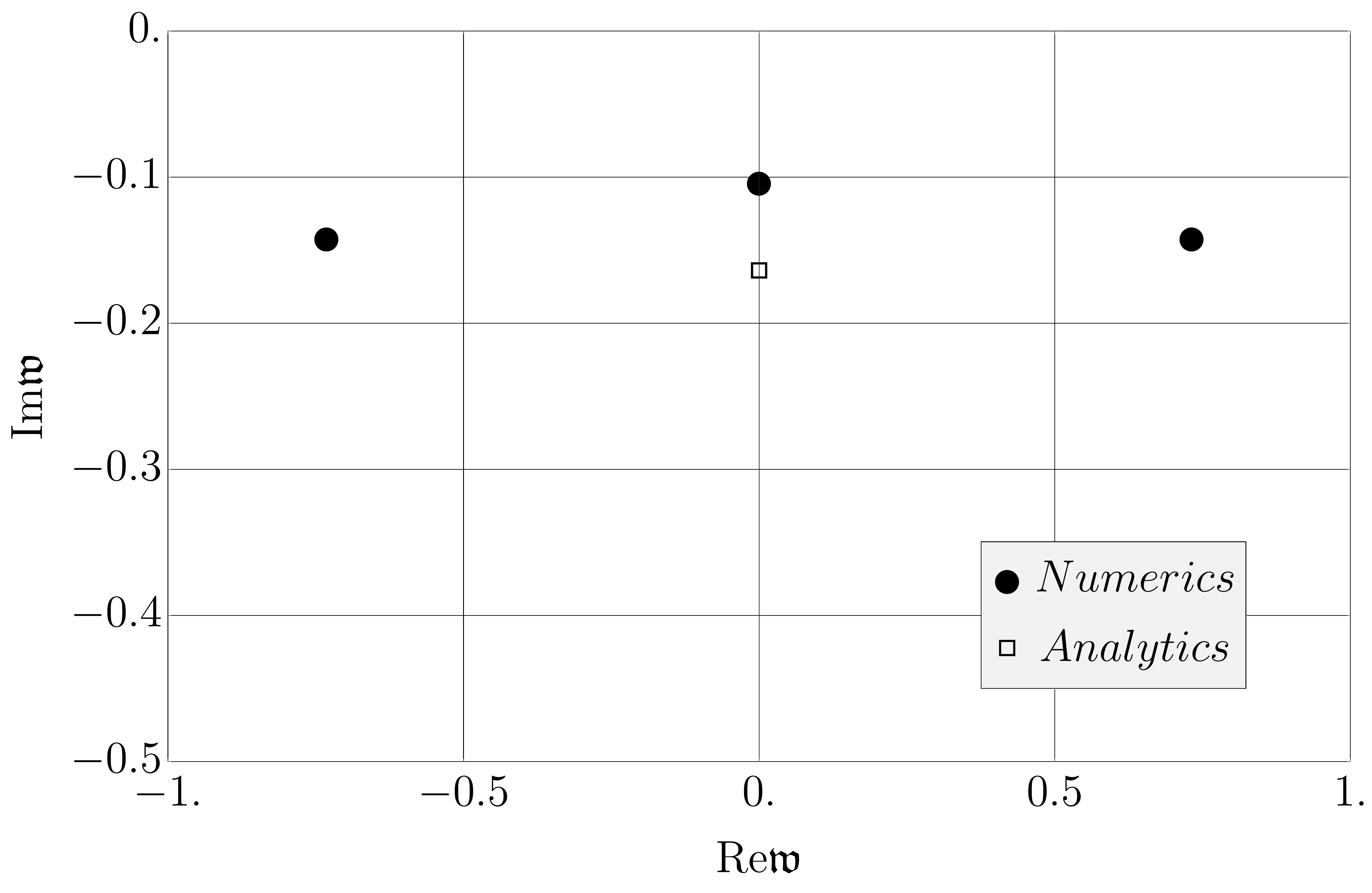}
\end{subfigure}
\\
\begin{subfigure}[b]{0.45\linewidth}
\includegraphics[width=1\linewidth]{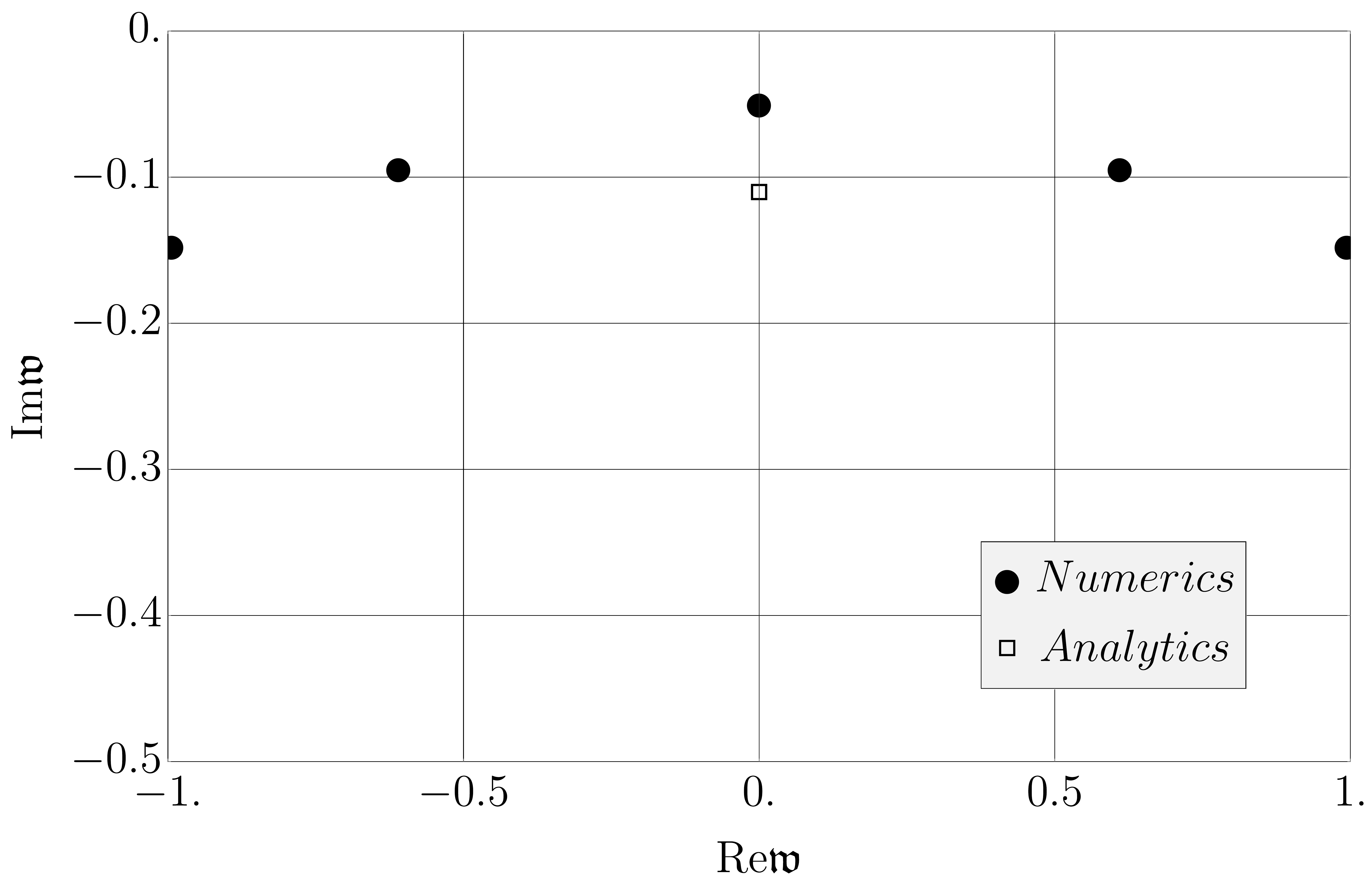}
\end{subfigure}
\qquad
\begin{subfigure}[b]{0.45\linewidth}
\includegraphics[width=1\linewidth]{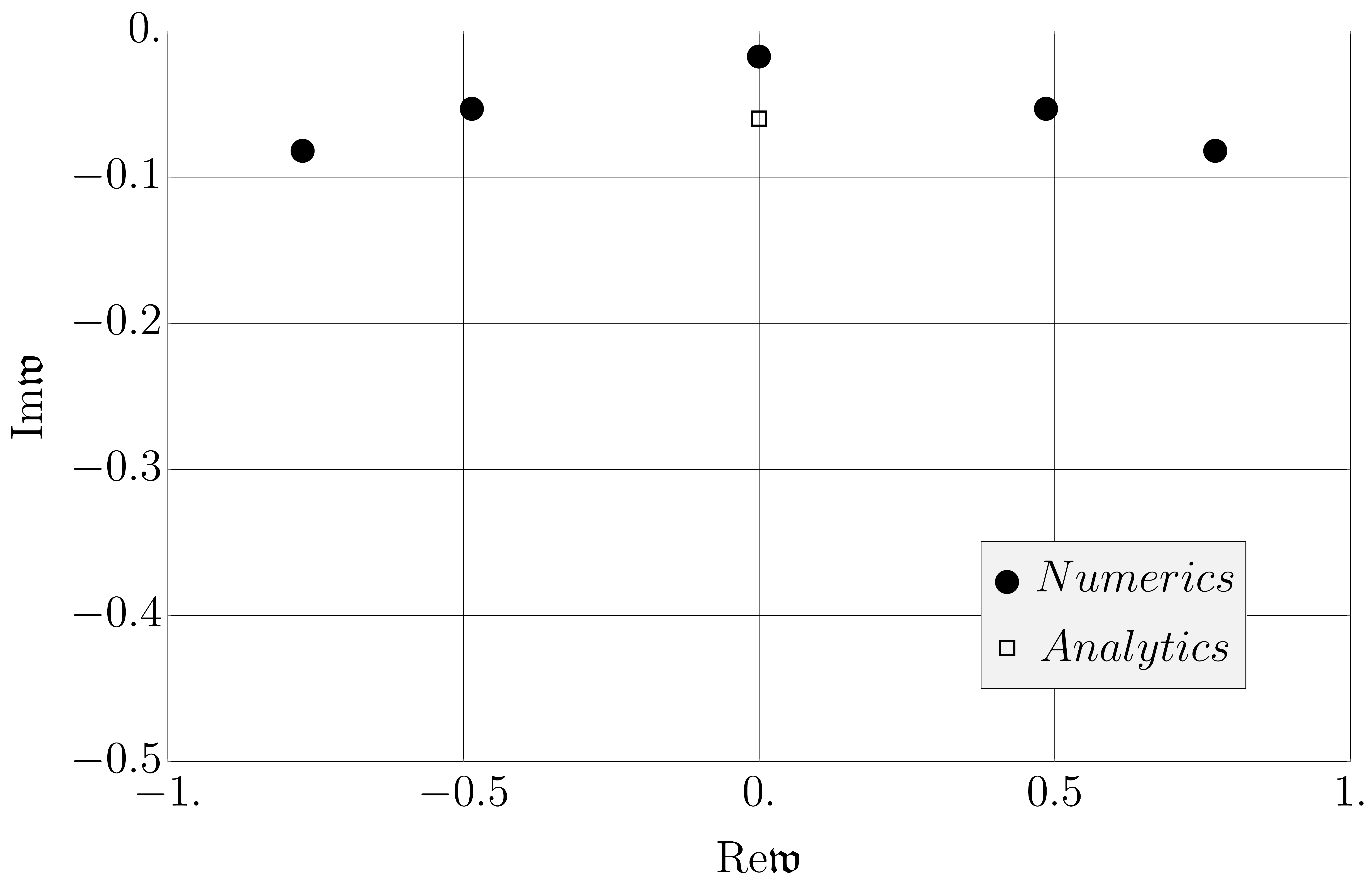}
\end{subfigure}
\caption{Quasinormal modes (shown by black dots), close to the origin, in the scalar channel of Gauss-Bonnet black brane metric perturbations, for increasing coupling constant and $\qfr=0.1$. From top left to bottom right: $\lgb =\{-2.8125, \,-4.8125,\, -7.3125,\, -13.8125\}$. The analytic approximation (\ref{eq:GB-scalar-gap}) to the gapped pole on the imaginary axis is shown by a white square.}
\label{fig:GB-Scalar-zoom}
\end{figure}

For $\eta/s <1/4\pi$ (corresponding to $\lgb > 0$), the poles in the two branches become more sparse relative to their $\lgb = 0$ distribution (see Figs.~\ref{fig:GB-Scalar-channel} and \ref{fig:N=GB-Scalar-3-poles}). In sharp contrast with the $\eta/s >1/4\pi$ case, here the branches lift up very slightly, almost infinitesimally, relative to their $\lgb=0$ positions. As shown in Figs.~\ref{fig:GB-Scalar-channel} and \ref{fig:N=GB-Scalar-3-poles}, an outflow of poles along the branches to complex infinity is observed and it is conceivable that the poles of the two branches are eventually completely pushed out of the finite complex plane. At the same time, there are still new poles rising up the imaginary axis. In the limit $\lgb \rightarrow 1/4$ ($\gammagb \rightarrow 0$) they are seen numerically to approach the positions (known exactly \cite{GBNesojen})
\begin{align}
\wfr = -i \left(4+2n_1 - \sqrt{4-3\qfr^2}\right) , \qquad \wfr = -i \left(4+2n_2 + \sqrt{4-3\qfr^2}\right)\,,
\label{eq:QNMScalar}
\end{align}
where $n_1$ and $n_2$ are non-negative integers.

The limit of vanishing shear viscosity $\lgb \rightarrow 1/4$ is difficult to explore numerically. However, the observed behavior is consistent with analytic results available at $\lgb = 1/4$. Indeed, exactly at $\lgb = 1/4$ the equations of motion can be solved in terms of hypergeometric functions and the quasinormal spectrum is determined exactly \cite{GBNesojen}. The only quasinormal frequencies at $\lgb = 1/4$ are the ones given by Eq.~\eqref{eq:QNMScalar}. This is consistent with the picture we observe numerically for $0 < \lgb < 1/4$.
\begin{figure}[ht]
\centering
\begin{subfigure}[t]{0.45\linewidth}
\includegraphics[width=1\linewidth]{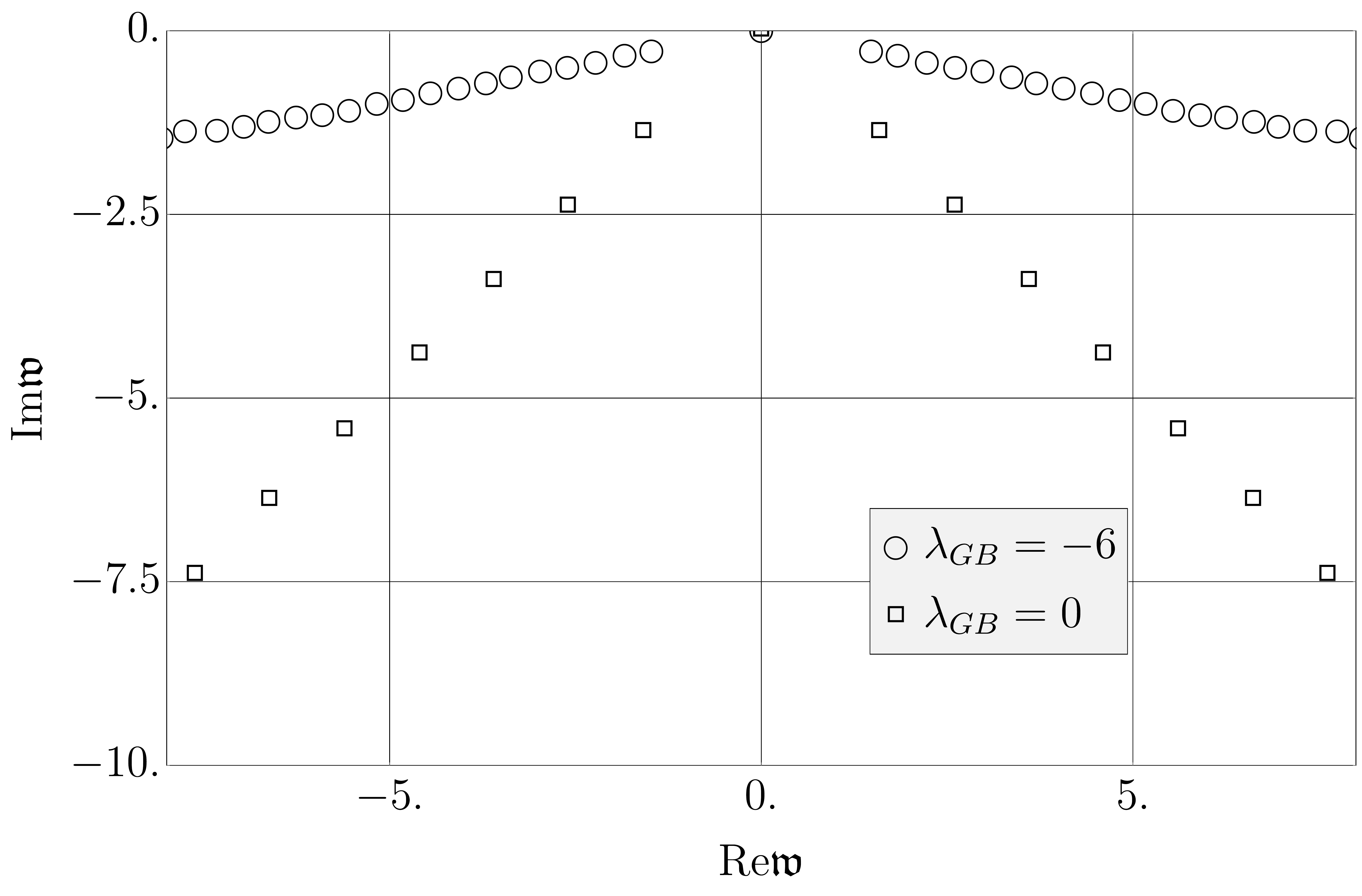}
\end{subfigure}
\qquad
\begin{subfigure}[t]{0.45\linewidth}
\includegraphics[width=1\linewidth]{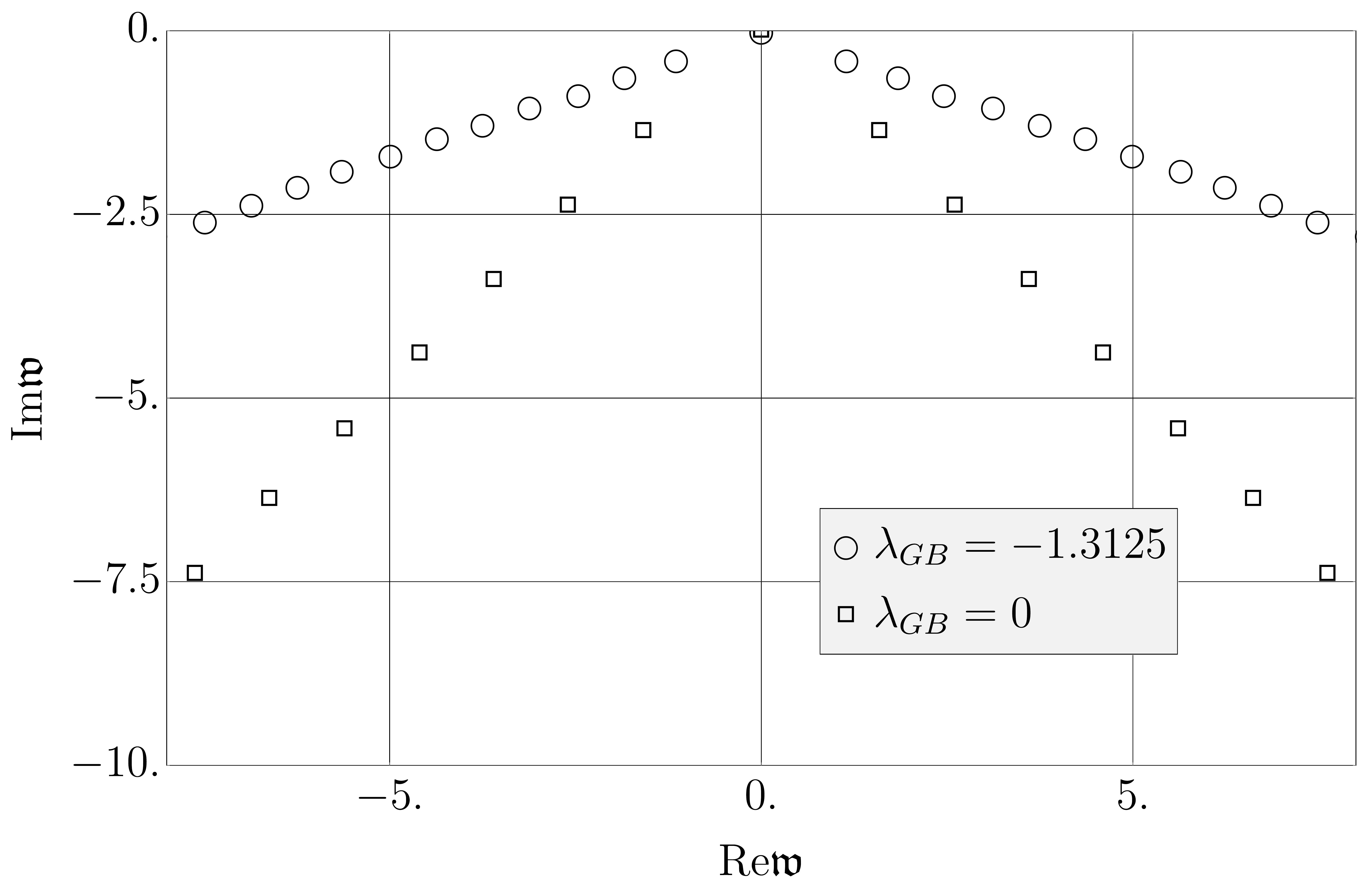}
\end{subfigure}
\\
\begin{subfigure}[b]{0.45\linewidth}
\includegraphics[width=1\linewidth]{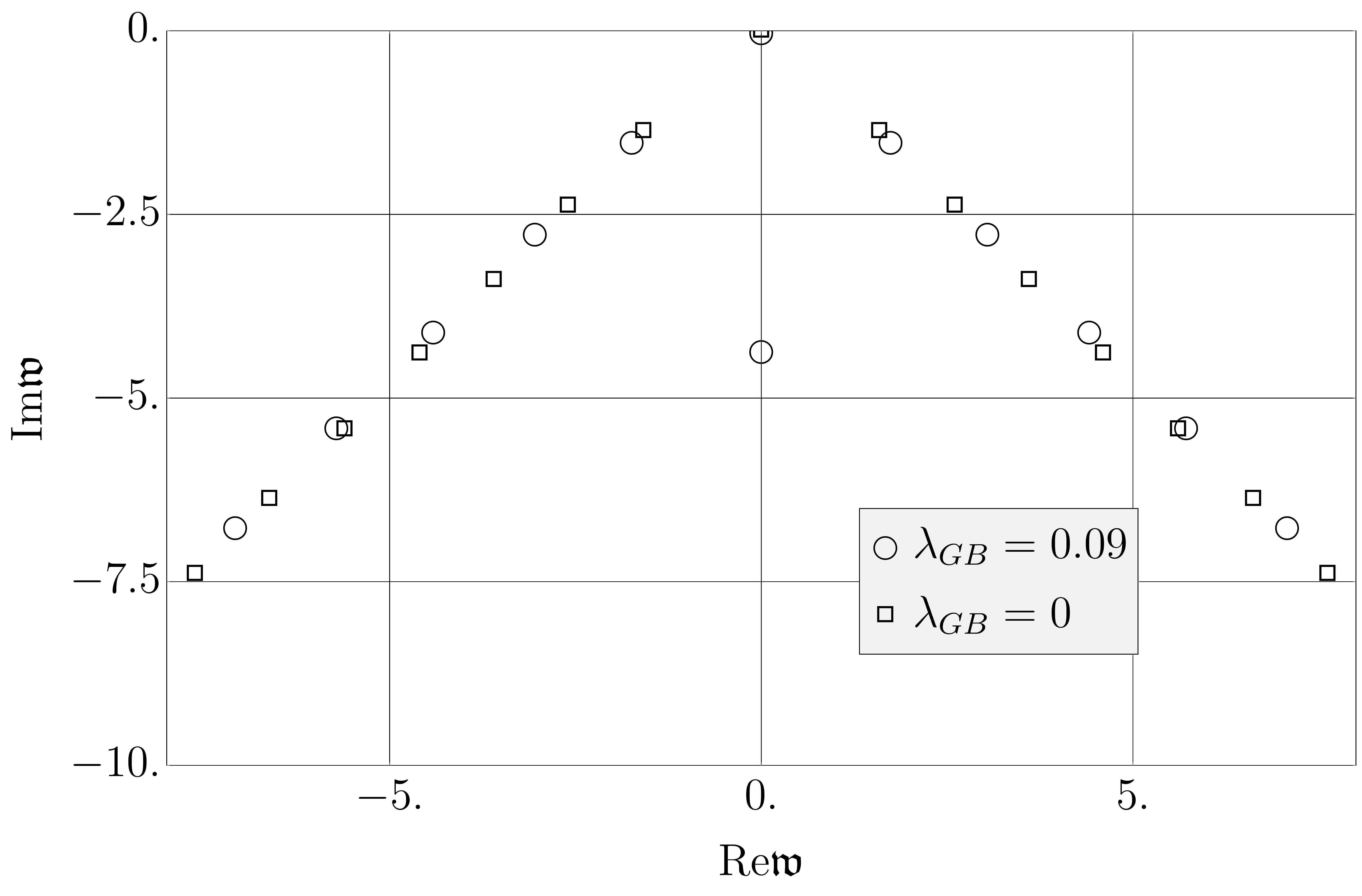}
\end{subfigure}
\qquad
\begin{subfigure}[b]{0.45\linewidth}
\includegraphics[width=1\linewidth]{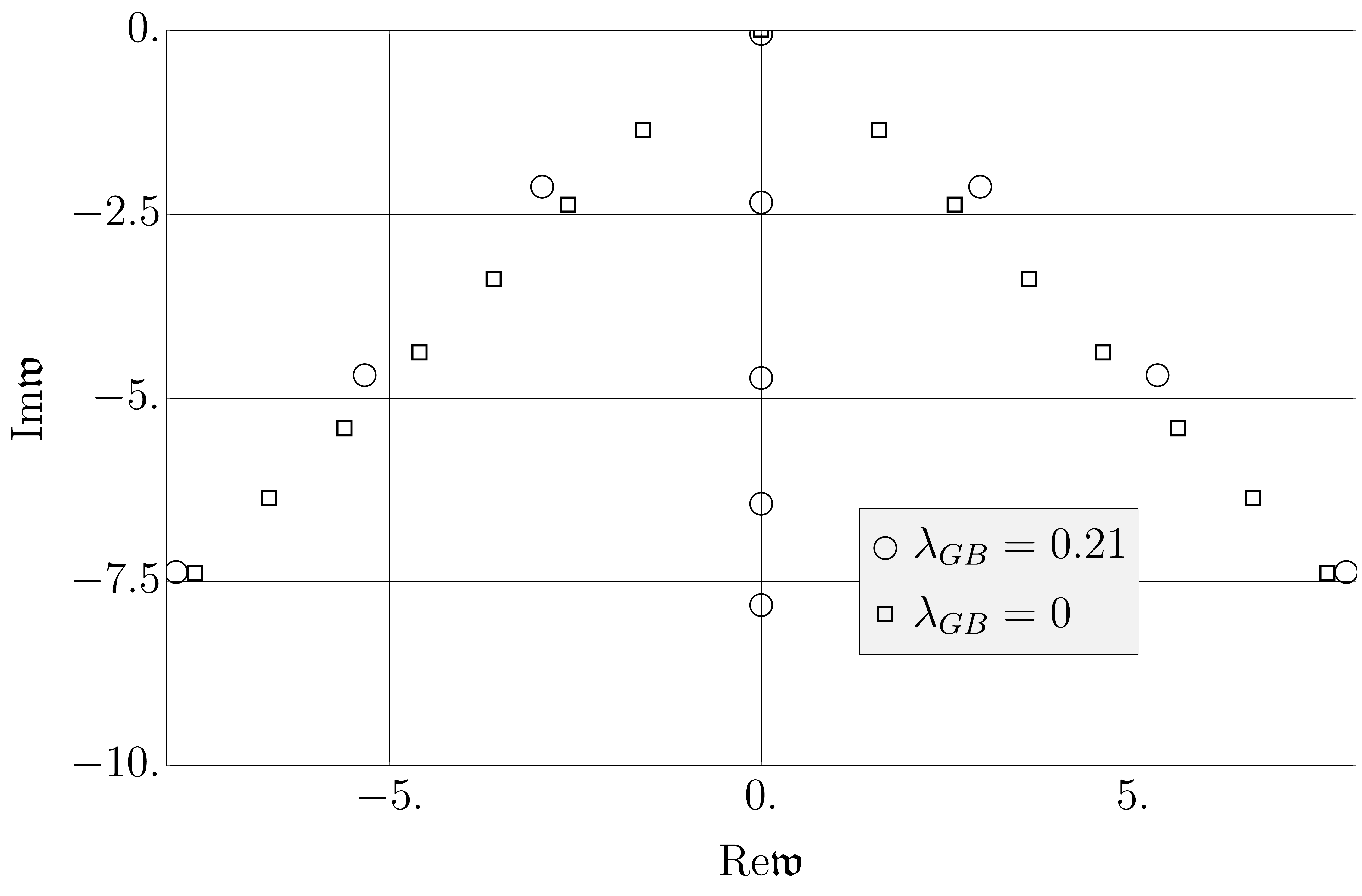}
\end{subfigure}
\caption{Quasinormal spectrum (shown by circles) of the shear channel metric perturbations in Gauss-Bonnet gravity for various values of the coupling $\lgb$ and $\qfr=0.1$. From top left: $\lgb = \{-6,\, -1.3125, \, 0.09,\, 0.21\} $. For comparison, the spectrum at $\lgb = 0$ is shown by squares.}
\label{fig:GB-Shear-channel}
\end{figure}
\begin{figure}[ht]
\centering
\includegraphics[width=0.45\linewidth]{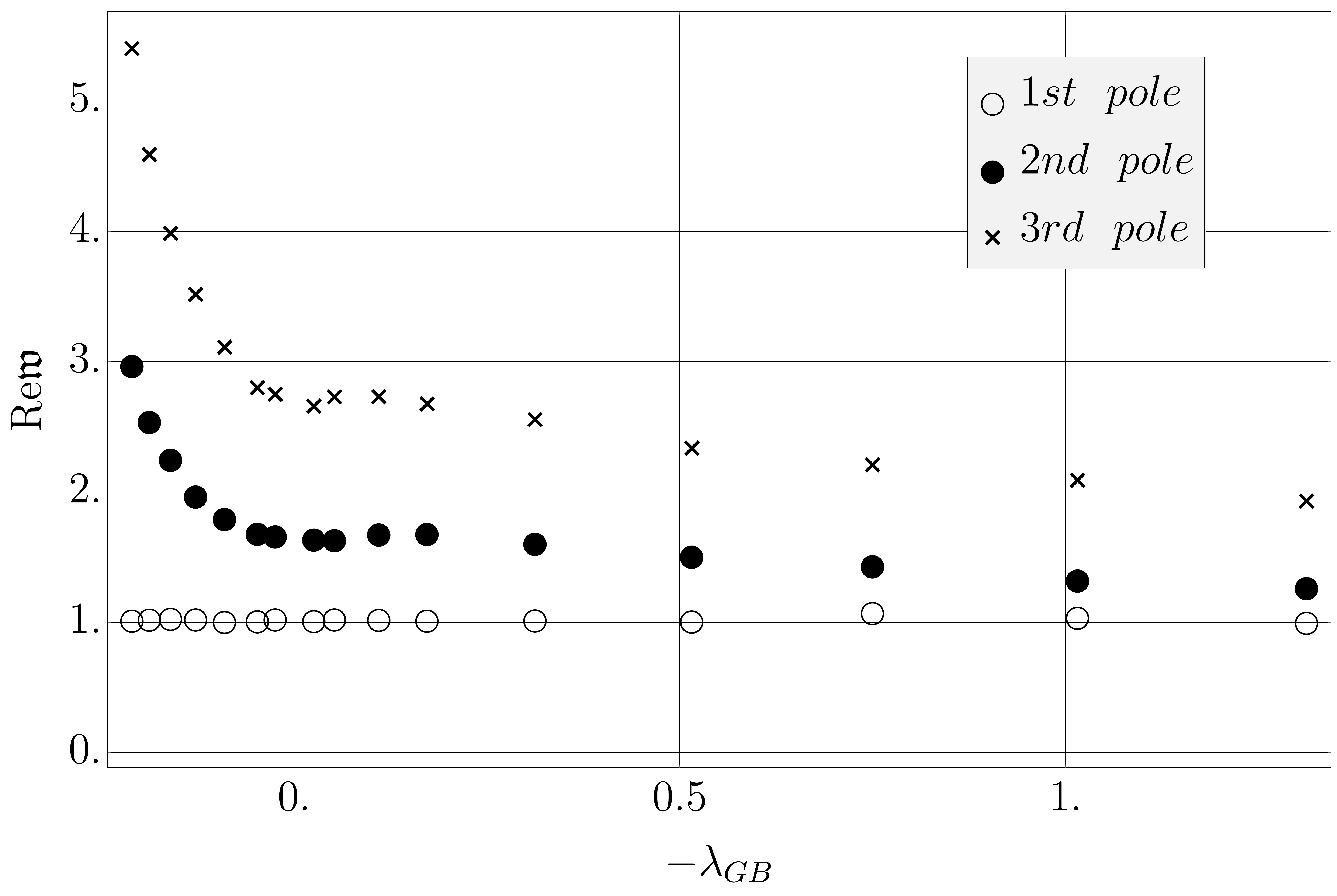}
\includegraphics[width=0.45\linewidth]{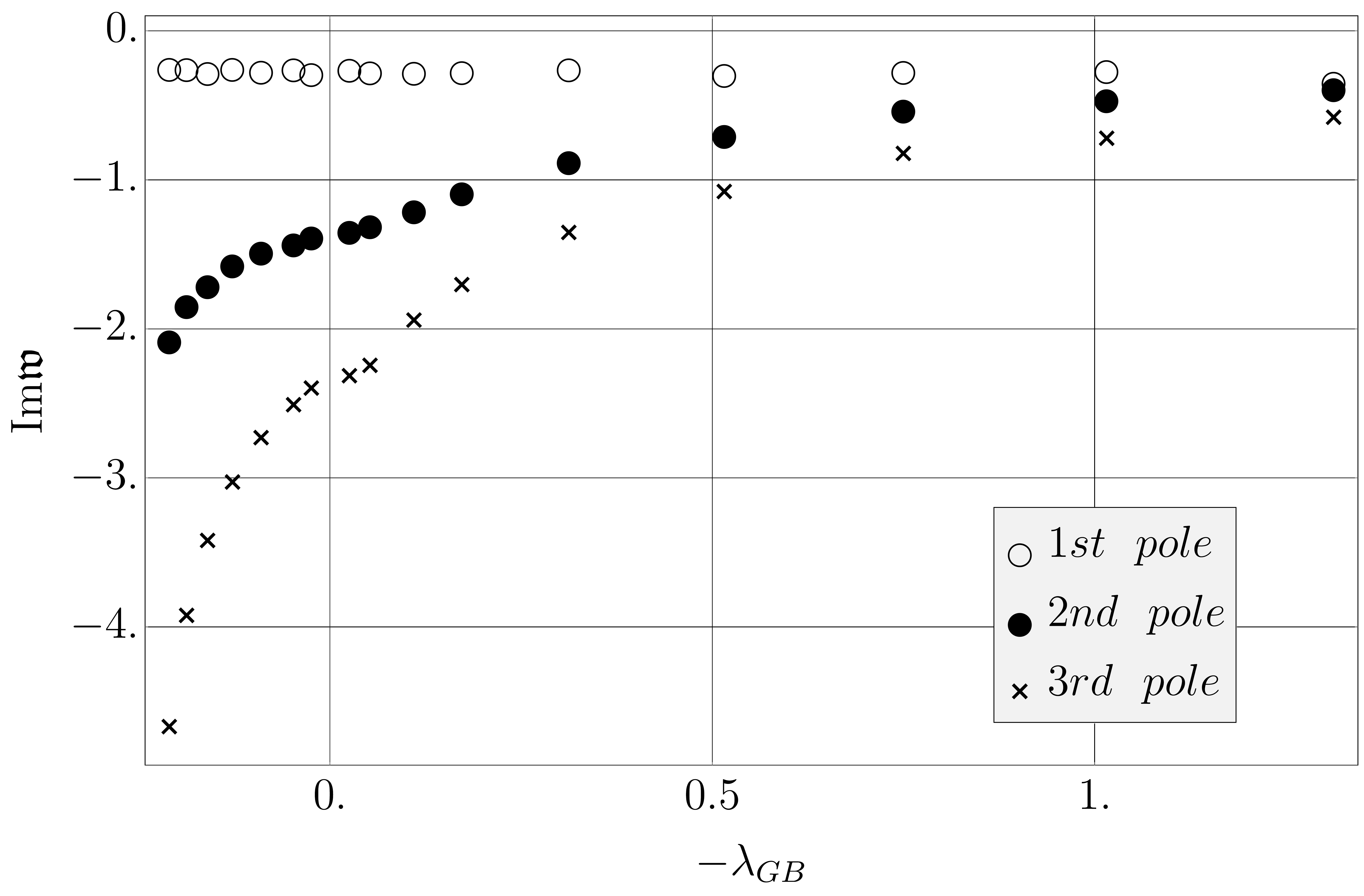}
\caption{Real (left panel) and imaginary (right panel) parts of the top three quasinormal frequencies in the symmetric branches in the shear channel of Gauss-Bonnet at $\qfr = 0.5$.}
\label{fig:GB-Shear-Poles-vs-gamma}
\end{figure}
\subsubsection{Shear channel}
The distribution of the poles in the shear channel is shown in Fig.~\ref{fig:GB-Shear-channel} and the coupling dependence of the real and imaginary parts of the top three poles in the symmetric branches can be seen in Fig.~\ref{fig:GB-Shear-Poles-vs-gamma}. The behavior of the poles in the symmetric branches is qualitatively similar to the one observed in the scalar channel. In the limit  $\lgb \rightarrow 1/4$, the new poles moving up the imaginary axis approach the $\qfr$-independent positions known analytically \cite{GBNesojen},
\begin{align}
\wfr = -2 i \left(1+n_1\right), & &\wfr = -2 i \left(3+n_2\right),
\label{eq:QNMShear}
\end{align}
where $n_1$ and $n_2$ are non-negative integers.
\begin{figure}[ht]
\centering
\begin{subfigure}[t]{0.45\linewidth}
\includegraphics[width=1\linewidth]{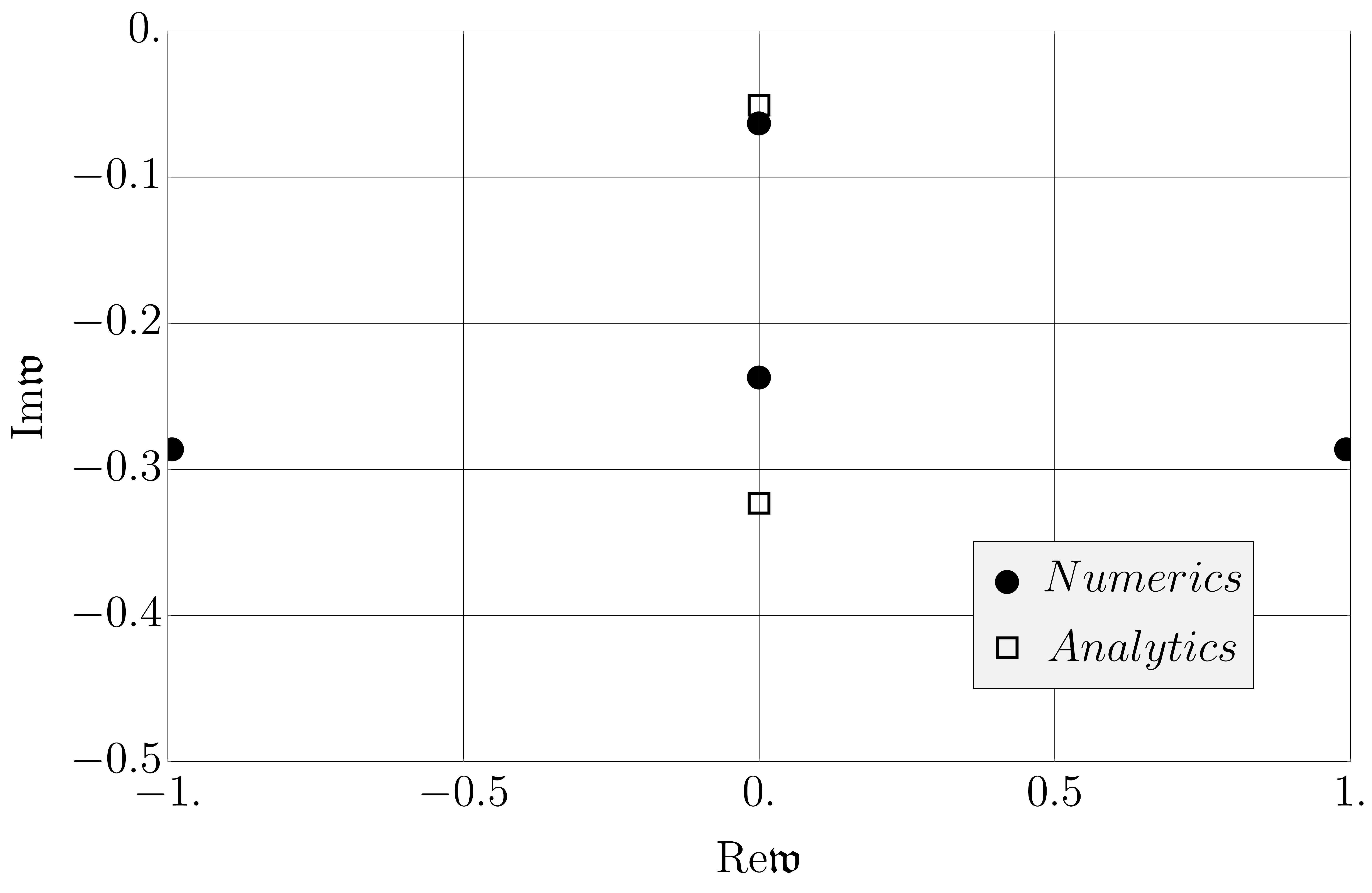}
\end{subfigure}
\qquad
\begin{subfigure}[t]{0.45\linewidth}
\includegraphics[width=1\linewidth]{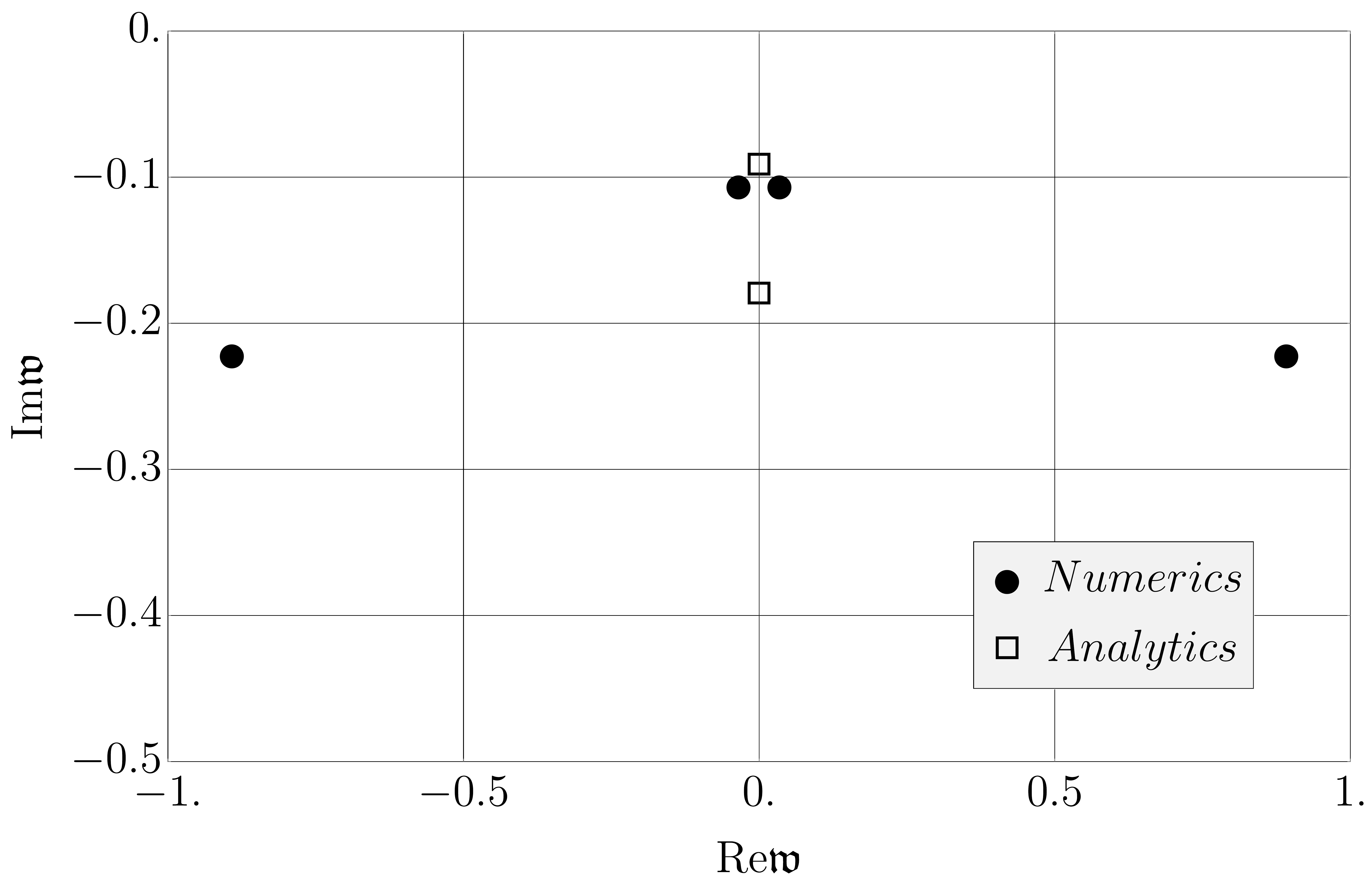}
\end{subfigure}
\\
\begin{subfigure}[b]{0.45\linewidth}
\includegraphics[width=1\linewidth]{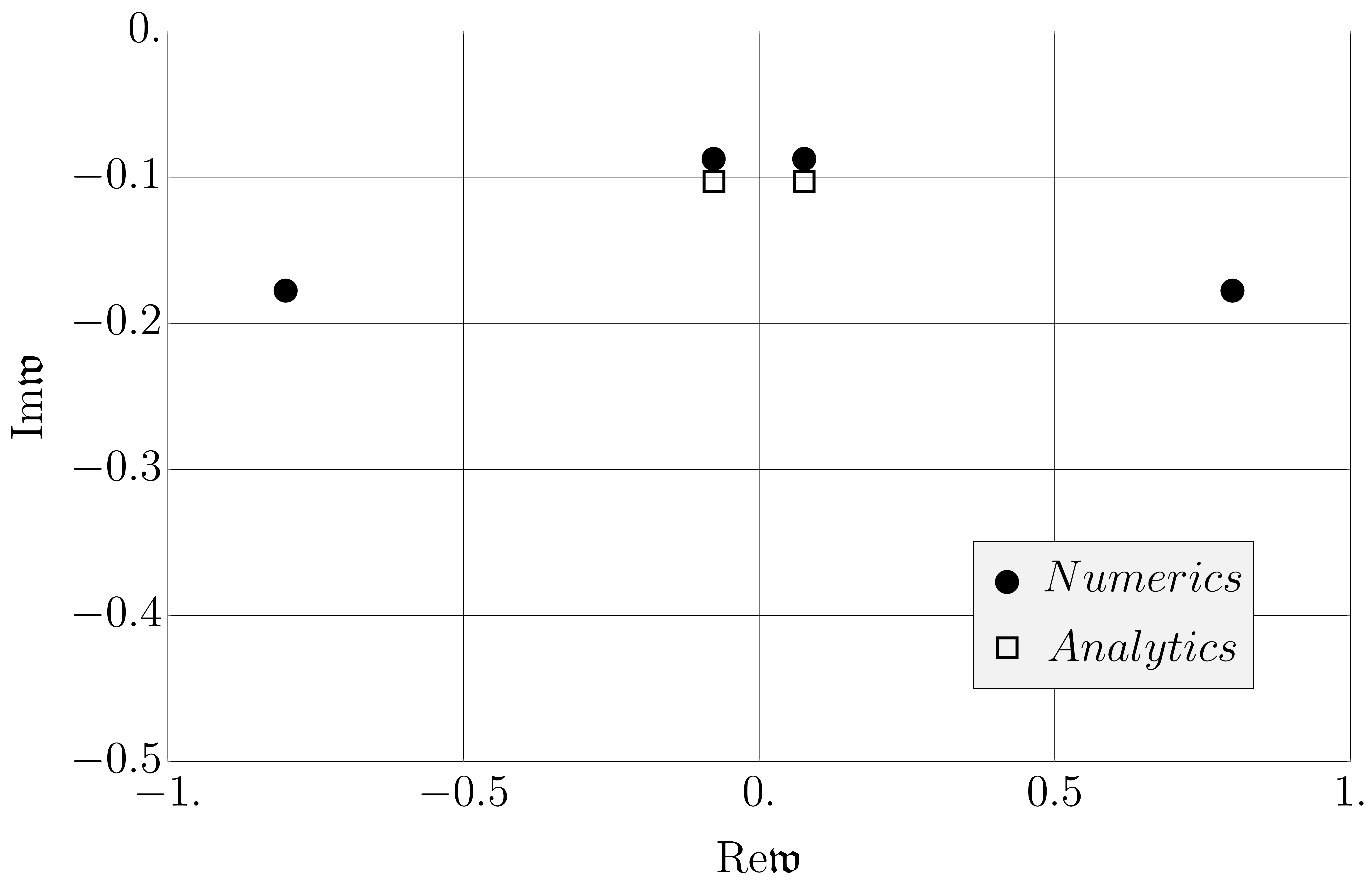}
\end{subfigure}
\qquad
\begin{subfigure}[b]{0.45\linewidth}
\includegraphics[width=1\linewidth]{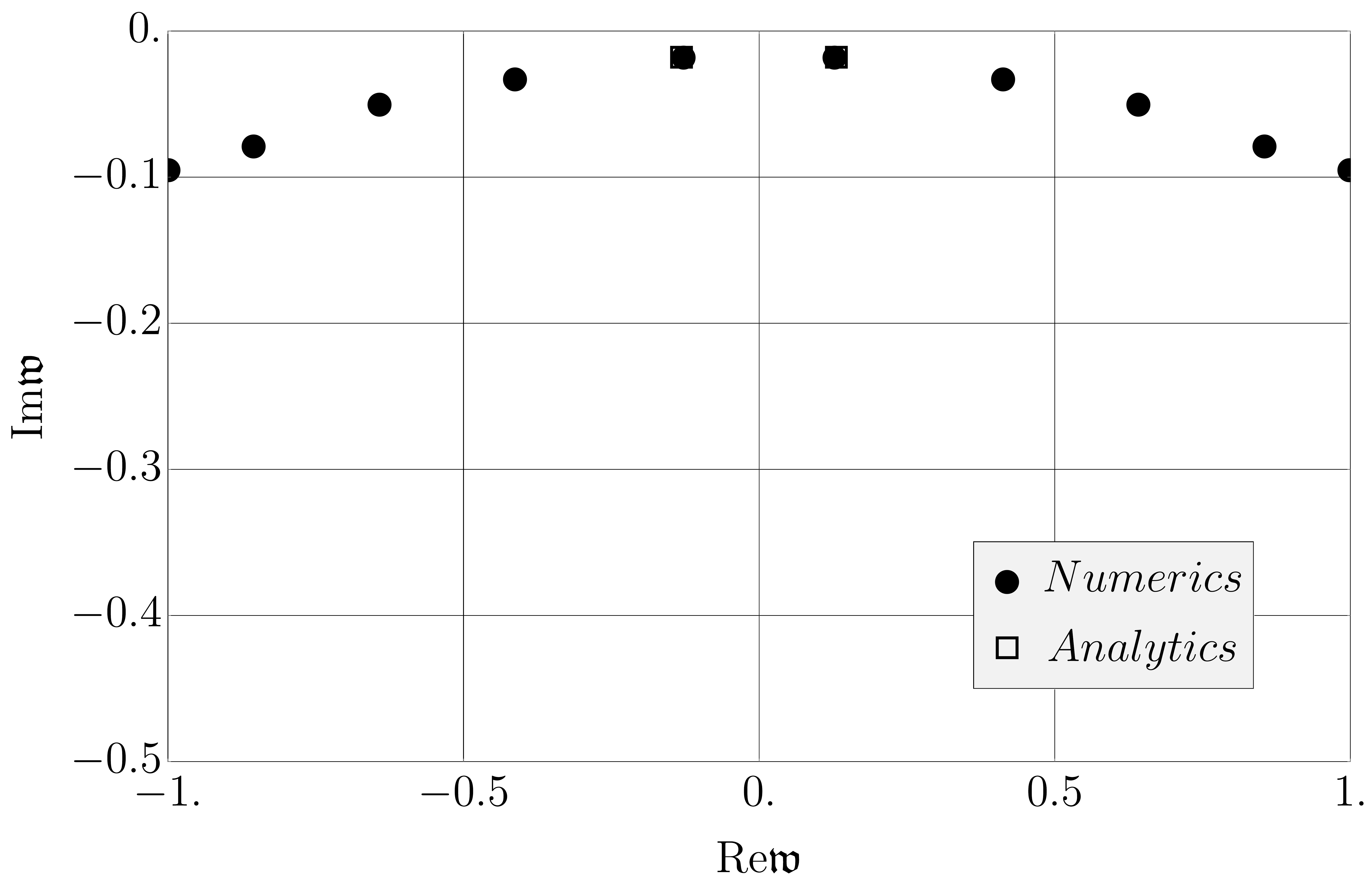}
\end{subfigure}
\caption{Quasinormal modes, close to the origin, in the shear channel of Gauss-Bonnet, for increasing coupling constant and $\qfr=0.1$. From left to right and top to bottom: $\lgb =\{-2.0000,\, -2.8125,\, -3.7500,\, -22.3125\}$.}
\label{fig:GB-Shear-zoom}
\end{figure}
\begin{figure}[ht]
\centering
\includegraphics[width=0.7\linewidth]{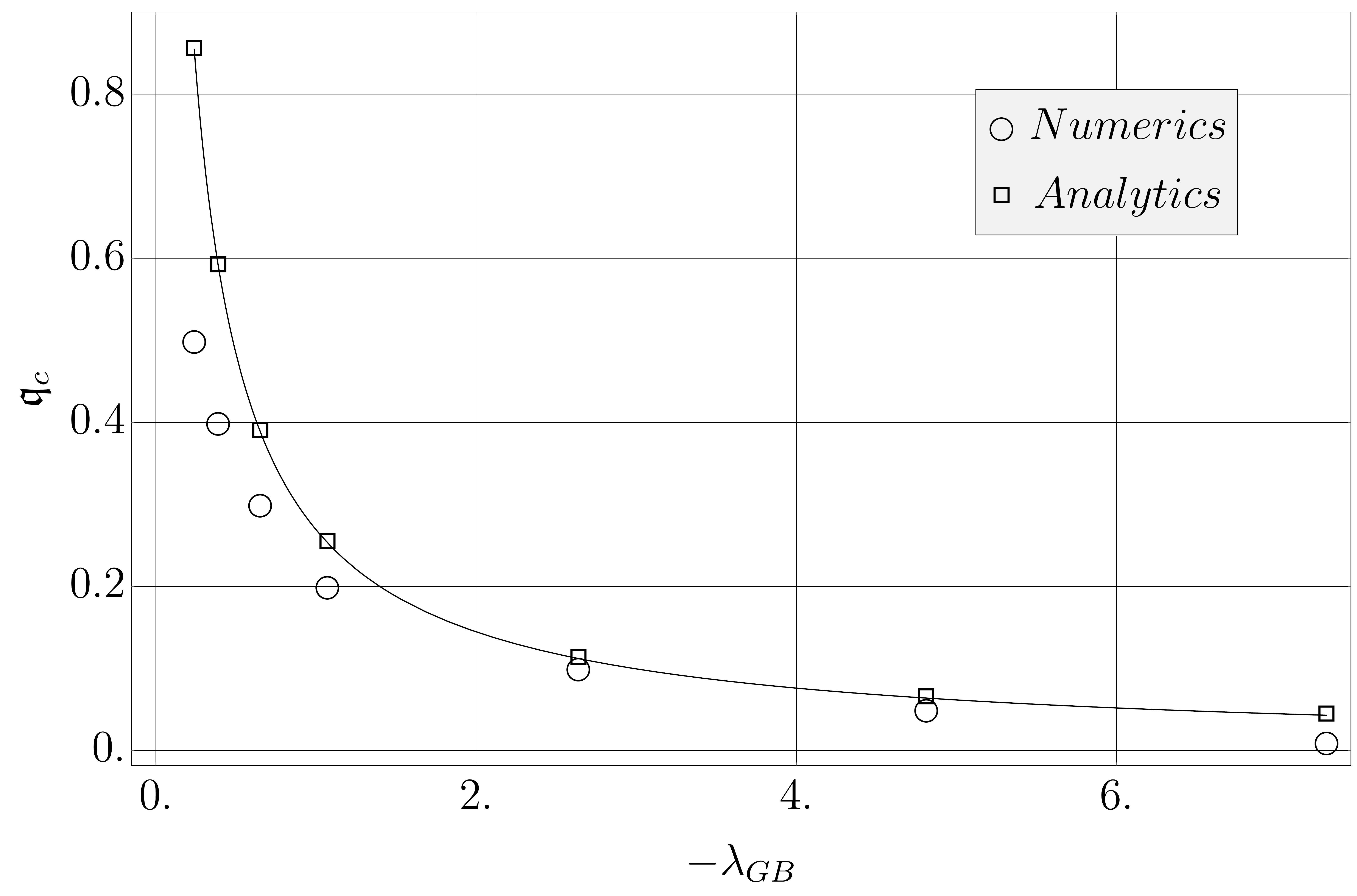}
\caption{Critical values of coupling $\lgb$, limiting the hydrodynamic regime, for the shear channel of Gauss-Bonnet.}
\label{fig:N=GB-Shear-critical}
\end{figure}

A characteristic feature of the shear channel is the presence of the hydrodynamic momentum diffusion pole on the imaginary axis. The dispersion relation for this mode is currently known analytically to quartic order in $q$ \cite{Grozdanov:2015kqa} and is given by
\begin{align}
\omega = - i \frac{\eta}{\varepsilon + P}\, q^2 - i  \left[ \frac{\eta^2\tau_\Pi}{(\varepsilon + P)^2}-\frac{\theta_1}{2(\varepsilon + P)}\right] q^4 + \cdots ,
\label{eq:disp-shear}
\end{align}
where the transport coefficients were defined in Section \ref{sec:SYM}. For Gauss-Bonnet gravity, solving the equation for the shear mode analytically, perturbatively in $\wfr\ll 1$, $\qfr\ll 1$ and non-perturbatively in $\ggb$ and imposing the Dirichlet condition $Z_2(0) = 0$, we find
\begin{align}
\wfr = &- i \frac{\ggb^2}{2}\, \qfr^2 - i \frac{\ggb^3}{16}\, \biggr[ \left(1+\ggb\right) \left(\ggb^2+ 5\ggb - 2 \right) \nn
& - 2 \ggb\, \ln \left[\frac{2 \left(1+\ggb\right)}{\ggb }\right] - 2 \left(2 \ggb^2 + \ggb - 1 \right)\biggr]\, \qfr^4+ \cdots .
\label{eq:wHydroThirdOrderGB}
\end{align}
The full set of non-perturbative first- and second-order hydrodynamic transport coefficients in Gauss-Bonnet theory was computed in \cite{Grozdanov:2015asa}. The coefficients relevant for the dispersion relation \eqref{eq:disp-shear} are given by
\begin{align}
&\eta = s \ggb^2/4 \pi\,, \label{eq:gb-visc} \\
&\tp = \frac{1}{2\pi T} \left[ \frac{1}{4} \left(1+\ggb\right) \left( 5+\ggb - \frac{2}{\ggb}\right) - \frac{1}{2} \ln \frac{2 \left(1+\ggb\right)}{\ggb } \right] . \label{eq:l0}
\end{align}
Thus, the value of the third-order coefficient $\theta_1$ in the Gauss-Bonnet theory can now be read off Eq.~\eqref{eq:wHydroThirdOrderGB}:
\begin{align}
\theta_1 = \frac{\eta}{8\pi^2 T^2} \ggb \left(2 \ggb^2 + \ggb - 1 \right).
\end{align}
In the limit of $\lgb \to 0$ ($\ggb\to 1$), this reproduces the corresponding result for $\mathcal{N}=4$ SYM theory found in Ref.~\cite{Grozdanov:2015kqa},
\begin{align}
\theta_1 = \frac{\eta}{4 \pi^2 T^2} .
\end{align}

The behavior of the momentum diffusion pole depends on whether $\eta/s$ is greater or less than $1/4\pi$. For $\eta/s < 1/4\pi$ ($0<\lgb <1/4$), the pole moves up the imaginary axis relative to its $\lgb = 0 $ position and approaches the origin. It completely disappears from the spectrum at $\lgb = 1/4$ \cite{GBNesojen}. For $\eta/s > 1/4\pi$ ($-\infty<\lgb <0$), its behavior is qualitatively similar to the one observed in $\mathcal{N}=4$ SYM: it moves down the imaginary axis and collides with the top new pole moving up the axis from complex infinity at which point the hydrodynamic description seemingly fails. Then the two poles move off the imaginary axis into the complex plane. For sufficiently large values of $|\lgb|$, this phenomenon happens in the range of small $\wfr$, $\qfr$ and thus can be approximated analytically (e.g. for $\lgb \sim -3$, the merger of the poles occurs at $|\wfr|\sim 0.1$, $\qfr \sim 0.1$). Solving the shear mode equation of motion perturbatively in $\wfr\ll 1$, $\qfr \ll 1$, we find a pair of quasinormal frequencies
\begin{align}
&\wfr_1 = \frac{-2 i + \sqrt{2 (\ggb -1) (\ggb +3) \ggb ^2 \qfr^2+4\ggb ^2 \qfr^2 \ln \left(\frac{2}{\ggb+1}\right)-4}}{\ggb  (\ggb +2) -3+2 \ln \left(\frac{2}{\ggb+1}\right)}, \label{eq:FullGBShearAnaly1} \\
&\wfr_2 = \frac{- 2 i -\sqrt{2 (\ggb -1) (\ggb +3) \ggb ^2 \qfr^2+4\ggb ^2 \qfr^2 \ln \left(\frac{2}{\ggb+1}\right)-4}}{\ggb  (\ggb +2) -3+ 2 \ln \left(\frac{2}{\ggb+1}\right)} \label{eq:FullGBShearAnaly2}
\end{align}
whose motion in the complex plane approximates the numerical observations quite well (see Fig.~\ref{fig:GB-Shear-zoom}). Expanding the above expressions for $\wfr_1$ and $\wfr_2$ to second order in $\qfr$, we find the standard hydrodynamic pole of Eq.~\eqref{eq:wHydroThirdOrderGB}
\begin{align}
\wfr_1 = - \frac{1}{2} i \ggb^2 \qfr^2 + \ldots\, \label{eq:w1ShearExpq}
\end{align}
and the new gapped pole
\begin{align}
\wfr_2 =  \wfr_{\mathfrak{g}}^{GB} + \frac{1}{2} i \ggb^2 \qfr^2 + \ldots\,, \label{eq:w2ShearExpq}
\end{align}
where the gap $\wfr_{\mathfrak{g}}^{GB}$ is identical to the one in Eq.~\eqref{eq:GB-scalar-gap}. The behavior of the poles is qualitatively the same as in the $\CN=4$ SYM theory. The diffusion pole moves down the imaginary axis while the new gapped pole moves up as $\lgb$ decreases from $0$ towards negative values. Then the two poles collide at some $\qfr$-dependent value of $\lgb^c = \lgb^c (\qfr)$ and move off the axis. An analytical approximation for this dependence (or more conveniently, for $\qfr_c = \qfr_c(\ggb)$) can be found from the condition $\wfr_1 (\qfr_c) = \wfr_2 (\qfr_c)$. We interpret this condition as the condition indicating inadequacy of hydrodynamic description for $\qfr > \qfr_c (\lgb)$. Equating the expressions (\ref{eq:w1ShearExpq}) and (\ref{eq:w1ShearExpq}), we obtain
\begin{align}\label{eq:qcShearGBqExp}
\qfr_{c} = \frac{2}{\ggb \sqrt{\ggb  (\ggb +2) - 3 + 2 \ln \left( \frac{2}{\ggb + 1 } \right)}} \sim \frac{1}{2 |\lgb|}.
\end{align}
Note that if we used instead the un-expanded Eqs.~(\ref{eq:FullGBShearAnaly1}), (\ref{eq:FullGBShearAnaly2}), we would find  $\qfr_{c}^{\text{(un-exp)}} = \qfr_{c}/\sqrt{2}$. The discrepancy is due to the additional $\qfr$ corrections not captured by Eqs.~(\ref{eq:w1ShearExpq}), (\ref{eq:w1ShearExpq}). The dependence $\qfr_c = \qfr_c (\lgb)$ obtained numerically as well as the analytic approximation \eqref{eq:qcShearGBqExp} are shown in Fig.~\ref{fig:N=GB-Shear-critical}.
\subsubsection{Sound channel}
\begin{figure}[ht]
\centering
\begin{subfigure}[t]{0.45\linewidth}
\includegraphics[width=1\linewidth]{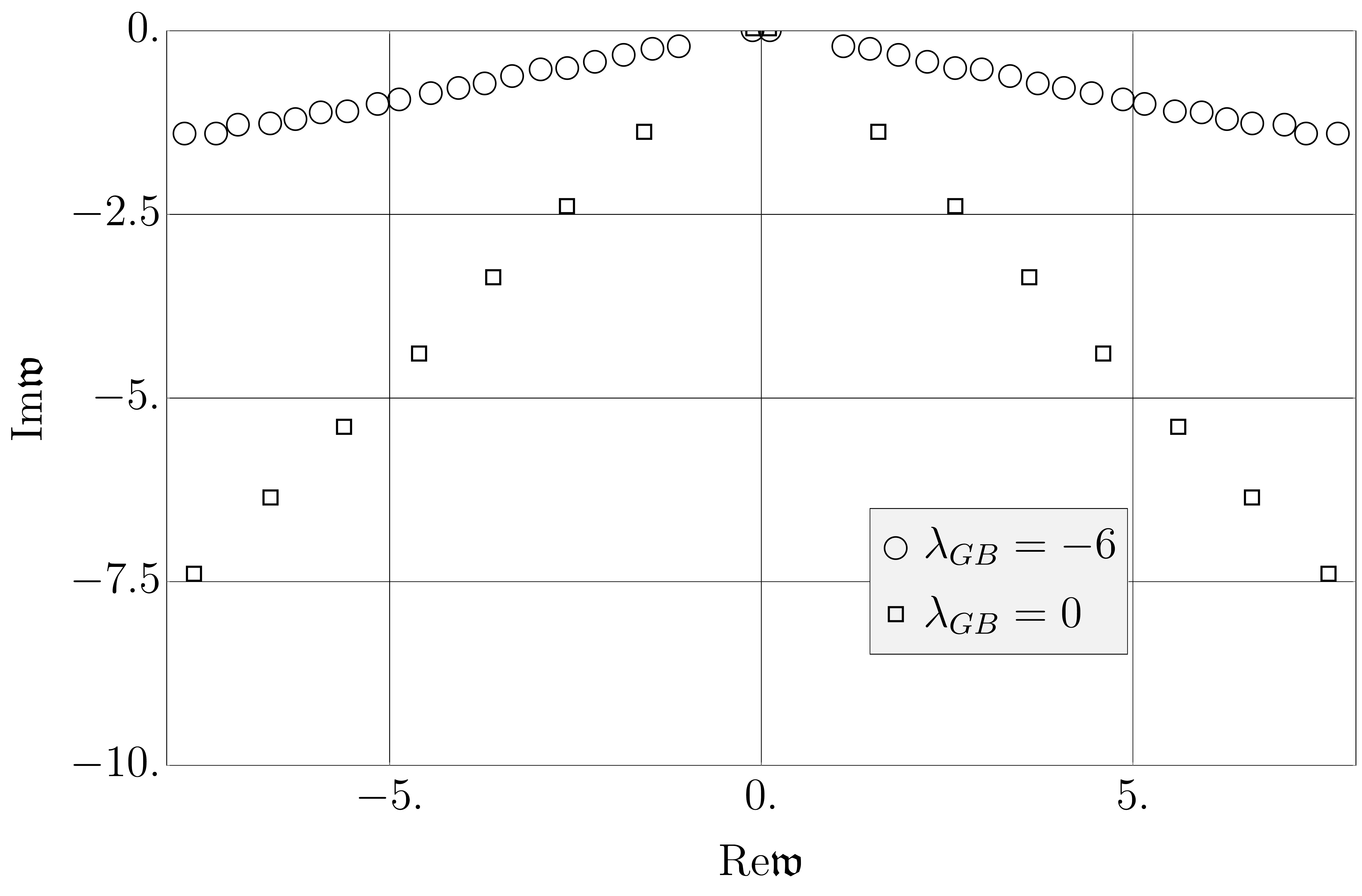}
\end{subfigure}
\qquad
\begin{subfigure}[t]{0.45\linewidth}
\includegraphics[width=1\linewidth]{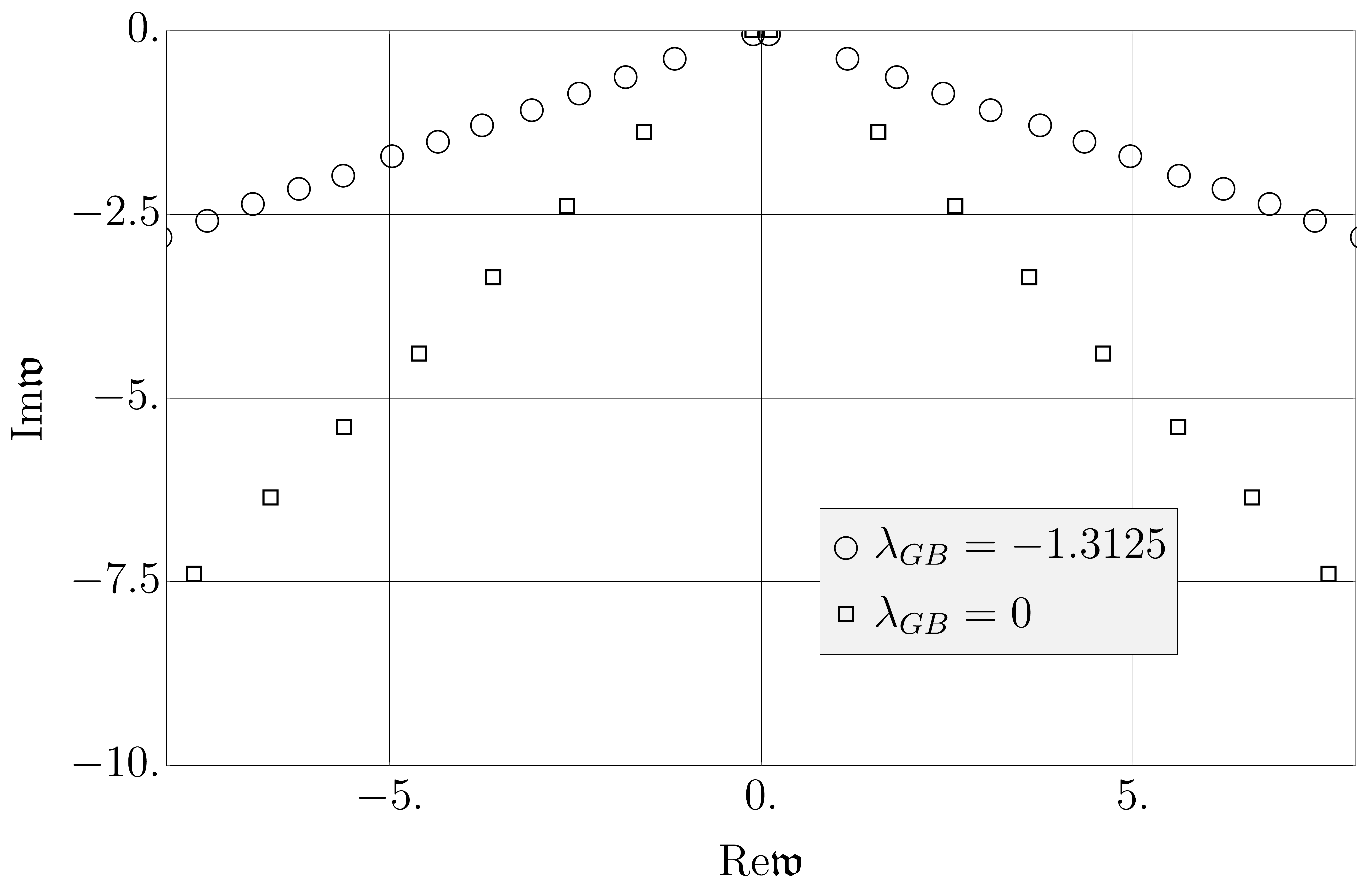}
\end{subfigure}
\\
\begin{subfigure}[b]{0.45\linewidth}
\includegraphics[width=1\linewidth]{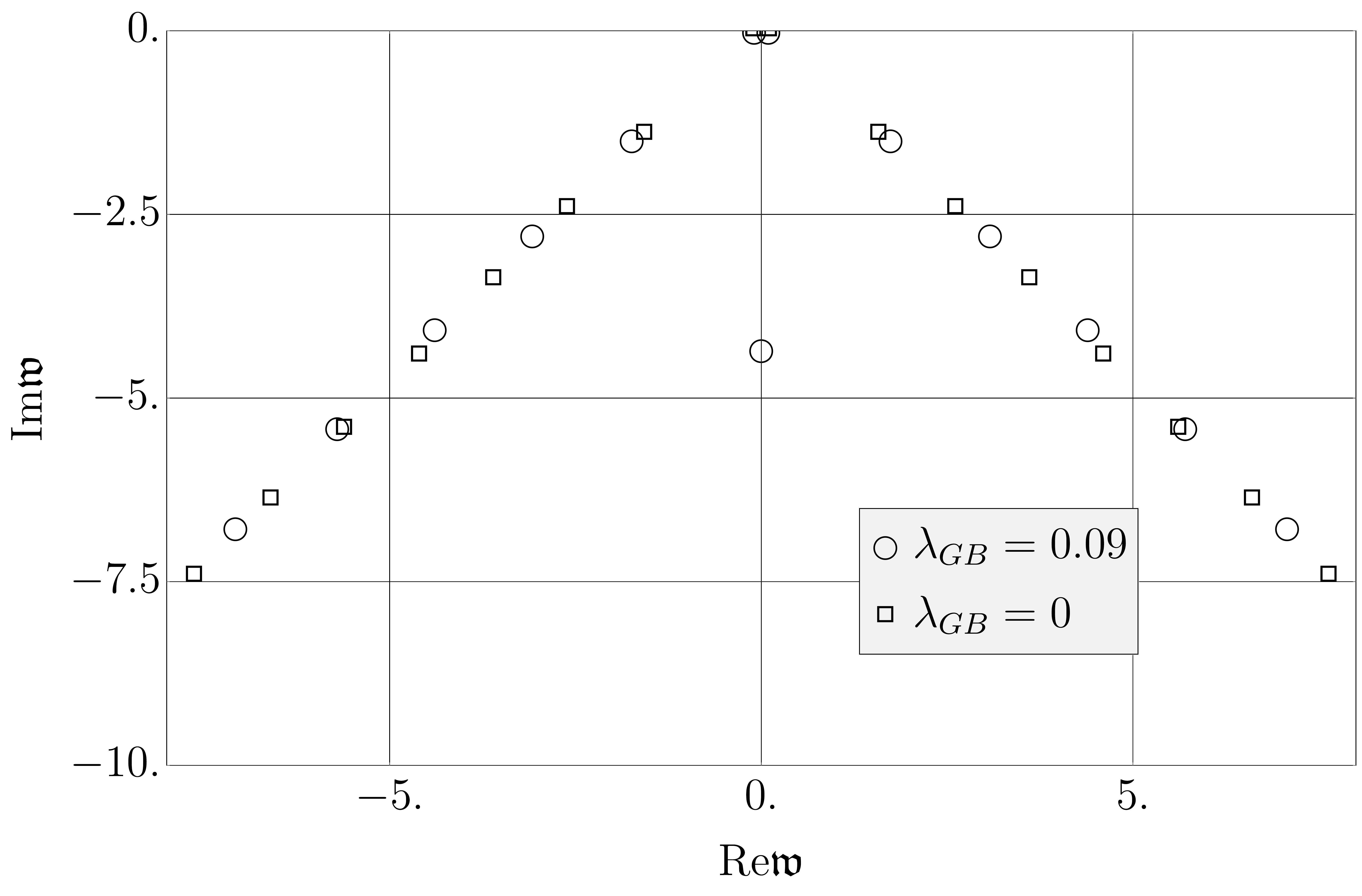}
\end{subfigure}
\qquad
\begin{subfigure}[b]{0.45\linewidth}
\includegraphics[width=1\linewidth]{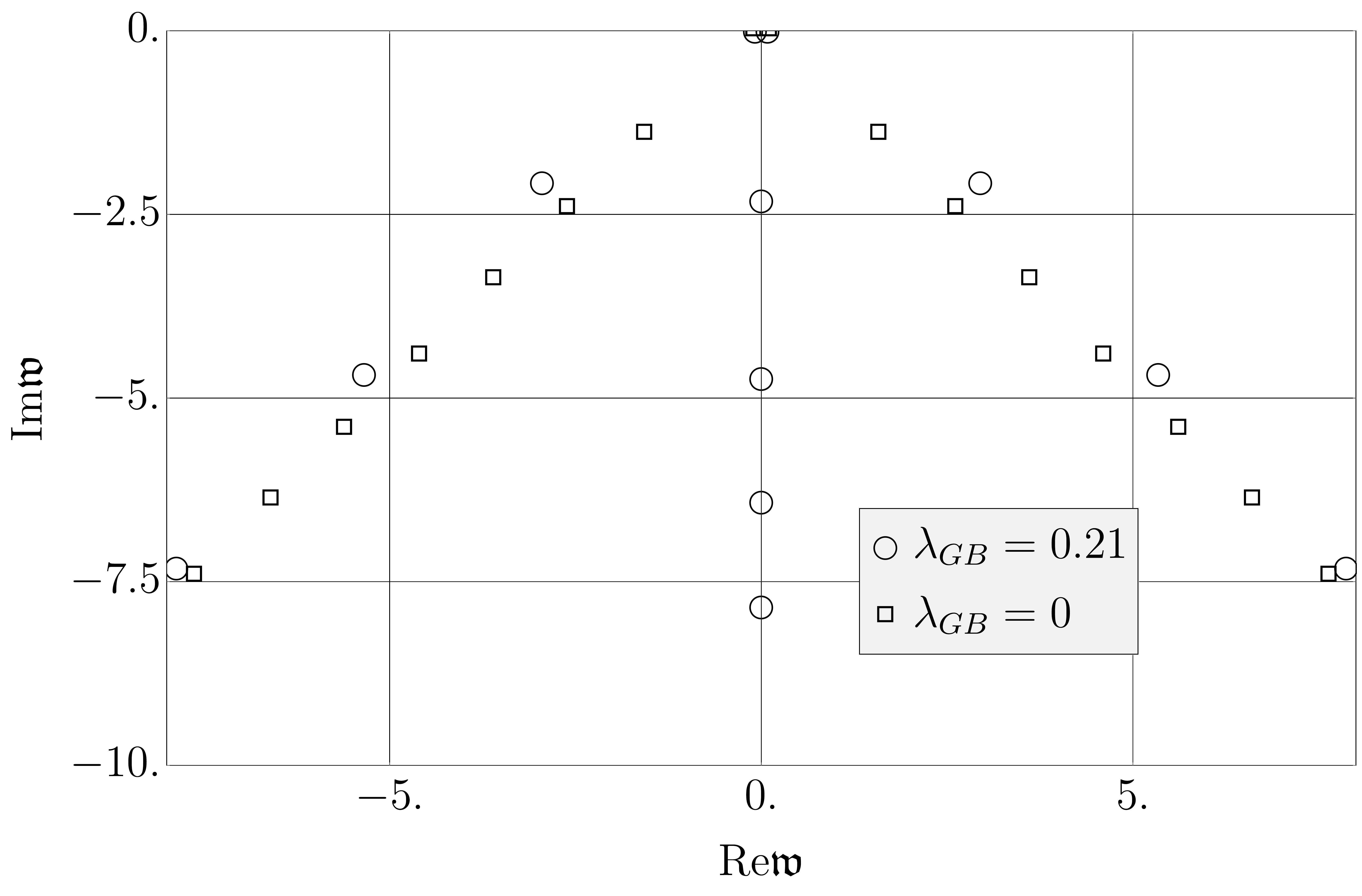}
\end{subfigure}
\caption{Quasinormal spectrum (shown by circles) of the sound channel metric perturbations in Gauss-Bonnet gravity for various values of the coupling $\lgb$ and $\qfr=0.1$. From top left: $\lgb = \{-6,\, -1.3125, \, 0.09,\, 0.21\} $. For comparison, the spectrum at $\lgb = 0$ is shown by squares.}
\label{fig:GB-Sound-channel}
\end{figure}

\begin{figure}[ht]
\centering
\includegraphics[width=0.45\linewidth]{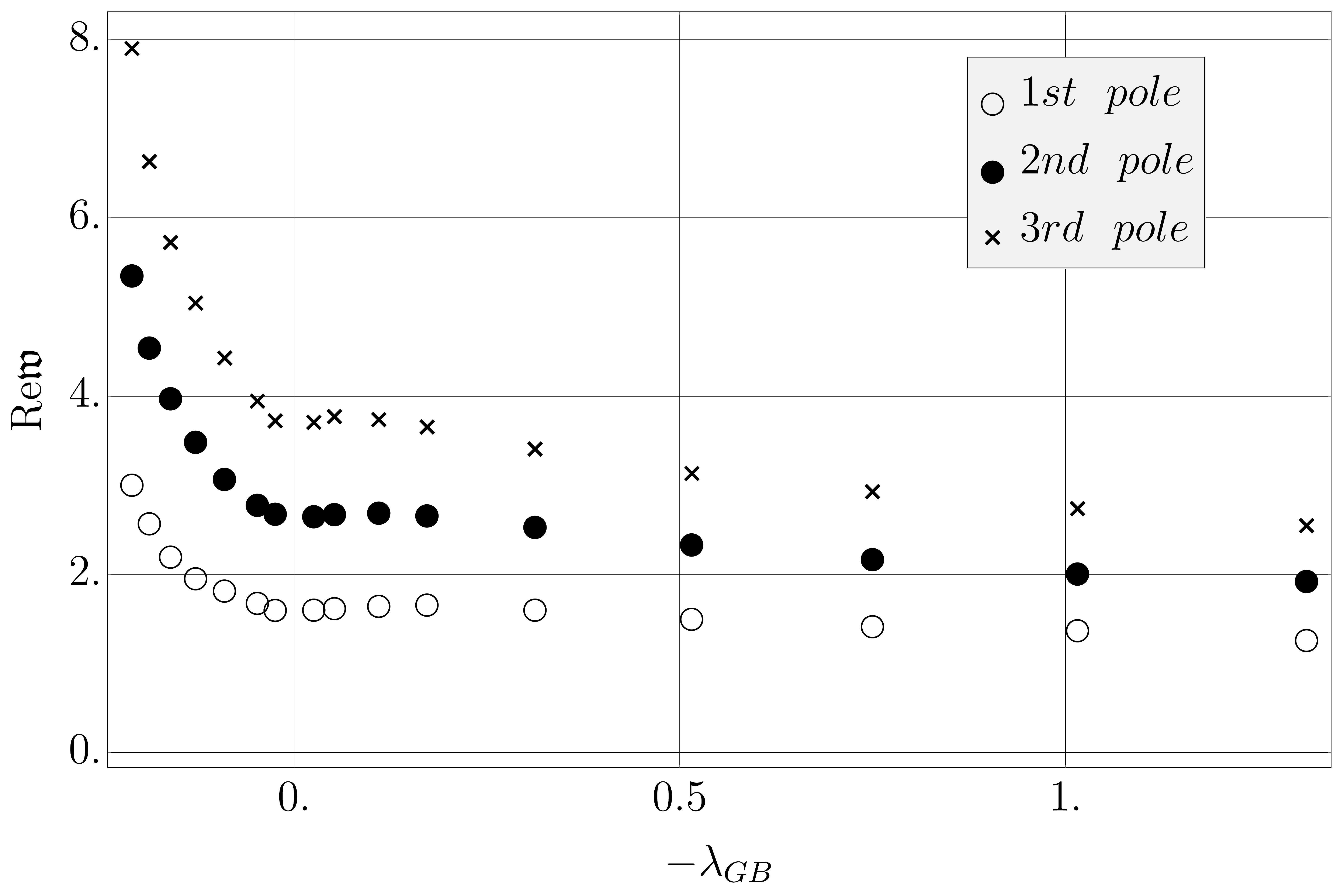}
\includegraphics[width=0.45\linewidth]{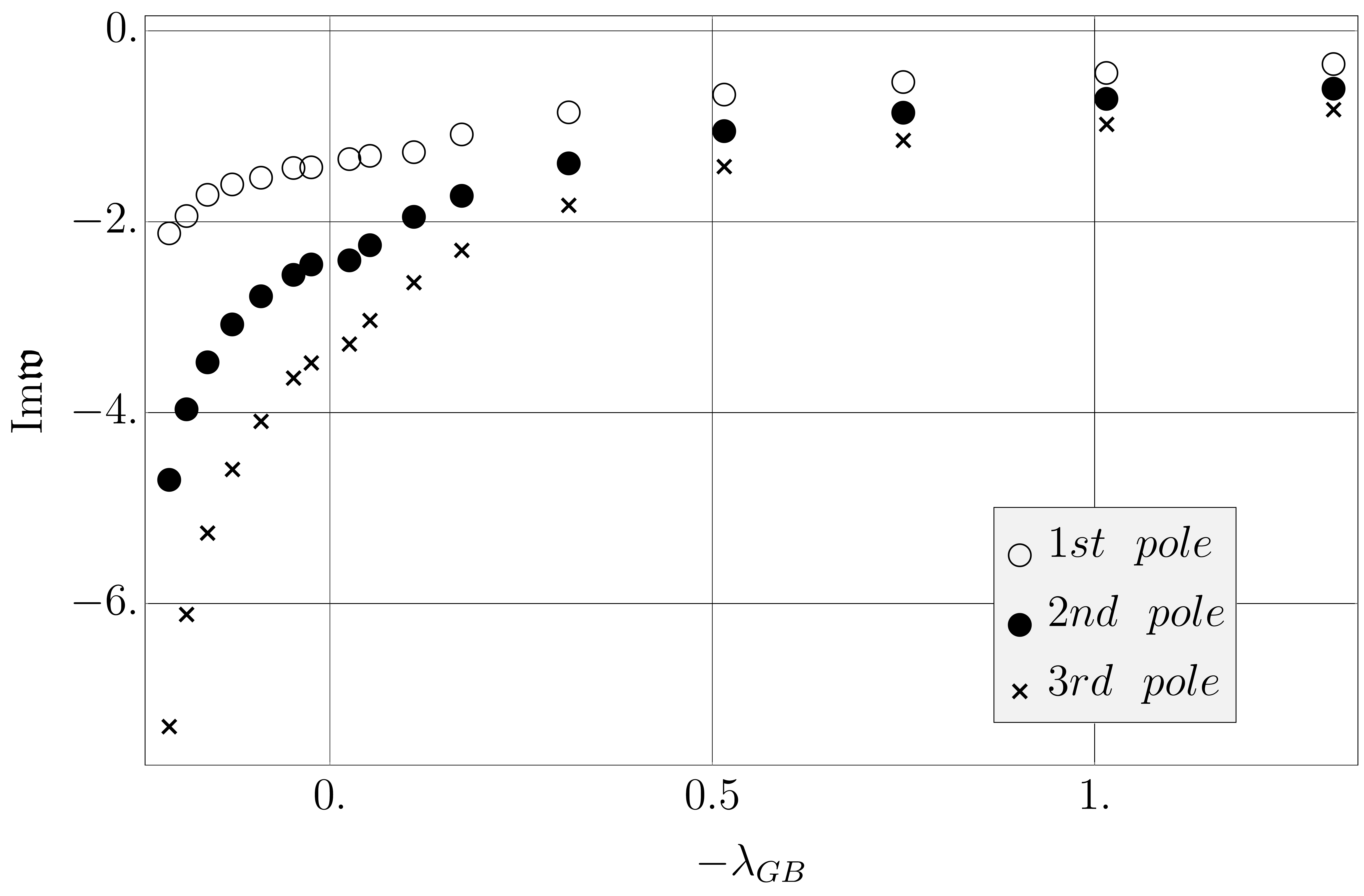}
\caption{Real (left panel) and imaginary (right panel) parts of the top three quasinormal frequencies in the symmetric branches in the sound channel of Gauss-Bonnet at $\qfr = 0.5$.}
\label{fig:GB-Sound-Poles-vs-gamma}
\end{figure}

The poles in the sound channel are shown in Fig.~\ref{fig:GB-Sound-channel} and the behavior of the real and imaginary parts of the three leading non-hydrodynamic poles in the symmetric branches is demonstrated in Fig.~\ref{fig:GB-Sound-Poles-vs-gamma}. We observe the same features of the coupling dependence of the spectrum as in the other channels. The two symmetric branches lift up from their $\lgb=0$ positions, moving swiftly towards the real axis and becoming more dense in the case of $\eta/s>1/4\pi$ ($\lgb <0$) and moving only slightly, becoming more sparse and apparently disappearing from the finite complex plane for $\eta/s<1/4\pi$ ($0<\lgb <1/4$).
There are new gapped poles rising up the imaginary axis regardless of the sign of $\lgb$. For $\eta/s<1/4\pi$ ($0<\lgb <1/4$), they reach the asymptotic values
\begin{align}
\wfr = -i \left(4+2n_1 - \sqrt{4+\qfr^2}\right) , & &\wfr = -i \left(4+2n_2 + \sqrt{4+\qfr^2}\right) ,
\label{eq:QNMSound}
\end{align}
where $n_1$ and $n_2$ are non-negative integers, in the zero viscosity limit $\lgb \rightarrow 1/4$. Here the modes \eqref{eq:QNMSound} are the exact quasinormal frequencies at $\lgb = 1/4$  \cite{GBNesojen}.
\begin{figure}[ht]
\centering
\begin{subfigure}[t]{0.45\linewidth}
\includegraphics[width=1\linewidth]{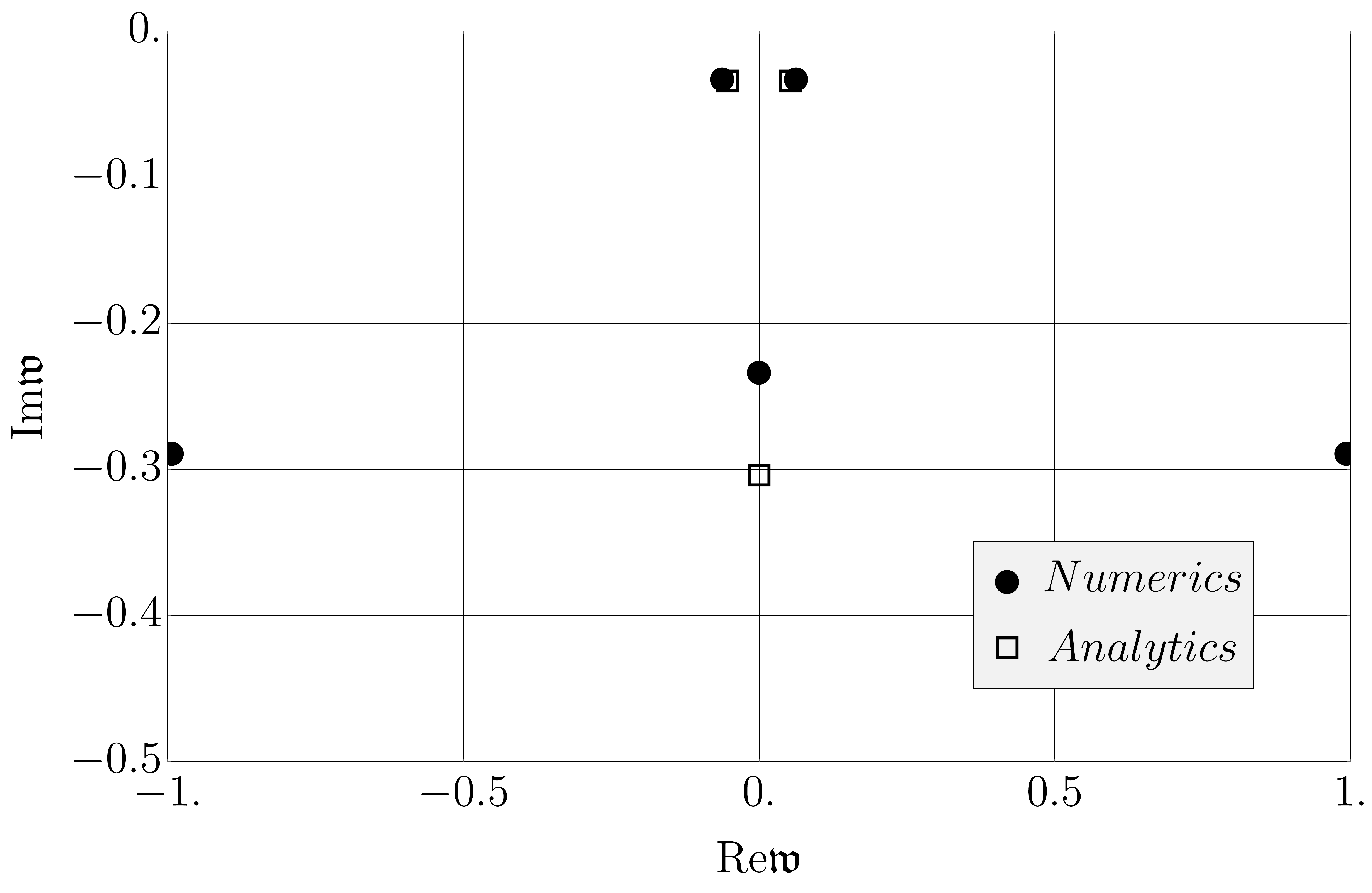}
\end{subfigure}
\qquad
\begin{subfigure}[t]{0.45\linewidth}
\includegraphics[width=1\linewidth]{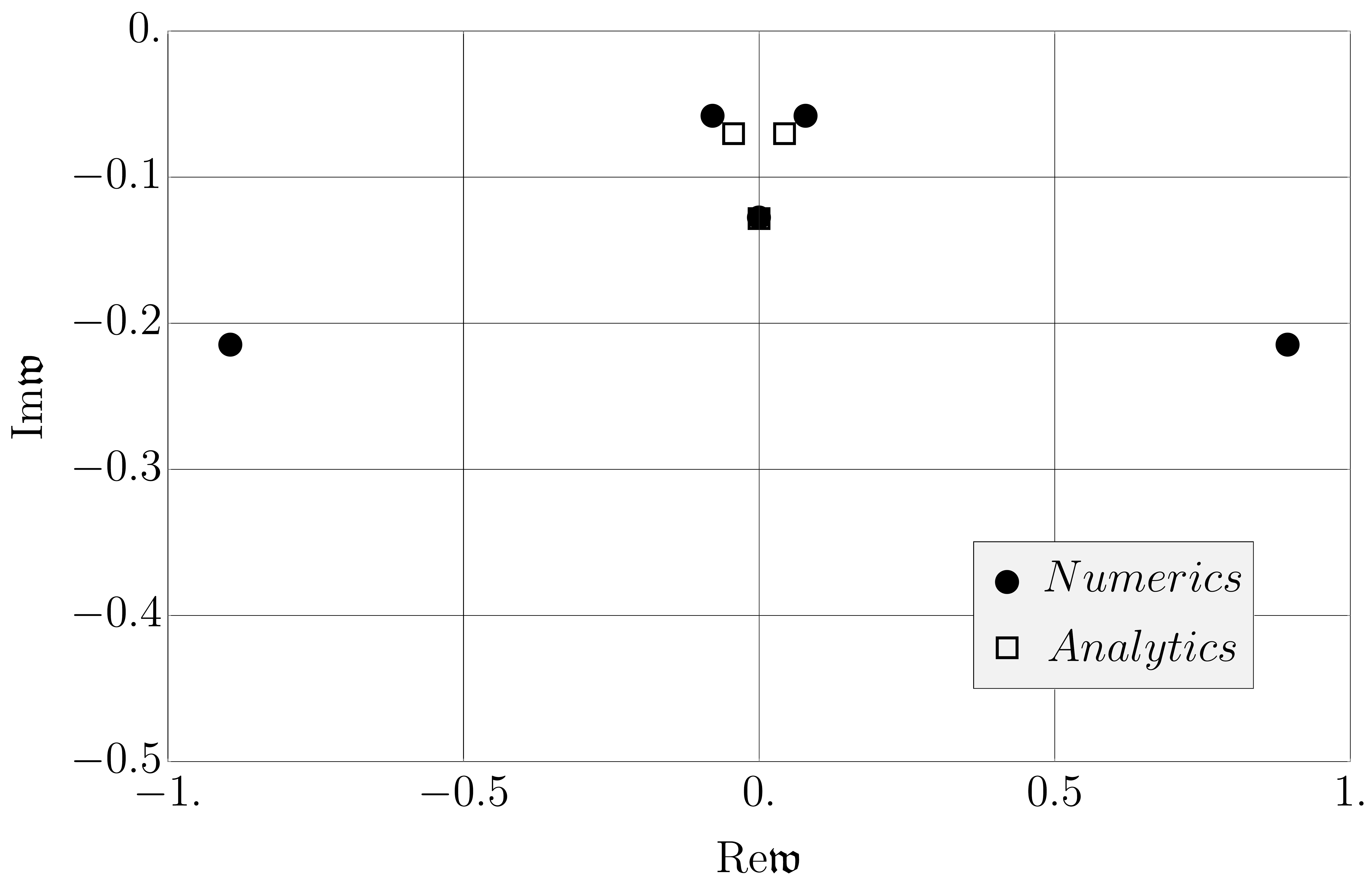}
\end{subfigure}
\\
\begin{subfigure}[b]{0.45\linewidth}
\includegraphics[width=1\linewidth]{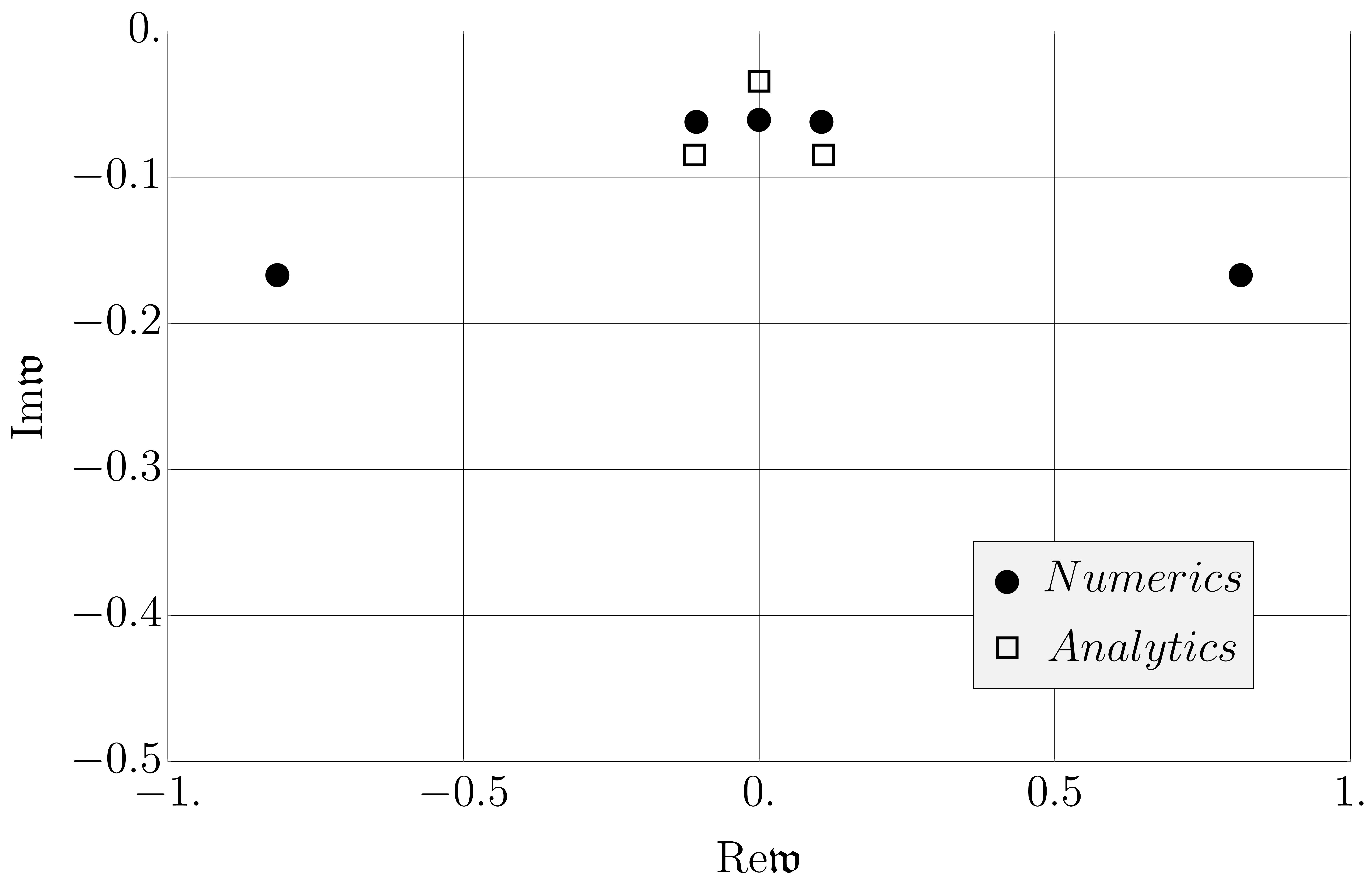}
\end{subfigure}
\qquad
\begin{subfigure}[b]{0.45\linewidth}
\includegraphics[width=1\linewidth]{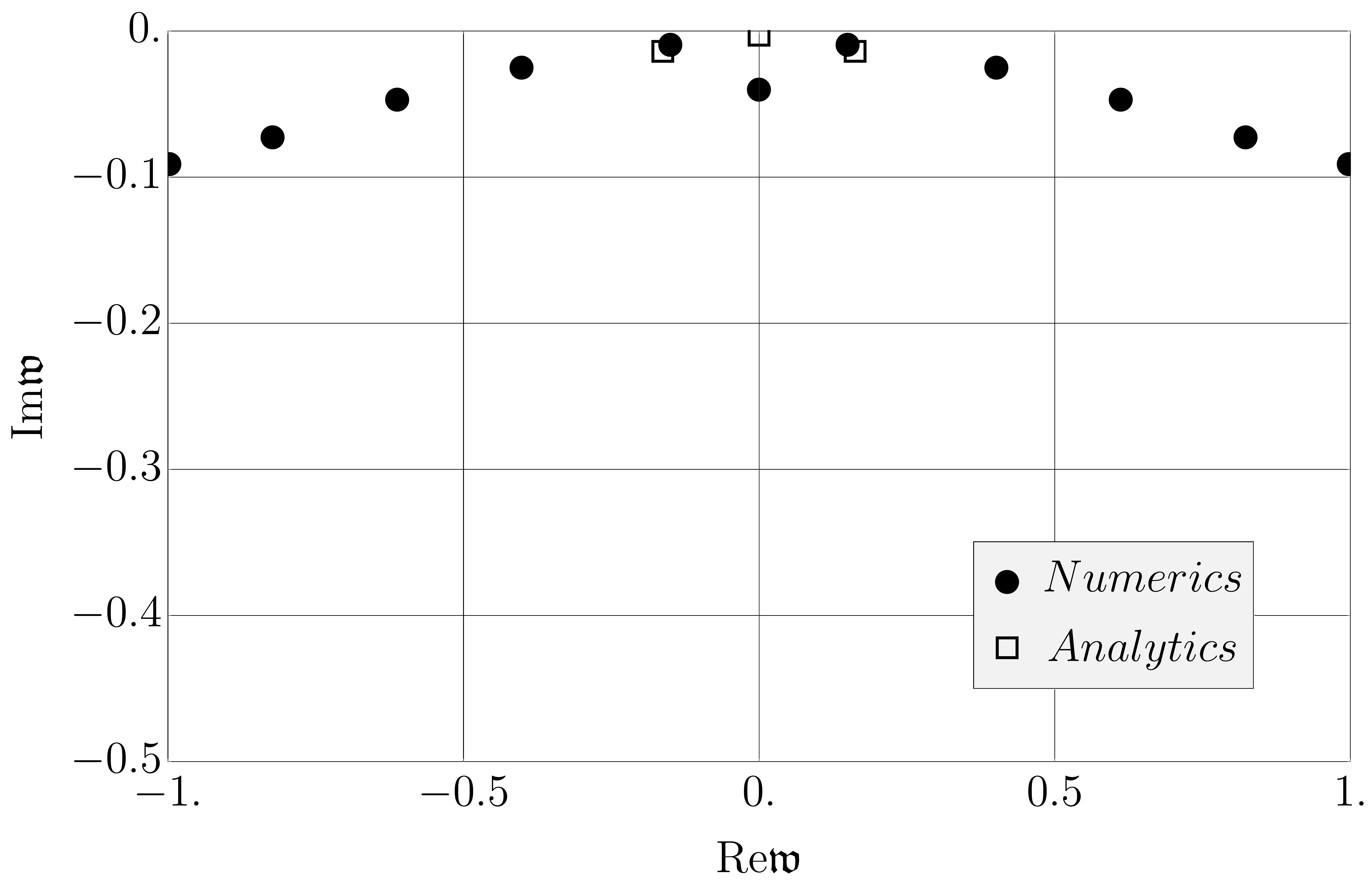}
\end{subfigure}
\caption{Quasinormal modes, close to the origin, in the sound channel of the Gauss-Bonnet theory, for increasing coupling constant and $\qfr=0.1$. From left to right and top to bottom: $\lgb =\{-2.0000,\, -2.8125,\, -3.7500,\, -24.7500\}$.}
\label{fig:GB-Sound-zoom}
\end{figure}
In the regime $\eta/s>1/4\pi$ ($\lgb <0$), the top new gapped pole moving up the imaginary axis with $|\lgb|$ increasing gradually approaches the level of the two symmetric sound mode poles and becomes aligned with them (see Fig.~\ref{fig:GB-Sound-zoom}). For larger values of $|\lgb|$, all three poles move closer to the real axis, with the sound poles now becoming parts of the symmetric branches. When the three poles are close to the origin, one can try to build an analytic approximation by solving the equation for $\CZ_3$ perturbatively in $\wfr \ll 1$, $\qfr \ll 1$. The Dirichlet condition ($Z_3 (0) = 0$) then gives the equation
\begin{align}
&9 \ggb ^2 \qfr^2 \wfr -3 \ggb ^2 \wfr ^3+2 \ggb  \qfr^2 \wfr -2 \qfr^2 \wfr  \ln (\ggb+1)-6 \ggb  \wfr ^3+6 \wfr ^3 \ln (\ggb+1 ) \nn
&-3 \qfr^2 \wfr+\qfr^2 \wfr  \ln 4+4 i \qfr^2+9 \wfr^3-3 \wfr ^3 \ln 4 -12 i \wfr ^2 = 0.
\label{eq:PolySoundGB}
\end{align}
The three roots, $\wfr_{1,2,3}$, can be found analytically, but the expressions are too cumbersome to present here. Their expansions in $\qfr$ to quadratic order are given by
\begin{align}
&\wfr_{1,2} = \pm \frac{1}{\sqrt{3}} \qfr - \frac{1}{3} i \ggb^2\qfr^2 + \ldots \, , \label{eq:FullGBSoundAnaly1}  \\
&\wfr_3 = \wfr_{\mathfrak{g}}^{GB} + \frac{2}{3} i \ggb^2 q^2 + \ldots\, ,
\label{eq:FullGBSoundAnaly2}
\end{align}
where the gap $ \wfr_{\mathfrak{g}}^{GB}$ is the same as in the scalar and shear channels (Eqs.~\eqref{eq:GB-scalar-gap} and \eqref{eq:w2ShearExpq}, respectively). The poles \eqref{eq:FullGBSoundAnaly1} correspond to the sound wave modes.

Defining the critical momentum $\qfr=\qfr_c(\ggb)$ as the one at which the hydrodynamic expansion no longer serves as an adequate description of the low-energy limit of the theory, we may choose the equation
$\im [\wfr_1 (\qfr_c) ]  =  \im [\wfr_2 (\qfr_c) ]  = \im [\wfr_3 (\qfr_c) ] $ to represent such a condition. Solving this for $\qfr_c(\ggb)$, we find exactly the same function \eqref{eq:qcShearGBqExp} as in the shear channel. Note, however, that the agreement between our numerical results and the analytic approximation is less satisfactory than in the shear channel (see Fig.~\ref{fig:GB-Sound-zoom}), apparently due to a stronger $\qfr$ dependence in the sound channel.
\begin{figure}[ht]
\centering
\begin{subfigure}[b]{0.45\linewidth}
\includegraphics[width=1\linewidth]{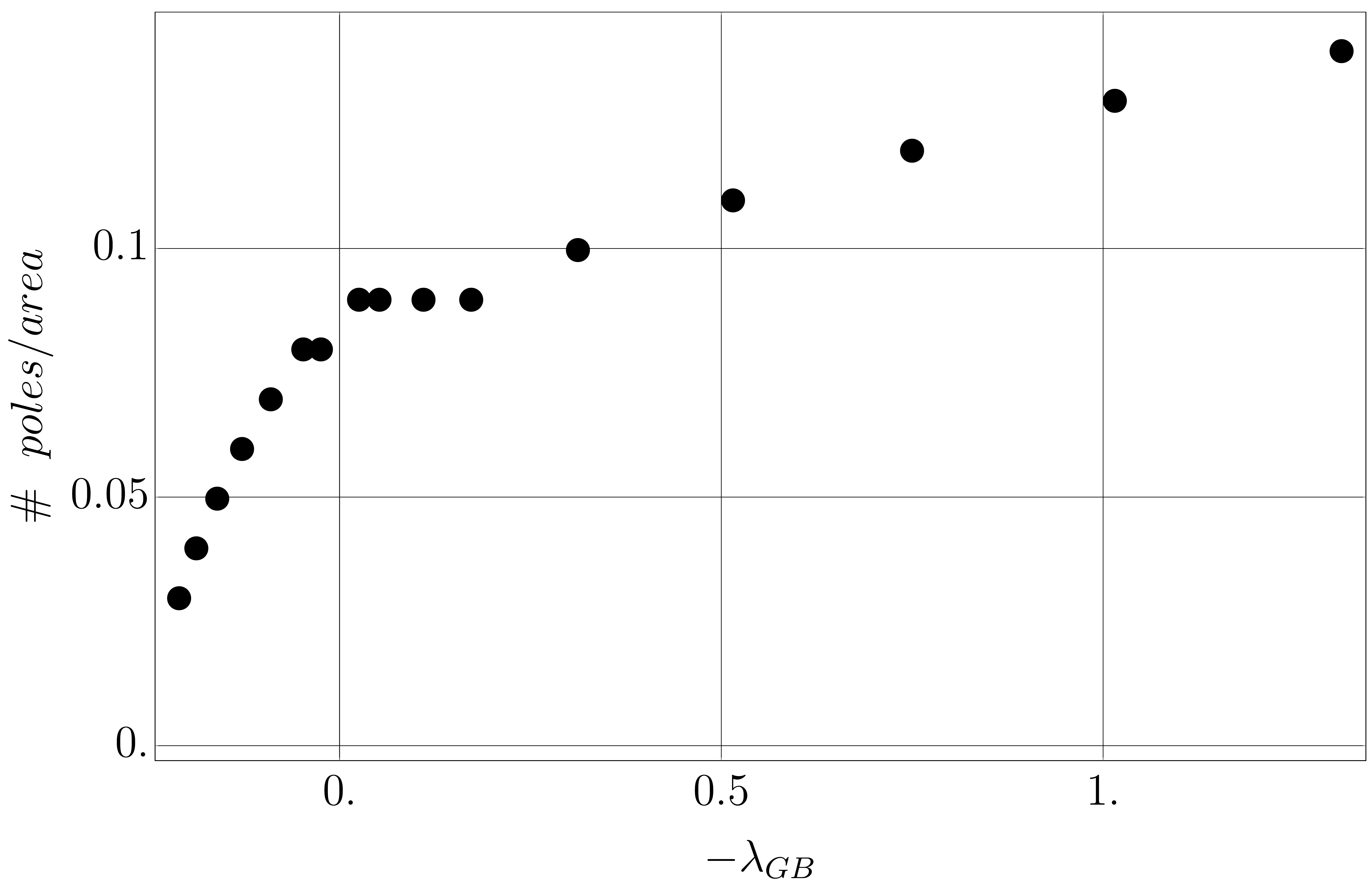}
\caption{Scalar}
\end{subfigure}
~
\begin{subfigure}[b]{0.45\linewidth}
\includegraphics[width=1\linewidth]{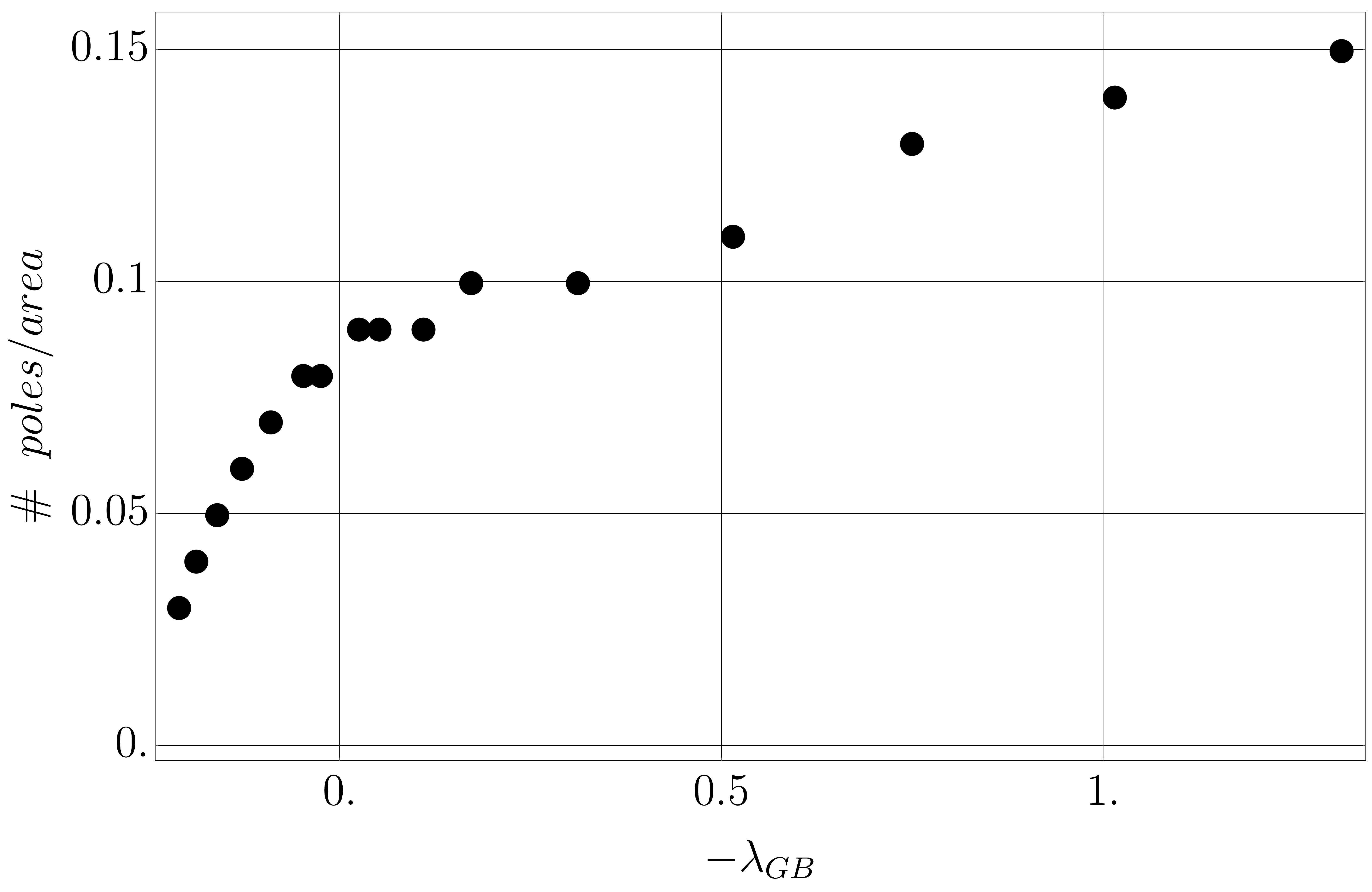}
\caption{Shear}
\end{subfigure}
\linebreak\linebreak
\begin{subfigure}[b]{0.45\linewidth}
\includegraphics[width=1\linewidth]{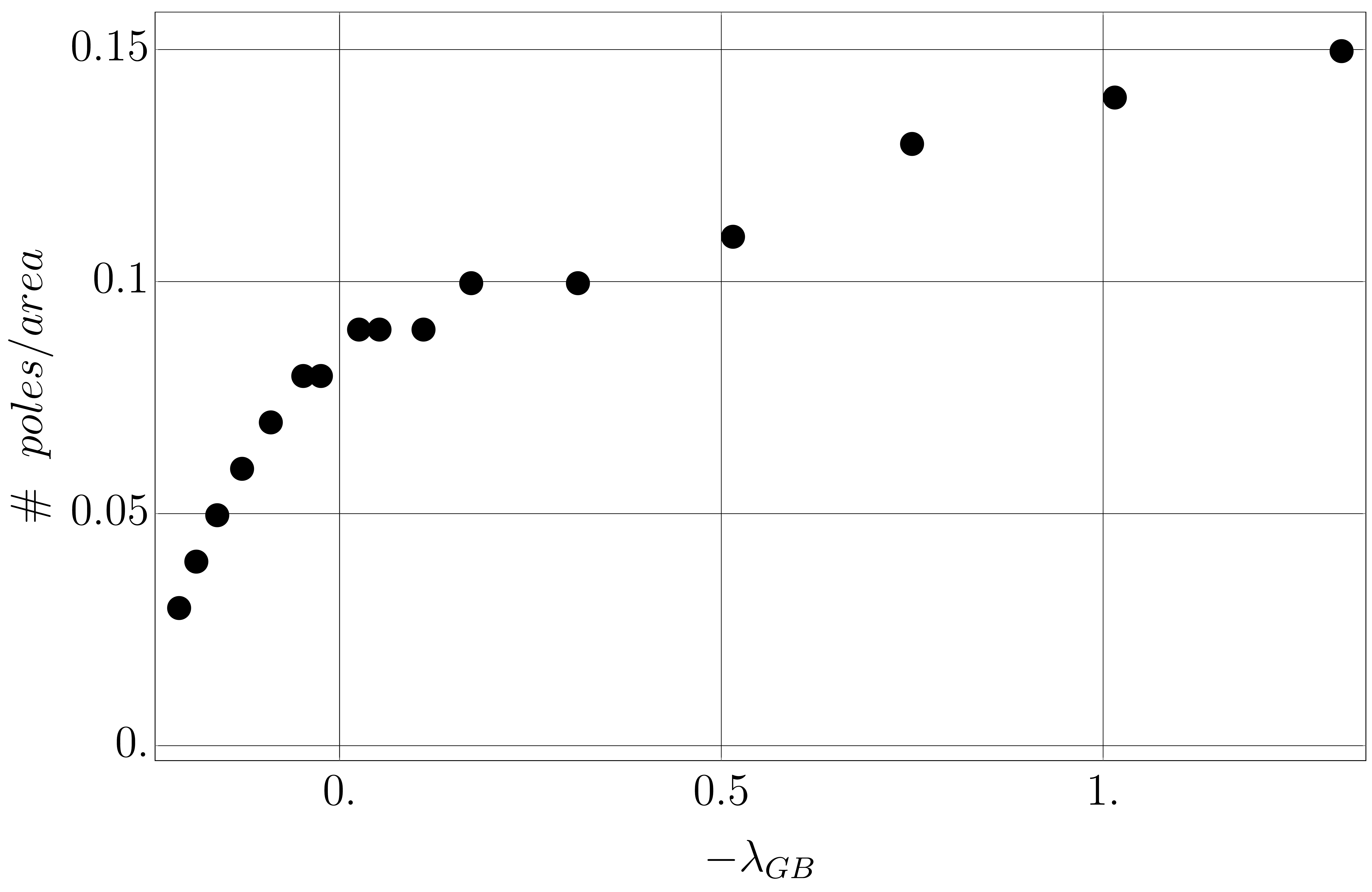}
\caption{Sound}
\end{subfigure}
\caption{Density of poles in the complex $\wfr$ plane plotted as a function of $-\lambda_{GB} \in (-1/4,\, 2 )$ at $\qfr = 0.5$.}
\label{figs:GBPoleDensity}
\end{figure}
\subsubsection{The density of poles and the appearance of branch cuts}
\label{sec:DensityBranch}
\begin{figure}[ht]
\centering
\includegraphics[width=0.7\linewidth]{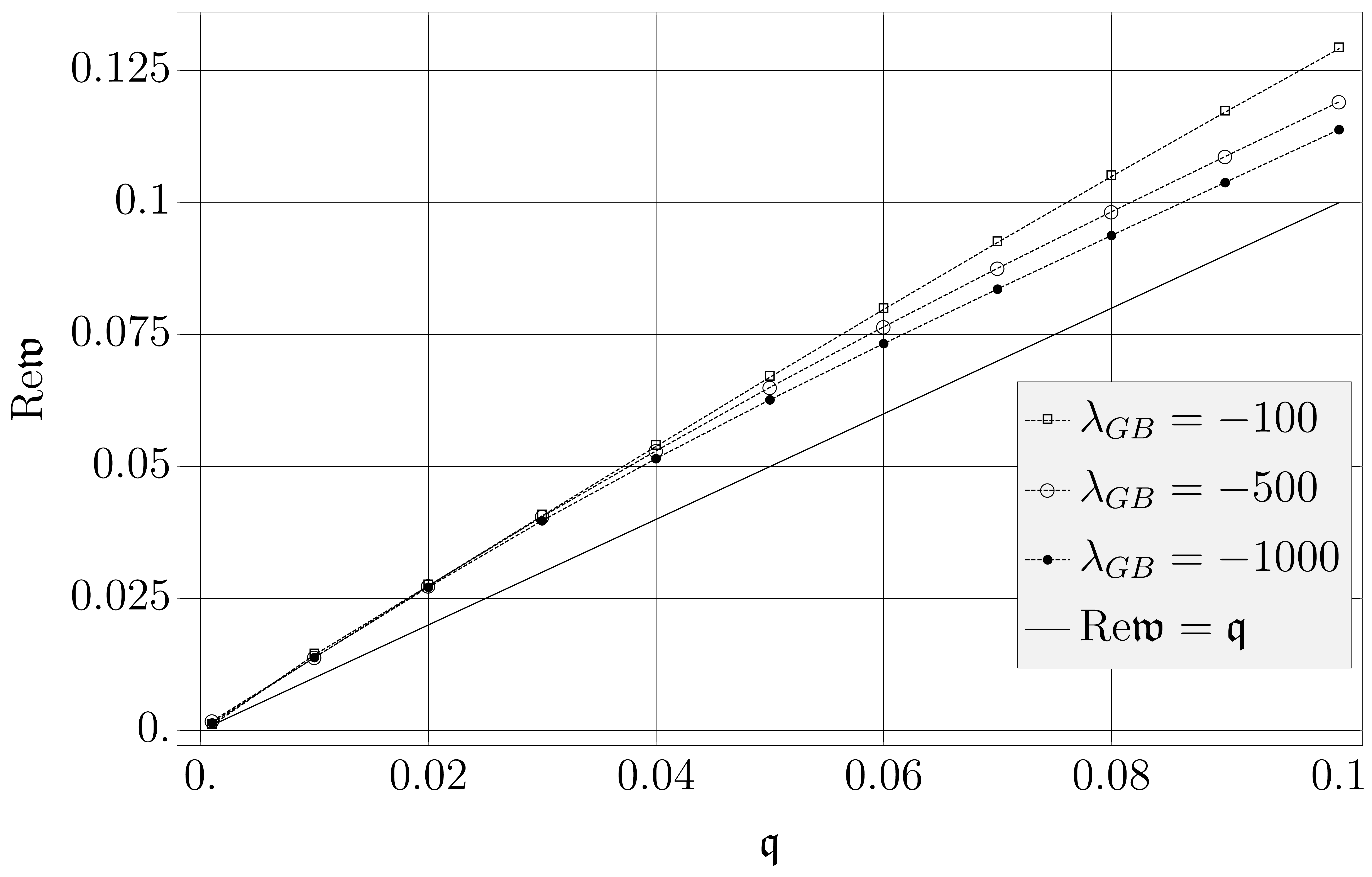}
\caption{Position of the first pole in the shear spectrum of the Gauss-Bonnet theory, for $\lgb=\{-100,-500,-1000\}$, as a function of momentum $\qfr$. The point indicates where the hypothetical branch cut would begin in the limit of large $|\lgb|$. The solid line corresponds to the expectation of where the position of $\mbox{Re}\, \wfr$ of the first pole should be in the limit of $\lgb \to -\infty$, i.e. $\mbox{Re} \,\wfr=\qfr$.}
\label{fig:GB-Shear-BranchCut}
\end{figure}

\begin{figure}[ht]
\centering
\includegraphics[width=0.7\linewidth]{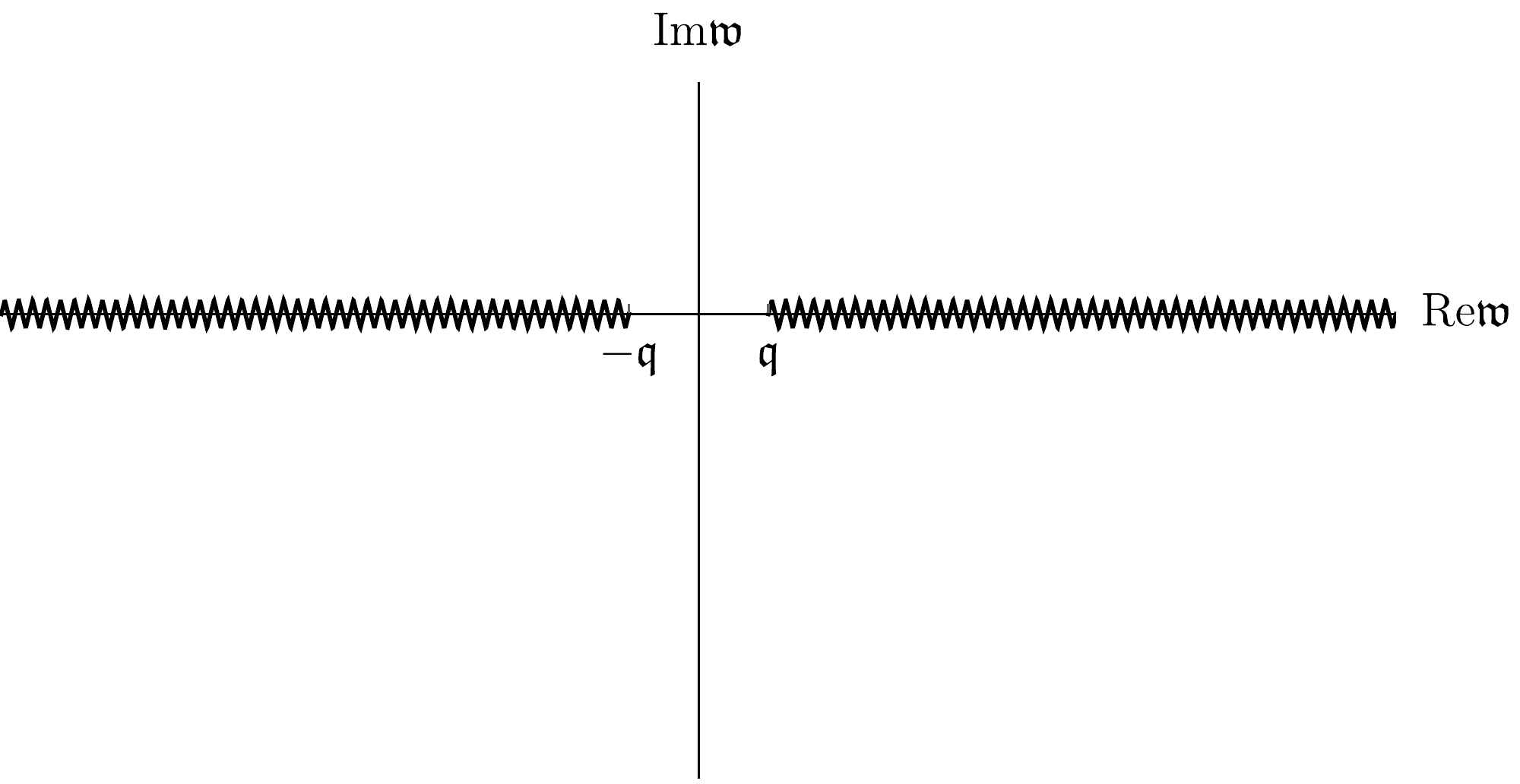}
\caption{The conjectured analytic structure of thermal correlators in holographic Gauss-Bonnet theory in the limit $\lgb \to -\infty$.}
\label{fig:GB-BranchCut}
\end{figure}

In Section \ref{sec:SYM} we observed that for $\CN=4$ SYM theory correlators, the density of poles in the two symmetric branches increases with 't Hooft coupling decreasing. In Gauss-Bonnet theory, the same phenomenon can be investigated in more detail since we are not constrained by infinitesimally small values of the higher derivative coupling. In all channels, the density of non-hydrodynamic poles in the two branches monotonically increases for $\eta / s > \viscb$ and decreases for  $\eta / s < \viscb$. Although the trend is apparent already from Figs.~\ref{fig:GB-Scalar-channel}, \ref{fig:GB-Shear-channel}, \ref{fig:GB-Sound-channel}, here we show the density of poles as a function of the coupling constant in Fig.~\ref{figs:GBPoleDensity}. The density is determined by selecting a region of the complex $\wfr$ plane, counting the number of poles in the symmetric branches inside that region and computing the resulting number density. The dependence in Fig.~\ref{figs:GBPoleDensity} is monotonic within the bounds of our numerical accuracy. The situation for $\lgb >0$  ($\eta / s < \viscb$) is clear: as $\lgb \rightarrow 1/4$ ($\eta / s \rightarrow 0$), the poles in the symmetric branches become less and less dense and in the limit they disappear from the finite complex plane altogether, as confirmed by analytic calculation at $\lgb = 1/4$. For  $\lgb <0$  ($\eta / s > \viscb$), the poles become more and more dense, the symmetric branches lift up toward the real axis with $|\lgb|$ increasing, and one may conjecture that in the limit $\lgb \to -\infty$ they merge to form branch cuts in the complex plane of frequency along $(-\infty , - \qfr]$ and $[\qfr, \infty)$. Numerically, we observe that  $\re[ \wfr]$ of the leading quasinormal mode in the (right) branch of poles monotonically approaches the line $\wfr = \qfr$ for large $|\lgb|$ (see Fig.~\ref{fig:GB-Shear-BranchCut}) which supports the conjecture that $\pm \qfr$ are the branch points of the correlator in the limit $\lgb \to -\infty$.
Note that all other poles (the ones not belonging to the symmetric branches at finite $\lgb$) in all channels either join the branches (in the sound and shear channels) or disappear due to vanishing residues (scalar and sound channels) in the limit $\lgb \to -\infty$. Thus, in that limit, the analytic structure of the correlator is represented by the branch cuts $(-\infty , - \qfr] \cup [\qfr, \infty)$ (see Fig.~\ref{fig:GB-BranchCut}). This resembles the zero temperature limit of the thermal correlator in a CFT dual to Einstein gravity with no higher derivative corrections. Such correlators are known analytically only for BTZ background. For example, the $\Delta = 2$ thermal correlator has the form \cite{Son:2002sd}
\begin{align}
G^R (\wfr,\qfr) \sim \left(\qfr^2-\wfr^2\right) \left\{ \psi \left[ 1 -\frac{i}{2}\left( \wfr - \qfr\right)\right] 
+ \psi \left[ 1 -\frac{i}{2}\left( \wfr + \qfr\right)\right] \right\}\,,
\label{eq:btz-corr}
\end{align}
where $\psi (z)$ is the logarithmic derivative of the Gamma-function with poles at $z=-n$, $n=0,1,2,...$. In the zero-temperature limit $\wfr\gg 1$, $\qfr \gg 1$, the poles merge forming two branch cuts running from the branch points $\omega = \pm q$ to infinity parallel to the imaginary axis. For large $z$, Binet's formula implies $\psi (z) \sim \ln z$ and thus in the limit of zero temperature the correlator (\ref{eq:btz-corr}) becomes $G^R \sim k^2 \ln{k^2}$, where $k^\mu = (-\omega,q)$. Similarly, the zero-temperature limit of the energy-momentum correlator in a $4d$ CFT dual to Einstein gravity is $G^R \sim (-\omega^2 + q^2)^2 \ln{(-\omega^2 + q^2)}$. This function has branch points at $\omega = \pm q, \infty$ joined by the branch cuts $(-\infty , - q] \cup [q, \infty)$.
\subsubsection{Coupling constant dependence of the shear  viscosity - relaxation time ratio in Gauss-Bonnet theory}
\begin{figure}[ht]
\centering
\includegraphics[width=0.7\linewidth]{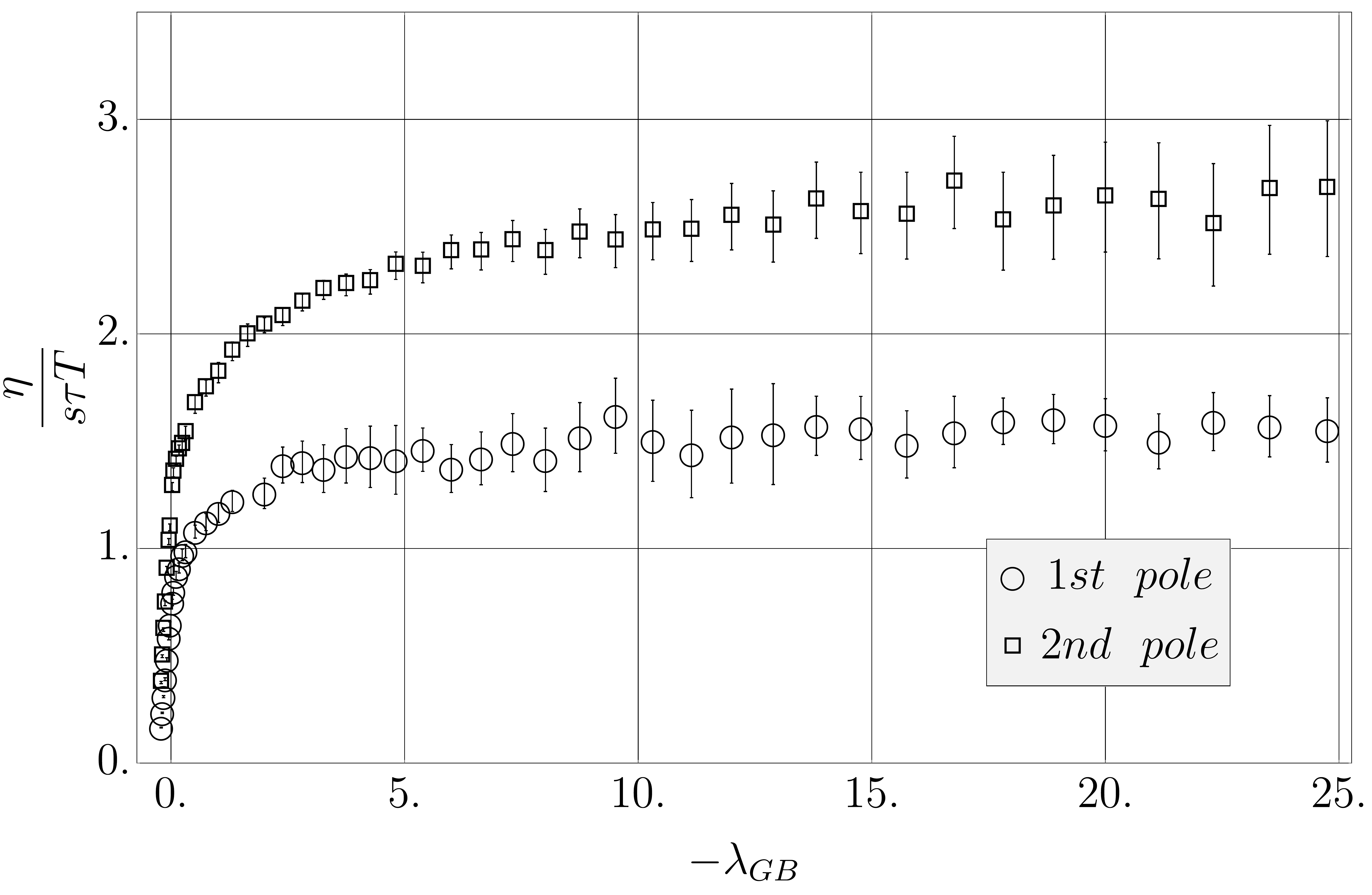}
\caption{The ratios $\eta/s T \tau_k$, for $k=\{1,\,2\}$, as functions of $\lgb$ in the shear channel of the Gauss-Bonnet theory. Here, $\tau_k$ are defined as $\tau_k = 1/|\mbox{Im}\, \omega_k|$ for the two smallest-in-magnitude non-hydrodynamical quasinormal frequencies $\omega_k$ at $\qfr=0$. The error bars correspond to resolution errors in the $\wfr$-plane.}
\label{fig:GB-Shear-etaOverStauT-vs-gamma}
\end{figure}

The coupling constant dependence of the ratio  $\eta / s \,\taur T$ in Gauss-Bonnet theory shows the same qualitative features as in $\CN = 4$ SYM discussed in Section \ref{sec:RatioN4}. In Fig.~\ref{fig:GB-Shear-etaOverStauT-vs-gamma}, we plot the ratios $\eta / (s \,\tau_k T )$, where $\tau_k$, $k=1,2$, are defined as $\tau_k = 1/|\mbox{Im}\, \omega_k|$ for the two smallest in magnitude non-hydrodynamical quasinormal frequencies $\omega_k$ at $\qfr=0$. We identify $\taur$ with $\tau_1$, $\omega_1$ being the fundamental frequency. The functions are monotonic, changing rapidly in the vicinity of $\lgb =0$ and flattening out in the region $|\lgb| \approx  3 -  6$. As in $\CN = 4$ SYM theory, the kinetic theory result (\ref{eq:rel-visc-rel}) seems to hold at intermediate coupling.
\subsubsection{Shear channel spectral function and quasiparticles at ``weak coupling''}
\label{sec:SpectFunGBShear}
\begin{figure}[ht]
\centering
\begin{subfigure}[t]{0.47\linewidth}
\includegraphics[width=1\linewidth]{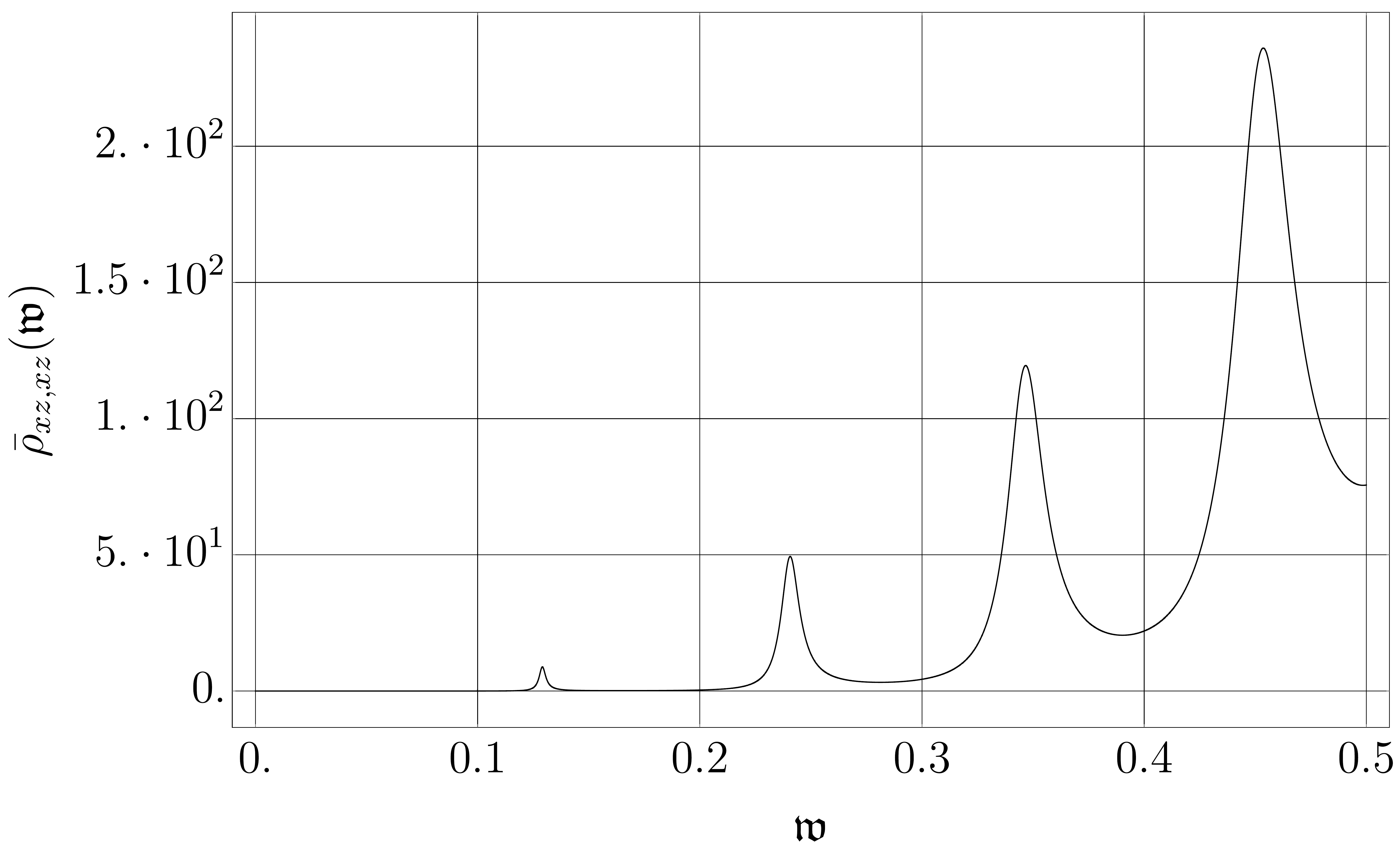}
\end{subfigure}
\qquad
\begin{subfigure}[t]{0.47\linewidth}
\includegraphics[width=1\linewidth]{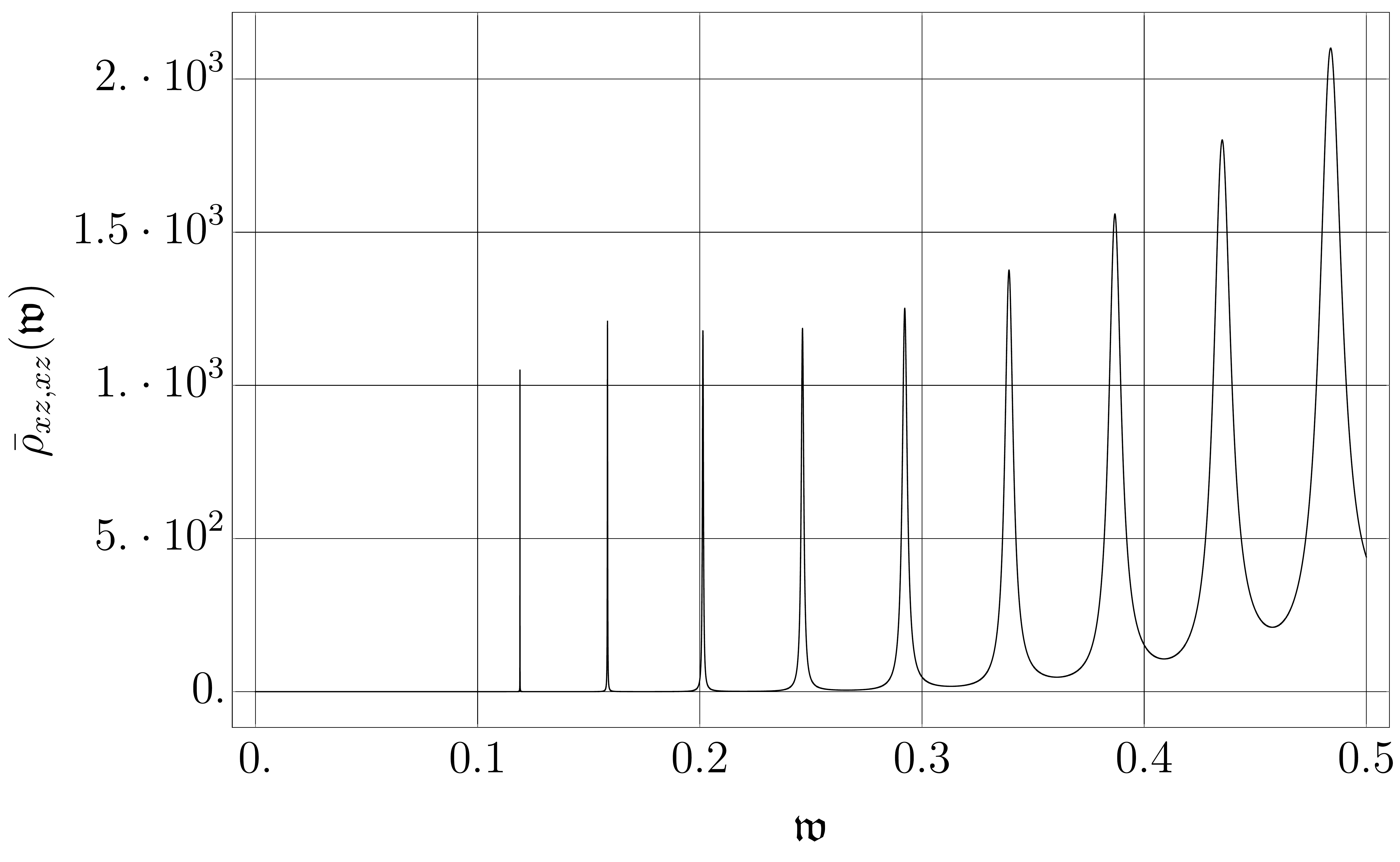}
\end{subfigure}
\caption{The dimensionless spectral function $\bar \rho_{xz,xz} \left(\wfr,\qfr,\lgb\right)$ in the shear channel of the Gauss-Bonnet theory for $\lgb = -100$ (left panel) and $\lgb = -500$ (right panel) at $\qfr = 0.1$.}
\label{fig:GB-Shear-SpectralFunctions}
\end{figure}

Since the non-hydrodynamic poles in the symmetric branches approach the real axis with $|\lgb|$ increasing (i.e. at weaker coupling), one may expect the corresponding spectral function to develop a structure resembling quasiparticle peaks. We check this by computing the spectral function in the shear channel. Choosing the spatial momentum along the $z$ axis, the shear channel retarded energy-momentum tensor correlator $G^{\text{R}}_{xz,xz}(\wfr,\qfr,\lgb)$ in Gauss-Bonnet theory can be computed as follows \cite{GBNesojen}:
\begin{equation}
\label{eq:corr-norm}
G^{\text{R}}_{xz,xz}(\wfr,\qfr,\lgb) =8 \pi^2 T^2 \wfr^2  \lim_{\varepsilon \rightarrow 0} \, {\cal C} (\varepsilon, \wfr, \qfr)\frac{\partial_u Z_2 (\varepsilon,\wfr,\qfr)}{Z_2(\varepsilon,\wfr,\qfr)}\,,
\end{equation}
where the function ${\cal C}$ is given by
\begin{equation}
{\cal C} (u,\wfr,\qfr) = \frac{\pi^2 T^2}{8\kappa_5^2}\, \frac{\bar{N} \bar{f} (1-\bar{f})}{ N_{GB}^5 u \left[ \bar{N} \bar{f} \qfr^2 - (1-\bar{f})^2 \wfr^2 \right]}\,,
\end{equation}
with
$$
\bar{f} = 1 - \sqrt{1- 4 \lgb (1-u^2)}\,, \;\; \qquad \;\; \bar{N} = N_{GB}^2\, \frac{1-4\lgb}{2\lgb}\,,
$$
and $Z_2(u)$ is the solution of the shear channel equation of motion obeying the incoming wave boundary condition at the horizon and normalized to one at the same $u = \varepsilon \to 0$. The solution $Z_2(u)$ can be written as $Z_2(u) = \CA_2 Z^I_2 (u) + \CB_2 Z^{II}_2 (u)$, where $Z^I_2 (u)$ and $Z^{II}_2 (u)$ are the two local Frobenius expansions at the boundary (see e.g. \cite{Kovtun:2006pf}). In terms of $\CA_2$ and $\CB_2$, the retarded Green's function \eqref{eq:corr-norm} is given by\footnote{Since the Frobenius expansion of $Z^{I}_2$ contains $Z^{II}_2$ multiplying $\ln u$, it is numerically more convenient to compute $\CB_2$ by subtracting off the logarithmic term, as was done in \cite{Kovtun:2006pf}. We find that $\CB_2 = \frac{1}{2} \lim_{u\to 0} \left( \partial^2_u Z_2 - 2 \CA_2 h \ln u \right) - \frac{3}{2} \CA_2 h$, where in the Gauss-Bonnet theory, $h = - 8 \lgb^4 \left( \qfr^2 - \wfr^2\right)^2 / \left( 1 - \sqrt{1-4\lgb} \right)^4$ \cite{GBNesojen}.}
\begin{align}
G^{\text{R}}_{xz,xz}(\wfr,\qfr,\lgb) =\frac{\pi^4T^4}{2 \kappa_5^2}\, \frac{\bar{N} \ggb \left(1-\ggb\right) \wfr^2 }{ N_{GB}^5 \left[ \bar{N} \left(1-\ggb\right) \qfr^2 - \ggb^2 \wfr^2 \right]} \, \frac{\CB_2}{\CA_2}.
\end{align}
The spectral function is then computed as
\begin{align}
\rho_{xz,xz} \left(\wfr,\qfr,\lgb\right) = - \im \, G^{\text{R}}_{xz,xz} \left(\wfr,\qfr,\lgb\right).
\end{align}
In Fig.~\ref{fig:GB-Shear-SpectralFunctions}, we plot the dimensionless spectral function
\begin{align}
\bar \rho_{xz,xz} \left(\wfr,\qfr,\lgb\right) \equiv \frac{ \kappa^2_5 }{ 4 \pi^2 T^4} \rho_{xz,xz} \left(\wfr,\qfr,\lgb\right)  ,
\end{align}
where $\kappa^2_5$ is the Newton's constant from the Gauss-Bonnet action \eqref{eq:GBaction} and $T$ the Hawking temperature \eqref{eq:GBTemperature}. As $|\lgb|$ increases and the symmetric branches of poles approach the real $\wfr$ axis, the appearance of quasiparticle-like peaks in the spectral function is clearly seen. As a result of the quasinormal modes now having $\left|\im [\wfr] \right| \ll \left|\re [\wfr] \right|$ at large $|\lgb|$, the peaks become sharp and very narrow. Since the density of poles increases with $|\lgb|$, the density of peaks increases as well. In the limit $|\lgb|\rightarrow \infty$, they presumably form a continuum.
\section{Generic curvature squared corrections to quasinormal spectra of metric perturbations}
\label{sec:r2}
In this Section, we comment on the quasinormal spectrum in the theory with general curvature squared terms in the action,
\begin{align}
S_{R^2} = \frac{1}{2 \kappa_5^2 } \int d^5 x \sqrt{-g} \left[ R - 2 \Lambda + L^2 \left( \alpha_1 R^2 + \alpha_2 R_{\mu\nu} R^{\mu\nu} + \alpha_3 R_{\mu\nu\rho\sigma} R^{\mu\nu\rho\sigma}  \right) \right],
\label{eq:R2Th}
\end{align}
where the cosmological constant is $\Lambda = - 6 / L^2$. For the special choice of the parameters $\alpha_1$, $\alpha_2$, $\alpha_3$ given by 
\begin{align}
\label{eq:gb-set}
\alpha_1  = \lgb/2\,, \qquad \alpha_2 = - 2 \lgb\,,  \qquad  \alpha_3 = \lgb/2\,,
\end{align}
the action \eqref{eq:R2Th} coincides with the Gauss-Bonnet action (\ref{eq:GBaction}). Generically, however, the action \eqref{eq:R2Th} leads to the equations of motion involving derivatives up to the fourth order. In this case the higher derivative terms in \eqref{eq:R2Th} are treated perturbatively and the parameters $\alpha_i$ are assumed to be infinitesimally small. We can find the corresponding quasinormal spectra by using a field redefinition and the known results for Gauss-Bonnet and $\CN = 4$ SYM theories. 

One may notice \cite{Brigante:2007nu} that the action \eqref{eq:R2Th} with $\alpha_3 = 0$ is equivalent via a field redefinition
\begin{align}
g_{\mu\nu} = \bar g_{\mu\nu} + \alpha_2 \bar R_{\mu\nu} - \frac{1}{3} \left(\alpha_2 + 2 \alpha_1 \right) \bar g_{\mu\nu} \bar R,
\label{eq:FieldRedef}
\end{align}
and an additional rescaling to the Einstein-Hilbert action with the same cosmological constant and modified Newton's constant (which does not enter the vacuum equations of motion).\footnote{See Ref. \cite{Grozdanov:2014kva} for a detailed description of this procedure where it was applied to the calculation of the second-order transport coefficients.} Consider now a gauge-invariant (with respect to infinitesimal metric perturbations) mode $Z\left(h_{\mu\nu}\right)$ that is linear in metric perturbations, $\delta g_{\mu\nu} = h_{\mu\nu}$. To linear order, the Ricci and Einstein tensors are invariant under diffeomorphisms, hence so is $g_{\mu\nu} R$. It therefore follows that $g_{\mu\nu}$ and $\bar g_{\mu\nu}$ transform identically under the diffeomorphisms and so
\begin{align}
Z\left(h_{\mu\nu}\right) = Z\left(\bar h_{\mu\nu}\right).
\end{align}
Hence, when $\alpha_3 = 0$, the quasinormal modes of $Z\left(\bar h_{\mu\nu}\right)$ are also those of $Z\left(h_{\mu\nu}\right)$, which means that the quasinormal mode spectrum of the AdS-Schwarzschild black brane (dual to thermal $\CN = 4$ SYM theory at infinite 't Hooft coupling) is exactly the spectrum of the theory defined by \eqref{eq:R2Th} with $\alpha_3 = 0$.

To include the $\alpha_3$ contributions, we can use the fact that the perturbative (in  $\alpha_i$) quasinormal spectrum generically has the form
\begin{align}
\omega^* = \omega_0^* + \alpha_1\ \tilde\omega_{1}^* + \alpha_2 \ \tilde\omega_{2}^* + \alpha_3 \ \tilde\omega_{3}^*,
\end{align}
where $\omega^*_0$ are the quasinormal frequencies of the AdS-Schwarzschild black brane. Moreover, the above discussion shows that $\tilde\omega_{1}^* = 0$, $\tilde\omega_{2}^* = 0$. Keeping in mind the identification (\ref{eq:gb-set}) and considering the linearized quasinormal spectrum in Gauss-Bonnet theory,
\begin{align}
\omega^*_{GB} = \omega^*_0 + \lambda_{GB}\, \tilde\omega^*_{GB},
\end{align}
we conclude that $\lambda_{GB} \tilde \omega_3^*/2 = \lambda_{GB} \tilde\omega_{GB}^*$. Hence, the quasinormal spectrum of a background defined by the action \eqref{eq:R2Th}  has the form
\begin{align}
\omega^* = \omega^*_0 + 2\,\alpha_3\,\tilde\omega^*_{GB}\,,
\end{align}
where $\omega^*_0$ is the corresponding frequency in the spectrum of AdS-Schwarzschild black brane with no higher derivative corrections included and $\tilde\omega^*_{GB}$ is the coefficient of the term linear in $\lgb$ in the corresponding spectrum of the Gauss-Bonnet theory. Thus, the coupling dependence of the relaxation time and other properties of the spectrum described in previous sections are qualitatively the same as the ones in the Gauss-Bonnet theory (and $\CN = 4$ SYM theory with large but finite 't Hooft coupling). In particular, one observes a qualitative difference between the regimes with  $\eta / s > \viscb$ and  $\eta / s < \viscb$ similar to the one described in the previous Section.
\section{Discussion}
\label{sec:discussion}
In this paper, we have studied the influence of higher derivative $R^2$ and $R^4$ terms on the quasinormal spectra of gravitational perturbations of black branes. In a dual QFT, this corresponds to changing the 't Hooft coupling or its analogue from infinite to large but finite value. Understanding, even qualitatively, how the physical quantities responsible for thermalization change from strong to weak coupling would be important both from a conceptual and a phenomenological point of view. We were looking for robust, model-independent qualitative features the higher derivative terms may bring about. Vulnerabilities of this approach are quite obvious. While $\CN = 4$ $SU(N_c)$ SYM is a well defined unitary theory, higher derivative corrections in its dual gravity description are only partially known even to leading order in $\gamma \sim \lambda^{-3/2}$ at infinite $N_c$ and those terms must be treated perturbatively. Moreover, as emphasized recently in \cite{Waeber:2015oka}, different physical quantities may have very different sensitivity to coupling corrections, and the smallness of the perturbative parameter $\gamma$ may not necessarily be a good indicator of the size of corrections. In contrast, the second order equations of motion of Gauss-Bonnet gravity can be treated fully non-perturbatively. However, the (hypothetical) dual field theory suffers from causality violation and even the bulk theory may need higher spin fields to mend the problems (the latter would imply that higher derivative corrections can only be treated perturbatively, i.e. the theory loses its special status with respect to Ostrogradsky instability). Unphazed by these uncertainties, we proceed to investigate coupling corrections in both theories and are encouraged to observe qualitatively similar results in both cases. Our findings are summarized at the end of Section \ref{sec:intro}.

One curious feature we find is the behavior of quasinormal spectrum leading to a breakdown of the hydrodynamic description at a coupling-dependent critical value $q_c$ of the spatial momentum. In both  $\CN = 4$ SYM and Gauss-Bonnet theories, the dependence on coupling implies that hydrodynamics has a wider applicability range at strong coupling. It may be interesting to investigate the convergence properties of the hydrodynamic derivative expansion at finite coupling, possibly along the lines of Refs.~\cite{Heller:2013fn,Heller:2015dha}.
 
Another qualitatively similar feature for both theories is the coupling dependence of the ratio of the shear viscosity to the product of relaxation time, entropy density and temperature. This quantity is (approximately) constant in kinetic theory at weak coupling. From the dual gravity with higher derivative corrections we find that this ratio changes rapidly in the vicinity of infinite coupling and then shows a very weak (essentially flat) dependence on coupling when the coupling is further decreased to large but finite values. Similar behavior is expected for other transport coefficients. Admittedly, corrections from the unknown higher derivative terms may influence the dependence at intermediate coupling. Yet, if correct, the observed tendency may help to explain certain phenomenological success and "unreasonable effectiveness" of kinetic theory methods far beyond their justified domain.
 
We also found that the behavior of coupling corrections to quasinormal spectrum and related quantities depends strongly on whether $\eta / s > \viscb$ or $\eta / s < \viscb$. In the regime of $\eta / s < \viscb$, the symmetric branches of quasinormal modes exhibit monotonically increasing $\left| \im\, \omega \right|$. Since this could lead to the relaxation time of the system $\taur$ decreasing below any possible lower bound (see Eq.~\eqref{eq:sachdev-const}), it is conceivable that this regime is pathological. Earlier work was focused on looking for possible pathologies (e.g. causality violation) in the ultraviolet sector of the theories having the regime   $\eta / s < \viscb$, and constraining higher derivative couplings accordingly. However, inconsistencies in this regime may exist in the infrared sector as well. As the qualitative behavior of the spectra critically depends on the sign of the correction to $\eta / s = \viscb$, we note that the relation between $\eta/s$ and the relaxation time $\taur$ raises the possibility that the bound on $\eta/s$ speculated upon\footnote{A number of strongly interacting many-body systems - from quark-gluon plasma and cold atoms \cite{Luzum:2008cw,Adams:2012th,Cremonini:2011iq} to dusty plasmas \cite{fortov-1,fortov-2} and rare gases and molecules in the vicinity of the critical point \cite{hohm}  have $\eta/s \gtrsim \viscb$.} in Ref.~\cite{Kovtun:2003wp} is related to a bound on relaxation time. In the holographic models considered in this paper, both $\eta/s$ and $\taur T$ are monotonic functions of the coupling. For $\eta/s$ decreasing below $\viscb$, the relaxation time also decreases below its value at infinite coupling. Is there a minimal relaxation time possibly correlated with the viscosity bound? Are there any universal constraints on the constant ${\cal C}$ in Eq.~(\ref{eq:sachdev-const})? Curiously, in 2006 Hod \cite{Hod:2006jw} suggested a universal bound on relaxation time in any system:
\begin{equation}
\taur \geq \tau_{min} = \frac{\hbar}{\pi k_B T}\,.
\label{eq:Hod-bound}
\end{equation}
For the black hole quasinormal spectrum, the inequality (\ref{eq:Hod-bound}) means that there exists at least one quasinormal frequency whose imaginary part lies in the strip $0 > \mbox{Im}\,  \omega \geq - \pi k_B T/\hbar$ in the complex frequency plane or, in terms of $\wfr = \omega/2\pi k_B T$, in the strip
\begin{equation}
0  > \mbox{Im}\, \wfr \geq - \frac{1}{2}\,.
\label{eq:Hod-bound-qnm}
\end{equation}
In the language of the kinetic theory linear collision operator spectrum, the bound implies
\begin{equation}
0 \leq \nu_{\min} \leq \nu_c =  \pi k_B T/\hbar\,,
\label{eq:Hod-bound-nu}
\end{equation}
see Fig.~\ref{fig:spectrum_kinetic}d. Apparently, the inequality (\ref{eq:Hod-bound-qnm}) holds for black holes (for black holes, the bound suggests that a black hole has (at least) one channel of slowly decaying perturbation modes which respect (\ref{eq:Hod-bound-qnm})). At first glance, however, the relaxation time bound is void of meaning since one expects the hydrodynamic modes to be always present in any system in the thermodynamic limit and they may relax arbitrarily slowly for sufficiently long wavelengths (in other words, gapless quasinormal frequencies corresponding to hydrodynamic modes are always present in the strip (\ref{eq:Hod-bound-qnm}) for sufficiently small spatial momentum $q$). Moreover, even if we regard the bound (\ref{eq:Hod-bound}) as the bound obeyed by the (non-hydrodynamic) relaxation time $\taur$ (and correspondingly, the inequality (\ref{eq:Hod-bound-qnm}) as the one for the fundamental non-hydrodynamic quasinormal frequency), it appears to be violated in all black brane channels (see e.g. tables of quasinormal frequencies in Refs.~\cite{Starinets:2002br,Nunez:2003eq,Kovtun:2005ev}). Nevertheless, we believe the question of whether holography or black hole physics implies an inequality of the type (\ref{eq:Hod-bound-nu}) or (\ref{eq:Hod-bound}) is an interesting one in view of its apparent validity for black holes and its possible connection to viscosity bound.
 
In this paper, we considered coupling constant corrections to equilibrium correlators in $4d$ CFTs. It would be interesting to consider non-conformal and time-dependent backgrounds. We also hope questions raised in this paper may stimulate additional work on weakly coupled thermal QFTs, via perturbation theory or kinetic theory, regarding the analytic structure of correlation functions and bounds of applicability of hydrodynamic description, with the goal to form a consistent qualitative picture interpolating between weak and strong coupling. 
\acknowledgments
We would like to thank J.~Casalderrey Solana, F.~Essler,  V.~Ker{\"a}nen, P.~Kleinert, R.~Konoplya, D.~Kovrizhin,  H.~Reall, A.~Schekochihin,  L.G.~Yaffe, J.~Zaanen and A.~Zhiboedov for illuminating discussions and S.~Hod for correspondence. A.O.S. is grateful to the Institute for Nuclear Theory at the University of Washington, Seattle, for its warm hospitality and to participants of the program INT-15-2c "Equilibration Mechanisms in Weakly and Strongly Coupled Quantum Field Theory" for useful discussions. S.G. is supported in part by a VICI grant of the Netherlands Organization for Scientific Research (NWO) and by the Netherlands Organization for Scientific Research/Ministry of Science and Education (NWO/OCW). N.K. is supported by a grant from the John Templeton foundation. The opinions expressed in this publication are those of the authors and do not necessarily reflect the views of the John Templeton foundation. This work was carried out on the Dutch national e-infrastructure with the support of SURF Foundation. The work of A.O.S. was supported by the European Research Council under the European Union's Seventh Framework Programme (ERC Grant agreement 307955).
\appendix
\section{The functions $\CG_1$, $\CG_2$ and $\CG_3$}
\label{sec:appendix-N4}

{\bf Scalar channel}
\begin{align}
\CG_1 =& -6 u \left(160 \qfr^2 u^3+129 u^4+94 u^2-25\right) \partial_u Z_1 \nn
&+  \frac{192 \qfr^4 u^5-\qfr^2 \left(851 u^6-789 u^4+75 u^2+30\right)+6 \left(-89 u^4+30 u^2+5\right) \wfr^2}{u \left(u^2-1\right)} Z_1.
\end{align}
\\
{\bf Shear channel}
\begin{align}
\CG_2 =&\, - \frac{2u}{\left(\wfr^2 - \qfr^2\left(1-u^2\right)\right)^2 } \biggr[ 640 \qfr^6 u^3 \left(u^2-1\right)^2 \nn
&-4 \qfr^4 u^2 \left(135 u^6-450 u^4-248 u^3 \wfr^2+495 u^2+200 u \wfr^2-180\right)\nn
&+\qfr^2 \wfr^2 \left(-462 u^6+1374 u^4+160 u^3 \wfr^2-1002 u^2+75\right)\nn
&+3\left(129 u^4+94 u^2-25\right) \wfr^4 \biggr]  \partial_u Z_2 \nn
& + \frac{3}{u \left(1-u^2\right) \left(\wfr^2 - \qfr^2\left(1-u^2\right)\right) } \biggr[ -64 \qfr^6 u^5 \left(u^2-1\right)\nn
&+\qfr^4 \left(425 u^8-880 u^6-64 u^5 \wfr^2+480 u^4-15 u^2-10\right)\nn
&+\qfr^2 \left(699 u^6-693 u^4+75 u^2+20\right) \wfr^2+2 \left(89 u^4-30 u^2-5\right) \wfr^4  \biggr] Z_2.
\end{align}
\\
{\bf Sound channel}
\begin{align}
\CG_3 =&\, - \frac{2u}{\left(3 \wfr ^2 -\qfr^2 \left(3-u^2\right) \right)^3}    \biggr[  32 \qfr^8 u^3 \left(35 u^6-291 u^4+753 u^2-585\right) \nn
&-3 \qfr^6 \left(3741 u^{10}-27911 u^8-2720 u^7 \wfr^2+60804 u^6+12992 u^5 \wfr^2-50112 u^4 \right.\nn
&\left. -12960 u^3 \wfr^2+16887 u^2-225\right)+3 \qfr^4 \wfr^2 \left(-19401 u^8+59832 u^6+4960 u^5 \wfr^2 \right. \nn
&\left. -53892 u^4-7200 u^3 \wfr^2+26094 u^2-1125\right)+9 \qfr^2 \wfr^4 \left(-1263 u^6+99 u^4 \right.\nn
&\left. +160 u^3 \wfr^2-3915 u^2+525\right)+81 \left(129 u^4+94 u^2-25\right) \wfr^6 \biggr]  \partial_u Z_3 \nn
&- \frac{1}{u \left(1-u^2\right) \left(3 \wfr ^2 - \qfr^2 \left(3-u^2\right)\right)^3} \biggr[192 \qfr^{10} u^5 \left(u^2-3\right)^3 \nn
&-\qfr^8 \left(u^2-3\right) \left(5811 u^{10}-41287 u^8-1728 u^7 \wfr^2+74004 u^6+5184 u^5 \wfr^2 \right.\nn
&\left. -35169 u^4+495 u^2+270\right)-3 \qfr^6 \left(11184 u^{13}-90072 u^{11}+17099 u^{10} \wfr^2 \right.\nn
&\left.+223952 u^9-106323 u^8 \wfr^2-16 u^7 \left(108 \wfr^4+12971\right)+185876 u^6 \wfr^2\right. \nn
&\left. +1728 u^5 \left(3 \wfr^4+34\right)-91107 u^4 \wfr^2+1800 u^3+2835 u^2 \wfr^2+1080 \wfr^2\right) \nn
&+3 \qfr^4 \wfr^2 \left(-68316 u^{11}+279504 u^9-40333 u^8 \wfr^2-319056 u^7+121158 u^6 \wfr^2\right.\nn
&\left. +36 u^5 \left(48 \wfr^4+2713\right)-81018 u^4 \wfr^2+3600 u^3+6075 u^2 \wfr^2+1620 \wfr^2\right)\nn
&-9 \qfr^2 \wfr^4 \left(21708 u^9-37140 u^7+7003 u^6 \wfr^2+12972 u^5-10017 u^4 \wfr^2 \right.\nn
&\left.+600 u^3+1755 u^2 \wfr^2+360 \wfr^2\right)-162 \left(89 u^4-30 u^2-5\right) \wfr^8  \biggr]Z_3
\end{align}
\section{The coefficients $A_i$ and $B_i$ of the differential equation (\ref{eq:eom_GB_ginv}) }
\label{sec:appendix-GB}

{\bf Scalar channel}
\begin{align}
A_1 = \, & -\frac{1}{u}   - u \left[ \frac{1}{\left(\ggb ^2-1\right) \left(1-u^2\right)^2 + 1 - u^2} + \frac{1}{\left(1-u^2\right) \sqrt{\ggb ^2-\left(\ggb^2-1\right) u^2}}\right], \\
B_1 = \, &\frac{(\ggb -1) (\ggb +1)^2 \left(3 \left(\ggb ^2-1\right) u^2-\ggb ^2\right)\left(-\ggb ^2+\left(\ggb ^2-1\right) u^2+U\right)}{4 u \left(\ggb^2-\left(\ggb ^2-1\right) u^2\right)^{3/2} \left(-\ggb ^2+\left(\ggb
   ^2-1\right) u^2+2 U-1\right)} \qfr^2 \nn
\, & + \frac{\left(\ggb ^2-1\right)^2 \left(-\ggb ^2+\left(\ggb ^2-1\right)u^2+U\right)}{4 u (U-1) \sqrt{\ggb ^2-\left(\ggb ^2-1\right) u^2} \left(-\ggb^2+\left(\ggb ^2-1\right) u^2+2 U-1\right)}\wfr^2,
\end{align}
\\
{\bf Shear channel}
\begin{align}
A_2 = \,&-\frac{2 \ggb ^4 (\ggb +1)  \left[\frac{1}{2} \left(1-\ggb ^2\right) \left(u^2-1\right) (U-2)+U-1\right]}{u (U-1) U^3 \left[\ggb ^2 (\ggb +1) (U-1)  \qfr^2  -\left(\ggb ^2-1\right) U^2 \wfr ^2\right]}  \qfr^2 \\
\,&-\frac{\left(1-\ggb ^2\right) \left(\ggb ^4+\left(1-\ggb^2\right)^2 u^4- 2 \left(1-\ggb ^2\right) u^2 \left(U-\ggb^2\right)-\ggb ^2 U\right)}{u (U-1) U \left[\ggb ^2 (\ggb +1)(U-1)  \qfr^2  -\left(\ggb ^2-1\right) U^2 \wfr ^2\right]}   \wfr^2, \\
B_2 = \,&  \frac{\ggb ^2 (\ggb +1) (U+1)}{4 u \left(u^2-1\right) U^2} \qfr^2 + \frac{\left(U^2+2 U+1\right) }{4 u \left(u^2-1\right)^2}  \wfr^2 ,
\end{align}
\\
{\bf Sound channel}
\begin{align}
A_3 =\,& \frac{3}{2 u} + \frac{3 (\ggb -1)  \left[\left(\ggb ^2-1\right) u^2-\ggb^2\right] \left[\left(\ggb ^2-1\right) u^2 (5 U-7)-5 \ggb ^2(U-1)\right]}{2 u (U-1) U^2 D_1} \wfr^2 \nn
\,&+ \frac{ \left(\ggb ^2-1\right)^2 u^4 \left(-3 \ggb ^2+5 U-7\right)+\ggb ^2 \left(\ggb ^2-1\right) u^2 \left(18 \ggb ^2-13 U+10\right) }{2 u (U-1) U^2 D_1} \qfr^2 \nn
\,&- \frac{ 15 \ggb ^4 \left(\ggb ^2-2 U+1\right) }{2 u (U-1) U^2 D_1} \qfr^2, \\
B_3 =&~ \frac{\left(\ggb ^2-1\right)^2}{D_{0}} \biggr\{ \,12 (\ggb -1)^2 \ggb ^2 (\ggb +1) \qfr^2 u^5-4 (\ggb -1) \ggb ^2 \qfr^2 u^3 \left(3 \ggb ^2-7 U+4\right) \nn
&+ \left(\ggb ^2-1\right)^3 \qfr^2 u^6 \left(3 (\ggb -1) \wfr ^2+\qfr^2\right) \nn
& -u^2 \ggb ^2 \left(\ggb ^2-1\right)  \left[\qfr^4 \left(\ggb ^2+2 U\right)+(\ggb -1) \qfr^2 \wfr ^2 \left(9 \ggb ^2-4 U\right)-6 (\ggb-1)^2 U \wfr ^4\right] \nn
&  + \left(\ggb ^2-1\right)^2 u^4 \left[\qfr^4 \left(3 \ggb ^2 (U-2)+U\right)+2(\ggb -1) \qfr^2 U \wfr ^2-3 (\ggb -1)^2 U \wfr ^4\right]\nn
& - 3 \ggb ^4 \left[\qfr^4 \left(\ggb ^2 (U-2)+U\right)+2 (\ggb -1) \qfr^2 \wfr^2 \left(U-\ggb ^2\right)+(\ggb -1)^2 U \wfr ^4\right]  \,\, \biggr\},
\end{align}
where we have defined
\begin{align}
D_1 &\equiv \left(\ggb ^2-1\right) u^2 \left(3(\ggb -1) \wfr^2+\qfr^2\right)+3 \ggb ^2 \left(\qfr^2 (U-1)-(\ggb -1)\wfr^2\right), \nn
D_0 &\equiv 4 (\ggb -1) u (U-1)^2 U^3 D_1 .
\end{align}
In the above expressions, we used $U^2 = u^2 + \ggb^2 - u^2 \ggb^2$, as well as the dimensionless frequency and momentum (\ref{eq:gothic}), where the Hawking temperature is given by Eq.~\eqref{eq:GBTemperature}.
\section{Numerical methods used}
\label{sec:Numerics}
In this Appendix, we briefly review the numerical methods used to compute quasinormal modes and spectral functions. The relevant equations of motion are \eqref{ScalarEqN4}, \eqref{ShearEqN4}, \eqref{eq:SoundEqN4} for the $\mathcal{N}=4$ and \eqref{eq:eom_GB_ginv} for the Gauss-Bonnet case. In all approaches, the first step involves computing the index of the corresponding regular singular point at the horizon, i.e. factoring out the singular part of the solution:
$$
Z_i \to (u-1)^{\nu_i} \tilde{Z}_i(u)\,,
$$
where  $\nu_{\pm} = \pm \frac{i}{2}\wfr$ in all channels for  $\CN=4$ SYM and Gauss-Bonnet theories, and $\tilde{Z}_i(u)$ is the Frobenius solution regular at the horizon. A similar factorization is done at the boundary.
\paragraph{Leaver's method}
In this method, originally introduced in \cite{Leaver:1990zz} and then used in \cite{Denef:2009yy,Edalati:2010hk,Edalati:2010pn,Davison:2011uk}, the regular function is represented by a series whose convergence radius  reaches both the horizon and the boundary,
\begin{equation}
\tilde{Z} = \sum_{n=0}^{N_{max}} a_n(\wfr,\qfr)\left(z-z_0\right)^n.
\label{eq:LeaversAnsatz}
\end{equation}
In principle, $z_0$ could be chosen arbitrarily, however, the optimal value is $z_0=1/2$. By inserting Eq.~\eqref{eq:LeaversAnsatz} into the equations of motion one gets a set of $N_{max}+1$ equations for $N_{max}+1$ unknowns. Written in a linear algebra language,
\begin{equation}
\sum_{n=0}^{N_{max}} M_{nm}(\wfr,\qfr) a_m (\wfr,\qfr) = 0.
\label{eq:LeaversEquation}
\end{equation}
For a given value of $\qfr$, the quasinormal modes correspond to frequencies $\wfr$ for which the system \eqref{eq:LeaversEquation} admits a non-trivial solution, i.e. the quasinormal spectrum is determined by the equation 
\begin{equation}
\det M_{nm} (\wfr,\qfr) = 0.
\label{eq:LeaversDet}
\end{equation}
In this work, $N_{max}$ has been chosen to be $150$. This method determines the quasinormal modes, i.e. the poles of the corresponding Green's functions but not the residues. The method is efficient in determining highly damped modes.
\paragraph{Integration}
When one is interested in the region of the complex frequency plane close to the origin, e.g. in hydrodynamic poles, or when the residue of the poles is needed, one could directly numerically integrate equations of motion, e.g. via  the Runge-Kutta method \cite{Kaminski:2009dh,Kaminski:2009ce}. Having already imposed the ingoing boundary condition at the horizon, there is still one boundary condition left. The numerical integration starts at the horizon (more precisely, at fixed small distance away from the horizon). In order to determine the starting values for the function and its derivative, one needs to iteratively solve the equations of motion by expanding the solution in series around the horizon. The freedom coming from the remaining boundary condition is encoded in a constant undetermined by this process which can be set to one without loss of generality. The quasinormal modes $\wfr$ are then determined by solving numerically the equation
\begin{equation}
\left.Z^*(\wfr)\right|_{u=\epsilon} = 0,
\end{equation}
where $Z^*$ is a numerical solution and $\epsilon$ is a boundary cut-off. This method was used in this work to determine the spectral functions in Section \ref{sec:SpectFunGBShear}. 
\paragraph{Spectral method}
In Ref.~\cite{Waeber:2015oka}, another numerical approach was used to compute quasinormal modes. It is based on the spectral method of solving 
ODEs. In this method, the regular function $\tilde{Z}$ is expanded in Chebyshev polynomials
\begin{equation}
\tilde{Z} = \sum_{n=0}^{N} c_n T_n(2u-1),
\label{eq:Chebyshev}
\end{equation}
where $T_n(x) = \cos(n \arccos z)$. Inserting \eqref{eq:Chebyshev} into the corresponding equation of motion and requiring that it is satisfied on the Chebyshev-Gauss-Lobatto grid, namely on $u_i=1/2(1-\cos(i\pi/N))$ for $i=0,1,\ldots,N$, the problem reduces to a linear-algebra problem
\begin{equation}
\hat{M}\cdot \underbar{c} = 0,
\end{equation}
where $\underbar{c}$ is the vector consisting of the $N+1$ coefficients $\left\{c_i\right\}$ and $\hat{M}$, an $(N+1)\times (N+1)$ matrix, consists of the evaluated equation of motion on the i-th grid point. A non-trivial solution exists only when $\det{\hat{M}}=0$. This determinant is a polynomial in $\wfr$ and its roots correspond to the quasinormal modes, easily computed numerically. One of the advantages of this method is the rapid convergence of the modes with $N$, requiring a relatively small matrix $\hat{M}$. 

In this paper, we have used all three described methods, where convenient, finding consistent results (within numerical resolution).
\bibliographystyle{JHEP}
\bibliography{refs}

\end{document}